\documentclass[letterpaper,12pt,openright]{report}
\usepackage[utf8]{inputenc}

\usepackage{graphicx}
\graphicspath{ {./Images} }

\usepackage{tabularx}

\usepackage{fancyhdr}
\usepackage{indentfirst}

\usepackage{epsf}
\usepackage{latexsym,euscript}

\usepackage{amsmath}
\usepackage{amsfonts}
\usepackage{amssymb, epsfig}
\usepackage{subcaption}
\usepackage[]{units}
\usepackage{mwe}
\usepackage{multirow}
\usepackage{verbatim}
\usepackage{geometry}
\usepackage{color,soul}
\usepackage{longtable}
\usepackage{booktabs}
\usepackage{float}

\usepackage{tikz}
\usepackage{dsfont}
\usepackage{braket}
\usepackage{blkarray}
\usepackage{comment}

\usepackage{lscape}
\usetikzlibrary{quantikz}
\usepackage{hyperref}
\usepackage{cleveref}

\usepackage{algorithm}
\usepackage{algorithmic}
\usepackage{simpler-wick}
\usepackage[titletoc]{appendix}

\usepackage[english]{babel}
\usepackage[babel]{csquotes}
\usepackage[natbib, maxnames=999, minnames=1, maxcitenames=2, mincitenames=1, maxbibnames=999, minbibnames=1, bibstyle=numeric, citestyle=numeric-comp, backend=bibtex, backref, hyperref, sorting=none,firstinits=true]{biblatex}

\AtEveryBibitem{\clearfield{pages}} 

\usepackage{tabularray}
\usepackage[export]{adjustbox}

\usepackage{microtype}
\usepackage[T1]{fontenc}

\usepackage{epigraph}

\bibliography{Bib_phs_thesis_EM}
\renewbibmacro{in:}{}

\begin{document}

\thispagestyle{empty}
\begin{center}
\vspace*{0.1 cm}
\Huge
\textbf{Investigating how to simulate lattice gauge theories on a quantum computer}

\vspace*{0.3 cm}
\LARGE
Emanuele Mendicelli
\vspace*{1.5 cm}

\Large
A Dissertation Submitted to\\
the Faculty of Graduate Studies\\
in Partial Fulfillment of the Requirements\\
for the Degree of

\vspace*{0.5 cm}

Doctor of Philosophy

\vspace*{0.8 cm}
\begin{center}
\includegraphics[width=0.4\textwidth]{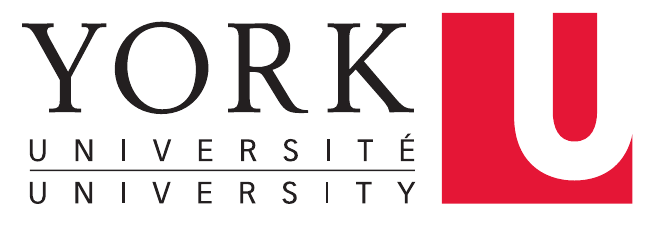}

\end{center}
\vspace*{0.5 cm}

Graduate Program in Physics and Astronomy\\
York University\\
Toronto, Ontario

\vspace*{3.0 cm}

August 2023

\vspace*{0.6 cm}

\textcopyright ~ Emanuele Mendicelli, 2023
\end{center}


\pagenumbering{roman}
\setcounter{page}{1}

\chapter*{Abstract}
\addcontentsline{toc}{chapter}{Abstract}
Quantum computers have the potential to expand the utility of lattice gauge theory to investigate non-perturbative particle physics phenomena that cannot be accessed using a standard Monte Carlo method due to the sign problem. Thanks to the qubit, quantum computers can store Hilbert space in a more efficient way compared to classical computers. This allows the Hamiltonian approach to be computationally feasible, leading to absolute freedom from the sign-problem. But what the current noisy intermediate scale quantum hardware can achieve is under investigation, and therefore we chose to study the energy spectrum and the time evolution of an SU(2) theory using two kinds of quantum hardware: the D-Wave quantum annealer and the IBM gate-based quantum hardware.


\chapter*{Acknowledgements}
\addcontentsline{toc}{chapter}{Acknowledgements}
\markboth{\MakeUppercase{Acknowledgements}}{\MakeUppercase{Acknowledgements}}

I would like to express my deepest gratitude to my PhD supervisor Professor Lewis for introducing and guiding me in my exploration of the field of quantum computing, enlightening me on many aspects of the research program. Over the past few years I have appreciated his unbounded patience, his ability to always find time, the clearness in dealing with my always twisted questions and most importantly, his scientific generosity devoted to creating opportunities and constantly encouraging me to move forward, from my first day, to summer schools and conferences, to my postdoc application and even further.\\ 
\\ 
I would like also to thank Professor Matthew C. Johnson and Professor Sean Tulin for having found the time to follow me through these years as part of my PhD research evaluation committee and by providing valuable advice and pointing me in good directions, furthermore I thank Professor Johnson for the extra support and advice during my postdoc applications period. \\ 
\\ 
During these years the challenge of engaging with quantum computing was eased by collaborating with Sarah Powell and Sarmed A Rahman by sharing together good notes, advices and two interesting works on quantum computing.\\ 
\\ 
Since my arrival I have enjoyed sharing an office with Brian Colquhoun, and off the lattice, being guided to the best concerts in town even if it meant walking to the venue during a crazy snow storm or while it was raining cats and dogs. After the social upheaval of Covid-19 I had the pleasure of enjoying the company of new colleagues Ali Hosseinzadehkavaki, William Parrott and Matthew Tsang with whom I actively shared liters of coffee in the pursuit of solving a problem about a wrong minus sign, what people call the scientific endeavour, and some good pub crawling. 
\begin{center}
***
\end{center}
There is something unique about living in Canada, especially when one comes from a monocultural European country, like Italy. In fact, the amazing complexity of a country so geographically extensive with such a diverse population made by Anglophones, Francophones, First Nations, Metis and constant influx of new immigrants, with different cultures, histories, cuisines and living in a vast territory all united by a strong sense of belonging to a country built by the contribution of so many. A society that it is not afraid to openly show that is naturally multicultural, a place that will inevitably foster a sense of belonging even to who, will be there temporarily, and for this I am once again grateful to Professor Lewis for this life opportunity.\\
\\
In this respect, Toronto gave me the amazing life experience that it will bring with me in the future, to grow up by sharing for four years in a quite large house with people from around the world, Ahmed al Baghdadi from Egypt, Dominique Monrose from Saint Lucia, Mallika Kumar from India, Yeraldin Molina from Venezuela, Samantha Baker from Jamaica and Zulqarnain Kazmi from Pakistan/Saudi Arabia, all with such a diverse and complex background but all united in their life journey to become permanent residents and future citizens, and to be permanently contributing to one of the most respectful and multicultural countries promoting inclusion in every possible way. You can image how amazingly complex our interactions were, mediated by diverse languages, cultures, religions and cuisines. We inevitably experienced deep joy in sharing delicious food, cheerful moments playing around and walking and biking in TO and tense time when someone was leaving for their next stage of life. I have directly and indirectly learned countless many life lessons from them and I am deeply grateful for their understanding, patience and friendship and I take this opportunity to wish them once again all the best in their future.\\ 
\\ 
I thank my family for the support even though life's hardships and distance have made everything a little stretched and twisted, but nevertheless it is always like a safe rock reachable whatever the ocean storm is.\\ 
\\  
I cannot be happier to have met Cortney here in Toronto, immediately I saw her brightness, and I am immensely grateful for her boundless, unconditional love and a truly $4\pi$ steradians support in every possible way during these years, even getting closer to get a minor in physics. I will always keep in my heart her patience while I was writing the thesis and continuously reminding me to "sit down and don't try to boil the ocean". We have a long road ahead and it can only be wonderful. Dulcis in fundo, a special thought goes to her cats Toby and Anders, "cagnolini" who with their cheerfulness manage to sweep away any sign of stress, and induce joy just by looking at them.


\newpage
\addcontentsline{toc}{chapter}{Contents}
\tableofcontents

\newpage
\addcontentsline{toc}{chapter}{List of Tables}
\listoftables

\newpage
\addcontentsline{toc}{chapter}{List of Figures}
\listoffigures


\clearpage
\pagenumbering{arabic}


\chapter{Introduction}\label{chap:Introduction}

\section{Lattice QCD and its methods}
Inside the Standard Model of particle physics, Quantum Chromodynamics (QCD) is a theory introduced to describe all the interactions mediated by the strong force. The theory assumes the existence of fundamental particles called quarks and gluons and describes their individual behaviour and their mutual interactions under the action of the strong force which is assumed to be mediated by gluons. Furthermore, their description assumes the existence of spinor fermion fields for quarks and vector fields for gluons. Aside from the common electric charge for quarks, both these fields have one new peculiar internal degree of freedom called a colour charge.\\
Up to this point, human knowledge was supported by extensive theoretical description, highly intensive numerical simulations and numerous precision experiments which gives us the possibility to state that QCD is the universally accepted theory for describing the strong interaction.\\
\\
Following the approach by Yang and Mills \citep{Yang:1954ek}, a theory can be built by requiring the fields describing the particles are gauge invariant under a chosen non-Abelian group. For the case of QCD the group is SU(3), a special unitary group of $3 \times 3$ matrices, Hermitian and with determinant equal to 1.  The theory can be described in the continuum space-time by the following Lagrangian density:
\begin{equation}
\label{L_QCD}
\mathcal{L}_{QCD} = - \dfrac{1}{4}G_{\mu \nu}^aG^{a\,\mu \nu} + \sum_f\overline{\psi_f}(i\,\gamma^\mu \,D_\mu-m_f)\psi_f
\end{equation}
where $\psi_f$ is a fermion quark field with mass $m_f$, where $f$ specifies the flavour of quarks known (up, down, charm, strange, top and beauty).\\
\\
$\gamma^\mu$ are the Dirac gamma matrices, $4 \times 4$  matrices with $\left\lbrace \gamma^\mu, \gamma^n \right\rbrace =2 \eta^{\mu \nu} I_4$, where $\eta^{\mu \nu}= \text{diag}(1,-1,-1,-1)$.\\
\\
$D_\mu = \partial_\mu - igT^aA_\mu^a$ is the gauge covariant derivative, where $T^a$ with $a$ running from 1 to 8 are eight generators of the Lie algebra of the SU(3) color group. The eight generators represented by $3 \times 3$ Gell-Mann matrices are subject to the following commutation relations $[T^a, T^b] = if^{abc}T^c$, where $f^{abc}$ are the structure constants characterizing the algebra of the group. $A_\mu^a$ with $a$ from 1 to 8 are the gluon boson gauge fields of the interaction and $g$ is the coupling strength between the fields $\psi_f$ and $A_\mu^a$.\\
\\
Finally,
\begin{equation}\label{eq:G_mn}
G_{\mu \nu}^a = \partial_\mu A_\nu^a - \partial_\nu A_\mu^a +gf^{abc}A^b_\mu A^c_\nu
\end{equation}
is the gluon field strength tensor.\\
\\
The gauge invariance of the QCD Lagrangian under the non-Abelian local gauge transformation is guaranteed by the following gauge transformation of its elements:
\begin{eqnarray}\label{con_local_transformation}
\psi(x) &\rightarrow& \psi'(x)=\Omega(x)\psi(x)\,, \\
\overline{\psi}(x)' &\rightarrow& \overline{\psi}(x)\Omega^\dagger (x)\,, \\
D_\mu &\rightarrow & D_\mu' =\Omega(x)D_\mu \Omega(x)^\dagger \,, \\
A_{\mu}(x) &\rightarrow & A_{\mu}(x)'=\Omega(x) A_{\mu}(x) \Omega(x)^\dagger +i \left( \partial_\mu  \Omega(x)  \right)  \Omega(x)^\dagger \,.
\end{eqnarray}
where $\Omega(x)$ belongs to the fundamental representation of SU(3), and can be written as $\Omega(x) \equiv e^{-iT^a\theta^a(x)}$ where $\omega^a$ are the 8 gauge transformation parameters.\\
\\
It is now important to consider the action of the theory because it is used to quantize the classical theory and secondly it is used in the lattice formulation of QCD. Its form is defined as:
\begin{equation}
\label{qcd_action}
S_{QCD} = \int dx^0 d^3x\left[- \dfrac{1}{4}G_{\mu \nu}^a(x)G^{a\,\mu \nu}(x) + \sum_f\overline{\psi_f}(x)\left(i\,\gamma^\mu \,D_\mu-m_f\right)\psi_f(x)\right].
\end{equation}
Since the exponential contains an imaginary term, the integrand is a highly oscillating function ill suited to be integrated numerically especially for a high-dimensional integral. It is usually rewritten by performing the Wick rotation, by changing the time variable from real to imaginary, $x_0\rightarrow -\mathrm{i} x_4$ and extending it to the gauge fields $A_0\rightarrow \mathrm{i} A_4$. Then we introduce the Euclidean gamma matrices $\gamma_\mu^E$ which are defined by the anticommutation relation $\left\lbrace \gamma_\mu^E,\gamma_\nu^E \right\rbrace =2\delta_{\mu \nu}$  and are related to the gamma matrices $\gamma_\mu$ according to $\gamma_0^E=\gamma^0$ and $\gamma_i^E=-i\,\gamma^i$. Substituting in the previous equation we obtain the QCD Euclidean action:
\begin{equation}
\label{qcd_action_eucli}
S_{QCD}^E = \int d^4x\left[\dfrac{1}{4}G_{a \mu \nu}^EG_{a \mu \nu}^E + \sum_f\overline{\psi}_f(x)\left(\gamma_\mu^E \,D_\mu^E+m_f\right)\psi_f(x) \right].
\end{equation}
The action is now in a form that can be numerically integrated, however since the time is now imaginary we cannot study its real-time dynamics and we are restricted to investigate only static properties.\\
\\
Many theoretical properties can be inferred by calculating the expectation value of an operator $O$ using the Feynman path integral formalism:

\begin{equation}
\langle O \rangle= \dfrac{\int D\psi D\bar{\psi} DA~ O[\psi, \bar{\psi},A]~ e^{-S_{QCD}^E[\psi,\bar{\psi},A]}}{\int D\psi D\bar{\psi} DA ~ e^{-S_{QCD}^E [\psi,\bar{\psi},A]}}
\end{equation}
The quantization of the theory is commonly performed using the Feynman functional-integral formalism in which all the possible n-points Green functions defined as the vacuum expectation value of time-ordered product of $n$ fields ${\phi}(x)$:

\begin{equation}
\bra{0}T[\hat{\phi}(x_1),\,\hat{\phi}(x_2),\dots,\hat{\phi}(x_n)]\ket{0}.
\end{equation}
whose value contributes not only to the calculation of the probability amplitude of a given process cross section that we are interested in, but can be used to extract the information on the spectrum as well as various other properties of the theory.\\
In this formalism Green functions are calculated simply as:

\begin{equation}
\bra{0}T[\hat{\phi}(x_1),\,\hat{\phi}(x_2),\dots,\hat{\phi}(x_n)]\ket{0}=  \dfrac{\int D\psi D\bar{\psi} \hat{\phi}(x_1),\,\hat{\phi}(x_2),\dots,\hat{\phi}(x_n)~ O[\psi, \bar{\psi},A]~ e^{-iS_{QCD}[\psi,\bar{\psi},A]}}{\int D\psi D\bar{\psi} DA ~ e^{-iS [\psi,\bar{\psi},A]}}
\end{equation}

These integrals can only be calculated analytically on rare occasions and, therefore a numerical approach has to be used. In attempting a numerical integration, we have to deal with an infinite number of degrees of freedom, since the fields are defined all over the space. We therefore have to discretize it and the time, which results in the introduction of a space-time lattice, that directly brings us toward defining a lattice gauge theory on a lattice, or in our specific case QCD on a lattice.
\\
The discussion regarding the problem of fixing the gauge that will lead to the presence of Faddeev-Popov ghost is quite an important aspect, but since in the lattice formulation there is no necessity to fix the gauge, we point the interested reader to the standard QCD textbooks \citep{Muta:2010xua, Rothe_Book, Cheng:1984vwu}, from which part of this discussion has been deduced.\\
\\
QCD is a theory rich with phenomena, the first striking one is related to its coupling strength whose value is not constant. In fact using arguments from the renomalization group it is possible to show that at the leading order of perturbation theory \citep{leader_predazzi_1996} its expression is:
\begin{equation}
\label{alfastrong}
 \alpha_s(Q)=\frac{12 \pi}{(33-2N_f) \log(Q^2/\Lambda^2)},
\end{equation}
where $N_f$ is the number of quark flavours used in the calculation, $Q$ is the momentum transferred between particles and $\Lambda$ is the QCD scale parameter, a constant to be determined experimentally which is necessary to match the theory prediction with the experiment results at a given energy scale.\\
\\
The strength of the strong force changes with the momentum transferred $Q$ inside a process giving the theory two main regimes, a perturbative regime studied analytically and a non-perturbative regime that uses mainly numerical computation on supercomputers.\\
In fact, for large values of the momentum transferred $Q$, or at short distances, the coupling constant gets smaller and the theory becomes asymptotically free when $Q \rightarrow 0$ meaning that the quarks and gluons are free particles and no bound states are present.
When $\alpha_s \ll 1$ in the perturbation regime the theory can be approximated using a perturbative expansion that uses the Feynman diagram approach to predict using analytical calculation the theory phenomena.\\
\\
Conversely, for small values of $Q$, or large distances, the coupling constant gets larger therefore quarks and gluons are confined in bound states commonly called hadrons. In this non-perturbative regime the main investigation approach is called lattice QCD (LQCD), in which the theory is discretized by first principles on a 4-dimensional lattice and studied by numerical simulations using supercomputers.\\
\subsection{Introducing the lattice}
\begin{figure}[H]
\centering
\includegraphics[width=0.9\linewidth]{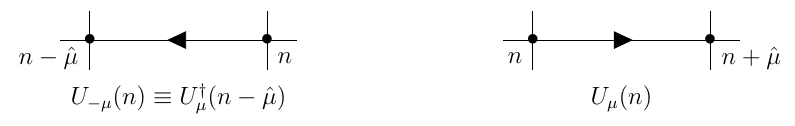}
\caption{Spatial representation of two link  link variable $U_{-\mu (n)}$ and $U_\mu (n)$.
\label{fig:LQCD_U_link}}
\end{figure}

In 1974 Kenneth Wilson introduced in his seminal paper \citep{PhysRevD.10.2445} the idea of discretizing a gauge theory from first principles, giving life to the field that is now referred to as lattice gauge theory (LGT). This idea paved the way to the possibility of studying the theory phenomena by numerical simulations, which was done for the first time by Michael Creutz in 1979 \citep{PhysRevD.21.2308}.\\
\\
In this approach the continuum space-time is discretized by using a 4-dimensional lattice with lattice spacing $a$, and the theory is defined from first principles on the lattice by direct discretizing its continuum version or by introducing new discrete definitions of the interaction. This freedom in choosing different definition for the discrete theory is only restricted by reproducing the continuum theory when $a\rightarrow0$.\\
The introduction of the lattice has the important role of regularizing the theory because it restricts the momentum in the range $\left( -\pi/a, \pi/a\right)$, de facto introducing a cutoff of order $1/a$ making the high and low energy divergencies of the theory finite. Therefore, any quantity calculated on the lattice is a function of the lattice spacing $a$, and to obtain the value corresponding to the continuum physics, when the lattice is removed $a\rightarrow0$, the theory has to be renormalized, by redefining the bare parameters of the theory like the particles masses, in such a way that the physical observables are independent of the lattice spacing $a\rightarrow0$ and assumes the known experimental value. The only evident negative effect of the lattice is the breaking of the rotational and Lorentz invariance which are restored in the continuum limit, which can limit the use of certain types of observables.\\
\\
The hypercubic lattice with linear sizes $L_\mu$ and lattice and spacing $a$ that discretizes the space-time is formally defined as:
\begin{equation}
\Lambda=\left\lbrace x_\mu =an_\mu ~ \vert ~ n_\mu = 0,1, \ldots L_\mu/a  \right\rbrace 
\end{equation}\\
The necessary prescription to perform the discretization of the action are:
\begin{eqnarray}
x_\mu &\rightarrow& n_\mu a\,, \\
\psi(x), \overline{\psi}(x) &\rightarrow& \psi(n a), \overline{\psi}(n a)\,, \\
\int d^4 x \ldots &\rightarrow& a^4 \sum_n \ldots, \\
\partial_\mu \psi(x) &\rightarrow& \dfrac{1}{2a} \left( \psi(n+\hat{\mu}) -  \psi(n-\hat{\mu}) \right) +\mathcal{O}(a^2) \,, \label{LQCD_derivative} 
\end{eqnarray}
which are very handy for showing that the lattice theory reproduces the continuum theory.\\
\\
The spinors $ \psi(n), \overline{\psi}(n)$ representing the fermion and the anti-fermion are placed on the nodes or sites of the lattice, while link oriented variable $U_\mu(n)$ is attached to each link connecting two sites, $n$, $n+\mu$ represented in Figure~\ref{fig:LQCD_U_link}.
$U_\mu$ is the fundamental gluonic variable representing the gauge field, the gluon vector potential field on the lattice and it is connected with the continuum form $A_\mu(x)$ by the relation $U_\mu=e^{ig a A_\mu(n)}$, that is the lattice version of the Schwinger line integral in the continuum theory $e^{i \mathrm{P}\int_x ^{x+a\hat{\mu}} dx_\mu A_{\mu}(x) }$.\\
\\
One of the most peculiar characteristics of the lattice approach is that it preserves the gauge invariance without the need to impose a gauge fixing. In fact to ensure that the lattice action will be gauge invariant under SU(3) local color rotations it is enough that the lattice fields $\psi(n)$, $ \overline{\psi}(n)$ and $ U_\mu$ transform under the action of an SU(3) element $\Omega(n)$ as:
\begin{eqnarray}\label{lattice_local_transformation}
\psi(n) &\rightarrow& \psi'(n)=\Omega(n)\psi(n)\,, \\
\overline{\psi'}(n) &\rightarrow& \overline{\psi}(n)\Omega ^\dagger (n)\,, \\
U_\mu(n) &\rightarrow& U'_\mu(n)=\Omega(n)U_\mu(n)\Omega ^\dagger(n+\hat{\mu})\,, \\  \label{LQCD_local_sl}
U_\mu ^\dagger(n) &\rightarrow& U^{' \dagger}_\mu  (n)=\Omega(n+\hat{\mu})U ^\dagger _\mu(n)\Omega ^\dagger(n)\,. 
\end{eqnarray}
To convince ourselves of the correctness of this approach it is enough to check that these invariance relations under the group element reproduce the ones of the continuum in \ref{con_local_transformation}. In particular to stress the connection by $A_\mu(x)$ and $U_\mu(n)$ it is enough to notice that by substituting $U_\mu=e^{ig a A_\mu(n)}$ in  Eq.~\ref{LQCD_local_sl}, expanding for small $a$ and performing the continuum limit one obtains the continuum transformation of $A_\mu(x)$ under the group element.\\
\\
Now that we have placed the necessary fields on the lattice and imposed the gauge invariance we can start building the action of QCD on the lattice, by working separately on the gauge and fermionic parts.\\
\\
Using an argument by Wilson, the gauge action part should contain only gauge invariant elements. Therefore the simplest gauge-invariant operator that we can build on the lattice is the trace of a path-ordered product of link variables around a lattice spatial square, where the path-ordered product of link variables is called a plaquette and it is defined as:
\begin{equation} \label{eq:qcd_pla}
U_{\mu\nu}(n) =U_\mu(n)U_\nu(n+\hat{\mu})U_{\mu}^\dagger(n+\hat{\nu}) U_{\nu}^\dagger(n)
\end{equation}
\begin{figure}[H]
\centering
\includegraphics[width=0.4\linewidth]{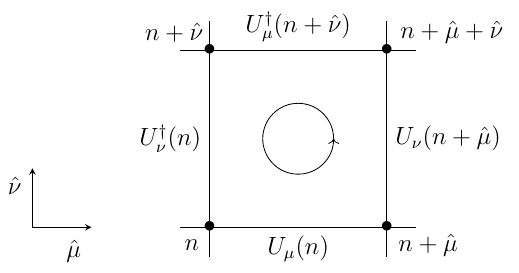}
\caption{The link variable composition of an elementary plaquette with link variables going counterclockwise as indicated by the circle with $\mu$ and $\nu$ expressing the two possible space directions. The plaquette $U_{\mu \nu}(n)$ located at site $n$ in the $\mu\nu-\rm plane$.
\label{fig:LQCD_U_pla}}
\end{figure}
It is important to notice that by substituting  $U_\mu=e^{iagA_\mu(n)}$ in the definition of the plaquette in Eq.~\ref{eq:qcd_pla} one gets:
\begin{equation}
U_{\mu\nu}(n) = \exp( i a^2 g \left\lbrace \partial_\mu A_\nu (n) -\partial_\nu A_\mu (n) + i g \left[ A_\mu(n), A_\nu(n) \right] \right\rbrace  +\mathcal{O}(a^3) ) = e^{ig a^2 G_{\mu \nu}(n) +\mathcal{O}(a^3)} 
\end{equation}
where $G_{\mu \nu}(n)$ is a discretized form of the field strength tensor $G_{\mu \nu}(x)$ in Eq.~\ref{eq:G_mn}.\\
\\
Therefore the Wilson gauge action invariant under the local transformations in Eq.~\ref{lattice_local_transformation} is a sum of all plaquettes as:
\begin{equation}\label{wilson_gauge_action}
S_G= \dfrac{3}{g^2} \sum_{n \in \Lambda}\sum_{\mu<\nu} \left[ 1-  \dfrac{1}{6} \mathrm{Tr}
\left( U_{\mu \nu} +U^\dagger _{\mu \nu}\right) \right] = \dfrac{3}{g^2}
\sum_{P}\left[ 1-  \dfrac{1}{6} \mathrm{Tr}
\left( U_{P} +U^\dagger _{P}\right) \right] .
\end{equation}
where the sum over $P$ extends over all distinct plaquettes on the lattice taken all in the same  counterclockwise (clockwise) direction.\\
Furthermore, it is easy to reproduce the continuum limit, by first expanding the term using the Baker-Campbell-Hausdorff formula for small lattice spacing $a$ and finally perform the limit $a\rightarrow 0$ obtaining the continuum expression of the gauge part present in Eq.~\ref{qcd_action_eucli}.\\
\\
The introduction of the fermion action on the lattice is not as smooth as in the case for the gauge field part of the action indicator, because there are subtleties. In fact, by naively discretizing the continuum fermionic action with the help of symmetric approximation for the derivative in Eq.~\ref{LQCD_derivative} leads to:
\begin{equation}
\label{naive}
S_{F, naive}\left[\psi,\overline{\psi},U\right]=a^4\sum_{n \in \Lambda} \overline{\psi}(n)\left(\sum_{\mu=1}^4\gamma_\mu\frac{U_\mu(n)\psi(n+\hat{\mu})-U_{\mu}(n-\hat{\mu})\psi(n-\hat{\mu})}{2a}+m\psi(n)\right)
\end{equation}
that is known as the naive fermionic action, because despite it is gauge invariant, it does not reproduce the correct continuum physics.\\
Essentially it introduces unphysical lattice artefacts called doublers, more fermion-like excitations in the momentum space than there should be.
In fact, by calculating the two-point fermionic correlation function it has more poles than there should be. Therefore, first the fermion fields are treated as Grassmann anticommuting numbers and, secondly in Wilson formulation to remove the doublers an ad hoc term is introduced in the naive fermionic action, in such a way that when we take the continuum limit $a\rightarrow0$ they disappear. This unfortunately is done at the expense of breaking the continuum chiral symmetry even for vanishing quark masses. It is possible to avoid the doublers in the strong interaction by defining lattice chiral symmetry in a way that accounts for the Nielsen–Ninomiya no-go theorem \citep{NIELSEN198120,NIELSEN1981219} and its recent generalization \citep{Le_2022}. Many different fermionic formulations can be obtained with their own advantages and disadvantages, and it is a lively research area.\\
\\
The fermion action with doublers removed using the Wilson term is:
\begin{equation}\label{wilson_gauge_action}
S_F^W= \sum_{f=1}^{N_f}  a^4\sum_{n \in \Lambda} \left\lbrace  \left( m^{(f)} + \dfrac{4}{a} \right) \overline{\psi}^{(f)}(n)\psi^{(f)}(n) -\dfrac{1}{2a} \sum_{\mu=\pm 1}^{\pm 4} \left(  \overline{\psi}^{(f)}(n+\hat{\mu}) \left[ \mathds{1} +\gamma_\mu \right]    \psi^{(f)}(n)   \right)     \right\rbrace 
\end{equation}
where we have defined $\gamma_{-\mu}=-\gamma_\mu$ with $\mu=1,2,3,4$.\\
\\
Now that we have the full LQCD in the Wilson formulation we can find the partition function for the LQCD:
\begin{eqnarray}\label{W_partition_function}
Z_{LQCD}&=& \int DU D\psi D\overline{\psi} ~ e^{-S_{LQCD} [U, \psi,\bar{\psi}]} \nonumber \\
&=& \int DU ~ e^{-S_{G}(U)} \int D\eta D\overline{\eta} ~ e^{-S_{F}^W(\eta,\overline{\eta})}  \nonumber \\
&=& \int DU ~ e^{-S_{G}} ~\prod_{f}^{N_f} \left( det[D] + m_f \right)
\end{eqnarray}
where in the second step we wrote the fermion integration measure as a measure Grassmann number, and in the last line the Matthews–Salam formula for the integration of a Gaussian integral with Grassmann numbers was used \citep{Gattringer:2010zz}, where $det[D]$ is the determinant of the Dirac operator.\\
\\
Now that we have the LQCD action all the theory information can be extracted by calculating gauge invariant operators called correlators:
\begin{equation}
\langle O\left( U, \psi, \bar{\psi} \right)  \rangle= \dfrac{1}{Z_{LQCD}} \int DU D\psi D\bar{\psi}~ O[U, \psi, \bar{\psi}]~ e^{-S_{QCD}^E[U,\psi,\bar{\psi}]}
\end{equation}
which are commonly calculated numerically, as described in the next section.\\
\\
In conclusion, the freedom to propose different definitions of the discrete QCD action makes the formalism very powerful because it can be adapted to conceptual and numerical needs. Therefore much more should be said for example about the other formulations of the fermion action with smaller errors or better computational cost compared to the Wilson fermion, but since we are not going to be using them in the discussion of this thesis, we point to the following well known LQCD textbooks \citep{montvay_munster_1994, Rothe_Book,Gattringer:2010zz, creutz_2023} from which this presentation has been adapted.\\
\\
The success of LQCD helps understand that the strong interaction is quite vast from the determination of the quark masses, Cabibbo–Kobayashi–Maskawa matrix for quark-mixing, decay constants, the $g_\mu -2$  and much more. For a review of the progress in the last 50 years see \cite{gross202250,ParticleDataGroup:2022pth}.
\subsection{The importance sampling Monte Carlo method}
In the lattice gauge theory approach to find the expectation value of an observables the following expression has to be calculated: 
\begin{equation}
\langle O \rangle= \dfrac{\int D[\mathcal{C}]~ O[\mathcal{C}]~ e^{-S[\mathcal{C}]}}{\int D[\mathcal{C}]~ e^{-S[\mathcal{C}]}}
\end{equation}
where the integration measure is on all the $\left\lbrace \mathcal{C}\right\rbrace $ possible configurations of the system.\\
This expression can be integrated analytically only for theories whose actions are simple, therefore for a generic theory a numerical approach has to be used. \\
\\
With the goal of choosing the most suitable numerical integration method, it is reasonable to estimate how numerically intensive the calculation is. For the sake of simplicity consider a 4-dimensional lattice with size $L$ and a theory with only variables on the lattice nodes and no variables on each link. For example a Potts model based on the $Z_3$ group, where the spin on each node can take 3 values. The number of all possible configurations is equal to the many possible ways we can organize spins on the lattice that is equal to $3^{L^4}$ and grows exponentially with the system volume. Considering a modest lattice size $L=10$ we have $3^{10^4} \approx 10^{4762}$, it follows that it is not possible to sum all of these terms, and consequently a more complex theory is even more numerically challenging.\\
\\
A clue for an efficient way is obtained by carefully observing the integrand, where one can notice that not all configurations contribute equally due to the Boltzmann weight $exp(-S)$. In fact, configurations with a high value for $S$ are exponentially suppressed, so their contribution to the sum is negligible, therefore an estimation of the integral can be obtained by just summing a few configurations with small actions. This is the concept behind the importance sampling \citep{andral2022attempt}.\\

A Monte Carlo using importance sampling will randomly generate a set of $N$ configurations ${\mathcal{C}}$ with a probability distribution given by $\exp\left( -S[\mathcal{C}_n]\right) $, and an observable expectation value $\langle O \rangle$ is approximated as:
\begin{equation}
\dfrac{1}{N} \sum_{n=1}^N O[\mathcal{C}_n] \pm \mathcal{O}\left( \dfrac{\sigma_O}{ \sqrt{N}}\right) 
\end{equation}
where the error is only proportional to $1/\sqrt{N}$ because $O[\mathcal{C}_n]$ are not statistically independent, because the configurations are not completely independent due to the way they are generated.\\
Essentially, the system is prepared in an initial configuration, and each new configuration candidate (obtained from the previous one by changing the value of all or a few spins) is always accepted if it has a lower or equal action and, if it does not have a lower action is accepted randomly with a low acceptance rate to recreate quantum fluctuations numerically. By repeating this process, a reasonable number of times, the system will reach the equilibrium, where each new accepted configuration does not have a significantly lower action, and these equilibrium configurations are the ones used to measure $O[\mathcal{C}]$.\\
\\
The subject is vast and fascinating, and thanks to it we have been able to understand many characteristics of the strong interactions over the last 50 years \citep{gross202250}. Here we wanted to emphasize the crucial role of using the  Monte Carlo, therefore the full discussion about it can be found in the following well known LQCD textbook \citep{creutz_2023, Rothe_Book, Gattringer:2010zz} from which this discussion has been adapted.\\
\section{The sign problem and the necessity to find new ways of computation}
The Monte Carlo method with importance sampling used to perform lattice gauge theory calculations stops working whenever the exponential of the action inside the partition function in Eq.~\ref{W_partition_function} becomes negative or complex. Therefore it cannot be interpreted as a probability distribution. If we cannot use importance sampling we are back to evaluating the full integral in traditional way in which all the points in the integration region are assumed to contribute equally to the final sum, and we have seen that this way of integrating scales exponentially with the size of the volume of the system. Therefore no integration can be done on a system of interest in particle physics.\\
\\
To solve the sign problem one could be tempted to perform a re-weighting, by taking the absolute value of integrand of the partition function, but as it was clearly shown by the authors in Ref.~\citep{Troyer:2004ge} the relative error associated with the observable measured using this trick increases exponentially with the volume of the system. Therefore, to achieve reasonable accuracy an exponentially larger number of Monte Carlo configurations has to be generated, a numerical task possible only for particular systems with mild sign problems.\\
\\
Nevertheless, many researchers are constantly trying to propose novel methods to avoid, but not solve, the sign problem and a recent review of the current state of the art of the computation methods that are being explored can be found in Ref.~\citep{Alexandru:2020wrj}. Moreover, applications of these methods to QCD suffer from residual mild sign problem that require non-polynomial computational resources and therefore broad applications on large systems remain daunting. Furthermore, for the analytical dual approaches in which the theory affected by the sign problem is analytically mapped to another without it, reviewed in Ref.~\citep{Gattringer:2016kco}, their applicability is still under investigation because for each theory a dual formulation has to be developed.\\ 
Consequently, at the moment there is no comprehensive method that can take the role of the Monte Carlo method and therefore other methods should be developed and explored.\\
\\
In the following subsections we discuss with some details the impact of the sign problem on the possibility to study interesting non-perturbative aspects of QCD using LQCD methods and in general of lattice gauge theories, and use them as motivation in regards to the necessity of developing and exploring new computational methods.
\subsection{Real-time evolution}
In constructing the lattice gauge theory method to make the action real and positive we have to perform the Wick rotation, which transforms time from real to imaginary, so that we can interpret the Boltzmann weight as a probability distribution and use the MC with importance sampling. This has no consequences in deriving static properties but completely rules out the possibility of studying any equilibrium quantity in real time. If instead we do not perform the Wick rotation we have a "dynamical" sign problem. Real-time standard evolution that uses the Hamiltonian formulation is not computationally feasible due to the exponential scaling of the computation with the size of the system. Therefore a serious limitation is present because we cannot visualize many processes, for example a particle propagating in space, how a strongly interacting matter evolves over real-time under the action of external parameters, or even more interesting, observing in real-time the process of a quark-antiquark pair string breaking that could give a hint on the mechanism behind quark confinement.\\
Finally, while we are well equipped to reconstruct system properties from initial and final static states, in general one can hope to learn and gain insight into the behaviour of the theory if it can be directly simulated in real time.

\subsection{Simulations with baryon chemical potential $\mu_B$}
The QCD action in Eq.~\ref{L_QCD} describes a system that has perfect symmetry between matter and antimatter, but if we are to consider a system like our reality in which there is more matter that antimatter, a term must be introduced that can tune an imbalance between quarks and antiquarks. This term is analogous to the chemical potential in the grand canonical ensemble of statistical mechanics. In this case a baryonic chemical potential quantifies the energy variation of the system as the number of baryons and antibaryons varies. Unfortunately, this term makes the QCD action complex and there is a sign problem, so we are not completely free to simulate the theory in the presence of a non-zero chemical potential. This excludes from LQCD simulations many interesting physical phenomena connected with the evolution of the early universe or the state of the matter inside the core of a neutron star, and, more importantly excludes the possibility of directly comparing the simulations with experiments  such as the ultra-relativistic heavy-ion collision currently at the Relativistic Heavy Ion Collider (RHIC) at the Brookhaven national laboratory (BNL) and at the Large Hadron Collider (LHC) at the European Organization for Nuclear Research (CERN), that are designed to investigate quark-gluon-plasma with baryon chemical potential tunable by the total center of mass energy.

\subsection{QCD phase diagram}
\begin{figure}[H]
\centering
\includegraphics[width=1.0\linewidth]{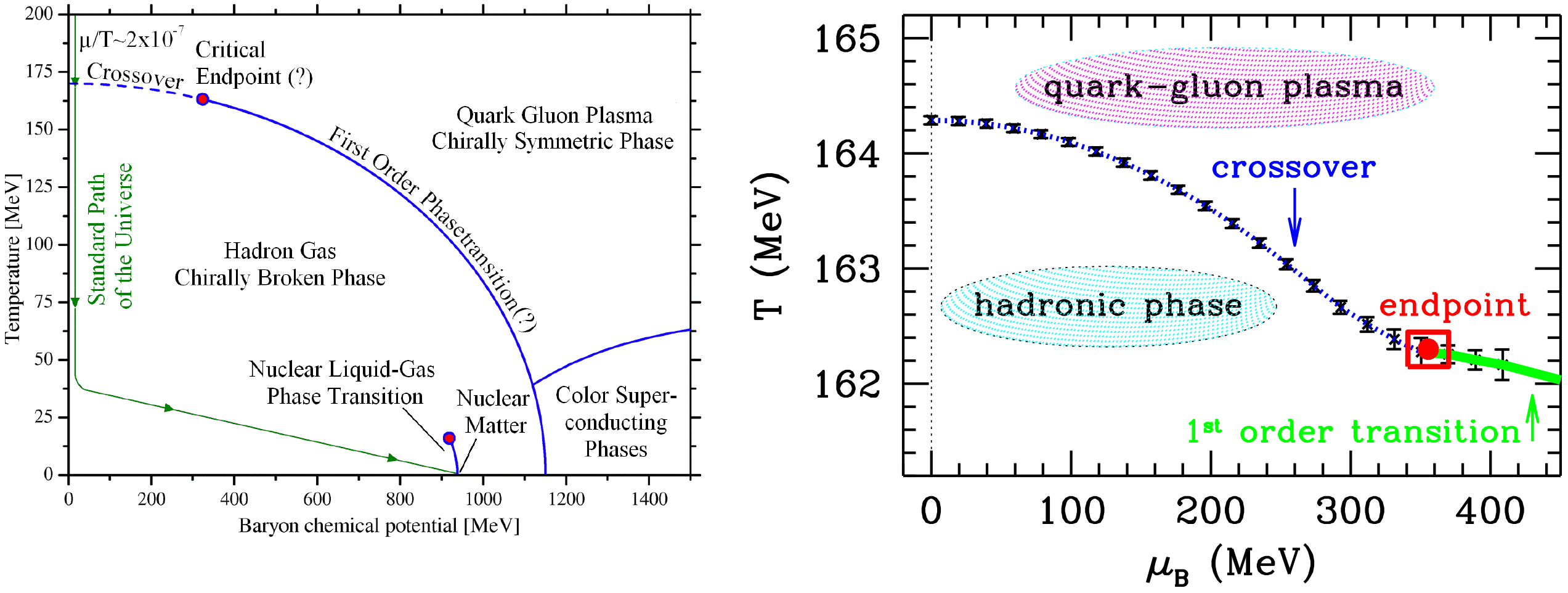}
\caption{Two examples of QCD $(T-\mu_B)$ phase diagrams displaying the conjectured QCD phases and the phases known from LQCD simulations. The left diagram displays most of the hypothetical QCD phases, and the green line represents the most accredited cooling down path of the early universe toward the formation of hadrons in a process called hadronization from calculation in Ref.~\citep{Fromerth:2002wb}. Figure taken from Ref.~\citep{Boeckel_2012}.\\
The right diagram displays the state of the art of the only known phases from LQCD simulations with the addition of the conjectured and notoriously debated second-order endpoint. Figure taken from Ref.~\citep{Fodor:2004nz}. As a reference a temperature of 1 MeV corresponds to $10^{10}K$ and for comparison the Sun's core temperature is $10^{-4} MeV$. The two diagrams clearly show that it is well accepted the absence of a phase transition in the region of temperature range 150-170 MeV for small value of the chemical potential $\mu_B$, therefore the two phases, hadron gas and quark gluon plasma can continuously transform into each other.}
\label{fig:QCD_phase_diagram}
\end{figure}

\begin{figure}[H]
\centering
\includegraphics[width=0.9\linewidth]{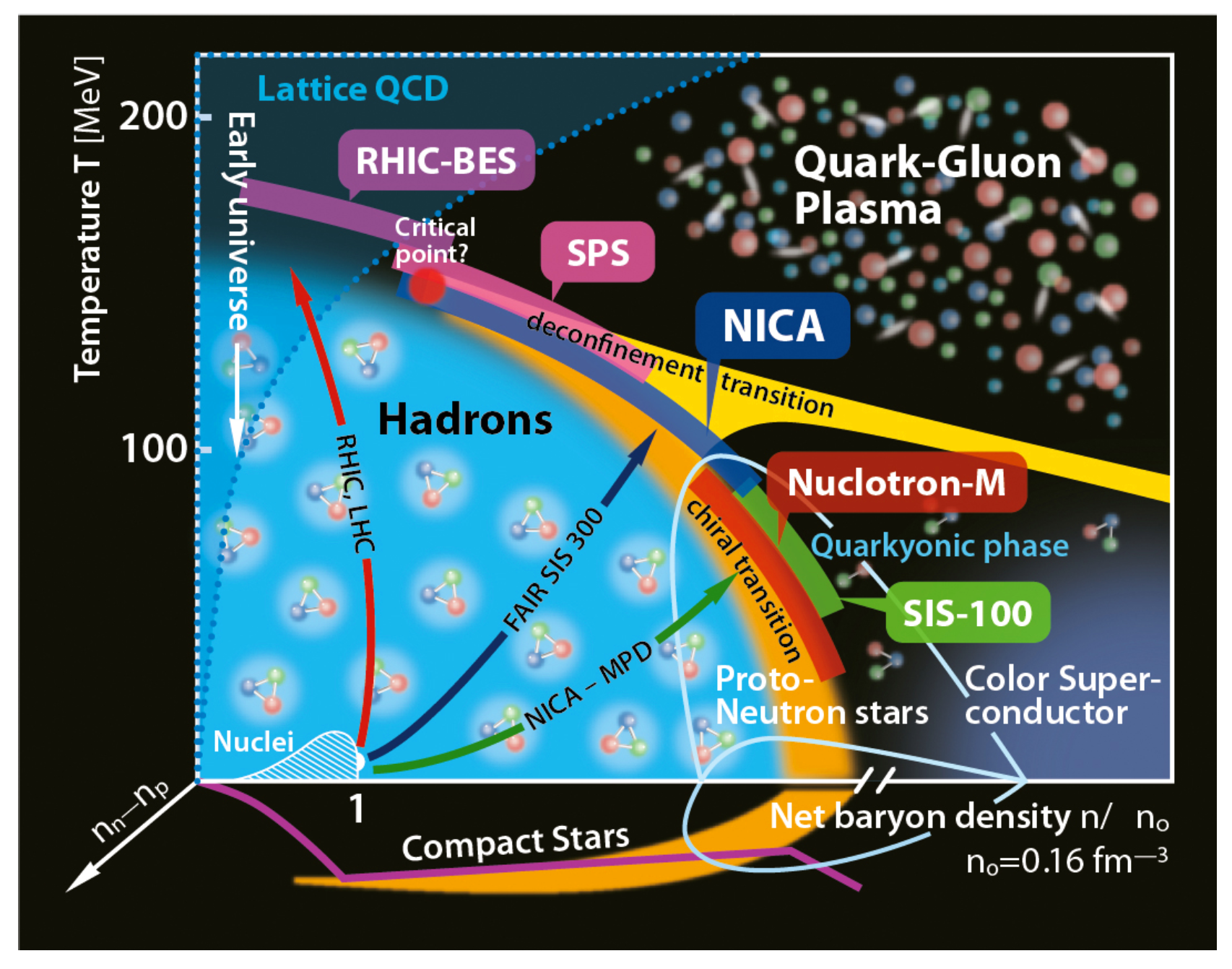}
\caption{QCD $(T-\mu_B)$ phase diagram displaying the regions investigated by current and future experimental facilities. The diagram shows the parameters region of the phase diagram where current experiments are ongoing or planned. The three arrows indicate the starting point of the heavy ion beams and the targeting area of the phase diagram that they plan to access. The experiment at RHIC using BES and LHC using SPS are ongoing with some results in \citep{Busza:2018rrf, An:2021wof}, while two accelerators are under construction \citep{Senger:2021cfo} with projects planned to directly investigate regions not accessible to LHC and RHIC, Nuclotron-based Ion Collider Facility (NICA) in Dubna, Russia and Facility for Antiproton and Ion Research (FAIR) in Darmstadt, Germany and their future projects Nuclotron-M at FAIR and SIS-100 and SIS-300 at NICA.Furthermore, the region where LQCD simulations are possible is highlighted in gray. As a reference a temperature of 1 MeV corresponds to $10^{10}K$ and for comparison the Sun core temperature is $10^{-4} MeV$.Figure taken from \citep{Sahoo:2021aoy}.}
\label{fig:QCD_phase_exp}
\end{figure}

The QCD $(T-\mu_B)$ phase diagram describes the different phases of the theory as a function of the temperature of the system $T$ and the baryonic chemical potential $\mu_B$ mainly in the non-perturbative regime. Compared to other diagrams, like for example $(T-\mu_I)$ where $\mu_I$ isospin chemical potential, it is the most studied QCD phase diagram because it is almost unsolved due to the numerical difficulty related to the sign problem and the long ongoing experiments at high energy. An overview of the current knowledge can be found in the reviews \citep{Guenther2021-ug, Fukushima:2010bq}, while \citep{Aarts:2023vsf} contains a review of how to investigate possible phase transitions from LQCD.\\
From the QCD properties it is known that at low temperatures quarks and gluons are confined, bound inside hadrons. While at high temperatures they are deconfined, that means that they are substantially free to form at first a gas of hadron, and a perfect liquid with low viscosity commonly called quark-gluon-plasma for even higher temperatures. For intermediate temperature and non-zero chemical potential should be present several phase transitions based on symmetries reasoning from idealized limits of the theory but not yet confirmed \citep{Stephanov:2006zvm}. This conceptual idea of phase diagram is shown in Figure~\ref{fig:QCD_phase_diagram}.
Due to the sign problem, LQCD simulations can only investigate the phase diagram along the T-axis in the absence of chemical potential $\mu_B=0$, and for very small values of $\mu_b$ using clever extrapolation methods, the exact region is highlighted in gray in Figure~\ref{fig:QCD_phase_exp}. Reliable LQCD simulations have determined that along the T-axis the transition between the hadron gas and the quark-gluon-plasma is a smooth process in which the hadron gas can continuously transform into a quark-gluon gas, which is called analytical crossover \citep{Aoki:2006we, Borsanyi_2020}. For the other phases shown in the left figure we do not have definitive knowledge of their existence, in fact those phases are conjectured based on studies using effective models of QCD. Therefore for reasons of universality, they should be present in the full QCD, but they have not yet been proven.\\
Therefore, further investigations are needed for these hypothesized phases both through simulations and from experiments, for example with heavy ion collision as shown in Figure ~\ref{fig:QCD_phase_exp}, some of which are currently under way at RHIC and LHC and future facilities are under construction, and a review of their results and the impact they can have in our understanding of strong interacting matter can be found here \citep{Busza:2018rrf, An:2021wof}.\\

Finally, from the leftmost diagram in Figure~\ref{fig:QCD_phase_diagram} we can see the negative impact of the sign problem in our understanding of QCD phases. In fact we can only get extra information about the phases of the system for points along the T-axis where $\mu =0$, and any other point in the plane is not accessible with LQCD simulations. So even if in the last 50 years many techniques have been developed to get around the problem of the sign when it is present in a mild form, such as when the chemical potential is small enough, such as the Taylor expansion, the Lefschetz thimbles, the Complex Langevin and the reweighting-method or performing the analytic continuation to simulations with imaginary chemical potential \citep{Bellwied:2015rza}) they are limited and a more general and direct numerical approach is needed to compare first-principle calculation with experimental results and, guide future experiments towards a range of parameters that are more interesting to study.\\
\subsection{Out-of-equilibrium process}
Within the lattice gauge theory approach based on the MC method only equilibrium configurations can be simulated. Therefore access to out-of-equilibrium processes are completely excluded. But there are many interesting out-of-equilibrium processes that we cannot simulate, for example the dynamics of a system close to a phase transition, the many interesting transient states of matter during the rapid expansion of the early universe, or processes immediately after a collision or a particle decay. These, in particular, exclude the possibility to oversee simulations directly with collision experiments performed in many accelerators around the world.\\
Even if this type of study is not directly influenced by the sign problem, simulating an out-of-equilibrium process requires an amount of computational resources that scales exponentially with the size of the system. Therefore classical calculations are limited to low-dimensional systems with little application to particle physics models, and other computational methods need to be explored.
\subsection{The end of Moore’s law}
The necessity to find a new way to simulate nature has other contributions that are relevant not only to highly intensive simulations in particle physics but in general to  society since everyday life activities require a constant growth in the computational power \citep{K_B_S_W_H_2011}.\\ 
Much of the LQCD advancements in understanding the strong force have come from the enormous advances in classical computational power over the past 50 years, and much more is expected in the future as the era of exascale computing unfolds \citep{Brower_2018, Joo:2019byq}, where a high-performance computing system will be able to do about $10^{ 18}$ floating point operations per second, but it is not clear that classical computing will continue to progress as rapidly as it has in the past.\\
In fact, in the unremitting effort to increase computing performance, the chip industries are approaching the end of Moore’s law, reaching the limit of miniaturization of transistors. Transistor are the building blocks of a computer central processing unit (CPU), effectively quantifying the computing power in terms of how many operations in parallel a computer can do. A Transistor is an electric driven switches that works by controlling the current flowing inside a wire connecting its two ends with an external voltage. Those ends called drain and source are made of silicon doped with other elements like arsenic, boron, phosphorus and gallium. Those elements have many more electrons so that the doped silicon behaves as a metallic material.\\
\begin{figure}[H]
\centering
  \includegraphics[width=1.0\linewidth]{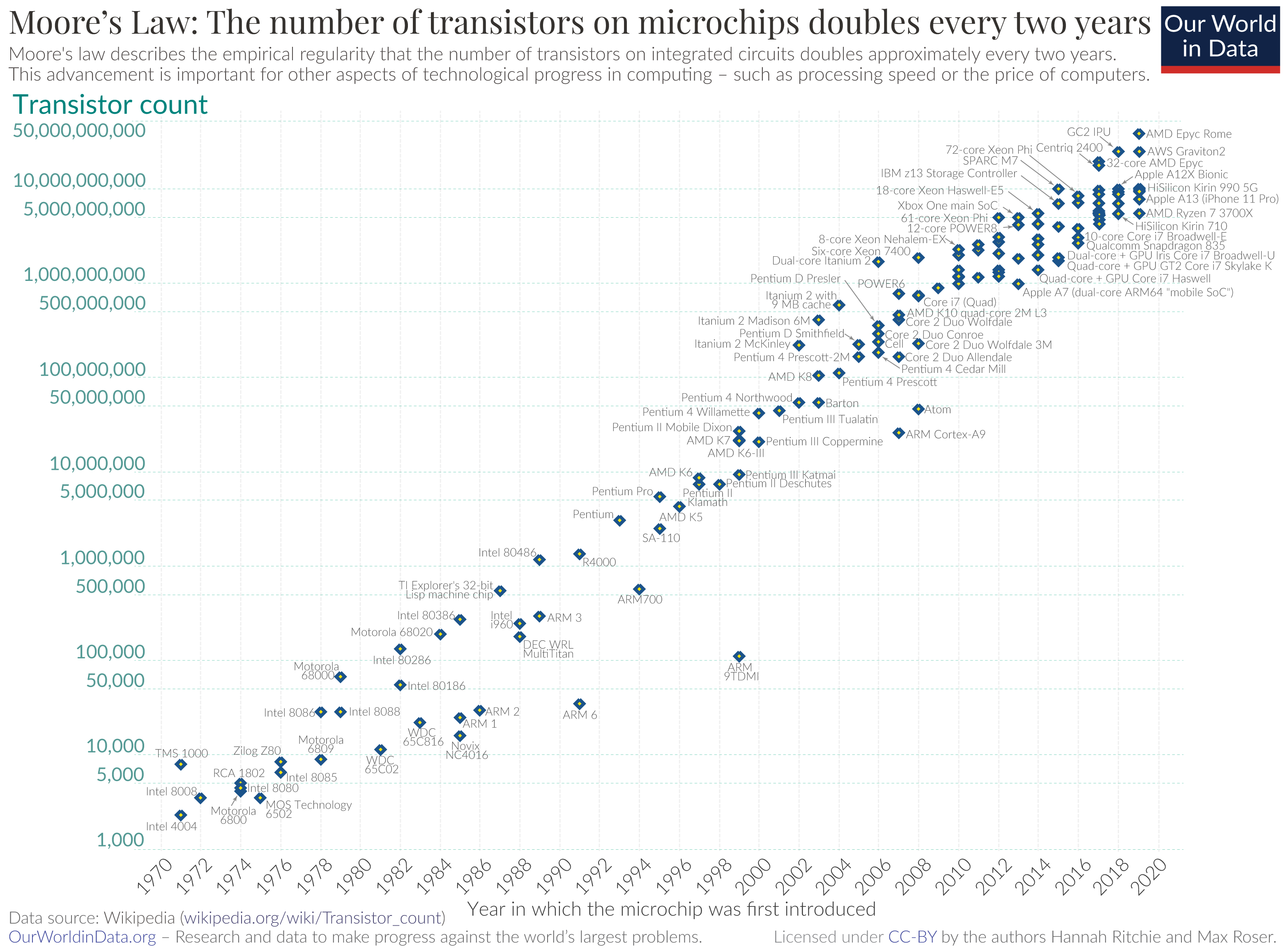}
\caption{
Prediction about the future number of transistors per integrated circuit by the empirical Moore’s law. In the right panel the exponential rise of the number of transistors inside an integrated circuit as a function of year, displayed in logarithm scale, figure taken from \citep{figure_Moore_law}. This has an effect on the single transistor size as it is shown in the left panel, where the transistor size is quantified by the wire widths, with pink and blue points actual transistor size, and future size prediction are shown by the black dashed lines, reaching the physical limits of atomic size between 2040 and 2050.}
\label{fig:Moore_law}
\end{figure}
Moore’s law describes empirically the growth of the number of transistors in an integrated chip doubles every 18-24 months \citep{Moore_Gordon_2006, Moore1995} as shown in the left panel of Figure~\ref{fig:Moore_law}. This exponential growth requires a progressive miniaturization of transistors that now are commonly around 7-10 nm in size with announcements for a 2 nm transistor by IBM \citep{IBM_2nm_chip}. If Moore’s laws empirical prediction continues to be followed, in the coming decades the size of a single transistor will get to the smallest possible size as shown in Figure~\ref{fig:Moore_law_Tsize}, reaching sizes comparable with few atoms making it not possible to realize due to completely different properties of the commonly used materials at that physical scale \citep{Michael_Frank_2005, Stojcev_Mile_K2004, Burg2021_ay, Lloyd_2000}.

\begin{figure}[H]
\centering
  \includegraphics[width=1.0\linewidth]{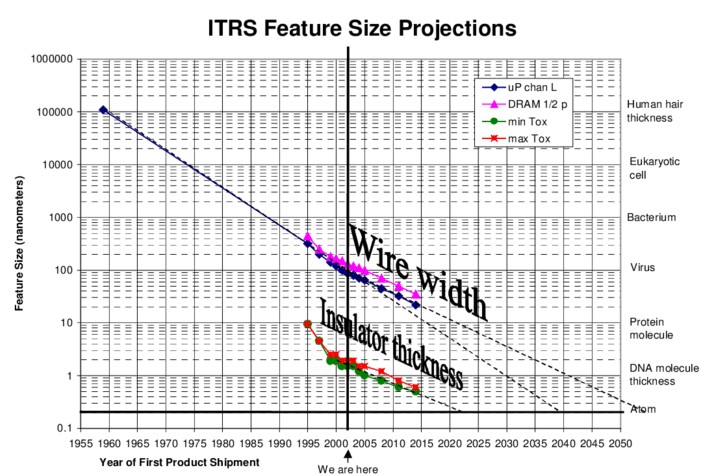}
\caption{
Prediction about the future single transistor size by the empirical Moore’s law. The exponential rise of the number of transistors inside an integrated circuit per year show in Figure~\ref{fig:Moore_law} has an effect on the single transistor size. The transistor size is quantified by the wire widths and displayed in logarithm scale as function of year. The pink and blue points are common time transistor size, while future size predictions are shown by the black dashed lines.  The graph shows that the physical limits of atomic size will be reached between 2040 and 2050. Figure taken from \citep{Michael_Frank_2005}.}
\label{fig:Moore_law_Tsize}
\end{figure}

In fact, in that future atomic dimension device quantum mechanics has a dominant role and quantum tunneling will make it difficult to prevent the electrons from moving freely between the drain and the source of the transistor. Consequently, the device will not be able to control the current flowing in it. Therefore it will no longer function as a transistor should. Hence new ways of increasing computational power should be found either in new technological innovations like for example with the introduction of carbon nanotube field-effect transistors \citep{Hills2019} or by considering new computational means that are not based on classical physics \citep{THOMPSON200620, Lloyd_2000}.
\section{Quantum computing, a new opportunity to explore}\label{quantum_explore}
In 1981 in a keynote speech \citep{Feynman1982} Richard Feynman contemplated the possibility of studying nature through simulations in a more natural way with an hypothetical computational hardware able to study a large system using an amount of resources proportional to the system volume and, that its internal functioning was directly based on the laws of quantum mechanics that are the ultimate laws describing nature in its microscopic essence. In subsequent work \citep{ Feynman1986} he began to advocate for the development of a quantum computer by actively proposing what and how its building blocks should be. As creator of the path integral formulation of quantum mechanics, (a formulation equivalent to the canonical approach to quantum mechanics that uses the Schrodinger equation to study the time evolution of a quantum system \citep{ RevModPhys.20.367, Feynman_Hibbs_1965}), he was deeply interested in the study of real time dynamics of a quantum system. In this respect he was concerned about the unreasonable attempts to study highly entangled quantum systems, such as those typical of particle or condensed matter physics, on a classical computer  because this would inevitably require an exponential increase in the amount of classical computational resources, making the task practically unfeasible.\\
\\
Why should we consider quantum computers? \\
\\
The reason is simple. In a quantum computer the fundamental computational element is represented by the qubit, the quantum version of a classical bit. While a bit can only take a 0 or 1 value at a time, the qubit being an object respecting the laws of quantum mechanics can be not only in the states $\left| 0\right>$ or $\left| 1\right>$ but simultaneously in any superposition of these two states, $a \left| 0\right> + b \left| 1\right>$ with $a$ and $b$ two complex numbers whose only requirement is that $|a|^2 +|b|^2=1$ needed to respect the probabilistic interpretation. \\
This means, that measuring a bit always return one value, $0$ or $1$, with absolute certainty, while a qubit can be found $100 \%$ in one of the two states or can be simultaneously in both states with a certain probability.\\
\\
To be more concrete, $N$ bits can represent up to $2^N$ numbers but only one of these can be stored in the memory at a time. Instead considering $N$ qubits the hardware can store a superposition of all states at one time.\\
This has an important consequence. If we now consider $N$ qubits these can be in any superposition, so a quantum hardware with $N$ qubits is able to encode a superposition of $2^N$ states using only $N$ qubits. This makes it possible to directly consider the Hamiltonian formulation of a theory, something that is not possible to study on a classical computer due to the computational exponential growing cost with the Hamiltonian size.\\
Indeed, the Hamiltonian formulation of a system with $N$ particles has a number of states that grows exponentially with the number of particles. If the particles are spin-1/2 it grows as $2^N$ and a classical computer requires $2^N$ memory slots to deal with the system. Therefore using only a few tens of particles makes the computation completely exorbitant in terms of memory resources and computational time. Exactly for this reason the lattice gauge theory approach was based on the  Monte Carlo importance sampling method.\\
Conversely, a quantum computer needs only $N$ qubits. This means that a quantum computer can make possible to do calculations in the Hamiltonain formalism. In fact, a quantum computer has the potential to be exponentially more efficient than a classical computer, because it can store a theory Hilbert space using a number of quantum resources that scale polynomially with the number of particles.\\
\\
It is this exponential efficiency that gives one the possibility to use the Hamiltonian formulation of a theory that is completely free of the sign problem limitations, hence quantum computers are expected to dramatically expand our computational approach to understand the fundamental laws of nature.\\
\\
In point of fact, a fault-tolerant quantum computer will offer a new possibility to study not only Quantum Chromodynamics but all non-perturbative phenomena of lattice gauge theories which are not possible to study numerically today using the standard Monte Carlo method due to the occurrence of the sign problem.\\
\\
Among the fields that will become computationally achievable are the studies of strong interacting matter in the presence of chemical potential, which is relevant to studying the matter of the early universe, the core of a neutron star, and it will certainly expand our understanding of the QCD phase diagram. The real-time dynamics of lattice gauge theory will become accessible and this will make it possible to simulate the dynamics of a heavy-ion collision currently at RHIC and LHC, creating the possibility to compare the results and, guide future experiments towards interesting parameter regions where a sign of possible physics beyond the Standard Model could materialize.\\
However, we should keep our enthusiasm tempered because those fascinating future studies are not yet possible on currently available quantum computers because they are not yet fault-tolerant, they are still noisy, lack error correction, and have small-scale quantum resources, we are in fact in the "Noisy Intermediate-Scale Quantum" (NISQ) era as John Preskill defined it \citep{ Preskill:2018jim}. Hence, as for the last 6-7 years and maybe for the next decades, only theories on small lattice sizes can be studied and error mitigation techniques have to be applied to reduce errors and in general quantum resources must be highly rationalized to keep the noise level as low as possible.\\
\\
These current technical limitations that force researchers to mainly study simplified theories on small lattices are nevertheless beneficial, because they stimulate a deeper understanding of the theory that could lead to the development of new approaches. Furthermore, the development of new quantum algorithms and their use on quantum hardware allows the research community to gain experience in quantum simulation, and ultimately take the opportunity to drive future hardware development toward the needs of the community, similarly to what happened with the development of supercomputers.

\subsection{The need for error mitigation techniques}
The main challenge of using the presently available NISQ era quantum hardware is the necessity of dealing with the noisy hardware due to the imperfect realization that generates results littered with errors and effectively hampering the usefulness of this new technology \citep{P_Shor_1996}. This is a temporary problem that should be solved with technological improvement, in fact from the theoretical point of view the quantum threshold theorem proves that once the hardware noise is reduced under a threshold level the quantum hardware can be as fault-tolerant as a classical computer \citep{Aharonov_et_all}.\\
\\
The main error sources affecting the NISQ hardware are the qubit interaction with the hardware, known as qubit decoherence; the imperfect action of gates, known as gate error; and errors made in reading the qubit state, measurement error.\\
\\
In contrast to the classical hardware where error correction can be implemented using some kind of data redundancy, basically by copying the data, anything similar to the classical case cannot be realized on quantum hardware, where copying a generic state is forbidden by the non-cloning theorem \citep{Wootters_Zurek,DIEKS1982271}. 
There have been a few error correction algorithms made available in the last few decades to take care of qubits decoherence like the Shor algorithm \citep{PhysRevA.52.R2493}, the bit-flip algorithm \citep{PhysRevA.32.3266}, Gottesman-Kitaev-Preskill algorithm \citep{Gottesman_2001}, the surface code \citep{Kitaev_2003} and a few others. But in general, these algorithms require hardware with a larger number of qubits and a greater qubit connectivity than those available today. For example, the Shor’s code needs 9 auxiliary qubits for each original qubit to keep free from decoherence.\\
\\
Therefore, while waiting for better and more powerful hardware with robust inbuilt  error correction capabilities, the user is left to troubleshoot by developing error mitigation techniques that can extend the usability of the current hardware by reducing the effects of the hardware errors on the measured data.\\

\subsection{A brief overview of the literature on quantum computing}
In the last 40 years since the first talk by Feynman and the independent work by Yuri Manin \citep{Manin_Yuri_1980} and Paul Benioff \citep{ Benioff:1979ce}, the scientific community has expressed a growing interest in the field and consequently produced a vast literature mainly dedicated to developing algorithms and speculating on how a quantum computer should be. Only in the last 6-7 years, the opportunity to use actual quantum computers, sometimes with free open access, has shifted the interest of researchers toward the direct implementation of LGT on the available quantum hardware, populating the literature with first results of quantum simulations using actual quantum hardware.\\
Hence, a succinct description of the historical literature is provided mainly to works that were considered useful to understand the growing field and motivate our interest in quantum computing, while a more detailed account of the literature related to the study of lattice gauge theory is provided, and the contribution of our two works \citep{ARahman:2021ktn, ARahman:2022tkr} to the literature is shown. Nevertheless, here there is no intention of giving a comprehensive and complete account of the whole quantum computing field not even of the part related to lattice gauge theories.\\
\\
In recent years many important contributions devoted to underlining the importance and the effectiveness of the development of quantum computers have been proposed, and here we give a chronological account. One of the first crucial works that showed how powerful a hypothetical quantum computer can be, is the one by Peter Shor \citep{Shor_Peter_1999} where the first account of how an algorithm for a quantum computer can perform exponentially better than a classical computer, for the concrete example of factorizing an integer in its prime factors. \\ 
Another important contribution was the one by Seth Lloyd that showed that a hypothetical quantum computer can efficiently simulate the dynamics of quantum mechanical system \citep{Seth_Lloyd_1996}, effectively confirming Feynman's conjecture.\\
\\
Regarding the properties that a hardware implementation should have to be considered a quantum computer David P. DiVincenzo proposed a list of criteria \citep{DiVincenzo_2000}, which guided researchers in developing actual hardware and are now commonly called the 7 DiVincenzo’s criteria.\\
\\
An important result toward the realization of quantum hardware was the proof of the quantum threshold theorem, that  ensures that a fault-tolerant quantum hardware can be obtained once the hardware noise is reduced under a threshold level making the quantum hardware as fault-tolerant as a classical computer \citep{P_Shor_1996, Kitaev:1997wr, Kitaev:1997wr, Aharonov_et_all}.\\
\\
For the development of the field of lattice gauge theory on quantum computers the work by Tim Byrnes and Yoshihisa Yamamoto was fundamental to developing the path for future studies by describing how to approach quantum simulations of $U(1)$, $SU(2)$ and $SU(3)$ \citep{Byrnes_Yamamoto}.\\
\\
A first comprehensive review of possible realizations of quantum hardware and on the possibilities and prospectives of the impact of quantum simulations in many fields of physics was reported by the authors of Ref.~\citep{Georgescu:2013oza}, and one of the first comprehensive textbooks on quantum computing by Nielsen and Chuang \citep{nielsen_chuang_2010} is now well known in the field as an important study reference.\\
\\
Strictly related to the field of lattice gauge theories using one dimensional lattice, the first study of a lattice gauge theory on a quantum computer was reported in \citep{Martinez:2016yna} a pioneering work where the real time dynamics of a one dimensional Schwinger model was studied on a trapped-ion quantum hardware, and after this milestone many other studies using a one dimensional lattice gauge theory using quantum hardware followed first using U(1) in \citep{Klco:2018kyo, Kokail:2018eiw, Mil:2019pbt, Surace:2019dtp, Yang:2020yer, deJong:2021wsd, Kim:2023sie}, Z(2) lattice gauge theory \citep{Carena:2022kpg, Pomarico:2023png, Charles:2023zbl},
$D_n$ \citep{Alam:2021uuq, Fromm:2022vaj}, SU(2) lattice gauge theory \citep{Klco:2019evd, Atas:2021ext} and SU(3) lattice gauge theory \citep{Ciavarella:2021nmj, Ciavarella:2021lel, Illa:2022jqb, Farrell:2022wyt, Atas:2022dqm}. We contribute to the field by performing the first study of part of the energy spectrum and the time evolution of SU(2) lattice gauge theory on a quantum annealer \citep{ARahman:2021ktn} and, the study of a SU(2) lattice on a gate-based quantum computer \citep{ARahman:2022tkr}.\\
\\ 
While for the case of lattice gauge theories on a (2+1) dimensional lattice no direct study on quantum hardware has been reported yet for more than a few plaquettes due to the shortage of quantum resources on the available quantum hardware, but instead many researchers have developed resource-efficient protocols that could be used in future hardware \citep{Celi:2019lqy, Kaplan:2018vnj, Paulson:2020zjd, Haase:2020kaj, Armon:2021uqr,  Kane:2022ejm, Gonzalez-Cuadra:2022hxt, Pardo:2022hrp, Bauer:2021gek}.\\
\\
Regarding the possibility to implement error correction techniques 
since those historical crucial algorithms are at the moment too expensive on the current hardware, many researchers have developed error mitigation techniques \citep{Geller_2013, Temme:2016vkz, Li:2016vmf, Klco:2018kyo, Funcke:2020olv, Alam:2021uuq, Urbanek:2021oej, Vovrosh:2021ocf, Atas:2022dqm, Ciavarella:2021lel, Farrell:2022wyt, Charles:2023zbl}, and we contribute to the field with the introduction of an error mitigation technique named Self-mitigation \citep{ARahman:2022tkr}.\\
\\
A few very interesting recent reviews of quantum computing for lattice gauge theory are already present in the literature \citep{Zohar:2015hwa, Ba_uls_2020, Alexeev:2020xrq, Humble:2022klb, Funcke:2023jbq, Bauer:2022hpo}.\\
\\
Finally, for a mindful motivating and historical account of the importance of quantum computing and its potential contributions to particle physics the many contributions by John Preskill to the quantum computing literature created in the last three decades are very inspiring \citep{ Preskill:1997dt, Preskill:1997ds, Preskill:1997uk, Preskill:1999he, Preskill:2012tg, Preskill:2018jim, Preskill:2018fag, Preskill:2022kgz}, and especially its latest report on the overall historical perspective of the entire industry over the past 40 years \citep{preskill2023quantum}.
%
%
%
%
%
%
\section{Structure of this thesis}
The origin of this thesis is grounded in the willingness to learn how to use the available quantum hardware to study particle physics phenomena that can be access by numerical calculation though direct computer calculation or by computer simulations. This approach of learning by doing using minimal lattice gauge theory models was contemporary shared by other research groups with a background in LQCD, in the interlude until better and more powerful quantum hardware able to host a full theory on large lattice size are developed.\\
\\
The thesis structure reflects the chronological development of the research, from the choice of the SU(2) theory as a first step toward the future goal of simulating SU(3), the gauge group on which QCD is based, to the strategies implemented and the results obtained in using two different open-access quantum hardware.\\
\\
In chapter 2 we present all the introductory material necessary to carry out the study of SU(2) pure gauge theory by presenting all the analytical derivation needed for the simulations and discussing the theory energy spectrum and time evolution calculated using a classical computer.\\
\\
In chapter 3 we first present the D-Wave quantum annealer and discuss how to use the hardware to find the ground state of a theory.  The numerical strategies and algorithms used to investigate the energy spectrum and the time evolution are described, and finally we discuss the numerical results for the energy spectrum and the time evolution of the SU(2) pure gauge lattice theory. This work was done in collaboration with Sarmed A Rahman, Randy Lewis and Sarah Powell and it was published in Phys. Rev. D \citep{ARahman:2021ktn}.\\
\\
In chapter 4 after presenting the IBM gate-base quantum hardware we investigated how to use the hardware by implementing it onto the time evolution of the theory and dealing with the hardware errors. This work was done in collaboration with Sarmed A Rahman, Randy Lewis and Sarah Powell and it was published in Phys. Rev. D \citep{ARahman:2022tkr}.\\
\\
Finally, in the concluding chapter, after a summary of the results of the entire thesis, a critical comparison between the two hardwares used is provided together with a general perspective for the future use of quantum computers in the field of lattice gauge theory.
%
%

\chapter{SU(2) lattice gauge theory}\label{chap:Chapter_1}
In this chapter we present the SU(2) pure gauge lattice theory studied, by presenting its Hamiltonian, the operators present therein and demonstrate how to find the physical states of the system used to represent it. In the last part we discuss the theory for a minimal lattice by deriving the physical states, the Hamiltonian representation and discuss the energy spectrum of the theory and its time evolution, pointing out some interesting physical phenomena like travelling excitation and the behaviour of the theory with truncation of the Hilbert space.\\
\\
The Hamiltonian for SU(2) lattice gauge theory was proposed by Kogut and Susskind in 1974 \citep{PhysRevD.11.395} and takes into account both the pure gauge part and the massive fermion part. In this thesis as a first fundamental step we concentrated our interest only in the pure gauge part leaving the massive fermion part for a future study, therefore only a derivation for the pure gauge part is presented in Appendix \ref{appendix:HSU2_derivation}.\\
In this respect, many recent articles addressed SU(2) lattice gauge theory in the context of quantum computing \citep{Byrnes_Yamamoto, Banerjee:2017tjn, Tagliacozzo:2012df, Zohar:2012xf, Stannigel:2013zka, Zohar:2015hwa, Zohar:2014qma, Mezzacapo:2015bra, Banuls:2017ena, Banerjee:2017tjn, Raychowdhury:2018tfj, Raychowdhury:2018osk, Kasper:2020akk, Davoudi:2020yln, Dasgupta:2020itb,Kasper:2020owz} but none of them used the available quantum hardware as they are limited and noisy.
The first simulation of SU(2) pure gauge theory on a quantum computer was done in 2019 by \citep{Klco:2019evd} and, it is important to mention that while we were working on \citep{ARahman:2021ktn}, the first work simulating the full SU(2) theory with fermions on a quantum computer appeared in the literature in 2021 \citep{Atas:2021ext}.\\
Our approach is mostly close to the treatment of \citep{Byrnes_Yamamoto,Klco:2019evd,Raychowdhury:2019iki} and here we follow the notation by Byrnes and Yamamoto in Ref.~\citep{Byrnes_Yamamoto}.\\
\\
When presenting a lattice gauge theory an important role is played by the space-time discretization. In our study that uses the Hamiltonian formalism the time is continuous while the space is discretized by using a single link joining two points in space as a primitive element on which the SU(2) gauge field $U$ resides. The link length $a$ is conventionally set to 1, $a=1$. These links are arranged in groups of four to form squares called plaquettes. A plaquette is the base element, and it is used to map the entire space by simply joining many of them together as shown in Figure~\ref{fig:lattice_explain}:
\begin{figure}[H]
\centering
\includegraphics[width=0.7\linewidth]{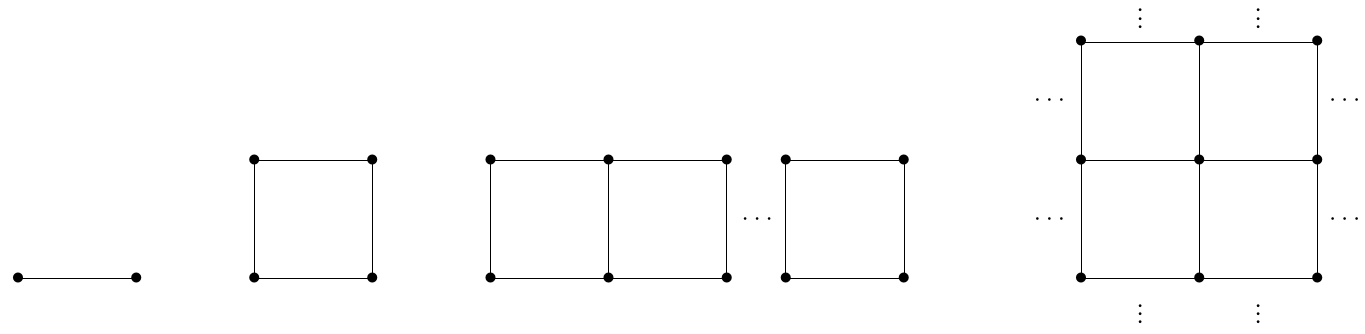}
\caption{The lattice notation is: a link is denoted by a line connecting two nodes indicated by black dots. From left to right we have a single link, a plaquette, a row of plaquettes and a square of plaquettes. A row of plaquettes can be used to map a 1-dimensional lattice while the square of plaquettes can be used to map a 2-dimensional lattice.\label{fig:lattice_explain}}
\end{figure}
In the works presented in this thesis we considered a minimal one-dimensional lattice made by a few plaquettes arranged in a row, with periodic or closed boundary conditions depending on the specific study. This choice was motivated by carefully considering the hardware noise and the shortage of quantum resource in the NISQ quantum hardware.

\section{SU(2) Hamiltonian and the operators}
We can start by first presenting the Hamiltonian and describing the action of the operators therein and successively formalizing their properties using the SU(2) group property. The interested reader can find in Appendix \ref{appendix:plaquette_derivation} a derivation of the pure gauge lattice theory Hamiltonian from the SU(2) pure gauge Lagrangian. \\
For clarity let's picture a generic row-lattice composed by four plaquettes with periodic boundary condition as shown Figure~\ref{fig:lattice}: 
\begin{figure}[H]
\centering
\includegraphics[width=0.6\linewidth]{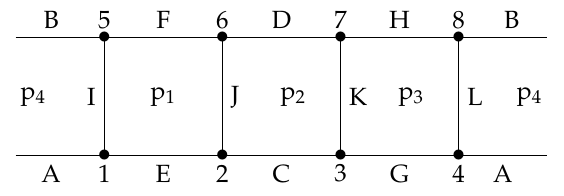}
\caption{A 4-plaquette lattice with periodic boundary condition in the longitudinal direction. The lattice is composed by 4 plaquettes made by the union of 12 gauge links which are joined by 8  nodes. 
Each plaquette is indicated by the label in the center of each square, each link by the letter at the middle point, and each node by the closest number.\label{fig:lattice}}
\end{figure}
The SU(2) pure gauge Hamiltonian considered is:
\begin{equation}\label{eq:H_c2}
\hat H = \frac{g^2}{2}\left(\sum_{i={\rm links}}\hat E_i^2-2x\sum_{i={\rm plaquettes}}\hat\square_i\right)
\end{equation}
where the only free parameter is the gauge coupling constant $g$, or the more convenient parameter $x\equiv 2/g^4$. The first term in $\hat H$, $\hat E_i$ is the square of the chromoelectric field operator whose action is to globally quantify the stored chromoelectric energy on the lattice by each link. The second term is the chromomagnetic term that contains the plaquette operator $\hat\square_i$ whose action is to add or subtract fluxes of energy on the $ith$ lattice plaquette. This Hamiltonian can be derived from the pure gauge SU(2) Lagrangian as it is shown in Appendix \ref{appendix:HSU2_derivation}.\\\\
\\
A more formal definition of the operators in connection with the SU(2) group properties is important to settle the discussion in a more insightful way. In this respect we adapt the discussion from \citep{Byrnes_Yamamoto}.\\
To discretize the space let's fix a reference frame at each lattice site $r$ and label the link as in Figure~\ref{fig:lattice_U_pla}:
\begin{figure}[H]
\centering
\includegraphics[width=0.4\linewidth]{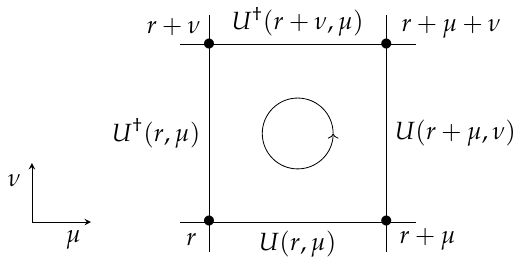}
\caption{A single plaquette where all the links are labelled starting from the lattice site r going counterclockwise by using the two unit vectors $\mu$ and $\nu$ expressing the two possible space directions.\label{fig:lattice_U_pla}}
\end{figure}
On each link of the lattice we define a gauge link operator as $2 \times 2$ matrix $\hat U(r,\mu)$ with the two end points oriented along the unit vector $\mu$  and the canonically conjugate operators $\hat E(r,\mu)^\alpha$, where for $\alpha=a,b,c$ we have the three color components of field respecting the SU(2) Lie algebra $[\hat E^a,\hat E^b]=i\epsilon^{abc}\hat E^c$.\\
The theory is quantized by imposing the following commutation relation at each lattice site:
\begin{eqnarray}
\left[ \hat{E(x)}^\alpha, \hat{U(x)} \right]  &=& \dfrac{1}{2} \sigma^\alpha \hat{U(x)}, \nonumber \\
\left[ \hat{E(x)}^\alpha, \hat{U(x)}^\dagger \right]  &=& -\dfrac{1}{2} \hat{U(x)}^\dagger \sigma^\alpha,
\end{eqnarray}
where $\sigma^\alpha$ $(\alpha=1,2,3)$ are the three SU(2) generators, the Pauli matrices.\\ 
The operators commute if they are at different lattice site.

The three $\hat E(x)^\alpha$ components can be used to define $\hat E^2 = (\hat E^a)^2 +(\hat E^b)^2 + (\hat E^c)^2$ as  the local gauge-invariant quadratic Casimir operator of SU(2).\\
\\
To further characterize these operators we need to understand how they act on the link state, but first we need to define a basis to represent a link state.
A possibility is to represent the state of each link composing the lattice by using one of the possible infinite countable SU(2) irreducible representations $j$ with $(j=0,\tfrac{1}{2},1,\tfrac{3}{2},2…)$ and two SU(2) projections $m, m^\prime$ with $(m, m^\prime = -j,-j+1,\ldots, j)$ to express the two link ends. The link state is indicated as $\left|j,m, m^\prime\right>$. This approach is called angular momentum formulation of a SU(2) lattice gauge theory. 
\\
Using the angular momentum formulation the operators $U$ and $\hat E^2$ act on link states as \citep{Byrnes_Yamamoto}:
\begin{eqnarray}
\hat{U}_{s,s^\prime}\left|j,m,m^\prime\right> &=& \sum_{J=\left|j-\tfrac{1}{2}\right|}^{j+\tfrac{1}{2}}\sqrt{\frac{2j+1}{2J+1}}
\sum_M\sum_{M^\prime}\left<J,M|j,m;\tfrac{1}{2},s\right>\left<J,M^\prime|j,m^\prime;\tfrac{1}{2},s^\prime\right>
\left|J,M,M^\prime\right>~~~~~~~~~
\end{eqnarray}
which has a non-vanishing contribution only for $M=m + s$ and $M^\prime=m^\prime + s^\prime$. This means that if the conditions on the projection are met the link operator is able to change the link state by a $\pm \frac{1}{2}$ unit of angular momentum.\\
Since $\hat E^2$ is the Casimir operator it acts on the link state as:
\begin{equation}
\hat E^2 \ket{j,m,m^\prime}= j(j+1)\ket{j,m,m^\prime}
\end{equation}
where $\ket{j,m,m^\prime}$ is an eigenstate of $\hat E^2$ with eigenvalue $j(j+1)$ which is interpreted as the chromoelectric energy stored in the link.\\
\\
The last operator present in the Hamiltonian is the plaquette operator. It is the simplest gauge invariant operator made by taking the trace of product of the four link operators placed around a plaquette as shown in Figure~\ref{fig:lattice_U_pla}:
\begin{equation}
\hat\square = \mathrm{Tr} [\hat{U}(r,\mu) \ \hat{U}(r+\mu,\nu) \ \hat{U}^\dagger(r+\nu,\mu) \ \hat{U}^\dagger(r,\nu)] \,.   
\end{equation}
Because the plaquette operator contains four link operators, it can change the state of each of the four links by $\pm \frac{1}{2}$ unit of angular momentum. The variation of each link state is acceptable only if the angular momentum composition rule holds at each of the four plaquette nodes.   
\\
\section{Deriving the states and the Hamiltonian representation}\label{section_H_representation}
Considering a generic 1-dimensional lattice made of a row of plaquettes as in Figure~\ref{fig:lattice}, we chose to characterize the states of the system using a global basis, in which the state of each link composing the lattice is expressed in the angular momentum basis $\left|j,m, m^\prime\right>$ and are normalized as $\left< l,n, n^\prime |j,m, m^\prime\right> = \delta_{l j} \delta_{m^\prime m} \delta_{n^\prime n}$ .\\
With this choice, a generic state of the system for a lattice like the one in Figure~\ref{fig:lattice} can be written as a direct product of each link state as:
\begin{equation}
\left|\psi\right> = \left|j_A,m_A,m_A^\prime\right>\left|j_B,m_B,m_B^\prime\right>\ldots\left|j_L,m_L,m_L^\prime\right> \,.
\end{equation}
The state so labelled is a physically allowed state of the system only if at each lattice node the SU(2) Gauss's law holds. This implies that in constructing the physical states one has to ensure that the links joining at each node respect the angular momentum composition rule.\\
It is important to note that even on a lattice made by a finite number of plaquettes, the Hilbert space is infinite because SU(2) is a continuous symmetry, therefore the associated Hamiltonian energy spectrum must be truncated at a certain energy value. However once the theory is truncated it has a finite number of states and it can be encoded using a finite classical or quantum computer memory. In this approach the full theory is approximated by successively increasing the energy truncation $j$. A larger $j$ truncation adds new states to those already present in the previous, smaller $j$ truncation. Therefore the truncation rationale can be evaluated by the agreement of the expectation value of an operator of interest between successive truncations.\\
\\
For future exploration not only of this specific Hamiltonian but also for an SU(2) non-abelian LGT with and without fermions one has to keep in mind that there are other ways to truncate the Hilbert space by using a different formulation. The authors of Ref.~\citep{Davoudi:2020yln} have thoroughly explored different formulations for SU(2) LGT reporting advantages and disadvantages of the most common formulations from the physics and computational point of view.
\\
To begin constructing the Hilbert space we have to identify the vacuum state, which could be defined as the state for which all the links are set to the $j=0$ SU(2) representation, that means that no chromoelectric energy flux is present on the lattice, that reads:
\begin{equation}
\hat{E}^2 \left|vacuum\right>= 0
\end{equation}
The excited states can be generated by applying any combination of plaquette operators to the vacuum. There is no limitation to how many plaquette operators can be applied, therefore to avoid dealing with an infinite number of states a truncation as to be introduced. The one adopted in this thesis is to limit the allowed $j$ value of the SU(2) irreducible representation for each of the link states composing the system state to a fixed $j_{max}$ value.\\
\\
Once the truncation value $j_{max}$ is chosen, the states generated can be used to find the Hamiltonian representation by simply calculating the contribution of the chromoelectric and chromomagnetic parts.
\\
The chromoelectric operator matrix elements are:
\begin{equation}\label{eq:HE}
\left<\psi_{ \rm final}\right|\sum_i\hat E_i^2\left|\psi_{ \rm initial}\right> = \sum_{i=A}^Lj_i(j_i+1) \delta_{\rm final, \rm initial} \,.
\end{equation}
which is diagonal, and the eigenvalues are the total chromoelectric energy stored on the lattice.
\\
The chromomagnetic matrix elements are more involved and have a specific contribution coming from each plaquette.
The contribution of a generic plaquette is derived in Appendix \ref{appendix:plaquette_derivation}, while here we show the result for  plaquette 1 of Figure~\ref{fig:lattice}:
\begin{align}
\left<\psi_{\rm final}\right|\hat\square_1\left|\psi_{\rm initial}\right> \nonumber 
 =& (-1)^{j_A+J_E+j_I}\sqrt{(2j_I+1)(2J_E+1)}\left\{\begin{array}{ccc} j_A & j_E & j_I \\ \tfrac{1}{2} & J_I & J_E \end{array}\right\} \nonumber \\
 & (-1)^{j_C+J_E+j_J}\sqrt{(2j_E+1)(2J_J+1)}\left\{\begin{array}{ccc} j_C & j_E & j_J \\ \tfrac{1}{2} & J_J & J_E \end{array}\right\} \nonumber \\
 & (-1)^{j_D+J_F+j_J}\sqrt{(2j_J+1)(2J_F+1)}\left\{\begin{array}{ccc} j_D & j_F & j_J \\ \tfrac{1}{2} & J_J & J_F \end{array}\right\} \nonumber \\
 & (-1)^{j_B+J_F+j_I}\sqrt{(2j_F+1)(2J_I+1)}\left\{\begin{array}{ccc} j_B & j_F & j_I \\ \tfrac{1}{2} & J_I & J_F \end{array}\right\} \nonumber \\
\label{eq:oneplaq}
\end{align}
where $j_i$ and $J_i$ are the j value for $\left|\psi_{\rm initial}\right>$ and $\left|\psi_{\rm final}\right>$, respectively.
Its contribution is non zero only if all the links in $\left|\psi_{\rm initial}\right>$ and $\left|\psi_{\rm final}\right>$ external to the acting plaquette have the same values, which is clear by looking to Eq.~\ref{eq:plaq_Istate}.
The four curly brackets are 6j symbols whose role is to enforce Gauss's law automatically at each node. Their numerical value is typically a square root of ratios and are provided in Appendix~\ref{appendix:plaquette_derivation}.
Notice that $\left|\psi_{\rm initial}\right>$ and $\left|\psi_{\rm final}\right>$ will never be the same state because
applying a plaquette operator necessarily changes each of those four gauge links by $\pm\tfrac{1}{2}$. Therefore all plaquette terms are off diagonal.\\
\\
Finally the Hamiltonian matrix elements for the lattice in Figure~\ref{fig:lattice} are:
\begin{equation}\label{eq:HE}
\left<\psi_{ \rm final}\right|\hat H\left|\psi_{ \rm initial}\right> = \sum_{i=A}^L j_i(j_i+1) \delta_{ \rm final, \rm initial} -2x \sum_{i=1}^4 \left<\psi_{ \rm final}\right|  \hat\square_i \left|\psi_{ \rm initial}\right> \,.
\end{equation}
There are three important observation.\\
The Hamiltonian chromoelectric and chromomagnetic contribution depends only on the $j$ SU(2) irreducible representation and not on the two projections at the end of each link $m$ and $m^\prime$. This observation motivated the change of the entire lattice state notation from\\$\left|\psi\right> = \left|j_A,m_A,m_A^\prime\right>\left|j_B,m_B,m_B^\prime\right>\ldots\left|j_L,m_L,m_L^\prime\right>$ to simply $\left|\psi\right> = \left|j_A,j_B, \ldots, j_L \right>$ and most important, enlighten toward the development of a simple algorithm to generate the allowed physical states given the lattice shape and the chosen $j_{max}$ truncation. This algorithm was used to generate the physical states for the Hamiltonian representations of the different lattice configurations present in this thesis. Its description in pseudocode and one calculation example was provided in Appendix \ref{appendix:state_algorithm}.\\
\\
The Hamiltonian has two regimes. In the strong coupling regime, $g^2 \to \infty$ $(x\to0)$, the chromomagnetic contribution is negligible and therefore the Hamiltonian closely coincides with the chromoelectric part therefore it is almost diagonal in the lattice states and these states are quasi-stationary. In the weak coupling regime, $g^2 \to 0$  $(x \to \infty)$, the chromomagnetic term dominates hence the eigenstates are now linear combinations of the chromoelectric states. Therefore many possible energetic transitions are possible between the chromoelectric states. In the intermediate case where $0<x<1$ the chromoelectric eigenstates are mildly affected by the chromomagnetic term creating a dynamics in which low-energy, single plaquette excitations, are traveling across the lattice.\\ 
\\
Since the Hamiltonian is not always diagonal in the lattice basis used, it is tempting to numerically diagonalize it. There are two good reasons not to succumb to this temptation: Numerical diagonalization hides the symmetries that are present and can be made evident by considering block diagonalization based on the system symmetries. Additionally, there are already quantum hardware built with the scope of finding the ground state in mind, as will be discussed in Chapter 2. Secondly, if we look toward the future when quantum computers will be fault-tolerant and have abundant quantum resources, there will be no need to diagonalize the Hamiltonian on a classical computer. Using quantum hardware, like the one discussed in Chapter 2 or implementing a quantum algorithm like the variational quantum eigensolver (VQE) \citep{Peruzzo_2014} to find the the eigenstates in a more efficient way and for a larger Hamiltonian than can be feasibly diagonalized using classical computer. For the present we should explore the potentiality of the available quantum resources by using only minimal assistance from the classical computer.

\section{2-plaquette $j_{max}=1/2$}
To make the theory more tangible, it is useful to consider a concrete example that summarizes the approach used to study the system dynamics on different lattices and with different $j$ truncations.\\
Let's consider the minimal case of a two plaquettes lattice with closed boundary conditions and work out how to derive the system states, discuss the time evolution and the effect of the truncation on the energy spectrum.   
The Hamiltonian for this case reads:
\begin{equation}\label{eq:H_2pla}
\hat H = \frac{g^2}{2}\left(\sum_{i={\rm links}}\hat E_i^2-2x\left(  \hat\square_L + \hat\square_R \right) \right)
\end{equation}
We adopt an energy truncation of $j_{max}=1/2$ implying that the link state can only have $j=0$, $j=1/2$ or a superposition of the two.\\
In this case the four allowed basis states are present in Figure~\ref{fig:four_states}:
\begin{figure}[H]
\centering
\includegraphics[width=0.7\linewidth]{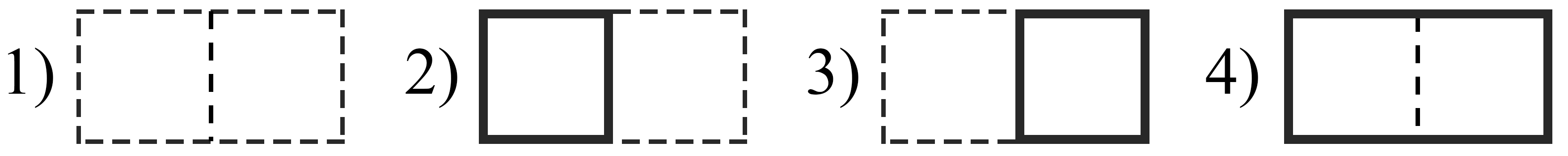}
\caption{The four possible states available on the 2-plaquette lattice after the $j_{max}=1/2$ truncation. Each solid line means a $j=1/2$ flux of energy is present, while the dashed line a $j=0$ meaning no energy is present. If all the links of a plaquette have $j=1/2$ we label the plaquette as switched ``on'', otherwise the plaquette is switched ``off''.}
\label{fig:four_states_chap_1}
\end{figure}
It is worth discussing how the states depicted in Figure~\ref{fig:four_states_chap_1} can be actually obtained, and because the system is minimal how this can be done without using the algorithm in Appendix \ref{appendix:state_algorithm}. We start with the vacuum and obtain the second state by applying the left plaquette operator $\hat\square_L$ on the vacuum, obtaining a $j=1/2$ energy flux on the left plaquette. If the left plaquette is now applied to the realized state, the second state, we obtain two states: one having all the links with $j=0$ that is the vacuum, and the other with $j=1$ flux in the left plaquette, which cannot be accepted because we fixed the truncation to $j_{max}=1/2$. Notice that these two $j$ values are simply obtained by the angular momentum composing rule, $\frac{1}{2} \otimes \frac{1}{2} = 0 \oplus 1$.\\
Similarly, the third state is obtained by applying the right plaquette operator $\hat\square_r$ on the vacuum obtaining a $1/2$ on the right plaquette.\\
Finally, the fourth state is obtained by applying both plaquettes on the vacuum. This should be done carefully by taking care of the angular momentum composition rule at the center vertical link, which now receives two contributions. 
We obtain two states: one in which both plaquettes contribute a $j=1/2$ in all the links, resulting with the vertical center link state having a $j=1$ value, which cannot be accepted due to the $j_{max}=1/2$ truncation. The other state is obtained by having a flux of $j=1/2$ in all the links except the vertical center link that has a $j=0$ value, which is fourth state.\\
\\
These four states in notation used so far $\left|j_a, j_b, j_c, j_d, j_e, j_f, j_g\right>$ are:
\begin{eqnarray}
&&1) \rightarrow \left|0, 0, 0, 0, 0, 0, 0\right> \quad   2) \rightarrow \left|\frac{1}{2}, \frac{1}{2}, \frac{1}{2}, \frac{1}{2}, 0, 0, 0\right>\\ \nonumber
&&3) \rightarrow \left|0, 0, 0, \frac{1}{2}, \frac{1}{2}, \frac{1}{2}, \frac{1}{2}\right>  \quad 4) \rightarrow \left|\frac{1}{2}, \frac{1}{2}, \frac{1}{2}, 0, \frac{1}{2}, \frac{1}{2}, \frac{1}{2}\right>
\end{eqnarray}
Since this notation is quite cumbersome we change it to one that closely represents a miniature of the entire lattice $\left|A_B^C D_E^F G\right>$, where the capital letters are $A=2 j_A +1$ and are used to have only integer values.\\
\\
Using these four states and the relations provided for the operators in the previous section the Hamiltonian representation is:
\begin{equation} 
\hat{H} = \frac{g^2}{2}
\left(
\begin{array}{cccc}
0 & -2x & -2x & 0 \\
-2x & 3 & 0 & -x \\
-2x & 0 & 3 & -x \\
0 & -x & -x & \frac{9}{2}
\end{array}
\right)
\begin{array}{c}
\left| 1_1^1 1_1^1 1\right> \\
\left| 2_2^2 2_1^1 1\right>\\
\left| 1_1^1 2_2^22 \right>\\
\left| 2_2^2 1_2^2 2 \right>\\
\end{array}
\end{equation}

\begin{figure}[H]
\centering
\begin{subfigure}{.5\textwidth}
  \centering
  \includegraphics[width=1.\linewidth]{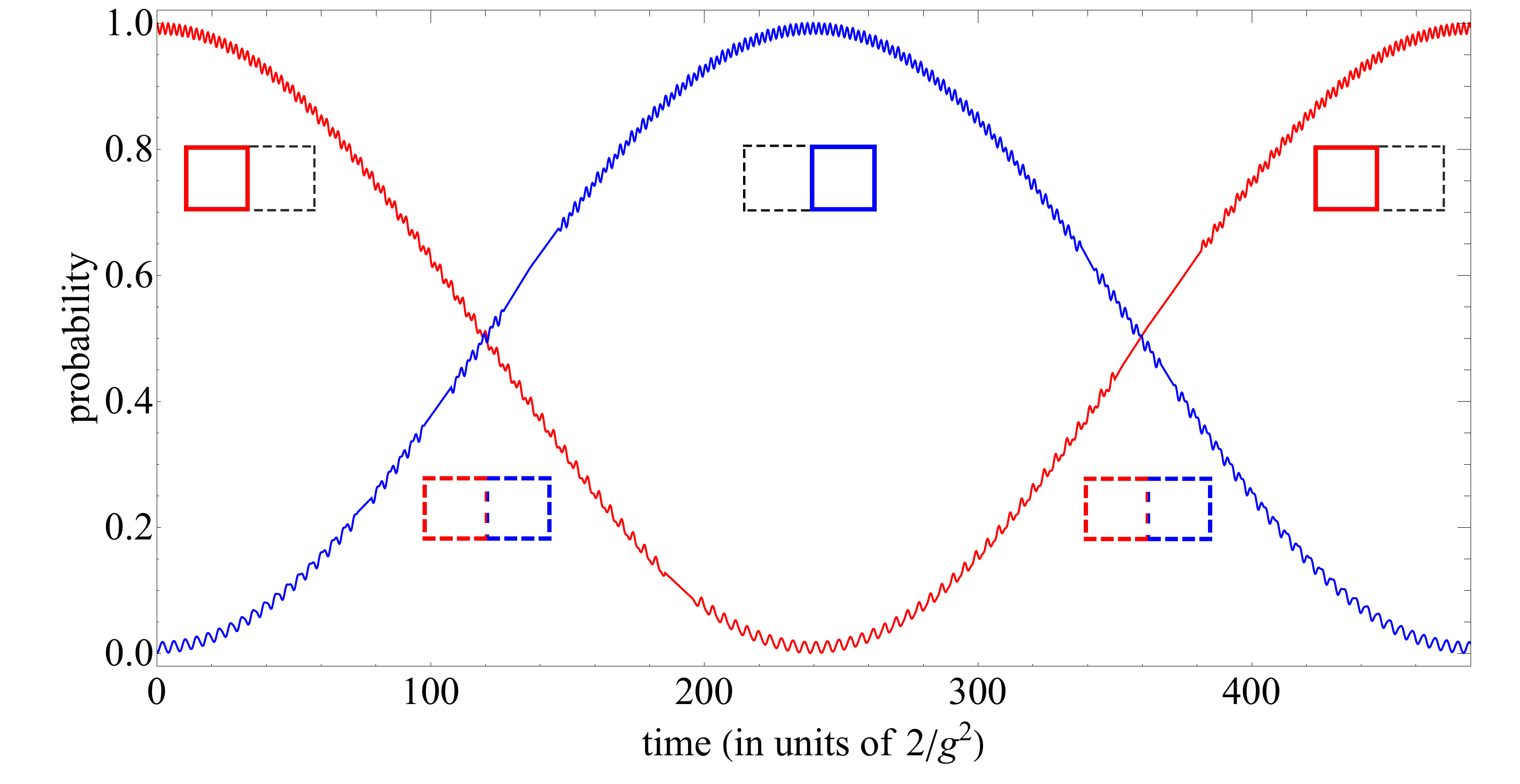}
  \label{fig:sub1}
\end{subfigure}%
\centering
\begin{subfigure}{.5\textwidth}
  \centering
  \includegraphics[width=1.\linewidth]{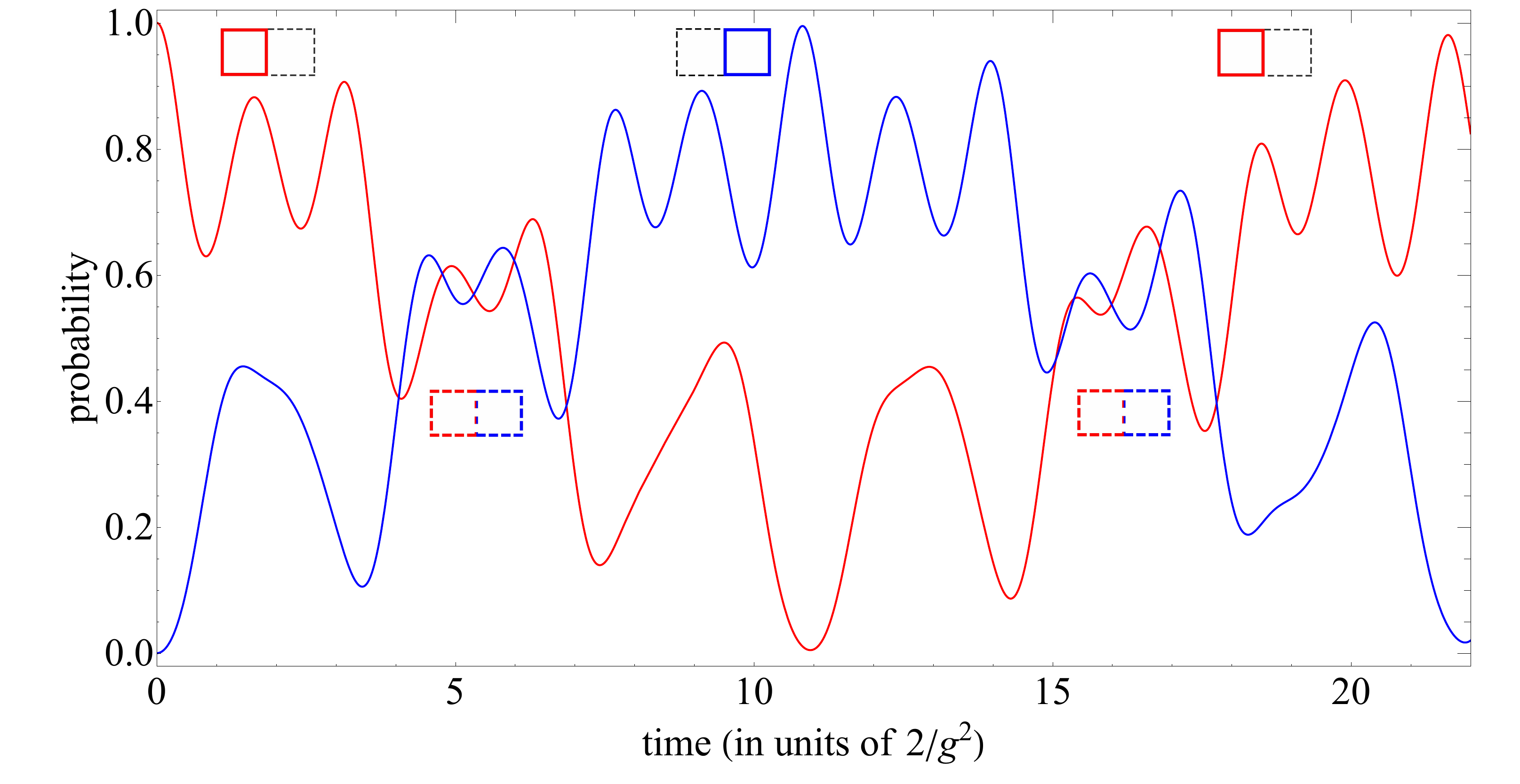}
  \label{fig:sub2}
\end{subfigure}
\caption{Exact calculation of the real time evolution for the 2-plaquette lattice for $x<1$. The left panel is at $x=0.1$ while the right panel at $x=0.6$. The red (blue) solid curve is the exact probability of the left (right) plaquette having an energy flux of $j=1/2$ over time. In both panels the small diagrams using the notation of Figure~\ref{fig:four_states} indicate the system state. The plots show the decrease of the excitation propagation time with the increasing of $x$ in the range $x<1$. }
\label{fig:Te_x<1}
\end{figure}

\begin{figure}[H]
\centering
\begin{subfigure}{.5\textwidth}
  \centering
  \includegraphics[width=1.0\linewidth]{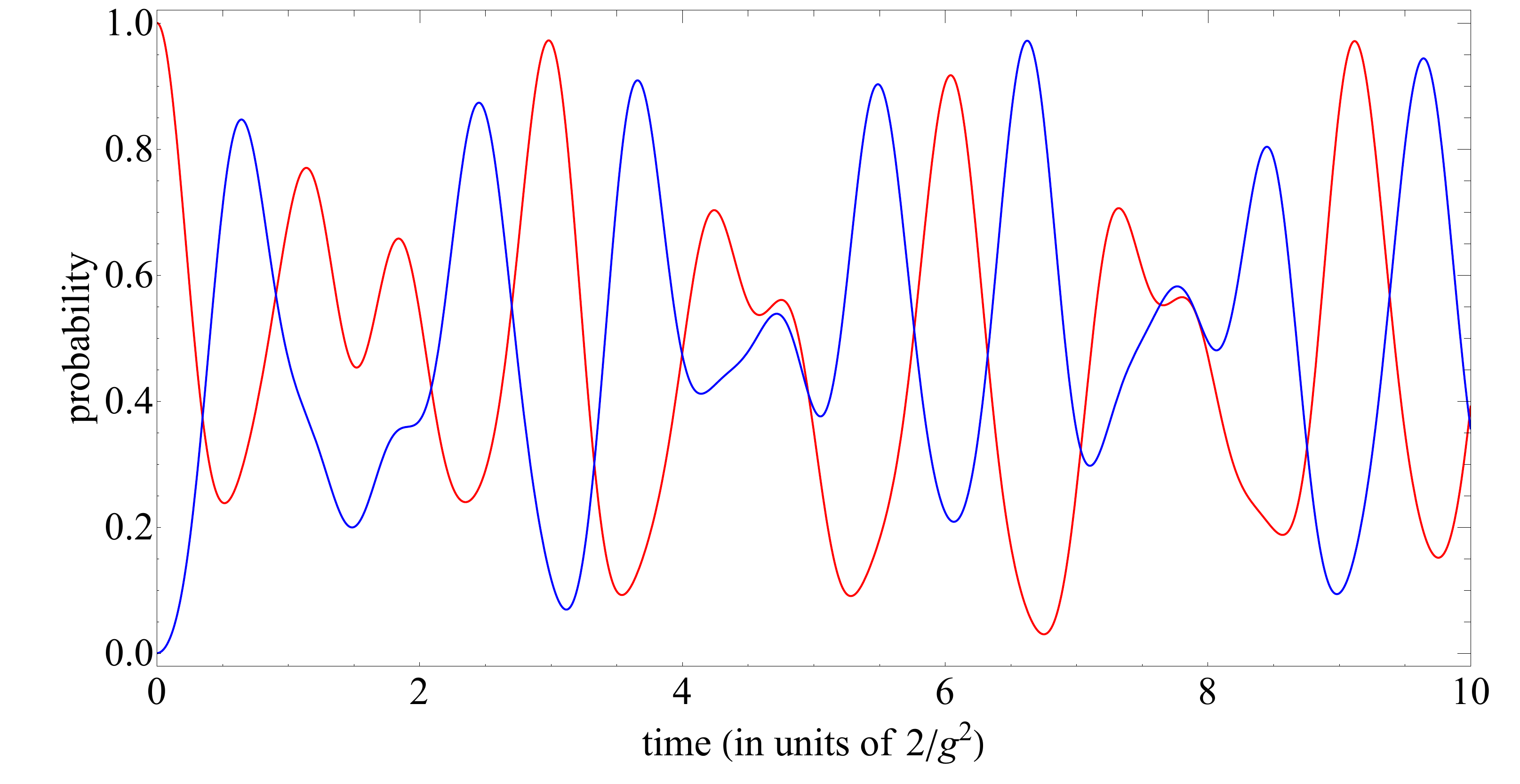}
  \label{fig:sub1}
\end{subfigure}%
\begin{subfigure}{.5\textwidth}
  \centering
  \includegraphics[width=1.0\linewidth]{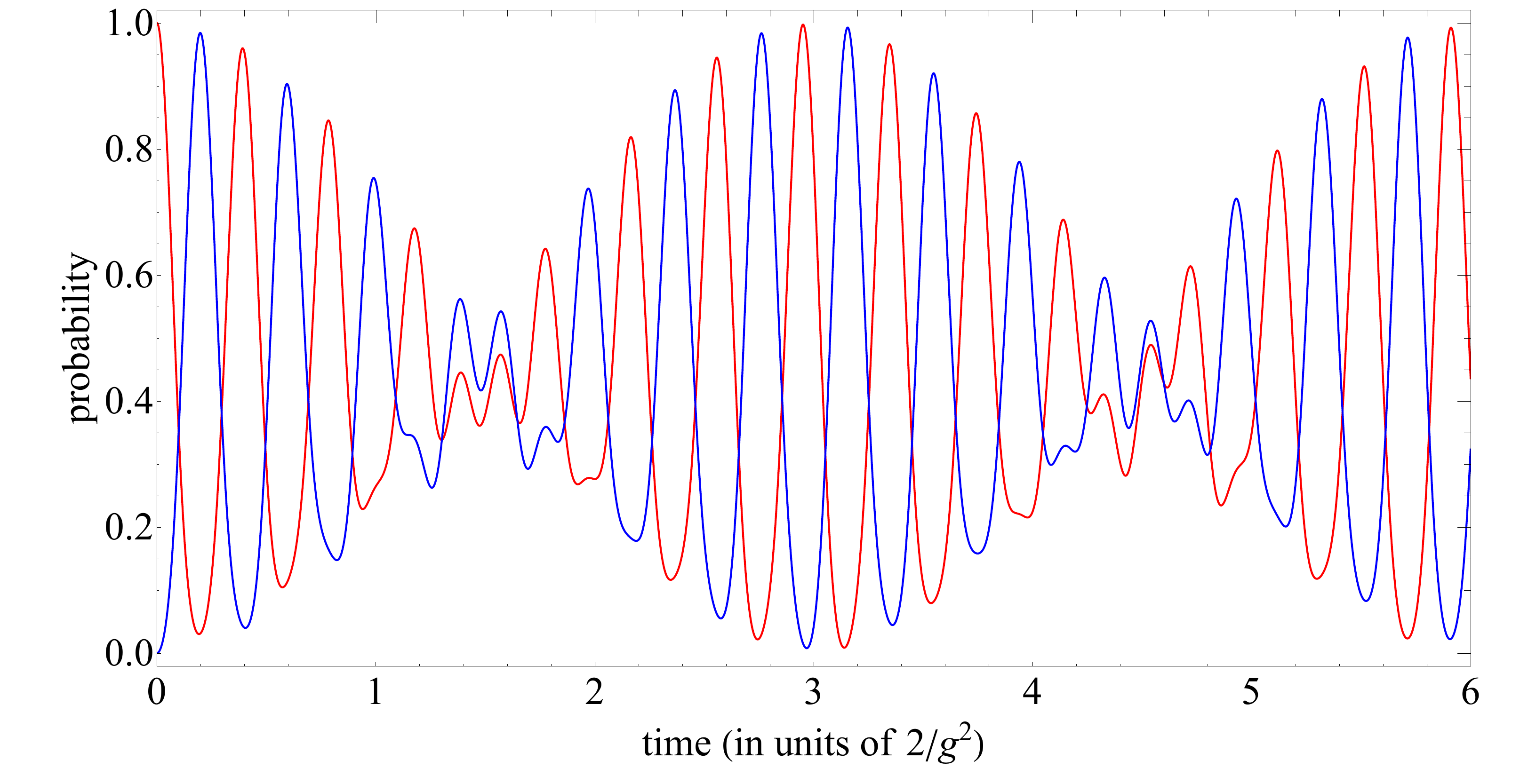}
  \label{fig:sub2}
\end{subfigure}
\caption{Exact calculation of the real time evolution for the 2-plaquette lattice for $x>1$. The left panel is at $x=1.5$ while the right panel is at $x=5.0$. The red (blue) solid curve is the exact probability of the left (right) plaquette having an energy flux of $j=1/2$ over time. The plots show the increase of higher frequencies with the increasing of $x$ in the range $x>1$.}
\label{fig:Te_x>1}
\end{figure}

This is enough to consider some physics phenomena. For example by considering the real-time evolution of the theory and calculating the probability that starting with the left plaquette switched ``on'', one finds that the left or the right plaquette will be  switched ``on'' after a certain time. \\
\\
The Hamiltonian has two regimes, one for large $x$ and one for small $x$.
For $x<1$ the Hamiltonian off-diagonal elements whose presence is due to the chromomagnetic operator are small. Therefore the dominant states are weakly coupled chromoelectric eigenstates. In this regime, the single-plaquette states will move across the lattice as shown in Figure~\ref{fig:Te_x<1}.
A traveling exitation is the interpretation of the phenomena present in the left panel of Figure~\ref{fig:Te_x<1} where for $x<1$ the probability of having an energy flux in the left plaquette (red curve) drops from $100\%$ at time $t = 0$ to close to 0$\%$ at time $t = 240$ while the probability of right plaquette (blue curve) does the opposite. Similarly in the right panel, the now larger value of $x$ superimposes some high frequencies.
That the excitation propagation time is shorter for larger $x$ is clearly visible by comparing the pseudo-period of the left panel at $x=0.1$ with the one present in the right panel at $x=0.6$.\\
The phenomena shown here on a larger lattice is interpreted as an only-energy particle a so called SU(2) glueball propagating into space.\\
\\
For $x>1$ the chromomagnetic contribution dominates mixing of the single-plaquette states. This is evident by the presence of many higher frequencies superimposed to the single-plaquette transitions in Figure~\ref{fig:Te_x>1}.

\begin{figure}[H]
\centering
\includegraphics[width=0.8\linewidth]{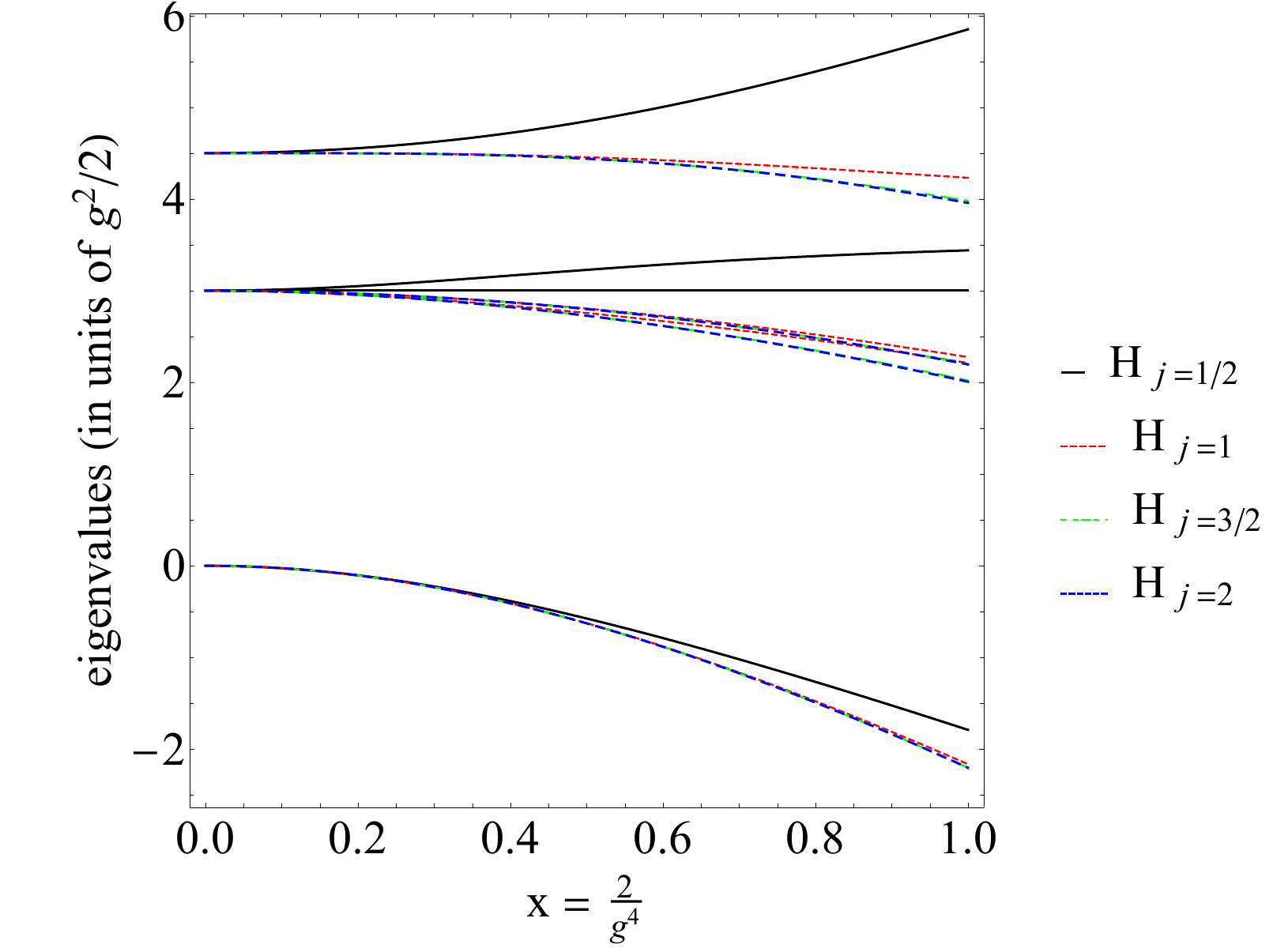}
\caption{The four lowest energy levels for the 2-plaquette lattice with closed boundary conditions Hamiltonian for various $j$ truncation versus $x$. In the range of $x$ considered rapid convergence is observed.}
\label{fig:E_2pla_1o2}
\end{figure}

\begin{figure}[H]
\centering
\includegraphics[width=0.9\linewidth]{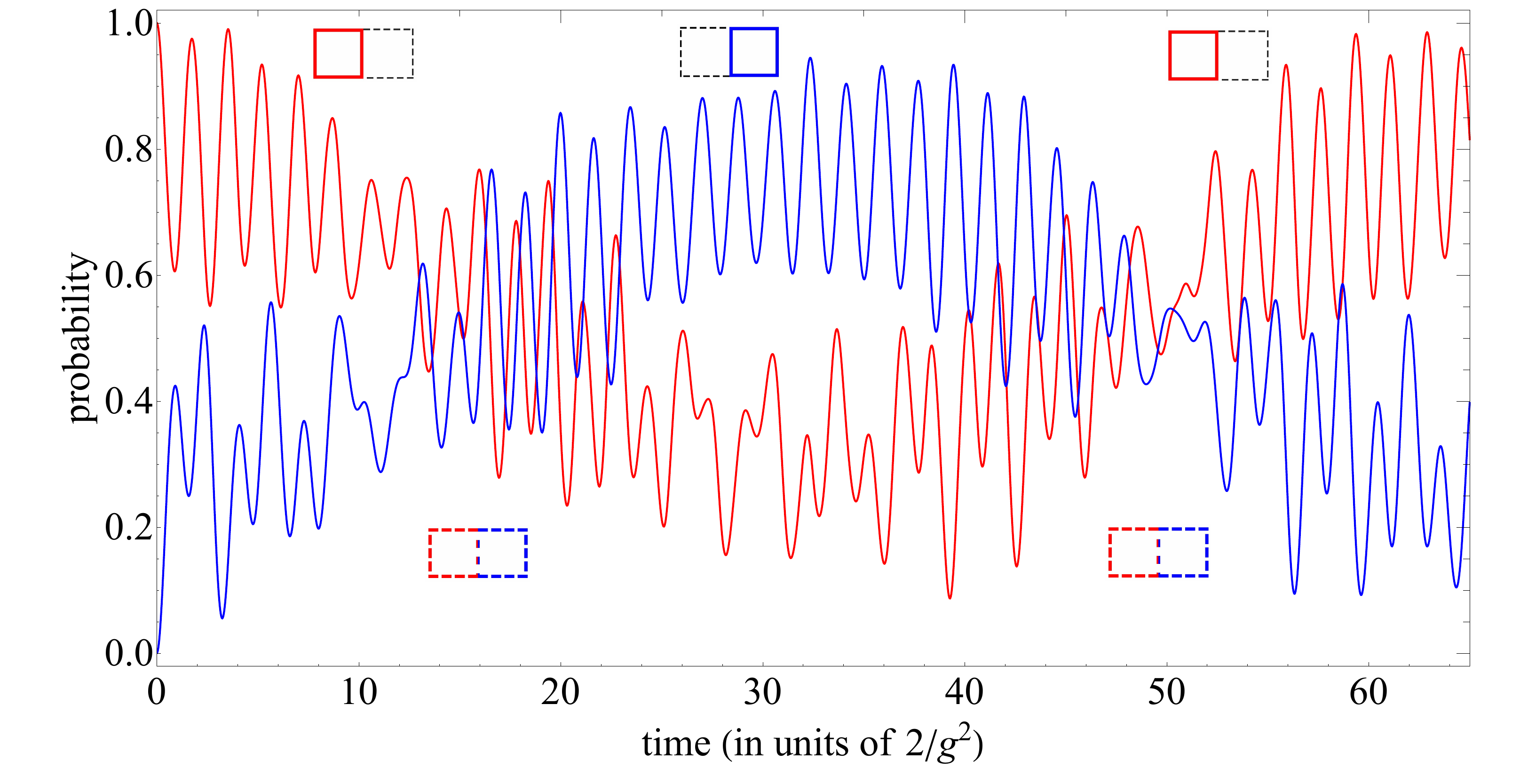}
\caption{Exact calculation of the time evolution for the 2-plaquette lattice for $x=0.6$ as in left panel of Figure~\ref{fig:Te_x<1} but coming from a $j=2$ truncation. The red (blue) solid curve is the exact probability of the left (right) plaquette having an energy flux of $j>0$ over time. The plot confirms that traveling exicitation are SU(2) phenomena and not an artefact of small $j$ truncation.}
\label{fig:to_jmax_2}
\end{figure}

At larger values of $x$, higher frequencies are present as can be seen moving from the left panel at $x = 1.5$ to the right panel at $x = 5.0$ in Figure~\ref{fig:Te_x>1}.\\
\\
Finally, in this section we should briefly discuss the effect of the $j$ truncation on the system by considering a relevant observable and investigate how it changes by considering larger $j$ values. In Figure~\ref{fig:E_2pla_1o2} it is shown how the four $j_{max}=1/2$ energy eigenvalues versus $x=2/g^4$ are changed by considering larger $j$ values.\\
The plot shows that the effect of the $j$ truncation is relevant for all four $j=1/2$ states and its effect grows with the value of $x$, but it is more prevalent on the excited states than on the ground state. What it is important to notice is the rapid convergence in the $x$ range considered among different $j$ values larger than $j=1/2$, and by $j=2$ precise results are clearly obtained.\\
\\
The impact of a larger $j$ truncation will have effect on the time evolution too. The plots in Figure~\ref{fig:E_2pla_1o2} will not look much different because a larger $j$ truncation has more states and many more off-diagonal elements contributing with more higher frequencies in the range $x>1$. Instead it is interesting to see that the traveling excitations in Figure~\ref{fig:Te_x<1} are still present, but the excitation propagation time is larger as can be seen in Figure~\ref{fig:to_jmax_2}. This is explained by noticing that a larger $j$ truncation has just more states, and in the range $x<1$ the off-diagonal elements are negligible, leaving all the larger number of chromoelectric eigenstates to contribute to the time evolution of a single state.

The plot in  Figure~\ref{fig:to_jmax_2} clearly shows the interesting physical phenomena of a travelling excitation is not an artefact due to a small truncation but instead it is clear distinct phenomena of the SU(2) pure gauge lattice theory considered.\\
A similar phenomena was found in a study conducted using $U(1)$ lattice gauge theory by the authors of Ref.~\citep{Lewis:2019wfx}.

\section{Toward the use of quantum computing}
To conclude this chapter we look forward to using quantum computing to reproduce the study presented in the last section by asking:\\
\\
Can a quantum computer be used to extract the energy spectrum and the time evolution of a SU(2) lattice gauge theory?\\
\\
We will answer this questions in the following two chapters by encoding the Hamiltonian representation and the time evolution operator on the two chosen quantum hardwares: the D-Wave quantum annealer \citep{D-Wave} and the IBM superconducting gate-based quantum hardware \citep{IBM}.\\
\\
Given the limitations of the noisy and small scale quantum hardware, two main simplifications will be adopted: first we will consider a minimal lattice made of a single row of plaquettes, limiting the study of the energy spectrum to lattices made from 2-plaquettes to 6-plaquettes and considering up to a $j=3/2$ truncation on D-Wave, leaving the larger values of $j$ to future studies. While for the real-time evolution on D-Wave and IBM a 2-plaquette lattice with a $j=1/2$ truncation is used by studying the probability of transition between excited states.


\chapter{SU(2) on the D-Wave quantum annealer}\label{chap:Chapter_2}
In this chapter \footnote{The results presented in this chapter were previously published in \href{https://doi.org/10.1103/PhysRevD.104.034501}{Phys. Rev. D 104, 034501} done in collaboration with  S. A Rahman, R. Lewis and S. Powell.} we present the main features of the D-Wave quantum annealer and show how the hardware can be used to extract the energy spectrum of SU(2) pure gauge theory for the case of 2-plaquette lattices with $j_{\rm max}=1/2, 1, \rm and \, 3/2$ and, 4 and 6 plaquettes lattices with $j_{\rm max}=1/2$ using periodic boundary conditions. The results are, overall, in perfect agreement with the exact classical results without using any error mitigation techniques. Then to study the time evolution of the theory we show how to implement a method called Kitaev-Feynman clock state and use it to calculate the real-time oscillation between two excited states, obtaining results in agreement with the exact classical result.  Finally, we conclude the chapter by discussing the overall results and looking forward to possible future extensions.\\
\section{The D-Wave quantum annealer}
The company D-Wave Systems Inc. \citep{D-Wave} founded in 1999 is based in British Columbia, Canada. In the last 24 years of activity it has developed 5 versions of a quantum annealer, which is a special type of quantum computer that exploits quantum fluctuations in the system to find the solutions to a given problem by using the adiabatic theorem \citep{Kadowaki_1998, farhi2000quantum, aharonov2005adiabatic, nature10012}.\\
\\
The D-Wave quantum annealer (DW) hardware  used for the result presented in this chapter is now the second-last version the D-Wave Advantage, that has 5760 physical qubits arranged into the hardware using the Pegasus architecture, where each qubit is connected to 15 other qubits. It is important to emphasize that many physical qubits have to be coupled together to form a logical qubit, which is the fundamental entity capable of performing a computation.\\
\\
The hardware is designed to find the ground state of an Ising-like Hamiltonian, better known in the quantum computing field as quadratic unconstrained binary optimization (QUBO):
\begin{equation}\label{eq:ising}
H(q) = \sum_{i=1}^Nh_iq_i + \sum_{i=1}^N\sum_{j=i+1}^NJ_{ij}q_iq_j
\end{equation}
where  $q_i$ can be 0 or 1, while $h_i$ and $J_{ij}$ are real valued coefficients chosen by the user.\\
\\
Therefore any problem that can be directly mapped into QUBO can be submitted to the hardware in matrix form, and the DW's Ocean software \citep{Ocean_2023} has the role of embedding the problems on the actual qubits present on the hardware by creating a graph that represents the problem as a connection between qubits.
The solution to a given problem is given in terms of a vector containing the state of the qubits and the energy associated with the qubits configuration.\\
\\
From the operative point of view, the DW hardware is able to solve the problem of finding the ground state by performing an adiabatic evolution starting from an initial Hamiltonian already present on the hardware to the one representing the user's problem. Whenever a problem is submitted, the DW hardware works on the following transverse-field Ising Hamiltonian:
\begin{equation}\label{eq:ising}
H(q,s) = A(s)\left[ \sum_{i=1}^N q_i^x\right] + B(s)\left[ \sum_{i=1}^Nh_i q_i^z  + \sum_{i=1}^N\sum_{j=i+1}^NJ_{ij}q_i^z q_j^z\right]
\end{equation}
where the first term multiplied by $A(s)$ is the initial prepared Hamiltonian with qubits aligned along $x$ due to the transverse field, so that initially all the qubits are arranged into an equal superposition with respect to the $z$ basis. This initial configuration is crucial to make every possible solution to the user's problem equally reachable. The second term multiplied by $B(s)$ is the problem submitted by the user with qubits aligned along $z$.
To find the ground state of the user's problem, the two annealing functions $A(s)$ and $B(s)$ are defined such that during the annealing time interval $t_a$, with the growing of the time parameter $s=t/t_a$, $A(s)$ slowly decreases from 1 to 0, while $B(s)$ slowly increases from 0 to 1, allowing the system to transition from the ground state of the initial Hamiltonian to the one from the user's problem \citep{MCGEOCH2020169}. The correctness of this process is guaranteed by adiabatic theorem, which under the general hypothesis assures that if the transformation from the initial to the final Hamiltonian is slow enough, the initial ground state will evolve into the ground state of the final Hamiltonian \citep{Kato_Tosio_1950, messiah2014quantum,PhysRevA.65.042308}. The annealing time can also be tuned starting from the default value of 20 microseconds.\\
\\
Since the connectivity of a qubit with the others is limited by the material connections present on the hardware's structure, a problem that requires a large qubit connectivity is embedded by connecting more qubits together using a chain term that forces those qubits to be in the same state by activating an extra Hamiltonian interaction of the form $\delta H = -J_{\rm chain}\sigma^z_j\sigma^z_k$ among those qubits. Those qubits connected by a chain are forming a logic qubit now having a larger connectivity to suit the needs of a specific problem. DW's feature of increasing the qubit's connectivity on a case-by-case basis is a powerful feature because the system has the flexibility to choose from a few qubits being connected to all-to-all qubits connectivity.\\
The down side of this feature is that the chain strength value $J_{\rm chain}$  has to be set by the user in a rational way. In fact, if the chain strength value is too small relative to the terms in the problem Hamiltonian, the system cannot hold the qubits forming a logic qubit to be aligned to the same state, causing a breaking of the chain and the incorrect encoding of the problem, while instead when it is too large, the chain term Hamiltonian changes the physics of the problem.\\ Therefore the user has to submit a few runs changing the value of the chain to be sure that the system is solving the correct problem, and in case of a chain breaking event repeat the run with a larger value of the chain, or accept the correction to the problem with chain breaking provided by the Ocean software.\\ 
\\
The DW applicability can appear to be rather limited but in reality there is a long list of problems that can be reformulated as an Ising Hamiltonian like many NP-complete and NP-Hard problems such as partitioning problems like number or graph partitioning, binary integer linear programming problem,covering and packing problems, optimization problem defined by inequalities like the Knapsack problem, or the well known traveling salesman problem and many more as shown by the author of \citep{Lucas_2014}, but not all of these problems have already been fully explored on DW.\\
\\
In the event that a problem or a physical system is representable only by a Hamiltonian that cannot be directly mapped into a Ising like Hamiltonian, the authors of Ref.~\citep{Teplukhin_2019, Teplukhin_2020, D0CP04272B, teplukhin2021sampling, Teplukhin2021} working in the field of computational chemistry developed an algorithm called Quantum Annealer Eigensolver (QAE) that maps the ground state variational problem onto an Ising problem, resulting in the possibility to use the DW hardware to find the ground state of any Hamiltonian by using the following functional:  
\begin{equation}\label{eq:F}
F \equiv \left<\psi\right|H\left|\psi\right> - \lambda\left<\psi|\psi\right>
\end{equation}
where $H$ is the Hamiltonian of the system under investigation and the parameter $\lambda$ is an extra Lagrange multiplier that has to be tuned by the user to avoid the trivial null solution $\left|\psi\right>=\left|0\right>$ and to find the minimum possible value for the eigenvalue and the corresponding eigenvector.\\
A similar approach can be used to extract the excited states by introducing, extra Lagrange multipliers to the function above to exclude the lower states from the state solution.\\
\\
Furthermore, to explore the possibility of having a solution vector with real number entries instead of the default 0 and 1, the same researchers proposed to use a number of qubits $(K)$ to construct a fixed-point representation such that each entry in the solution vector is:
\begin{equation}
a_i = -q_K^i + \sum_{k=1}^{K-1}\frac{q_k^i}{2^{K-k}} \,
\end{equation}
where the possible values of each vector entry are spaced within $[-1,1)$ with $K$ as:
\begin{equation}\label{eq:K_values}
\begin{array}{c}
K=1 \\[3pt]

K=2 \\[3pt]

K=4 \\[3pt]
\ldots
\end{array}
\quad
\begin{array}{cccccccccccccccc}
& & & & & & & -1 & 0\\[3pt]
& & & & & & -1 & -\frac{1}{2} & 0 & \frac{1}{2} \\[3pt]
-1 & -\frac{7}{8} & -\frac{3}{4} & -\frac{5}{8} & -\frac{1}{2} & -\frac{3}{8} & -\frac{1}{4} & -\frac{1}{8} & 0 & \frac{1}{8} & \frac{1}{4} & \frac{3}{8} & \frac{1}{2} & \frac{5}{8} & \frac{3}{4} & \frac{7}{8}\\[3pt]
& & & & & & & & \ldots
\end{array}
\end{equation}
meaning that for $K=1$ there are only two possible values -1 or 0, for $K=2$ there are four values, and so on.\\
This approach extends the usability of DW's hardware to work with real numbers with a set precision tunable by the user, at the cost of using extra $K$ qubits. This implies that a problem represented by an $N \times N$ matrix requires a number of logic qubits equal to $N K$.\\
In the next section we will see the details of how to use the QAE algorithm but, in short, the user has to submit the matrix representing the problem, then choose the number of qubits $K$ to set the precision, and then in each run one has to tune the value of $\lambda$ to find the ground state of the system and, in case of a chain break, tune the chain strength.\\
It is important to notice that the precision of the calculation cannot be improved by simply increasing $K$, because its value is limited by the total available logical qubits and, most practically, by the fact that the more qubits are used the more chains are needed to represent the problem on the hardware and this effectively increases the chances of chain break making it difficult or impossible to perform a calculation.\\
This limitation can be extended by reducing the size of the matrix by performing a block diagonalization and by using an algorithm that improves the precision without increasing the number of qubits $k$ like the one we proposed in \citep{ARahman:2021ktn} the adaptive QAE (AQAE), both tools are discussed in details in the next section.\\
\\
Despite these challenges, over the last few years many studies from various fields have been conducted using the DW hardware without the need of using extra tools. The most common application is for optimization problems \citep{Boixo_2014, neukart2017traffic, birdal2021quantum, Inoue_2021, Irie_2021}, while other applications can be seen in life sciences \citep{perdomoortiz2012finding, Nazareth_Spaans, Li_2018, Gonzalez_Calaza_2021} and in finance for portfolio optimization in stock market
\citep{Rosenberg_2016, cohen2020picking, phillipson2020portfolio, Grant_2021}. At the publishing of this paper there are more than 250 results listed on an updated list of featured applications of DW hardware \citep{Dwave_list_art}.
\\
Regarding the field of particle physics, the first use of DW was done to solve a Higgs-signal-versus-background optimization problem in \citep{nature24047}, followed by a study of vertex reconstruction of hadronic interaction in \citep{Das:2019hrw} without using any extra tools. We contribute to the literature with the first use of DW for studying lattice gauge theory with the study of SU(2) energy spectrum and time evolution in \citep{ARahman:2021ktn}, which was followed by a study of SU(3) time evolution and dynamics of neutrino entanglement in \citep{Illa:2022jqb}, a study of the hadronic spectrum of 1+1 quantum chromodynamics in \citep{Farrell:2022wyt}, and the study of LGT on dihedral group in \citep{Fromm:2022vaj}. All of these studies used the QAE algorithm,  instead a recent paper \citep{Kim:2023sie} studied a U(N) LGT on 2-dimensional lattice imposing geometric constrains  to extract the system equilibrium configurations effectively performing an importance sampling.

\section{SU(2) energy spectrum on D-Wave}
\begin{figure}[H]
\centering
\includegraphics[width=0.6\textwidth]{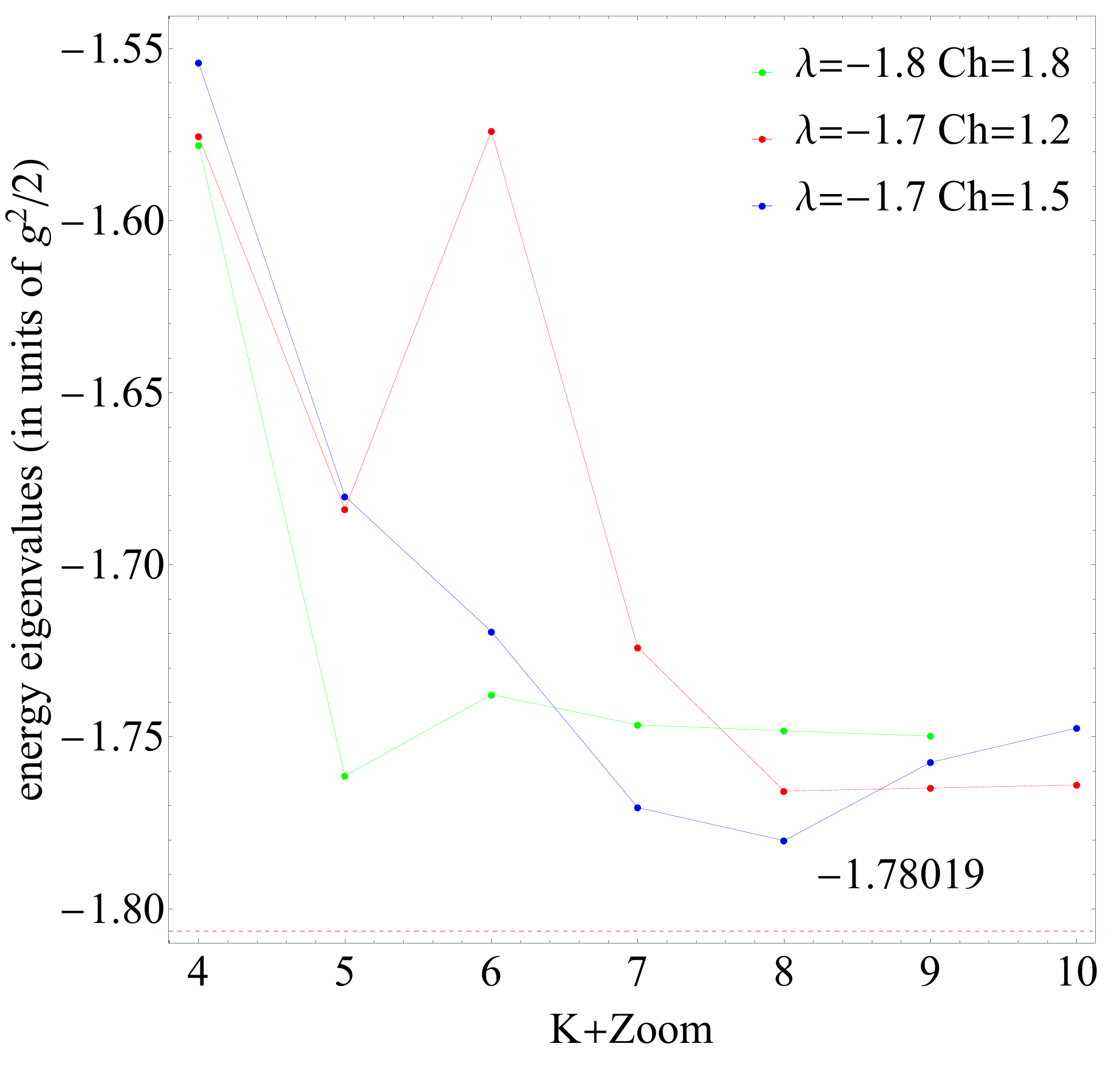}
\caption{Ground state energy as a function of $K+\rm Zoom$ for the case of 2-plaquettes and $j_{\rm max}=1$ at $x=0.9$ for different values of $K$ qubits used to represent the real number. Three datasets for different value of $\lambda$, chain strength $(\rm Ch)$ and $10^4$ reads obtained using the AQAE algorithm on DW hardware Advantage are shown, while exact eigenvalue is indicated by the dashed red line. The best obtained value is the lowest blue point indicated by its energy value $-1.78019$.  
}\label{fig:Fig_x_0.9__2_pla_jmax_1_chain}
\end{figure}

\begin{figure}[H]
\centering
\includegraphics[width=0.7\textwidth]{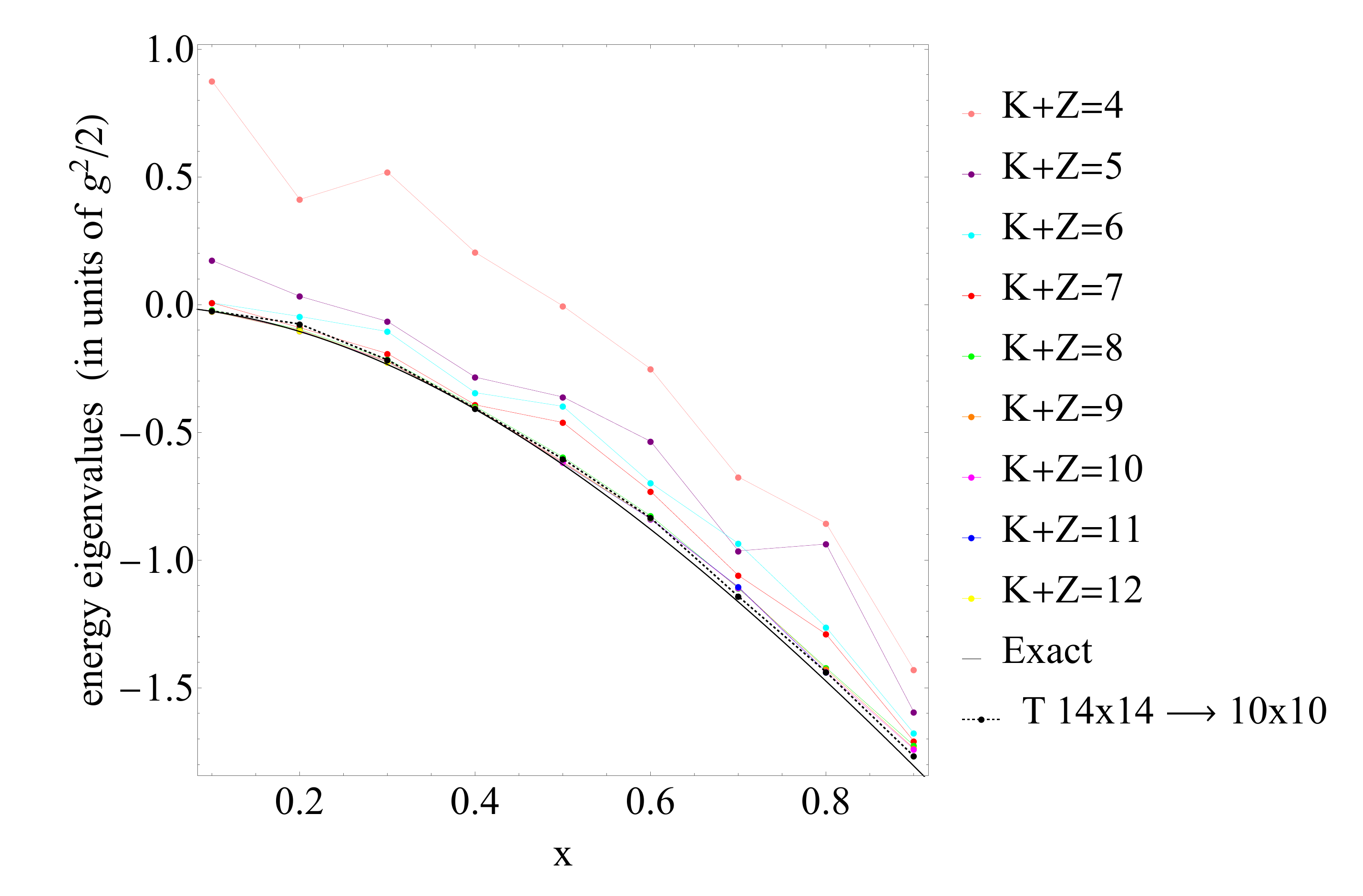}
\caption{Ground state energy versus gauge coupling $x$ for the case of 2-plaquettes with $j_{\rm max}=1$. The colored points with line as a guides for the eye are obtained using AQAE algorithm on the DW hardware Advantage from $10^4$ reads for different precision iteration $(K+Z)$ starting from $K=4$ and $Z=0$ then increasing $Z$ by 1 in each subsequent step. The black points joined by a dotted line are data obtained using AQAE on DW Advantage with $K=4$, $Z=3$ and $10^4$ reads after having truncated the matrix from $14 \times 14$ to $ 10 \times 10$.} \label{fig:Fig_2_pla_jmax_1_AQAE}
\end{figure}

The SU(2) pure gauge Hamiltonian considered is the one presented in chapter~\ref{chap:Chapter_1}:
\begin{equation}\label{eq:H_dw}
\hat H = \frac{g^2}{2}\left(\sum_{i={\rm links}}\hat E_i^2-2x\sum_{i={\rm plaquettes}}\hat\square_i\right)
\end{equation}
Here we consider the theory with an even number of plaquettes 2, 4 and 6 and periodic boundary conditions (p.b.c) along the longitudinal direction. The five separate physical systems considered are 2 plaquettes with $j_{\rm max}=1/2, 1$ and $3/2$ and, 4 and 6 plaquettes with $j_{\rm max}=1/2$.\\
The Hamiltonian representation for these systems can be obtained by using the method illustrated in Appendix~\ref{appendix:state_algorithm}, here for a question of space only the Hamiltonian for the case of 2 plaquettes with $j_{\rm max}=1/2$ is shown:
\begin{equation}\label{eq:2pla_periodic}
H = 
\frac{g^2}{2}\left(
\begin{array}{cccc}
0 & -2x & -2x & 0 \\
-2x & 3 & 0 & -\frac{x}{2} \\
-2x & 0 & 3 & -\frac{x}{2} \\
0 & -\frac{x}{2} & -\frac{x}{2} & 3
\end{array}
\right)
\begin{array}{c}
\left|1_1^11_1^1\right> \\
\left|2_2^22_1^1\right> \\
\left|2_1^12_2^2\right> \\
\left|1_2^21_2^2\right>
\end{array} \,
\end{equation}
where the states on the right of the Hamiltonian are shown in Fig.~\ref{fig:four_states_pbc}:
\begin{figure}[H]
\centering
\includegraphics[width=0.7\linewidth]{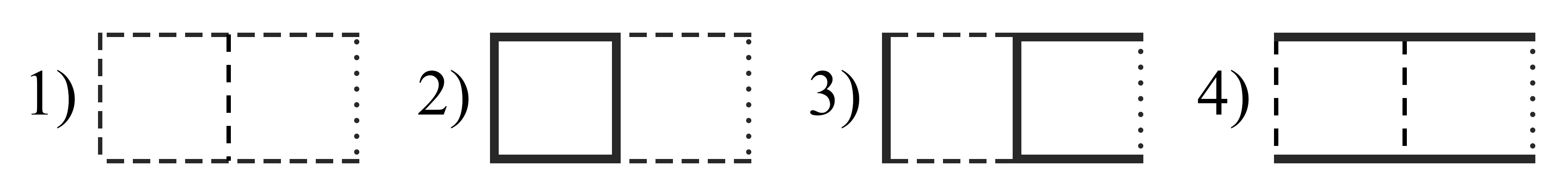}
\caption{The four possible basis states available on the 2-plaquette lattice with periodic boundary conditions in the longitudinal direction after the $j_{\rm max}=1/2$ truncation. Each solid line means a $j=1/2$ flux of energy is present, a dashed line a $j=0$ meaning no energy is present, while the dotted line on the right indicates the lattice side that is periodic.}
\label{fig:four_states_pbc}
\end{figure}
As we discussed in the previous Section~\ref{quantum_explore}, in the NISQ era rationalizing the use of quantum resources by limiting the number of qubits is crucial in order to reduce noise. In this direction a possibility comes from block diagonalization of Hamiltonian using the symmetries of the system, so that it can be divided into smaller blocks and each one of them can be studied separately in the quantum hardware, reducing the number of qubits needed and effectively improving the performances of the hardware.\\
\\
To make it tangible, let's focus on the 2 plaquette Hamiltonian, where we can observe that the two states $\left|2_2^2 2_1^1\right>$ and $\left|2_1^1 2_2^2\right>$ can be interchanged leaving the Hamiltonian invariant, therefore we may substitute these states with linear combinations $\frac{1}{\sqrt{2}}(\left|2_2^2 2_1^1\right> \pm \left|2_1^1 2_2^2\right>)$ that consequently  has the effect of substituting their corresponding rows and columns with new ones obtained by adding and subtracting the original one.\\
Performing this operation on the 2-plaquette Hamiltonian divides it into two separate blocks a $3 \times 3$ and a $1\times1$:
\begin{equation}\label{eq:3x3}
H = 
\frac{g^2}{2}\!\!\left(
\begin{array}{ccc|c}
0 & \!\!\!-2\sqrt{2}x\!\!\! & 0 & 0 \\
\!\!\!-2\sqrt{2}x\!\!\! & 3 & -\frac{x}{\sqrt{2}} & 0 \\
0 & -\frac{x}{\sqrt{2}} & 3 & 0 \\
\hline
0 & 0 & 0 & 3
\end{array}
\right)
\begin{array}{c}
\left|1_1^11_1^1\right> \\
\tfrac{1}{\sqrt{2}}\left(\left|2_2^22_1^1\right>+\left|2_1^12_2^2\right>\right) \\
\left|1_2^21_2^2\right> \\
\tfrac{1}{\sqrt{2}}\left(\left|2_2^22_1^1\right>-\left|2_1^12_2^2\right>\right)
\end{array}
\end{equation}
This approach can be generalized and applied to any Hamiltonian matrix of any dimension by exploiting the symmetries present on the basis states like top-to-bottom reflection,
left-to-right reflection and, spatial translation in the longitudinal direction. Therefore once all the physics states allowed for a particular system are found, those two or multiple states that are symmetric under one of the possible symmetries, are replaced by symmetric and antisymmetric combinations.\\
The result of block diagonalizing for the five physical systems considered, 2 plaquettes with $j_{\rm max}=1/2, 1$ and $3/2$ and 4 and 6 plaquettes with $j_{\rm max}=1/2$, is summarized in Table~\ref{Tab:systems_sizes}, where for each physics system it is shown the block decomposition and the groundstate block size.
\begin{table}[H]
\centering
\begin{tabular}{ccccc}
\hline
\hline
$N_{\rm plaq}$ & $j_{\rm max}$ & Size of $H$ & blocs decomposition & vacuum block  \\
\hline
2 & 1/2 &  4$\times$4  & $(3 \times3)$ $(1 \times1)$ 	&  $(3 \times3)$  \\
4 & 1/2 & 16$\times$16 & $(6 \times 6)$ $3~(3 \times 3)$ $(1 \times 1)$	& $(6 \times 6)$   \\
6 & 1/2 & 64$\times$64 & $(13 \times13)$ $2~(11 \times 11)$ $2~(9 \times9)$ $(7 \times7)$ $(3 \times3)$ $(1 \times1)$	& $(13 \times 13)$ \\
2 &  1  & 27$\times$27 & $(14 \times14)$ $(5 \times 5)$ $2~(3 \times 3)$ $2~(1 \times 1)$	&  $(14 \times 14)$ \\
2 & 3/2 & 95$\times$95 & $(36 \times 36)$ $(20 \times 20)$ $(14 \times 14)$ $(12 \times 12)$ $(7 \times 7)$ $(2 \times2)$ 	& $(36\times 36)$ \\
\hline
\end{tabular}
\caption{ Results of block diagonalization for the five physics systems considered. Each system is labelled by the lattice size expressed in plaquettes number, $N_{\rm plaq}$ and, the gauge truncation $j_{\rm max}$. In the third column the full Hamiltonian block decomposition is displayed and in the last column the groundstate block size.}
\label{Tab:systems_sizes}
\end{table}
Now that we have decomposed the Hamiltonian representation in blocks for each system  we are interested in, we can discuss in detail how to use the QAE algorithm on the DW hardware.\\
As we already discussed, the QAE algorithm implements the variational method using the following functional:
\begin{equation*}
F \equiv \left<\psi\right|H\left|\psi\right> - \lambda\left<\psi|\psi\right>
\end{equation*}
where $\lambda$ has to be tuned by the user to find the minimum eigenvalue and eigenvector.\\
To make the procedure more clear let's show how to find the ground state for a fixed value of the gauge coupling $x$ for the 2-plaquette lattice with $j_{\rm max}=1$, which it is contained in the $14 \times 14$ block.\\
\\
Given this block the matrix representation of the functional $F$ has to be entered using a python code working within the DW Ocean environment.\\
In all the results presented in this chapter the default annealing time of 20 microseconds was used. It is possible to use a longer time but it increases the chances of depolarization of the qubit causing the chain to break, while a shorter time may not be enough to allow the system to transit to the final system.\\
The procedure is used to extract the ground state and is made by two consecutive steps:\\
\\
The first step consists of running the code many times on the DW quantum simulator to scan the value of $\lambda$ to find the optimal value for which DW returns the minimum possible energy. Those runs are done at a fixed value of $K$ qubits used to represent each real entry in the matrix. Those runs are repeated a few times by increasing $K$ only to check that there is no significantly smaller value for the energy of the system. The results of this first step are shown in Figure~\ref{fig:Fig_x_0.9__2_pla_jmax_1} for the case of 2-plaquettes with $j_{\rm max}=1$ and $x=0.9$:
\begin{figure}[H]
\centering
\includegraphics[width=0.6\textwidth]{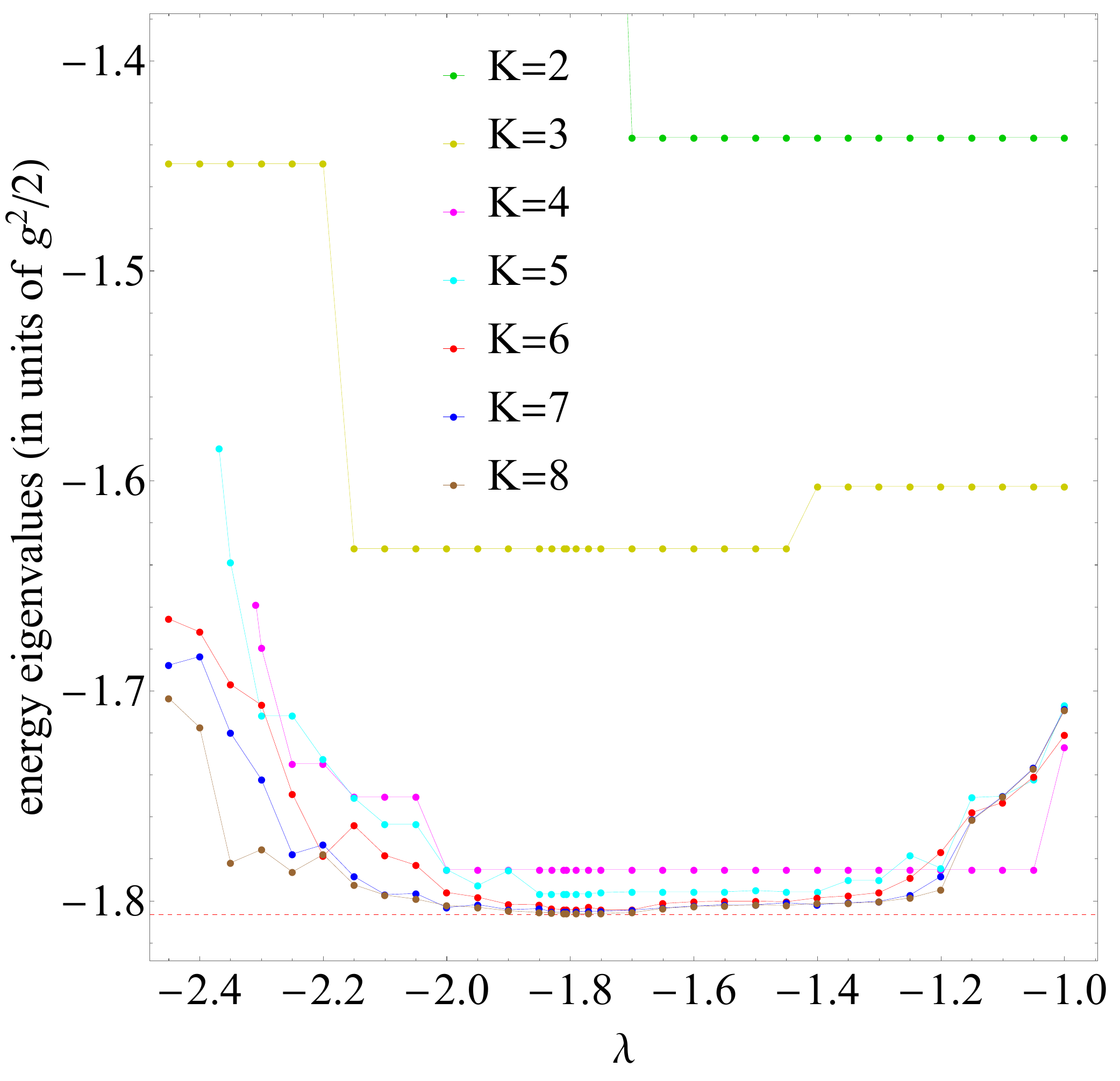}
\caption{Ground state energy versus the penalty parameter $\lambda$ for the case of 2-plaquettes and $j_{\rm max}=1$ at $x=0.9$ for different values of $K$ qubits used to represent the real number and $10^4$ reads. The data points are obtained using the QAE algorithm on a DW classical simulated annealer, while exact energy is indicated by the dashed red line.  
} \label{fig:Fig_x_0.9__2_pla_jmax_1}
\end{figure}
In Figure~\ref{fig:Fig_x_0.9__2_pla_jmax_1} it is evident that for $K\leq 5$ is not enough to approach the exact results. For $K=6$ the result is close to the exact eigenvalue, while for $K=8$ and $7$ a broad range of $\lambda$ values reproduce the exact eigenvalue. From this plot the range of optimum lambda ($\lambda_{opt}$) values that correspond to the minimum energy of the system is extracted.\\
\\
The second step is performed by running the same code using a DW quantum hardware, which requires to set the chain strength. A value close to $\lambda_{\rm opt}$ is used for $\lambda$ and the best value from the first step is used for $K$. The second step is repeated few times by reducing the chain strength value until no lowering of the system energy is observed and no chain breaks happened. If the precision obtained on the DW hardware is not satisfactory, one can increase $K$ and repeat the second step as shown in Figure~\ref{fig:Fig_x_0.9__2_pla_jmax_1_chain} where the best estimation of the groundstate energy is $-1.78019$.
There is one main observation, in our study we always used the DW simulator to find $K$ and $\lambda_{\rm opt}$ because we needed to maximize our one minute hardware time per month. However, in the case one has enough hardware time the calculations can be all performed on a quantum hardware, by executing the first step for a initially set value of the chain strength, once the best value for $K$ and $\lambda_{\rm opt}$ are found the chain strength can be reduced to find a possible lower energy solution.\\
\\
To extract precise results for a matrix block larger or equal to a $14 \times 14$ the QAE requires a large number of qubits $K$ to represent the real number limiting the precision achievable and the matrix size that can be studied on the DW hardware. With the goal of extending the QAE usability we developed an adaptive QAE (AQAE) algorithm in \citep{ARahman:2021ktn} which improves the precision without increasing the number of qubits used. The AQAE runs the QAE several times with the same value for $K$ and at the end of each run gives a user the possibility to find a better solution by re-running the code reusing the best eigenvector found in the previous run as a starting point for the new solution search. The new search is centred around the previous minimum with the entries in the new solution vector now taking one of the $2^K$ values around the previous entry value, therefore the precision in resolving the minimum is doubled and each step, effectively performing a zoom in around the minimum. Extra details and the code implementation can be found in the appendix of our published articles \citep{ARahman:2021ktn}.\\
The AQAE not only rationalizes the use of qubits and reduces the the possibility of chain breaks but creates the possibility of using DW for much larger matrices without reducing the qubits deployed for precision.\\
The effectiveness of the AQAE can be seen in Figure~\ref{fig:Fig_2_pla_jmax_1_AQAE} where the groundstate of 2-plaquettes with $j_{\rm max}=1$ contained in a $14 \times 14$ block is extracted using only $K=4$ precision qubits.
The plot clearly shows a good agreement with the exact result for $x\leq 0.4$ already using $K=4$ with 4 zoom step $(k+Z=8)$. While for $x\geq 0.4$ the data $k+Z\geq8$ are slightly above the exact curve but they are in good agreement with the data coming from the $10 \times 10$ truncation of the $(14 \times 14)$ block, therefore some systematic hardware error should be present.\\
The result can be improved if needed by choosing a $K$ value larger than 4 and rerunning the AQAE algorithm.\\
\\
In the next sections the QAE and the AQAE with the procedure just described are used to systematically extract the smallest eigenvalue and the corresponding eigenvector of each block present in the block decomposition for each of the five physical systems.
\subsection{Eigenvalues and eigenvectors using QAE and AQAE algorithm}
\begin{figure}[H]
\centering
\includegraphics[width=0.7\linewidth]{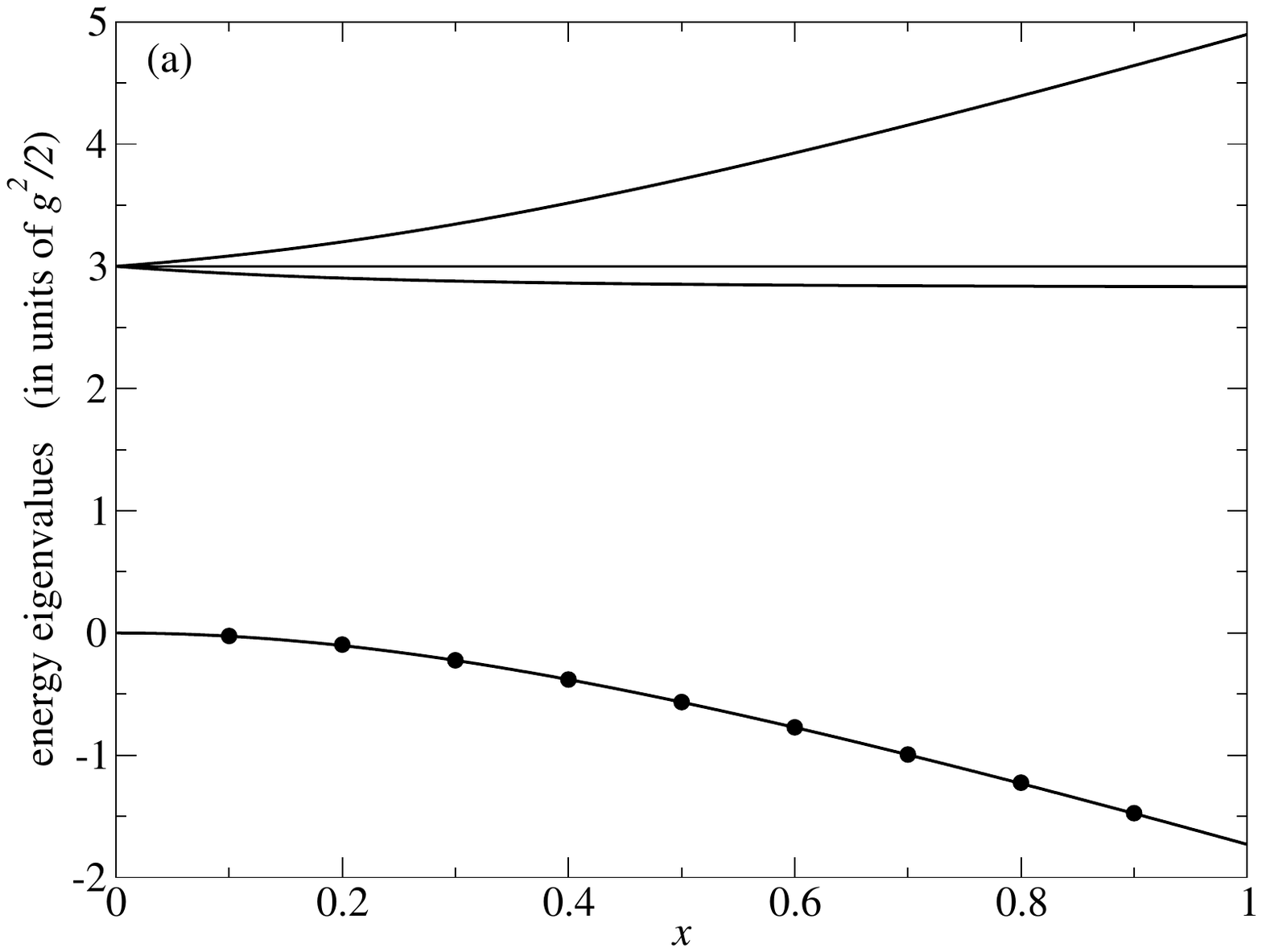}
\caption{Smallest energy eigenvalues for each block of a 2-plaquette lattice with $j_{\rm max}=1/2$ as a function of the gauge coupling $x$. The solid curves are the exact eigenvalues calculated by diagonalization on a classical computer. Data points are obtained using DW quantum hardware Advantage. Figure reproduced from Ref.~\citep{ARahman:2021ktn}.}
\label{fig:E_2pla_j_1o2_dw}
\end{figure}

\begin{figure}[H]
\centering
\includegraphics[width=0.7\linewidth]{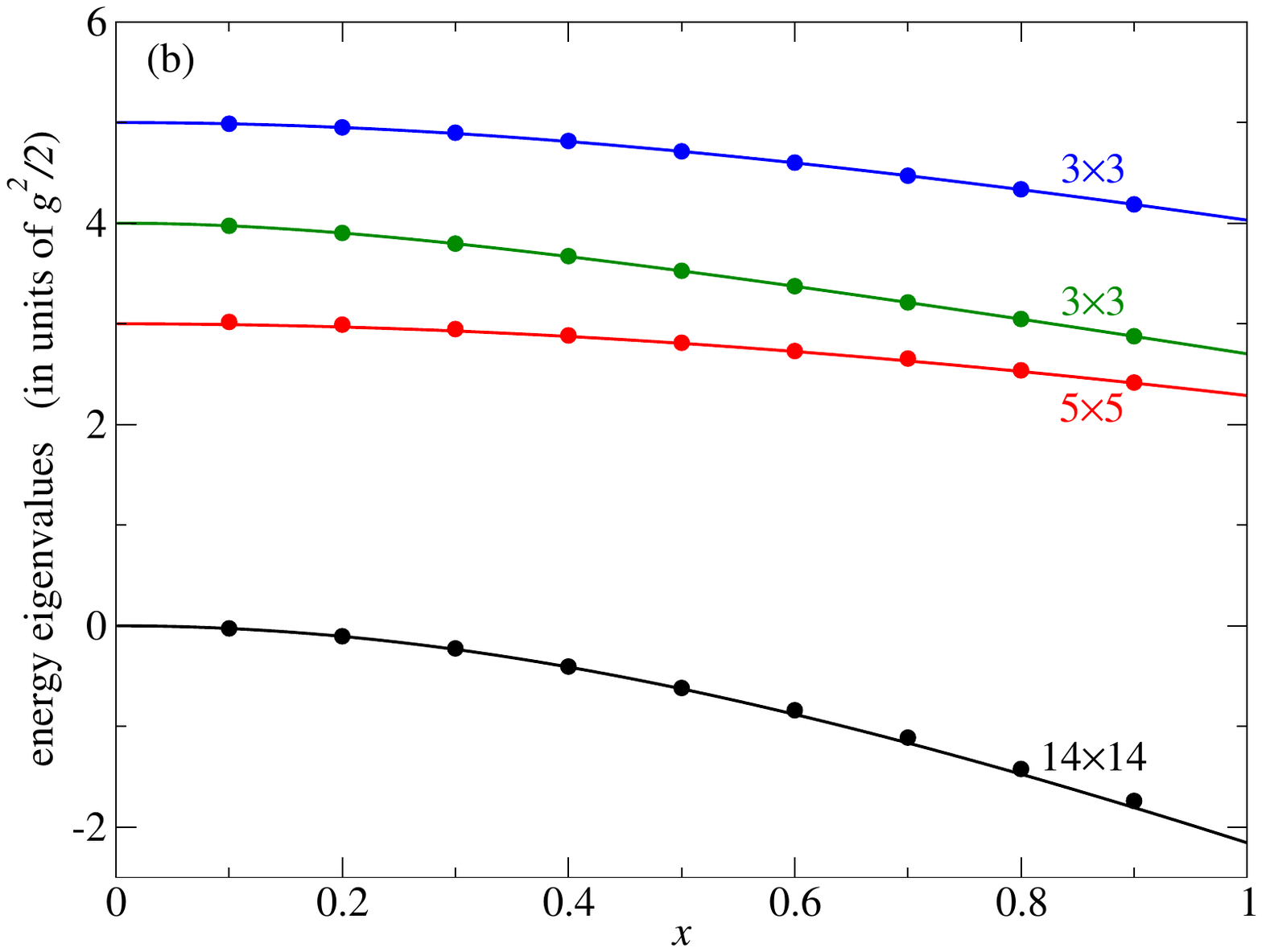}
\caption{Smallest energy eigenvalues for each block of a 2-plaquette lattice with $j_{\rm max}=1$ as a function of the gauge coupling $x$. The solid curves are the exact eigenvalues calculated by diagonalization on a classical computer. Data points are obtained using DW quantum hardware Advantage. Figure reproduced from Ref.~\citep{ARahman:2021ktn}.}
\label{fig:E_2pla_j_1_dw}
\end{figure}

\begin{figure}[H]
\centering
\includegraphics[width=0.7\linewidth]{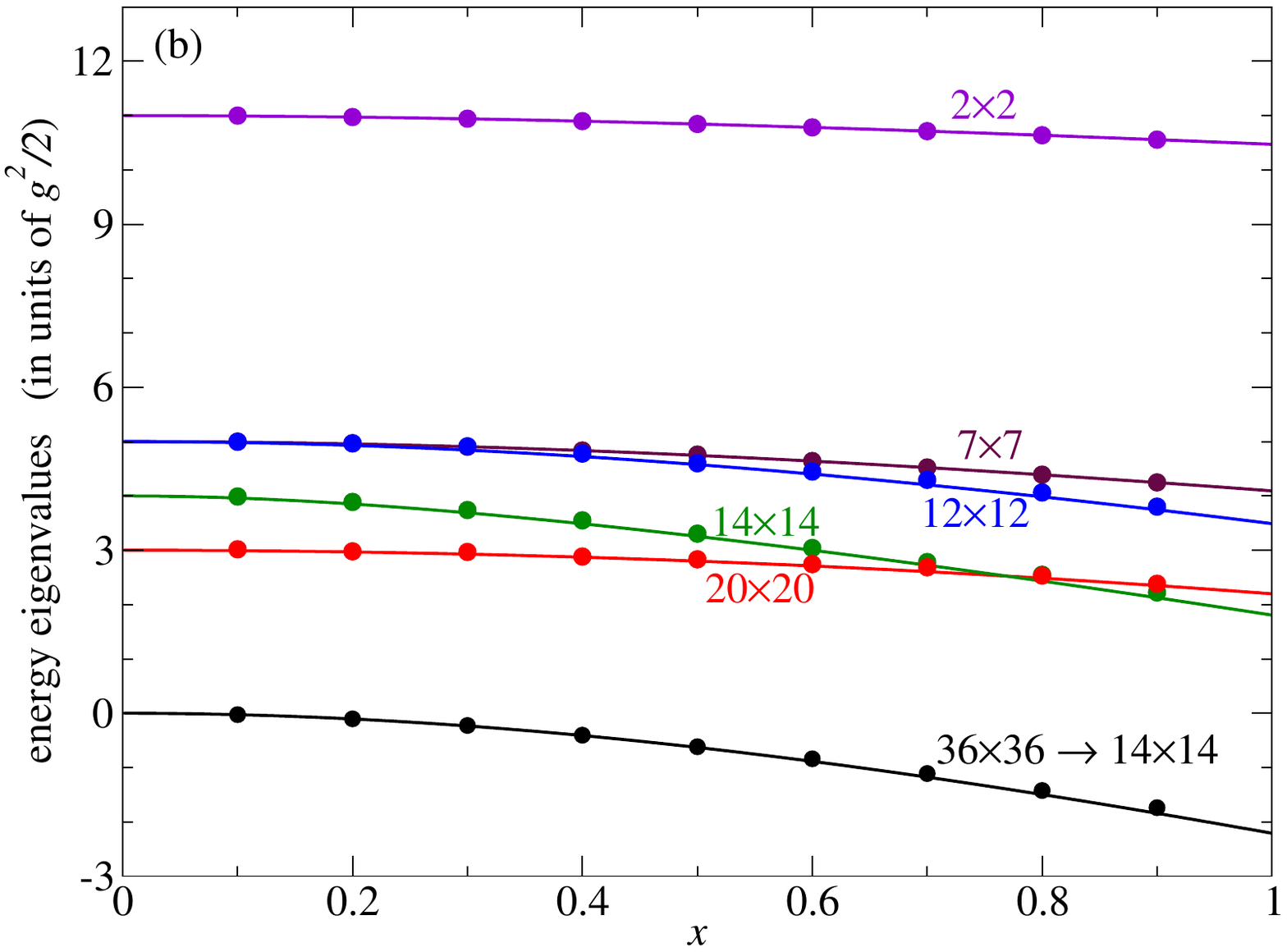}
\caption{Smallest energy eigenvalues for each block of a 2-plaquette lattice with $j_{\rm max}=3/2$ as a function of the gauge coupling $x$. The solid curves are the exact eigenvalues calculated by diagonalization on a classical computer. Data points are obtained using DW quantum hardware Advantage.  Figure reproduced from Ref.~\citep{ARahman:2021ktn}.}
\label{fig:E_2pla_j_3o2_dw}
\end{figure}

\begin{figure}[H]
\centering
\includegraphics[width=0.7\linewidth]{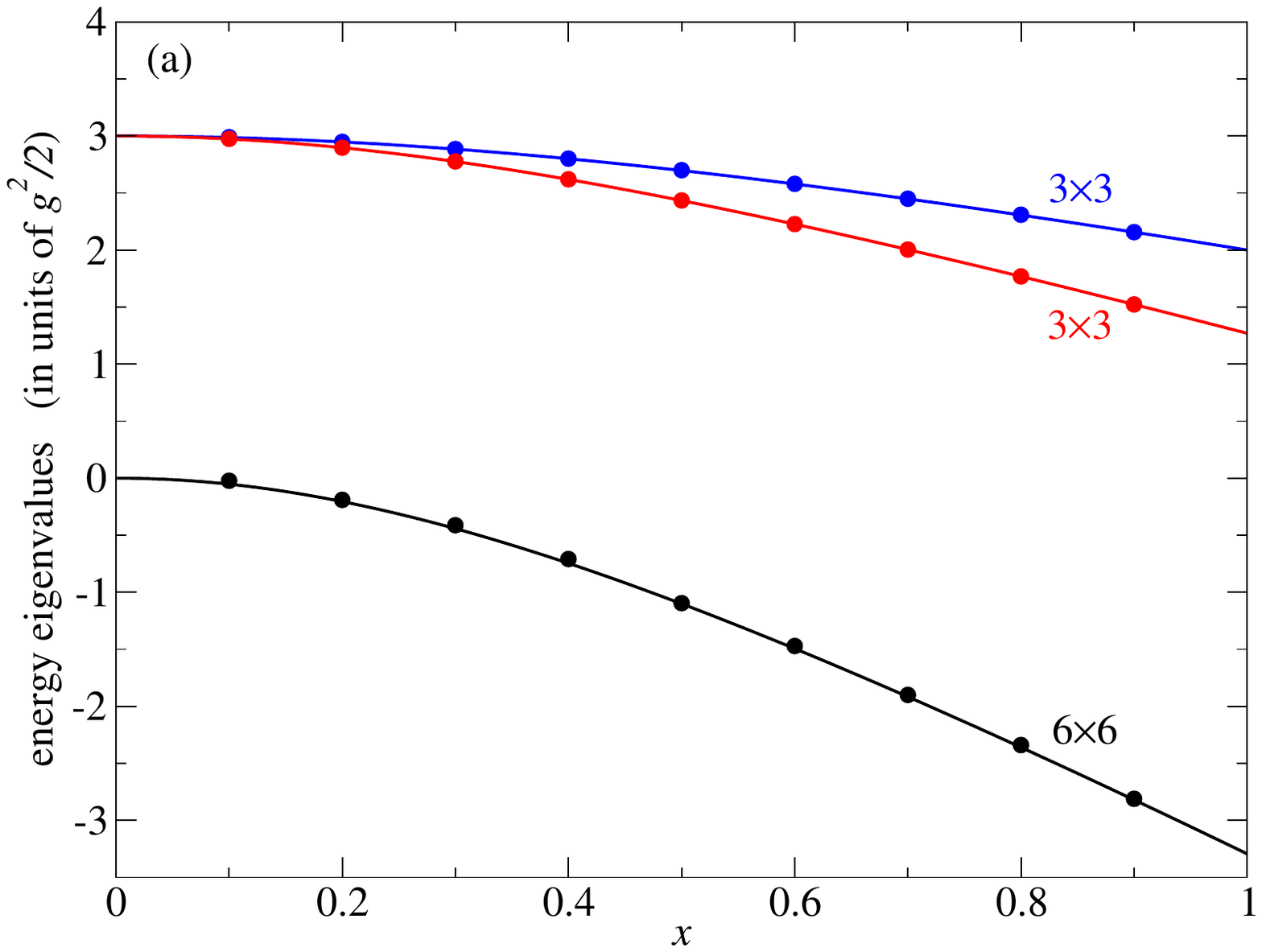}
\caption{Smallest energy eigenvalues for each block of a 4-plaquette lattice with $j_{\rm max}=1/2$ as a function of the gauge coupling $x$. The solid curves are the exact eigenvalues calculated by diagonalization on a classical computer. Data points are obtained using DW quantum hardware Advantage.  Figure reproduced from Ref.~\citep{ARahman:2021ktn}.}
\label{fig:E_4pla_j_1o2_dw}
\end{figure}

\begin{figure}[H]
\centering
\includegraphics[width=0.7\linewidth]{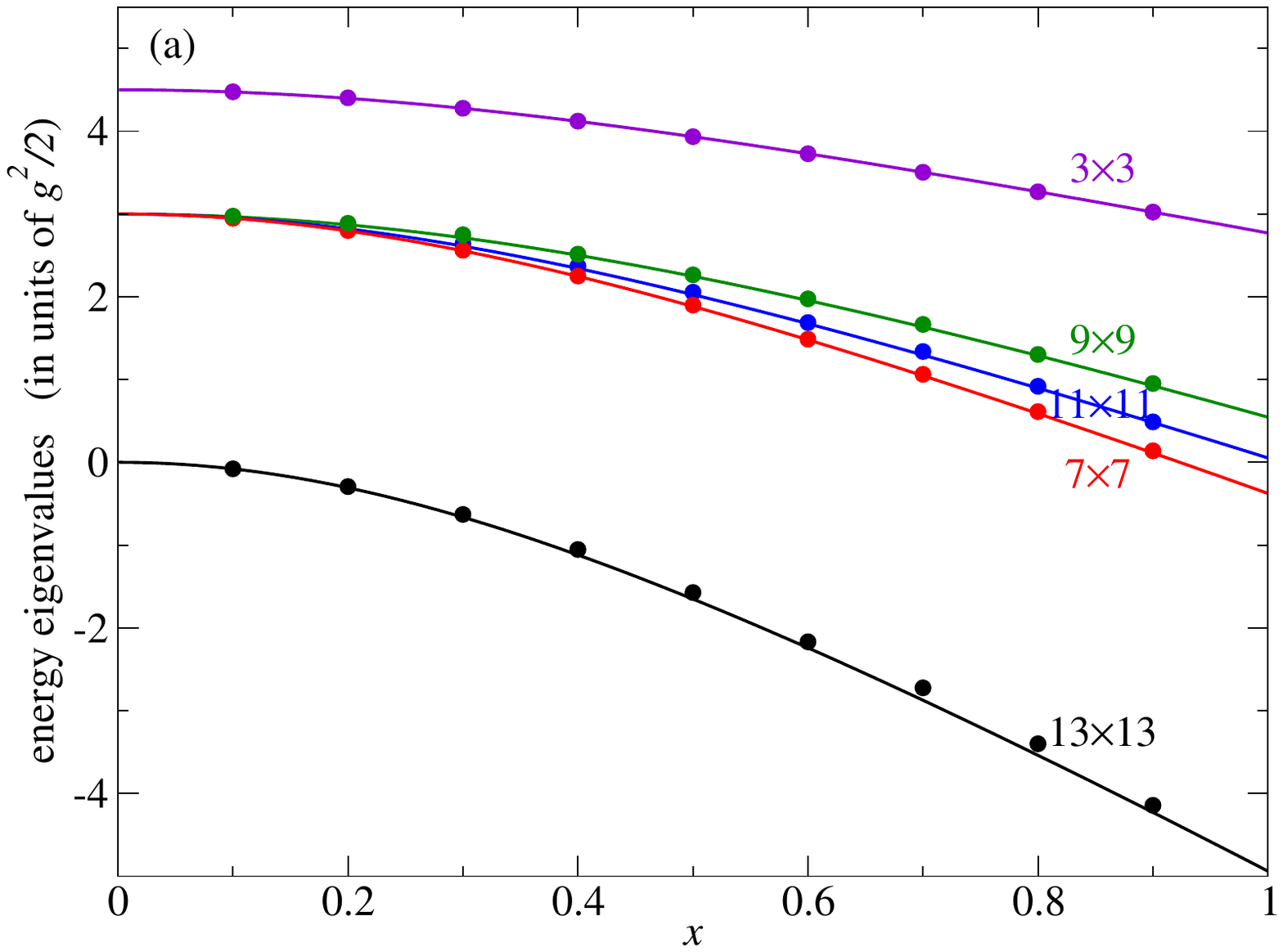}
\caption{Smallest energy eigenvalues for each block of a 6-plaquette lattice with $j_{\rm max}=1/2$ as a function of the gauge coupling $x$. The solid curves are the exact eigenvalues calculated by diagonalization on a classical computer. Data points are obtained using DW quantum hardware Advantage.  Figure reproduced from Ref.~\citep{ARahman:2021ktn}.}
\label{fig:E_6pla_j_1o2_dw}
\end{figure}

\begin{figure}[H]
\centering
\includegraphics[width=0.7\linewidth]{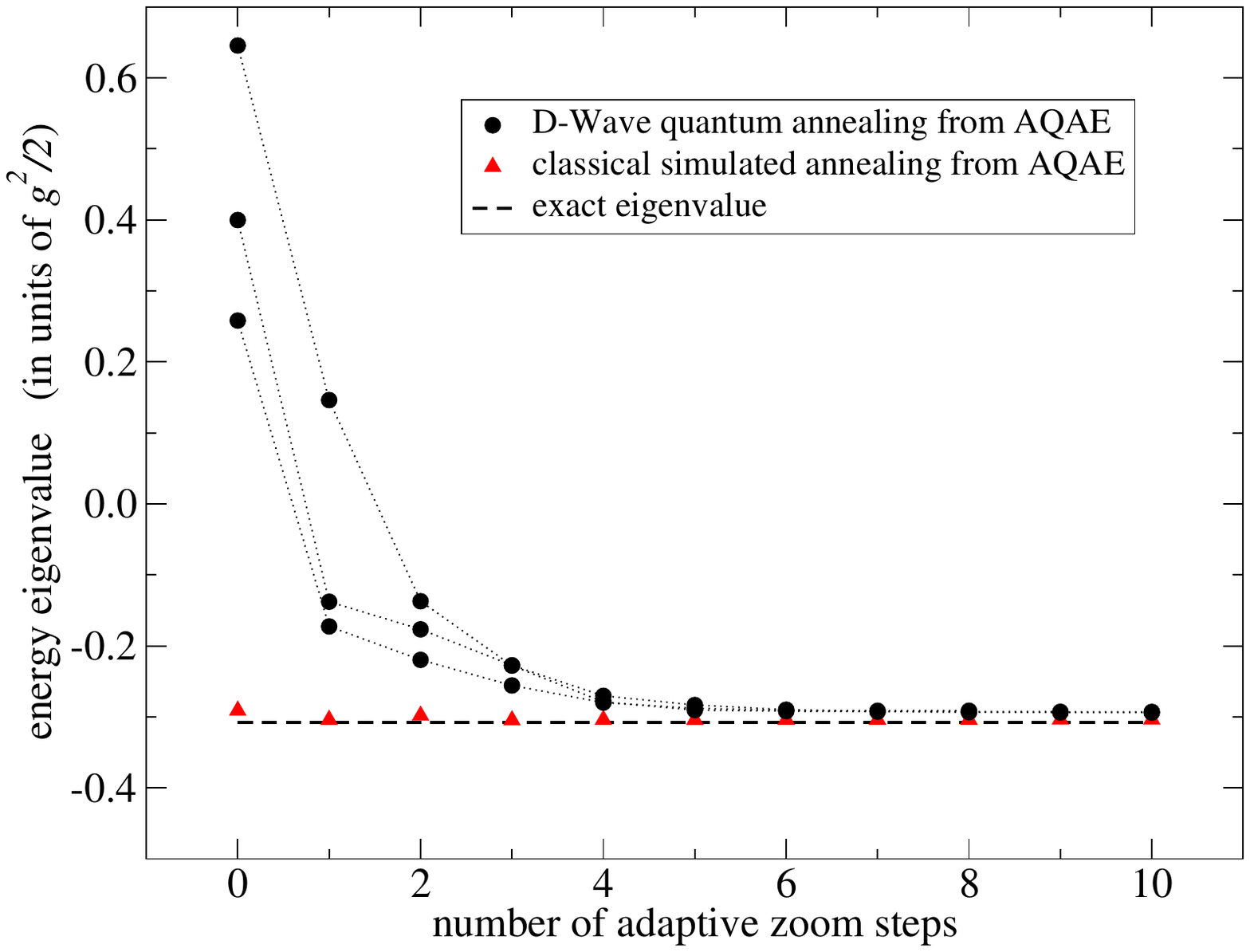}
\caption{AQAE efficiency to extract a precise result for the case of 6-plaquette lattice with $j_{\rm max}=1/2$ at $x=0.2$. Each set of black points joined by a dotted line is data from the same run with increasing zoom steps using the AQAE on the DW Advantage hardware. The red triangles are data obtained using the classical simulator using AQAE, while the dashed black line is the exact eigenvalue obtained on a classical computer.  Figure reproduced from Ref.~\citep{ARahman:2021ktn}.}
\label{fig:AQAE_6_pla}
\end{figure}

We begin this study by considering the simplest possible system made of two plaquettes with $j_{\rm max}=1/2$ whose Hamiltonian in Eq.~\ref{eq:3x3} which is made by a $3 \times 3$ block and a $1 \times 1$ block, with the result shown in Figure~\ref{fig:E_2pla_j_1o2_dw}.
The results in Figure~\ref{fig:E_2pla_j_1o2_dw} shows the four  eigenvalues of the 2-plaquette lattice with $j_{\rm max}=1/2$ as a function of the gauge coupling $x$. The data points displayed only for the ground state are coming from the QAE DW hardware Advantage using  $10^3$ reads, annealing time of 20 microseconds, and  $K=7$ qubits for precision for a total of 21 logical qubits. The results are in good agreement with the exact results displayed as a solid line and no visible discrepancy is observed with $x$. There are no data points for the other three eigenvalues because the QAE algorithm was set to extract only the smallest eigenvalue and eigenvector of each block.\\
\\
It is valuable to investigate how the theory changes when longer lattice and larger gauge field truncation $j_{\rm max}$ are considered, therefore extra simulations where done and the results for the eigenvalues obtained using the default annealing time of 20 microseconds are shown in Figures~\ref{fig:E_2pla_j_1_dw}, \ref{fig:E_2pla_j_3o2_dw}, \ref{fig:E_4pla_j_1o2_dw}, and \ref{fig:E_6pla_j_1o2_dw}.\\
\\
The first extension of this study is shown in Figure~\ref{fig:E_2pla_j_1_dw}  where the 4 lowest eigenvalues of the 2-plaquette lattice with $j_{\rm max}=1$ as a function of the gauge coupling $x$ are displayed. For the two $3 \times 3$ blocks and the $5 \times 5$ block the QAE with $10^3$ and $10^4$  reads respectively and $K=7$ for a total of 21 and 35 logical qubits respectively. For the $14 \times 14$ block the QAE was not able to find a reasonably precise result using $10^4$ reads, therefore the AQAE with $10^4$ reads, $K=4$ and between 4 to 9 zoom steps for a total of 56 logic qubits was used for each value of $x$. 
The results are in good agreement except for the $14 \times 14$ block, where the data points are slightly deviated from the exact curve for growing value of $x$ reaching a discrepancy of $4\%$ at $x=0.9$.
The second extension of this study is shown in Figure~\ref{fig:E_2pla_j_3o2_dw}  where the 6 lowest eigenvalues of the 2-plaquette lattice with $j_{\rm max}=3/2$ as a function of the gauge coupling $x$ are displayed. All the blocks, from the $2 \times 2$ to the $14 \times 14$ are obtained using the AQAE with $10^4$ reads, $K=4$ and between 4 to 9 zoom steps using a range of logical qubits from 8 for the $2 \times 2$ case to 56 for the $14 \times 14$ case. The data points of the lowest eigenvalues coming from the $36 \times 36$ block are not displayed because they are not resolved in a way to be better than the data coming from the $14 \times 14$ block, since $36 \times 36$ block contains the $14 \times 14$ block, the data of this later block are displayed instead.  
Considering the quantum resources available, the study can be further extended by fixing $j_{\rm max}=1/2$  to quantify how  the theory changes on a longer lattice, the results of this extension are showed in Figures~\ref{fig:E_4pla_j_1o2_dw} and \ref{fig:E_6pla_j_1o2_dw}.\\
\\
The first lattice extension is shown in Figure~\ref{fig:E_4pla_j_1o2_dw}  for the 4-plaquette lattice case. For the two $3 \times 3$ blocks $10^3$ and reads while for the $6 \times 6$ block $10^4$ reads were necessary while using the QAE. The number of precision qubits was $K=7$ for a total of 21 logical qubits for the $3 \times 3$ blocks and 42 for the $6 \times 6$.\\
The second lattice extension is shown in Figure~\ref{fig:E_6pla_j_1o2_dw}  for the 6-plaquette lattice case. All the blocks, from the $3 \times 3$ to the $13 \times 13$ are obtained using the AQAE with $10^4$ reads, $K=4$ and between 4 to 9 zoom steps using a range of logical qubits from 8 for the $2 \times 2$ case to 52 for the $13 \times 13$ case. The results are in good agreement for all the blocks except for the $13 \times 13$, where the data points are slightly deviating from the exact curve for growing value of $x$.
Over all the DW quantum hardware Advantage has shown good performances by producing accurate data points by implementing the QAE and, for precise result the AQAE which reduces the number of qubits needed and reduces the chances of chain breaks, allowing to obtain plots where statistical error bars are smaller than the data symbols. The AQAE effectiveness in reducing the statistical errors can be appreciated in an example of calculation shown in Figure ~\ref{fig:AQAE_6_pla},  where three different runs are obtained for the 6-plaquette with $j_{\rm max}=1/2$ at $x=0.2$ as a function of the zoom steps in the AQAE with $10^4$ reads and $K=4$. The black points joined by dotted lines are three separate datasets obtained on the DW Advantage hardware showing a rapid convergence between them and toward the exact result, represented by a black dashed curve, already after 4 zoom steps. The remaining discrepancy with the exact result is due to systematic hardware error.


\subsection{Investigating how the theory changes with a larger $j_{\rm max}$ truncation and lattice size.}
\begin{figure}[H]
\centering
\includegraphics[width=0.7\linewidth]{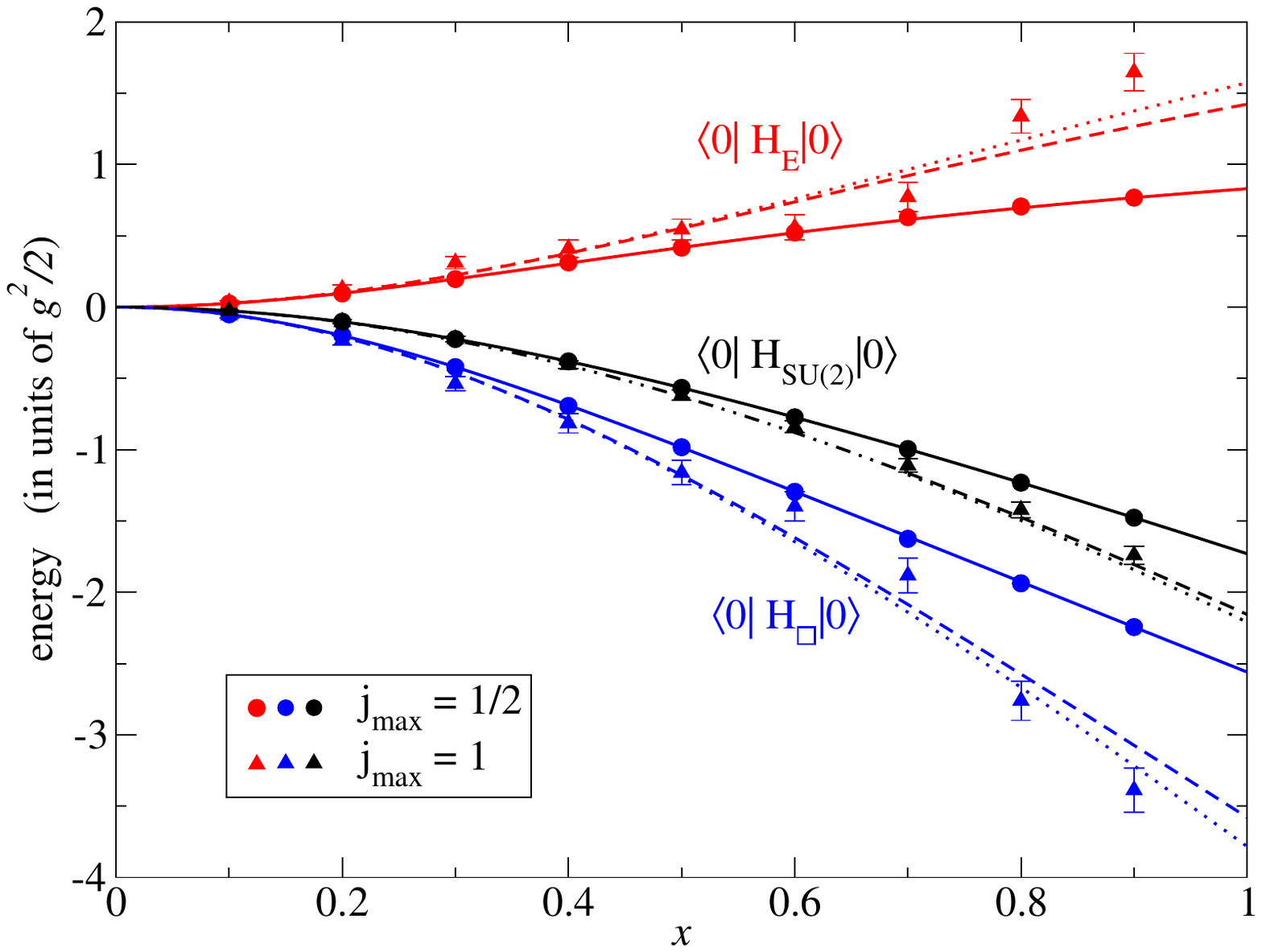}
\caption{Vacuum expectation value of the SU(2) Hamiltonian and its operators for 2-plaquette lattice with p.b.c. for different value of $j_{\rm max}$ as a function of the gauge coupling $x=2/g^4$. The solid, dashed, and dotted curves are the exact classical calculation for $j_{\rm max}=1/2$, $1$, and $3/2$ respectively. The circles and triangles are data point obtained from the Advantage DW quantum hardware for $j_{\rm max}=1/2$ and $1$ respectively.  Figure reproduced from Ref.~\citep{ARahman:2021ktn}.}
\label{fig:0_H_0_jmax}
\end{figure}

\begin{figure}[H]
\centering
\includegraphics[width=0.7\linewidth]{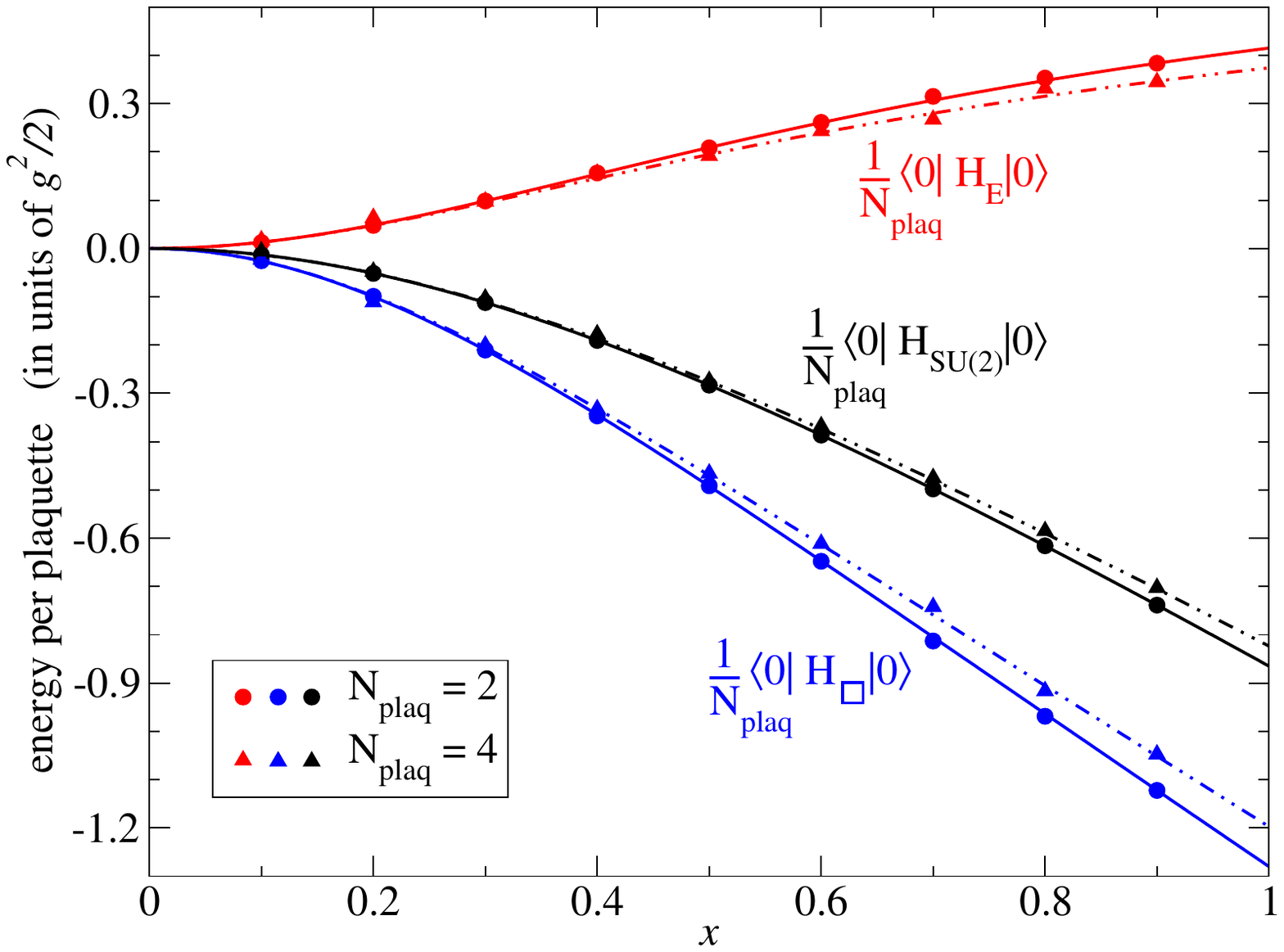}
\caption{Vacuum expectation value of the SU(2) Hamiltonian and its operators on different lattice sizes and for fixed $j_{\rm max}=1/2$ as a function of the gauge coupling $x=2/g^4$. The solid, dashed, and dotted curves are the exact classical calculation for $N_{\rm plaq}=2$, $4$, and $6$ respectively. The circles and triangles are data point obtained from the Advantages DW quantum hardware for for $N_{\rm plaq}=2$ and $4$ respectively.  Figure reproduced from Ref.~\citep{ARahman:2021ktn}.}
\label{fig:0_H_0_latt}
\end{figure}

It is interesting to show how precise the simulations done on DW are, to the point that they can be used to characterize how the theory changes when considering a higher value for the gauge truncation $j_{\rm max}$ or a larger lattice size. This can be done by using any state, but it is particularly meaningful to use the state with the lowest possible energy, that traditionally is used to extract the vacuum expectation value of observables that describe important properties of the system like spontaneous breaking of the system’s symmetries,  one of the most iconic examples is the Higgs field that has a non zero vacuum expectation value breaking the electroweak symmetry which gives masses to the particles.\\
This can be done easily on DW without doing new simulations, because the result of each simulation is an energy and the corresponding eigenstate. Therefore, to calculate the vacuum expectation value it is enough to consider the corresponding vectors to the ground state energies used to create the previous plots.\\
\\
In our case the only operator considered is the Hamiltonian, therefore we consider its vacuum expectation value and the one of the chromoelectric and the chromomagnetic operator: $\left<0\right|H_{\rm SU(2)}\left|0\right> $, $\left<0\right|H_E\left|0\right> $ and$ \left<0\right|H_\square\left|0\right>$. The calculation can simply be done by recalling that the matrix representation of $H_E$ is given by the diagonal element in the Hamiltonian while the values of $ H_\square $ are given by the off-diagonal elements.\\
How the vacuum expectation value of those three operators changes with the gauge truncation $j_{\rm max}$  is shown in Figure~\ref{fig:0_H_0_jmax} while their change with the lattice size is shown in Figure~\ref{fig:0_H_0_latt}.\\
The data points in Figure~\ref{fig:0_H_0_jmax} show a particularly good agreement with the exact classical calculation for the case $j_{\rm max}=1/2$ for the three operators, while for the case $j_{\rm max}=1$ a visible deviation from the exact curve growing with the value of the gauge coupling $x$ is present for the chromoelectric and chromomagnetic term, but not for the full Hamiltonian. Finally, even if the data points for $j_{\rm max}=3/2$ are not displayed due to the fact that DW did not resolve them from the one of $j_{\rm max}=1$, it is valuable to notice that $<0|H|0>$ for exact calculation of $j_{\rm max}=3/2$ represented by a dotted curve is particularly close to the one of $j_{\rm max}=1$, dashed line, and the DW data for $<0|H|0>$,  the black triangles are in agreement with both curves. This clearly shows that in the gauge coupling range $0<x<1$, the truncated theory has a rapid convergence toward the full untruncated theory already at small value of $j_{\rm max}$.\\
\\
To better visualize how the vacuum expectation value depends on the lattice size, it is convenient to consider an energy density obtained by normalizing the energy by the number of plaquettes as shown in Figure~\ref{fig:0_H_0_latt}.
In Figure~\ref{fig:0_H_0_latt} all data points are in good agreement with the exact curves. The exact curves for $N_{\rm pla}=4$ and 6, dashed and dotted, are perfectly superimposed, and the triangles, data points for $N_{\rm pla}=4$ are in agreement with both curves. This clearly shows that the normalized vacuum expectation values are local quantities independent of  the lattice size.\\
\\
By globally observing both plots in Figure~\ref{fig:0_H_0_latt} and Figure~\ref{fig:0_H_0_jmax} one can notice that the chromoelectric and chromomagnetic terms have larger error and discrepancy compared to the total energy.\\
\\
This can be explained by simply recalling that in the variational method the eigenvalues are extracted with higher precision compared to the eigenvector. In a more quantitative way, given a perturbation $\epsilon$, the eigenvalues are corrected to an order $O(\epsilon)$  while the eigenvalues to an order $O(\epsilon^2)$ \citep{sakurai_napolitano_2020}.\\
\\
In conclusion in the range of gauge coupling considered $0<x<1$, the theory on a 2 plaquette lattice shows rapid convergence toward the full untruncated theory already at $j_{\rm max}=1$, while for the case $j_{\rm max}=1/2$ the theory is local, showing no visible dependence on the lattice size. 
\section{SU(2) time evolution on D-Wave}
\begin{figure}[H]
\centering
\includegraphics[width=0.5\linewidth]{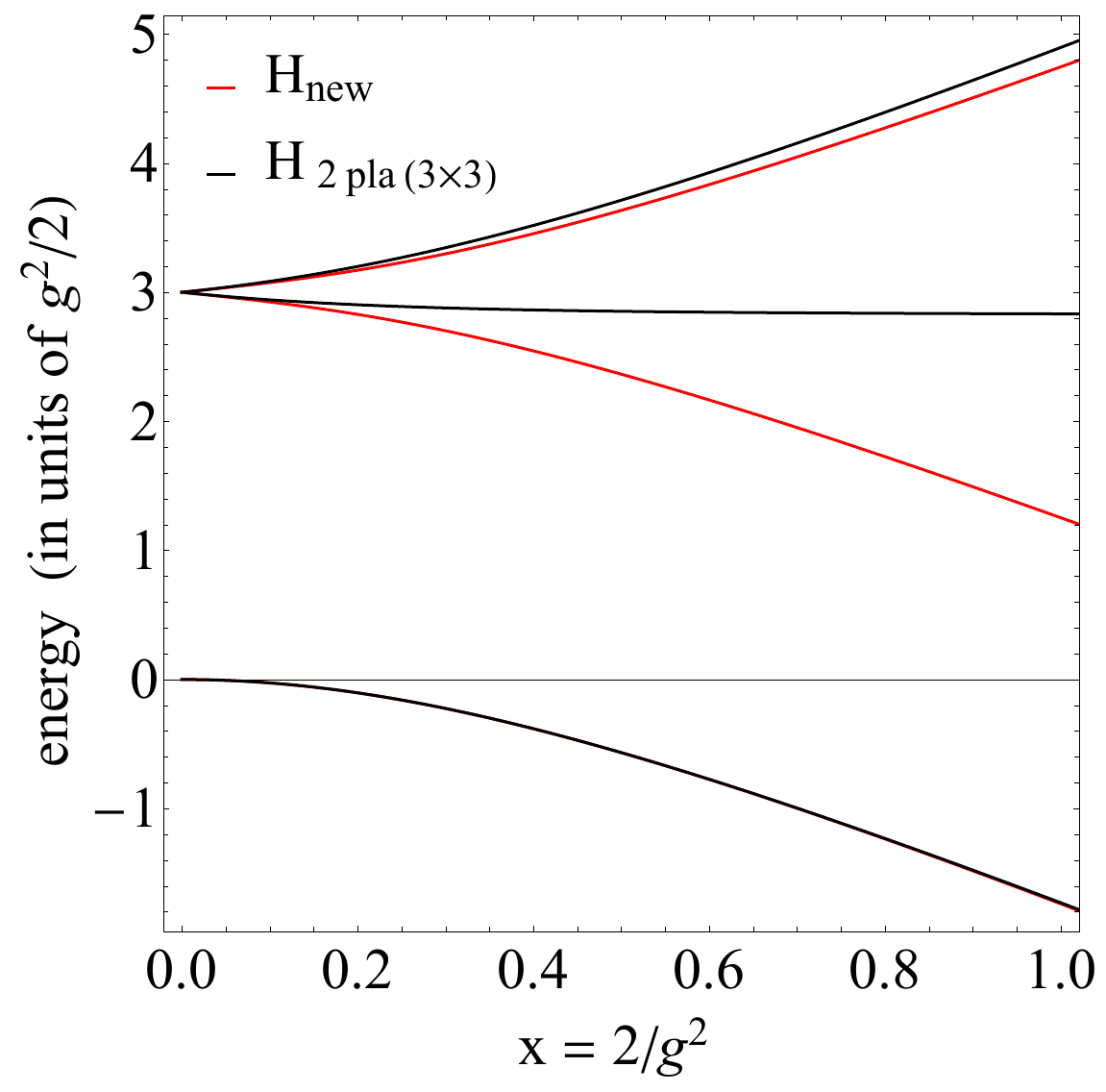}
\caption{Energy spectrum of the 2 plaquette Hamiltonian with $j_{\rm max}=1/2$ as a function of the gauge coupling x showing how the eigenvalues of the $3 \times 3$ block change due to the heavy state present in the new Hamiltonian Eq.~\ref{eq:h_heavy_state}. The three black solid lines are the eigenvalues of the $3 \times 3$ block while the red solid lines are the lowest three eigenvalues of the Hamiltonian with the extra heavy state. The plot shows eigenvalues of the two Hamiltonians are in close agreement in the gauge coupling window $0<x<0.2$.}
\label{fig:energy_s_heavy_state}
\end{figure}

The real-time evolution of the model can be directly accessed in the Hamiltonian approach using the time evolution operator $e^{-iHt}$, but for DW there is no simple way to encode it on the hardware. One interesting solution is to rethink the time evolution problem as a ground-state eigenvalue problem using the so called Kitaev-Feynman clock states \citep{Feynman_85,McClean_2013, Tempel_2014, Kitaev_2022}. In a nutshell, this 
consists of using an operator with a ground state that contains the entire time evolution sequence from the initial to the final state, allowing one to calculate in parallel the entire time evolution of the system. To illustrate this approach we follow the discussion in \citep{McClean_2013}.\\
\\
Let's suppose that we would like to solve the Schr\"odinger equation
\begin{equation}
{i \partial_t \left|\Psi(t)\right> = H(t)\left|\Psi(t)\right>}
\end{equation}
with the goal of finding the state of the system at a later time $t_f$. The solution will have the general form $\left|\Psi(t_f)\right> = U(t_f,t_0)\left|\Psi(t_0)\right>$  where $U(t_f,t_0)$ is the time evolution operator from the initial time to the final time.\\
\\
We can now divide the time evolution process into $N$ steps of size $\epsilon$, such that:
\begin{equation}
{\left|\Psi(t_f)\right> = U(t_f,t_0)\left|\Psi(t_0)\right> =
  U(t_f,t_{f-1}) \enspace \dotsc  U(t_2,t_1)  \enspace U(t_1,t_0)\left|\Psi(t_0)\right>}
\end{equation}
and let's keep track of each time evolution step with a clock state set $\lbrace\left|t\right>\rbrace$ such that:\\
\begin{equation*}
\begin{array}{cc|ccc|c|cc|c|cc}

\left|0\right> & & & 
\left|1\right> &&
\dotsc &
\left|t+\epsilon\right>  & & 
\dotsc & &
\left|t_f-1\right>\\

  & & &    &&   &   & &   & &  \\
 
\left|\Psi(t_0)\right> & & & 
U(t_1,t_0)\left|\Psi(t_0)\right> && 
\dotsc & 
U(t+\epsilon, t-\epsilon)\left|\Psi(t-\epsilon)\right> & &
\dotsc & &
U(t_f,t_{n-1})\left|\Psi(t_{n-1})\right>

\end{array}
\end{equation*}
when the clock is $\left|0\right>$ the system is in the state $\left|\Psi(t_0)\right>$ and at a later clock time $\left|t_f-1\right>$ the system is in the state  $\left|\Psi(t_f)\right>$.\\
It is useful to create a connection between the clock states set $\lbrace\left|t\right>\rbrace$ and the corresponding states of the physical system $\left|\Psi (t)\right>$ by entangling each corresponding pairs and then sum all pair creating the following state:
\begin{equation}
\left|\Theta\right> =\frac{1}{\sqrt{N}} \bigg( \left|\Psi(t_0)\right>\otimes\left| 0 \right>    + \dotsc +  \left|\Psi(t_f-1)\right>\otimes\left| t_f-1 \right>  \bigg)  =\frac{1}{\sqrt{N}} \sum_{t=t_0}^{t_f-1} \left|\Psi(t)\right>\otimes\left| t \right>
\end{equation}
This state contains the entire time evolution from $t_0$ to $t_f$, hence the name history state.
\\
\\
The operator which maps the sequence of unitary time operators $\lbrace U_i \rbrace$ to the the sequence of clock states $\lbrace\left|t\right>\rbrace$ and has the history state as a ground state is called the clock Hamiltonian and was originally proposed by Feynman. One of its possible forms is:
\begin{equation}
H_c = C_0 + \frac{1}{2}\sum_{t=0}^{t_f -1}\bigg(I\otimes\left|t\right>\left<t\right|-U_t\otimes\left|t+\epsilon\right>\left<t\right| \nonumber -U_t^\dagger\otimes\left|t\right>\left<t+\epsilon\right|+I\otimes\left|t+\epsilon\right>\left<t+\epsilon\right|\bigg)
\label{eq:clockH}
\end{equation}
where $U_t=e^{-i\epsilon H_t}$ is the time evolution operator from time $t$ to time $t+\epsilon$ and $C_0$ is used to specify the initial state at time $t=0$.\\
Therefore the history state can be found by using the QAE algorithm using as a functional the following:
\begin{equation}\label{eq:timefunctional}
{\cal L} = \sum_{t,t^\prime}\left<t^\prime\right|\left<\Psi_{t^\prime}\right|H_c\left|\Psi_t\right>\left|t\right>
         - \lambda\left(\sum_{t,t^\prime}\left<t^\prime\right|\left<\Psi_{t^\prime}|\Psi_t\right>\left|t\right>-1\right)
\end{equation}
where $\left|\Psi_t\right>$ is the state of the system at time $t$, $\left|t\right>$ is the state of the clock at time $t$ and the Lagrange's multiplier is needed for the solution to have a unit norm. This  functional was obtained using a discrete-time variational principle that authors of \citep{McClean_2013} called time-embedded discrete variational principle (TEDVP).\\
\\
The computational cost of using the clock Hamiltonian in terms of logical qubits is given by the product of the dimension of the time evolution operator, the number of time steps chosen and number of precision qubits $K$. 
\\
This method can be used for example to study the real-time evolution between the two excited states $1$ and $2$, and show how to extract from the time evolution their energy gap.\\
\\
To perform this study one has to first deal with the fact that DW hardware does not handle complex numbers, which are inevitably present in the time evolution operator $U_t=e^{-i\epsilon H_t}$. One possible solution to this is to use a basis that makes the Hamiltonian imaginary so that the time evolution operator is a real matrix.\\   
For the case of two-plaquette lattice Hamiltonian in Eq.~\ref{eq:2pla_periodic} there is no change of basis that converts the Hamiltonian into a purely imaginary matrix, therefore to  find a new basis in which $H_{3x3}$ is a imaginary matrix a possible solution is the introduction of a fictitious heavy state:
\begin{equation} \label{eq:h_heavy_state}
H_{\rm new} = \frac{g^2}{2}\left(\begin{array}{ccc|c} 0 & -2\sqrt{2}x & 0 & \textcolor{red}{0} \\ -2\sqrt{2}x & 3 & -\tfrac{x}{\sqrt{2}} &\textcolor{red}{0} \\

0 & -\tfrac{x}{\sqrt{2}} & 3 & \textcolor{red}{-2\sqrt{2}x}
\\
\hline
 \textcolor{red}{0} & \textcolor{red}{0} & \textcolor{red}{-2\sqrt{2}x} & \textcolor{red}{ 6} \end{array}\right)
 \begin{array}{c}
\left|1_1^11_1^1\right> \\
\tfrac{1}{\sqrt{2}}\left(\left|2_2^22_1^1\right>+\left|2_1^12_2^2\right>\right) \\
\left|1_2^21_2^2\right> \\
\textrm{heavy state}
\end{array}
\,
\end{equation}
where the red column and row are due to the introduced heavy state and are needed for finding a purely imaginary representation of the Hamiltonian.\\
The effect of this heavy state on the energy spectrum is shown in Figure~\ref{fig:energy_s_heavy_state}, where it is clear that its effect is negligible in the range of gauge coupling $0<x<0.2$. Therefore in this range $H_{\rm new}$ can be used to study the time evolution of the original Hamiltonian. 

\begin{figure}[H]
\centering
\includegraphics[width=0.7\linewidth]{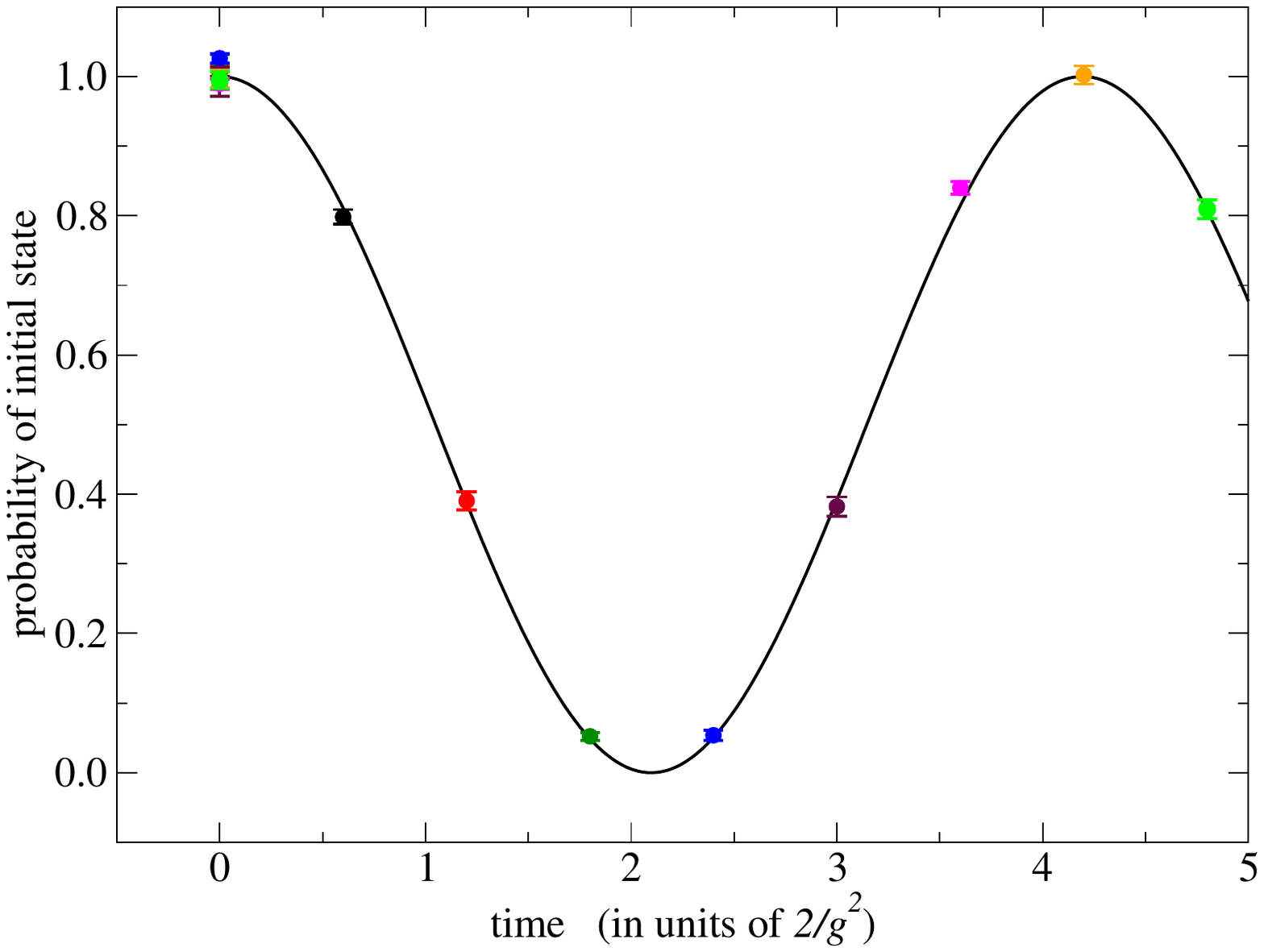}
\caption{Probability oscillations between the two state $\left|1_2^21_2^2\right>$ and $\tfrac{1}{\sqrt{2}}\left(\left|2_2^22_1^1\right>+\left|2_1^12_2^2\right>\right)$ at gauge coupling $x=0.1$. The solid line is the exact calculation $cos^2(\omega t)$ and, the eight coloured pairs of data points are obtained from eight separate runs with two time step $t=0$ and $t>0$ having different step-size $\epsilon$ by using DW quantum hardware Advantage with $K=7$, $5 \times 10^4$ reads, $\lambda=0.0$ and $\rm chain strength=0.5$ as simulation parameters. Figure reproduced from Ref.~\citep{ARahman:2021ktn}}
\label{fig:TE_non_stoquastic}
\end{figure}

\begin{figure}[H]
\centering
\includegraphics[width=0.7\linewidth]{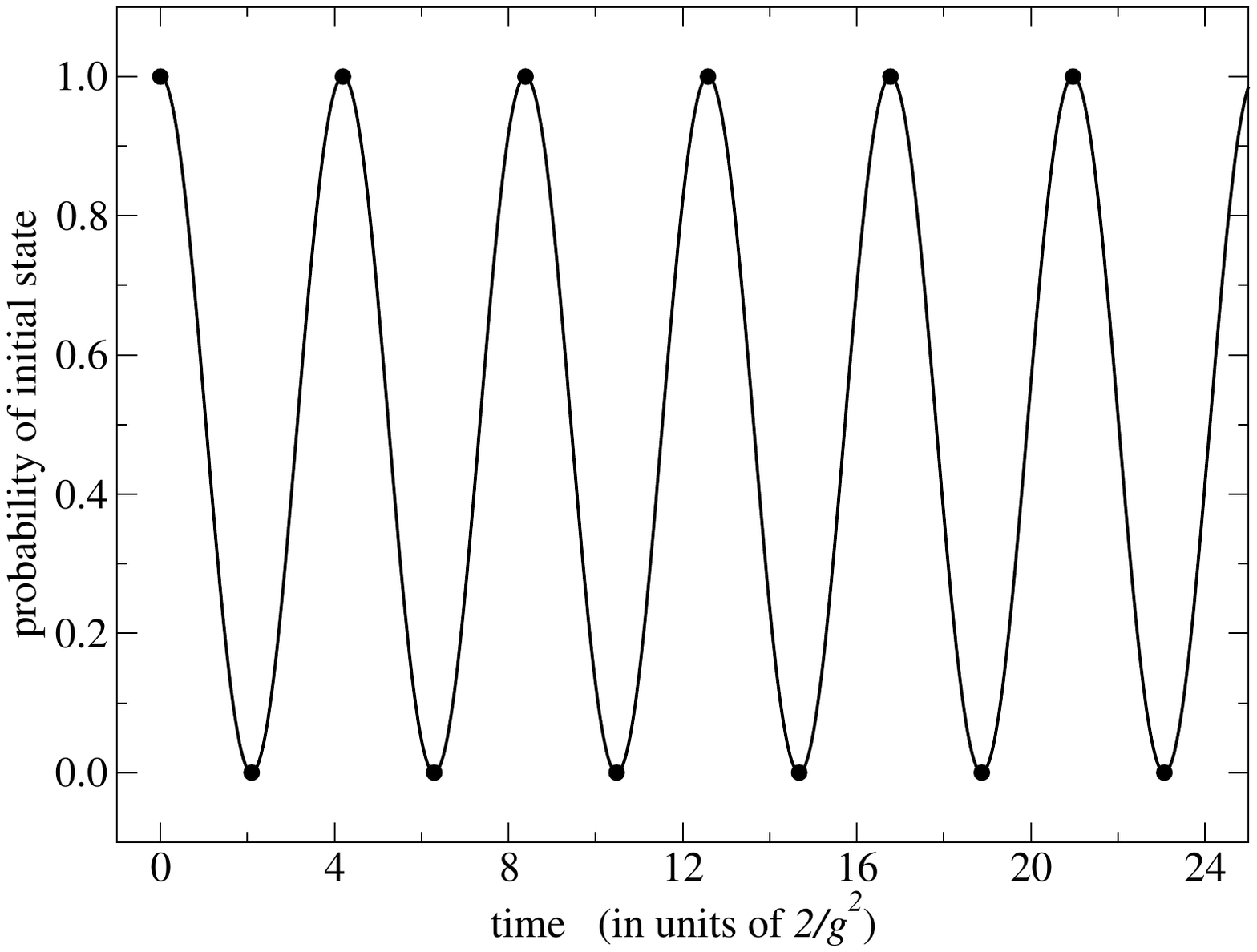}
\caption{Probability oscillations between the two states $\left|1_2^21_2^2\right>$ and $\tfrac{1}{\sqrt{2}}\left(\left|2_2^22_1^1\right>+\left|2_1^12_2^2\right>\right)$ at gauge coupling $x=0.1$ . The solid line is the exact calculation $cos^2(\omega t)$ and, the twelve black points are obtained from one run with 12 time steps with step-size $\epsilon=\pi/\omega$ using DW quantum hardware Advantage with $K=2$, $10^4$ reads, $\lambda=0.12$ and $\rm chain strength=0.3$ as simulation parameters. Figure reproduced from Ref.~\citep{ARahman:2021ktn}.}
\label{fig:TE_stoquastic}
\end{figure}

By construction $H_{\rm new}$ can be written up to a constant in a purely imaginary form as: 

\begin{equation}\label{eq:Hnew_imagi}
P^{-1}H_{\rm new}P = \frac{g^2}{2}\left(\begin{array}{cccc} 
\textcolor{blue}{3} & \textcolor{blue}{-ih_-} & 0 & 0 
\\
\textcolor{blue}{ih_-} & \textcolor{blue}{3} & 0 & 0 
\\
0 & 0 & \textcolor{red}{3} & \textcolor{red}{ -ih_+ }
\\
0 & 0 & \textcolor{red}{ih_+} & \textcolor{red}{3}
\end{array}\right)
\end{equation}
where $h_\pm = \frac{1}{2}\sqrt{18+33x^2\pm\sqrt{65x^4+1116x^2+324}}$ and, the left top block in  blue contains the two physical excited states while the right bottom block in red contains the ground state and the added heavy state. The constant part of the Hamiltonian can be dropped because we are only interested in extracting the energy difference of the two states. The time evolution operator of the blue block needed for the clock Hamiltonian is obtained by direct calculation of $e^{-iH_{\rm blue}t}$ and it is represented by a real matrix:
\begin{eqnarray}\label{T_ev_two_states}
U_t =e^{-i\epsilon H_{\rm blue}}= \left(\begin{array}{rr} \cos(\omega\epsilon) & -\sin(\omega\epsilon) \\
      \sin(\omega\epsilon) & \cos(\omega\epsilon) \end{array}\right)
 \ , \ \ \ 
\omega=h_-g^2/2 \,
\end{eqnarray}
assuring that clock Hamiltonian is now real. \\
\\
The results of using the DW hardware and the described method to observe the real-time transition from these two exited states $[\left|1_2^21_2^2\right> \leftrightarrow \tfrac{1}{\sqrt{2}}\left(\left|2_2^22_1^1\right>+\left|2_1^12_2^2\right>\right)]$ are shown in Figure~\ref{fig:TE_non_stoquastic} and Figure~\ref{fig:TE_stoquastic}.

In Figure~\ref{fig:TE_non_stoquastic} the time evolution is obtained with eight different simulations using two-time steps at once, one at $t=0$ and one at later time $t=\epsilon$ and they are indicated by pairs with the same colours. Each run used $5 \times 10^4$ reads, $\lambda=0.0$, $\rm chain strength=0.5$ and  $K=7$ qubits for precision for a total of 28 logical qubits. Although  the number of logic qubits is not large compared to other simulations in the previous section, the time interval cannot be extended to a longer time because the DW hardware works best only with stoquastic matrices, which have only off-diagonal nonpositive elements. The clock Hamiltonian is not stoquastic for a generic time step $\epsilon$, therefore the systematic error grows with the size $\epsilon$ present in the off-diagonal positive elements \citep{Ozfidan_2020}.\\
\\
The real time evolution can be extended to a longer period of time by choosing $\epsilon=\pi/\omega$ that makes all the off-diagonal element nonpositive whose result is shown in Figure~\ref{fig:TE_stoquastic}, where a clock Hamiltonian with 12 time step was used to obtain the time evolution of 12 steps in one run with $10^4$ reads, $\lambda=0.12$, $\rm chain strength=0.3$ and  $K=2$ qubits for precision for a total of 48 logical qubits.
From both Figure~\ref{fig:TE_non_stoquastic} and Figure~\ref{fig:TE_stoquastic} the energy gap between the two states can be extracted by measuring the oscillation period and calculating the frequency $\omega$, obtaining $E_2-E_1=2\omega$.\\
\\
The same approach was previously used in \citep{Jalowiecki2020} to observe six Rabi oscillations using the DW hardware 2000Q a previous version of DW Advantage. In our result twelve time steps have been reached showing significant improvement between two subsequent hardware versions.

\section{Conclusions and looking forward}
In this chapter we have presented the main features of D-Wave quantum annealer (DW), and the results obtained by using it to study the energy spectrum and the time evolution of a SU(2) lattice gauge theory (LGT).\\
Our work was inspired by the first study of SU(2) LTG using a minimal lattice done on a IBM gate-based quantum computer \citep{Klco:2019evd}, which we tried to extend in many ways: we worked on larger lattices made by 2,4, and 6 plaquettes; implemented a larger energy truncation $j_{\rm max}=1/2$, $1$ and $3/2$; and extended the time evolution to a larger time range, of about 10 times.  With the result from DW we showed that in the range of gauge coupling $0<x<1$, the energy spectrum swiftly converges already for a small energy truncation $j_{\rm max}$ and in the case of systems with the same $j_{\rm max}=1/2$ and longer lattices, their spectrum rapidly convergence after few plaquettes.\\
\\
Although DW was designed to mainly solve Ising-type problems, these results were possible using the QAE algorithm developed in \citep{Teplukhin_2019, Teplukhin_2020, D0CP04272B, teplukhin2021sampling, Teplukhin2021}, which not only gave us the possibility to study the energy spectrum and the time evolution of a LGT, but it also paved the way to exploring other LGTs.\\
\\
It should be noted that the accuracy of our results was obtained only by rationing the number of qubits used, without using many of the available tools present on DW Ocean software such as error-correction, chain break reduction, tuning of the annealer time and the possibility of choosing the problem’s embedding. Using these extra tools might improve the use of the quantum resources but would not change the scaling of resources with lattice size. Rationing the use of quantum resources is important, because even if DW had a large number of physical qubits, the logical qubits are limited, so we rationed the use of logical qubits by the system symmetries to block diagonalizing the Hamiltonian representation of each case, resulting in hardware manageable matrix sizes and, leaving only for a few problematic cases the need to further truncate the matrix. \\
\\
For additional rationing, a crucial choice was the introduction of the adaptive QAE algorithm, which allowed us to work with large matrix blocks without increasing the number of qubits used to make the calculation precise.\\
\\
The time evolution of the theory was possible by an intelligent use of the Kitaev-Feynman clock state which converts the time evolution problem into a ground state, implementing for our case the TEDVP developed in \citep{McClean_2013}. Despite the DW limitations associated with the use of this method, we were able to study the transition between two excited states and extract their energy splitting. Additionally, the study extended the time range in which the DW hardware can operate compared to the results of a previous Rabi oscillations study using the previous generation of DW hardware 2000Q \citep{Jalowiecki2020}.\\
\\
It is important to emphasize that our work did not try to use quantum resources in such a way that we could achieve polynomial scaling while working with the entire physical Hilbert space of the theory, but we tried to achieve this by extracting physical information by block diagonalizing the theory.\\
When asked the much more fundamental question of how polynomial scaling can be achieved in study of LGT on DW hardware, we believe this goal can only be addressed by changing the way users interact with qubits, giving them more freedom and flexibility to directly access the qubits state.\\
\\
Ultimately, the work presented in this chapter was the first to use the DW to study a LGT, and represents an important proof of concept that has inspired other research groups to further investigate the capabilities of  DW's hardware. In fact, the approach has been used by the authors of Ref.~\citep{Illa:2022jqb} to study the energy spectrum of harmonic and anharmonic oscillators, and the time evolution of a SU(3) LGT on a single plaquette, and neutrino flavor dynamics in beam-beam collisions. For the latter two studies, an important extension has been achieved while using the Kitaev-Feynman clock states by allowing complex values in the QUBO elements by expressing the functional to be minimized in its real and imaginary part, with the important result of making it easier to investigate the time evolution of a generic theory.\\
A further application of DW has been done in \citep{Fromm:2022vaj} by studying the energy spectrum and time evolution of a 2 plaquette LGT using dihedral groups instead of the more common SU(N). Finally a recent study was done in \citep{Kim:2023sie} to find the configurations of a U(N) LGT paving the way toward importance sampling of LGT on DW.\\
\\
Looking to the future use of DW, new perspective will be available in the near future as part of the company’s "Clarity" road map announced in 2021 at the annual D-Wave conference, where it was unveiled that the company is focused on the development of a gate-based quantum annealer, extending the functionality of the present hardware technology to use gates to directly operate on qubits, similarly to what is already possible with other quantum computers.\\
\\
This future gate-based quantum annealer will give users more flexibility in encoding their problem on the hardware and controlling the qubits, expanding the applicability of the hardware to other fields of knowledge. For the studying of lattice gauge theories there will be the possibility to use larger lattices for the systems already studied and, secondly the opportunity to consider particle physics systems having more complex dynamics. In fact, having a gate-based quantum annealer will certainly reduce the number of qubits needed to encode an operator, because what now it is done by submitting the full matrix representation, it will likely be done by encoding it using a few gates. Reducing the number of qubits needed for encoding a problem, automatically shortens the length of the chain which has the effect of reducing the chances of breaking the chain, making the hardware even more reliable. Ultimately, this advance could lead to a polynomial scaling of the quantum resources needed to study a lattice gauge theory with lattice size growth, achieving the definitive goal of outperforming a classical computer.\\
\\
In conclusion, as discussed by DW researchers in \citep{Ozfidan_2020}, the current available DW annealer implements a stoquastic traverse-field Hamiltonian making calculation free from the sign problem only with problems represented by a stoquastic Hamiltonian. Therefore a future breakthrough may come by extending the hardware to non-stoquastic systems with the introduction of interactions of the form $\sigma_i^x \sigma_j^x$ or $\sigma_i^y \sigma_j^y$. This will not only extend the usability of the hardware to non-stoquastic Hamiltonians, but will also set the scene for the introduction of a universal quantum annealer, which will be able to accommodate any Ising-like Hamiltonian without restrictions. Such a hardware will be equivalent to an adiabatic quantum computer \citep{Albash_2018} for which it was proved to be polynomially equivalent to a conventional quantum computer \citep{aharonov2005adiabatic}.

\chapter{SU(2) on IBM gate-based quantum computers}\label{chap:Chapter_3}

In this chapter \footnote{The results presented in this chapter were previously published in \href{https://doi.org/10.1103/PhysRevD.106.074502}{Phys. Rev. D 106, 074502} done in collaboration with  S. A Rahman, R. Lewis and S. Powell.} we present the main features of IBM quantum computers and show how they can be used to calculate the real-time evolution of the SU(2) pure gauge lattice theory for the case of a 2 and 5 plaquettes lattice with $j_{ \rm max}=1/2$ using non-periodic boundary conditions, and in particular the case of a travelling energy excitation across the lattice. These results are in good agreement with the exact classical result thanks to the use of a remarkable error mitigation technique that we introduced, called self-mitigation. Then we present an extension of this approach for the case of 2-plaquette lattices with $j_{\rm max}=3/2$ whose simulations are under preparation. In conclusion, we provide an overview of the results and discuss potential directions for future research.\\

\section{The IBM quantum computer}
\begin{figure}[H]
\centering
\includegraphics[width=1.0\linewidth]{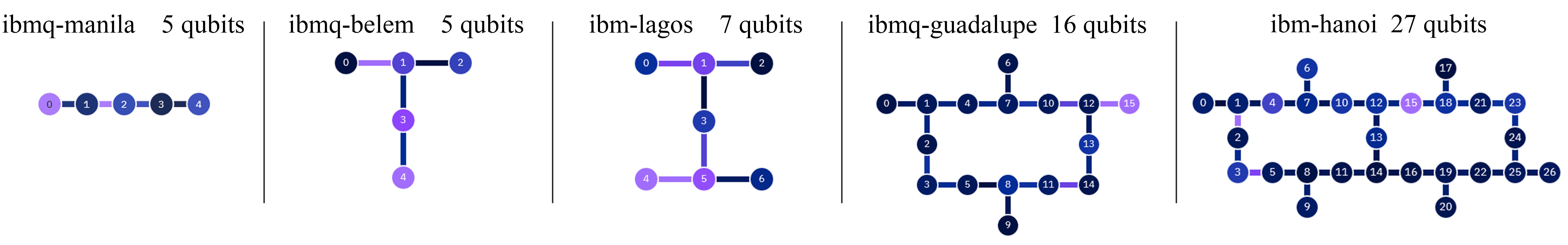}
\caption{Example of qubit layout and connectivity of some of the available IBM quantum hardware. Each qubit is represented by a circle and the interqubit connection by a bar. In these specific figures taken from the IBM quantum experience, the color on the qubits indicated the readout error associated with the measurement of each qubit, while the color of the bars the error associated to each $CX$. The error range is indicated in a scale from black to light violet, with black for the lowest error rate and light violet for the highest error rate, which are based on the hardware calibration at the time of the picture was taken. The hardware layout were taken from \citep{IBM_hardware}.}
\label{fig:IBM_hard_geometry}
\end{figure}

\begin{figure}[H]
\centering
\includegraphics[width=0.9\linewidth]{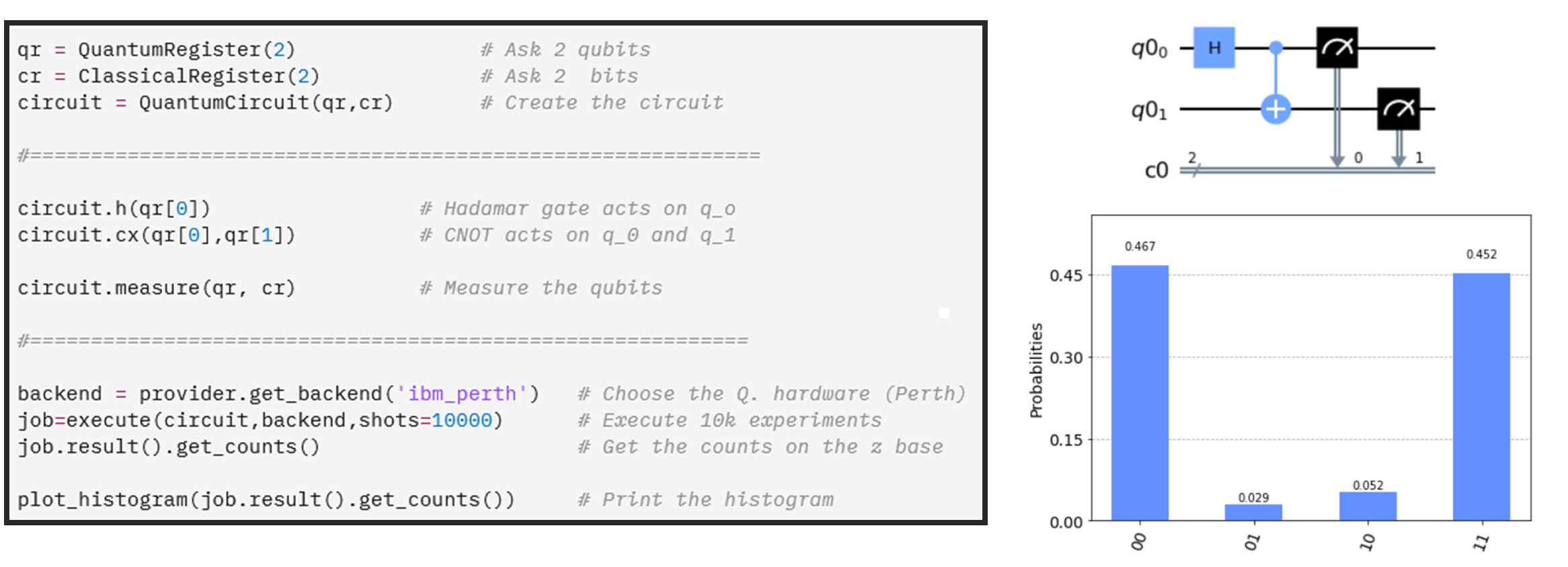}
\caption{Job execution for creating and measuring the Bell state $\frac{1}{\sqrt{2}}(\vert 00 \rangle + \vert 11\rangle)$ on the IBM hardware {\tt ibm\_perth}. On the left the coding necessary to encode the operation on a two-qubit circuit, submit 10000 experiment and express in probabilities the measurements of the state of two qubits in the z-basis. On the top right the corresponding circuit, where each single horizontal line represents a qubit while the bottom double line represents the two classic memory channels used to store the result of the measurements, and after the $H$ and $CX_{01}$ the two speedometer like symbols are the measurement operators with vertical arrows pointing to the  classical memory channels used to store the results. On the bottom right the bar diagram displays the probabilities for each element of the computational basis.}
\label{fig:IBM_Bell_circuit}
\end{figure}

Since 2016 the multinational technology company IBM has made their quantum computer hardware available with the IBM Quantum Experience. While they started with few hardware ranging from 1 to 5 qubits, at the time of writing 23 different computers ranging from 5 to 433 qubits are available. Of the 23, 9 computers with 5 to 7 qubits are available to the public free of charges \citep{IBM_hardware}.\\ 
\\   
The IBM quantum computers are universal gate-based quantum computers, which means that the qubit state can be directly manipulated by using operators called gates. Native gates, are a specific set of gates that are implemented on the hardware and form a universal basis for all possible operators. Therefore, any operator can be expressed as a combination of native gates, making the hardware a universal quantum computer.\\
The software that allows the user to interact with the hardware provides a variety of default gates outside the native get set, and the most common operators which are used in this chapter are listed in Table~\ref{tab:IBM_gates_table}.
\begin{table}[H] 
\centering 
\begin{tabular}{c c c}      
\hline
\hline                                                                
Operator  & Gate & Matrix Representation \\
\hline\\
        Pauli X
        & \begin{tikzcd}
		&\gate{X} & \qw
		\end{tikzcd}
		&$\begin{pmatrix}0&1\\1&0\end{pmatrix}$ \\ [4ex]
		%
		%
        Pauli Y
        & \begin{tikzcd}
		&\gate{Y} & \qw
		\end{tikzcd}
		&$\begin{pmatrix}0&-i\\i&0\end{pmatrix}$ \\ [4ex]
		%
		%
        Pauli Z
        & \begin{tikzcd}
		&\gate{Z} & \qw
		\end{tikzcd}
		&$\begin{pmatrix}1&0\\0&-1\end{pmatrix}$ \\ [4ex]
		%
		%
        $RX(\theta)$
        & \begin{tikzcd}
		&\gate{RX(\theta)} & \qw
		\end{tikzcd}
        & $\begin{pmatrix}cos(\theta/2)&-i \, sin(\theta/2)\\-i \, sin(\theta/2)&cos(\theta/2)\end{pmatrix}$  \\[4ex]
        %
		%
        $RY(\theta)$
        & \begin{tikzcd}
		&\gate{RX(\theta)} & \qw
		\end{tikzcd}
        & $\begin{pmatrix}cos(\theta/2)&-sin(\theta/2)\\sin(\theta/2)&cos(\theta/2)\end{pmatrix}$  \\[4ex]
        %
		%
        $RZ(\phi)$
        & \begin{tikzcd}
		&\gate{RZ(\phi)} & \qw
		\end{tikzcd}
        & $\begin{pmatrix}e^{-i\phi/2}&0\\0&e^{i\phi/2}\end{pmatrix}$  \\ [4ex]
        %
		%
        Hadamard H
        & \begin{tikzcd}
		&\gate{H} & \qw
		\end{tikzcd}
		&$\frac{1}{\sqrt{2}}\begin{pmatrix}1&1\\1&-1\end{pmatrix}$ \\ [4ex]
		%
		%
        Phase gate $S=\sqrt{Z}$
        & \begin{tikzcd}
		&\gate{S} & \qw
		\end{tikzcd}
		&$\begin{pmatrix}1&0\\0&i\end{pmatrix}$ \\ [4ex]
		%
		%
        Controlled-x
        $CX_{01}$
        & \begin{tikzcd}
		&\ctrl{1} & \qw \\
		&\targ{} & \qw
		\end{tikzcd}
        & $
        \begin{pmatrix}
        1&0&0&0\\
        0&0&0&1\\
        0&0&1&0\\
        0&1&0&0
        \end{pmatrix}$\\ [4ex]\\
        %
		%
        Controlled-x
        $CX_{10}$
        & \begin{tikzcd}
		&\targ{} & \qw \\
		&\ctrl{-1} & \qw
		\end{tikzcd}
        & $
        \begin{pmatrix}
        1&0&0&0\\
        0&1&0&0\\
        0&0&0&1\\
        0&0&1&0
        \end{pmatrix}$\\ [4ex]\\
        %
		%
        SWAP gate
        & \begin{tikzcd}
		&\swap{1} & \qw \\
		&\targX{} & \qw
		\end{tikzcd}
        & $
        \begin{pmatrix}
        1&0&0&0\\
        0&0&1&0\\
        0&1&0&0\\
        0&0&0&1
        \end{pmatrix}$\\ [1ex]\\

\hline
\hline
    \end{tabular}
\caption{List of the most common available gates which are used in this chapter, an extended list of the default IBM gates present in the Qiskit library can be found here \citep{IBMGATES}. The first eight gates listed are single qubit gates: Pauli X Pauli Y, Pauli Z, the rotation gates RX, RY and RZ, the Hadamard $H$ and the $S$ gate, the last three act on a pair of qubits. The first two are Controlled-x gates $CX_{ij}$ and they applies a Pauli-X gate on the $j$ qubit when the $i$ qubit is in the state $\vert 1 \rangle$, and leave it  unchanged when the $i$ qubit is in the state $\vert 0 \rangle$, while the last one, the SWAP gate, as the name suggest swaps the states of two qubits.}
\label{tab:IBM_gates_table}
\end{table}
The state of the qubit is represented in the z-basis, often called the computational basis, as $\vert 0 \rangle=\begin{pmatrix} 1\\0 \end{pmatrix}$ and $\vert 1 \rangle=\begin{pmatrix} 0\\1 \end{pmatrix}$ and any superposition of these two states, therefore the Pauli gates $X, Y, Z$ and rotation gates $RX, RY, RZ$ have a well known action on the qubit state, as can be confirmed by performing the matrix multiplication.\\
\\
The Hadamard gate $H$ when acting on a qubit rotates it in the x-basis, therefore creating a superposition of the two qubit’s states, in fact $H \vert 0 \rangle =\frac{1}{\sqrt{2}}\left( \vert 0 \rangle + \vert 1 \rangle  \right)  = \vert + \rangle$ and $H \vert 1 \rangle =\frac{1}{\sqrt{2}}\left( \vert 0 \rangle - \vert 1 \rangle  \right)  = \vert - \rangle$.\\
The phase gate $S$ is a phase shift gate of $\pi/2$ and it is the square-root operator of the Pauli Z gate,  $\sqrt{Z}$.\\
\\
The controlled-x gate $CX_{c t}$, often denoted as $CNOT$, is a 2-qubit gate, where the subscript $c$ labels the control qubit and $t$ labels the target qubit. The target qubit state is flipped by applying to it an $X$ gate when the control qubit state is in $\vert 1 \rangle$ and, and leaves the target qubit unchanged when the control qubit is in $\vert 0 \rangle$. The control not gates play an important role in quantum computation because they are entangling gates. In fact, if the control qubit is in a superposition, the control and the target qubit get entangled.\\
\\
Finally, the $\rm SWAP$ is a 2-qubit gate that swaps the state of two qubits.\\
\\
The measurement is done by default in the z-basis, and to measure in other bases it is enough to project the qubit’s state before the measurement, in a such a way that the result of the measurement of the z-basis returns the corresponding value in the chosen basis. For example, to measure in the x-basis it is enough to enter a Hadamard gate before the measurement.\\
\\
It is important to consider that each hardware has only a specific set of native gates, the most common is  $\left[ CX, I, RZ,\sqrt{X}, X \right] $, therefore any time a non-native gate is used an internal algorithm called transpiler rewrites it in native gates.\\
In this respect under more general hypothesis about the set of native gates that it is assumed to be able to generate a dense subset of SU(d), the Solovay-Kitaev theorem guarantees that any unitary operator can be expressed as a combination of elements of a set of native gates with accuracy $\epsilon$ using a finite sequence of gates, whose number is estimated by $O(m\log^c(m/\epsilon))$ where $m$ is the number of $CNOT$ gates and $c$ is a constant \citep{A_Yu_Kitaev_1997, Solovay_2000}. A recent review and an algorithmical proof can be found here \citep{dawson2005solovaykitaev}, and for a more detailed treatment and the history behind see \citep{nielsen_chuang_2010}. This fully justifies the fact that the IBM hardware is a universal quantum computer.\\
\\
The qubits are arranged on the hardware with a certain geometry and therefore their connectivity varies depending of the hardware, from 1 to maximum 3 as can be seen in Figure~\ref{fig:IBM_hard_geometry}. This means a qubit can communicate directly with at most three other qubits, therefore a qubit can be directly entangled, with a $CX$ gate, only with the qubits to which it is connected.
The other hardware with a larger number of qubits $33, 65, 127, 433$, whose layout is too large to be shown, have the same maximum connectivity  equal to 3 and, the qubit layout is obtained by repeating in space the 16 qubits layout present in Figure~\ref{fig:IBM_hard_geometry}. 
\\
\\
In case an operator requires connectivity larger that 3, one can use a $\rm SWAP$ gate, which is a non-native gate made from the combination of three $CX$ gates, making it one of the noisiest gates.\\
\\
The hardware is still noisy so the gates have a non-negligible error, the error rate changes on each hardware hour by hour depending on the hardware calibration schedule. This information is publicly accessible by choosing any hardware on the IBM quantum experience \citep{IBM_hardware}.\\
But in general the $CX$ has an error of order of $10^{-2}$, the error in measuring the qubit is of order of $10^{-2}$, while the single qubit gates have an error of the order of $10^{-4}$. \\
\\
A job can be executed on a chosen hardware by using the IBM Qiskit library written in python. The operation is written in gates arranged to form a circuit, whose action on the qubits is obtained by measuring the results of a certain number of experiments executed on the hardware in terms of probabilities of obtaining a state in the z-basis.\\
An example of this process is shown in Figure~\ref{fig:IBM_Bell_circuit} where the IBM hardware \\{\tt ibm\_perth} was used to create and measure the Bell state $1/\sqrt{2}(\vert 00 \rangle + \vert 11\rangle)$ starting from the initial state in which both qubits are in the zero state $\vert 0 \rangle_1 \vert 0 \rangle_0$ by applying $H$ on the first qubit, $H_0 \vert 0 \rangle_1 \vert 0 \rangle_0 = \vert 0 \rangle_1 1/\sqrt{2} \left( \vert 0 \rangle_0 + \vert 1 \rangle_0  \right)$ and then applying a $CX$ on the pairs $CX_{01} 1/\sqrt{2} \left( \vert 0 \rangle_1 \vert 0 \rangle_0 + \vert 0 \rangle_1 \vert 1 \rangle_0  \right)$ obtaining  $1/ \sqrt{2}(\vert 00 \rangle + \vert 11\rangle)$.\\
From the 10000 experiments the measurements obtained are $[\vert 00 \rangle :4666, \vert 01 \rangle :294, \vert 10 \rangle :517, \vert 11 \rangle : 4523  ]$  we can see that the result is close to what was expected with close to half the states in $\vert 00 \rangle$ and half in $\vert 11 \rangle$, but there is a small probability for the states $\vert 01 \rangle$ and $\vert 10 \rangle$ that should not be there. These are present due to what we generally call hardware errors.\\
\\
In the next section we discuss some of the most common error mitigation techniques that we have used to mitigate the hardware error present in our circuits.\\
\\
Since 2016 there have been many particle physics studies that used the IBM quantum simulator and/or hardware, therefore here we limit our references to an essential list of the work related to the field of lattice gauge theory whose simulations were done on an actual quantum hardware. This is inevitably naïve list because many works that used the actual quantum hardware benefited from previous studies that focused in encoding the theory on gate-base quantum hardware without using the actual hardware or that used only the quantum simulator, but an interested reader can find more in the references inside the following listed articles.\\
Works on U(1) lattice gauge theory \citep{Klco:2018kyo, deJong:2021wsd}, Transverse Ising model \citep{Gustafson:2019vsd}, recent studies using Z(2) lattice gauge theory \citep{Carena:2022kpg, Pomarico:2023png, Charles:2023zbl}, SU(2) lattice gauge theory \citep{Klco:2019evd, Atas:2021ext, ARahman:2022tkr}, SU(3) lattice gauge theory \citep{Ciavarella:2021nmj, Ciavarella:2021lel,   Farrell:2022wyt, Atas:2022dqm}.

\section{Error mitigation techniques}
The main error sources affecting the IBM hardware are: the qubit interaction with the environment, known as qubit decoherence; the imperfect action of gates, known as gate error; and errors made in reading the qubit state, measurement error.\\
In our case, the study of the time evolution of the theory requires a circuit with a simple structure but it contains many gates, in particular many CNOT gates which are the noisiest, resulting in noisy results, therefore to extract the signal a comprehensive set of error mitigation techniques was used.

\begin{itemize}
\item Mitigation of measurement error:

To mitigate the error that the hardware makes when it reads the state of a qubit a method called mitigation of measurement error \citep{Bravyi_2021} is commonly used and present inside the IBM’s Qiskit library, but in our study we preferred to develop our own implementation. 
\\
In general, for a circuit that uses $n$ qubits, the measurement results can be found in any of the $2^n$ orthogonal states present in the computational basis. 

To quantify the hardware error the method first creates each of these $2^n$ states with a dedicated circuit on the hardware, then it uses the result of their measurement to construct a $2^n\times2^n$ calibration matrix, whose entries are the probabilities that each state once it is measured, has a superposition with another state.\\
In case the hardware is error-free, for example a quantum simulator, the calibration matrix is diagonal, meaning that the states once measured have no superposition with each other. The calibration matrix is then used to correct the readout error present in the measurements of the physics circuit, by effectively rotating the results of a measurement affected by the hardware error to their best possible error reduced version.\\
\\
To better visualize how this process works let’s consider the case of a 2-plaquette lattice, where 2 qubits are needed. The 4 mitigation circuits needed are shown in Figure~\ref{fig:4_circuits_readout}. 
\begin{figure}[H]
\centering
\includegraphics[width=0.9\linewidth]{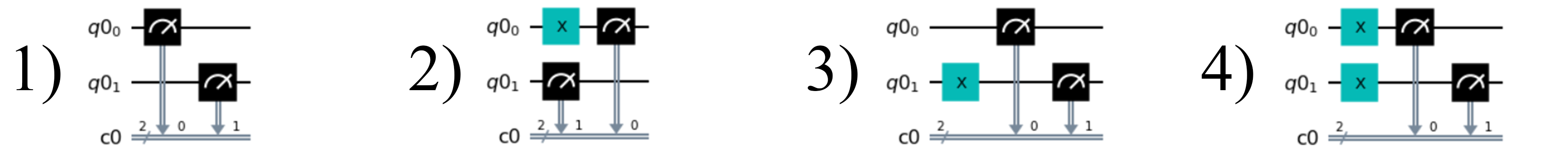}
\caption{The four circuits respectively representing the states $\left|00\right>, \left|01\right>, \left|10\right>, \left|11\right>$ needed by the mitigation of readout error when the circuit considered uses two qubits, as for the 2-plaquette case.}
\label{fig:4_circuits_readout}
\end{figure}
Once these four circuits are measured and their result it is used to create the $4 \times 4$ calibration matrix, this is applied through a linear fitting procedure called sequential least squares programming to mitigate the readout error in the measurements of the physical circuit, by extracting its most probable measurement if the readout error was not present.\\
\end{itemize}
The gate error is more complex to mitigate therefore two error mitigation techniques called randomized compiling and self-mitigation are used and in case extra mitigation is required these two techniques can be combined with zero-noise extrapolation.
\begin{itemize}

\item Randomized compiling also known as Pauli-twirling:\\
This is a technique developed in \citep{Wallman_2016} to transform the CNOT coherent noise into incoherent noise, by preventing the CNOT noise from forming a particular pattern. This is done by creating new physics circuits in which each original CNOT gate is equivalently substituted by a CNOT gate “surrounded” by Pauli and identity gates. It can be shown that there are 16 possible ways to rewrite a CNOT gate as  indicated in Figure~\ref{fig:randomized_c}:
\begin{figure}[H]
\centering
\includegraphics[width=0.9\linewidth]{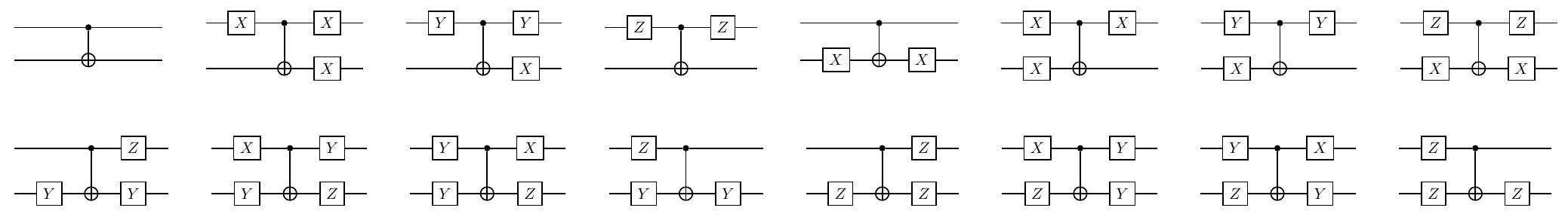}
\caption{The 16 possible circuits equivalent to a single CNOT gate obtained by surrounding a CNOT gate with combinations of Pauli and identity gates.}
\label{fig:randomized_c}
\end{figure}
This means that to actively transform the CNOT noise from coherent to incoherent, the physics circuit has to be run a sufficient number of times so that each CNOT gate present in the original physics circuit can be randomly substituted with as many as possible combinations of the 16 possible identities, this make this method particularly computationally expensive. The final physics measurement is obtained by averaging the measurements of these runs.
\item	Self-mitigation:\label{item_Self_mitigation}\\
This method developed in \cite{Rahman:2022rlg} and inspired by a similar approach present in \cite{Urbanek:2021oej} belongs to a class of error mitigation techniques in which an extra circuit is used to estimate the hardware errors.\\
In these approaches, the error-estimation circuit should be as similar as possible to the gate structure of the physics circuit and should have a known exact result. The estimation of the hardware error from the error estimation circuit is obtained by measuring how far the final state is from the known exact result. This knowledge is then used to mitigate the hardware error affecting the physics circuit by using a proportional relation between the error of the physics circuit and the error-estimation circuit.\\
For the case of time evolution, self-mitigation uses as error-estimation circuit identical to the physics circuit as shown in Figure~\ref{fig:Fig_self_mit}.  
\begin{figure}[H]
\centering
\includegraphics[width=0.6\linewidth]{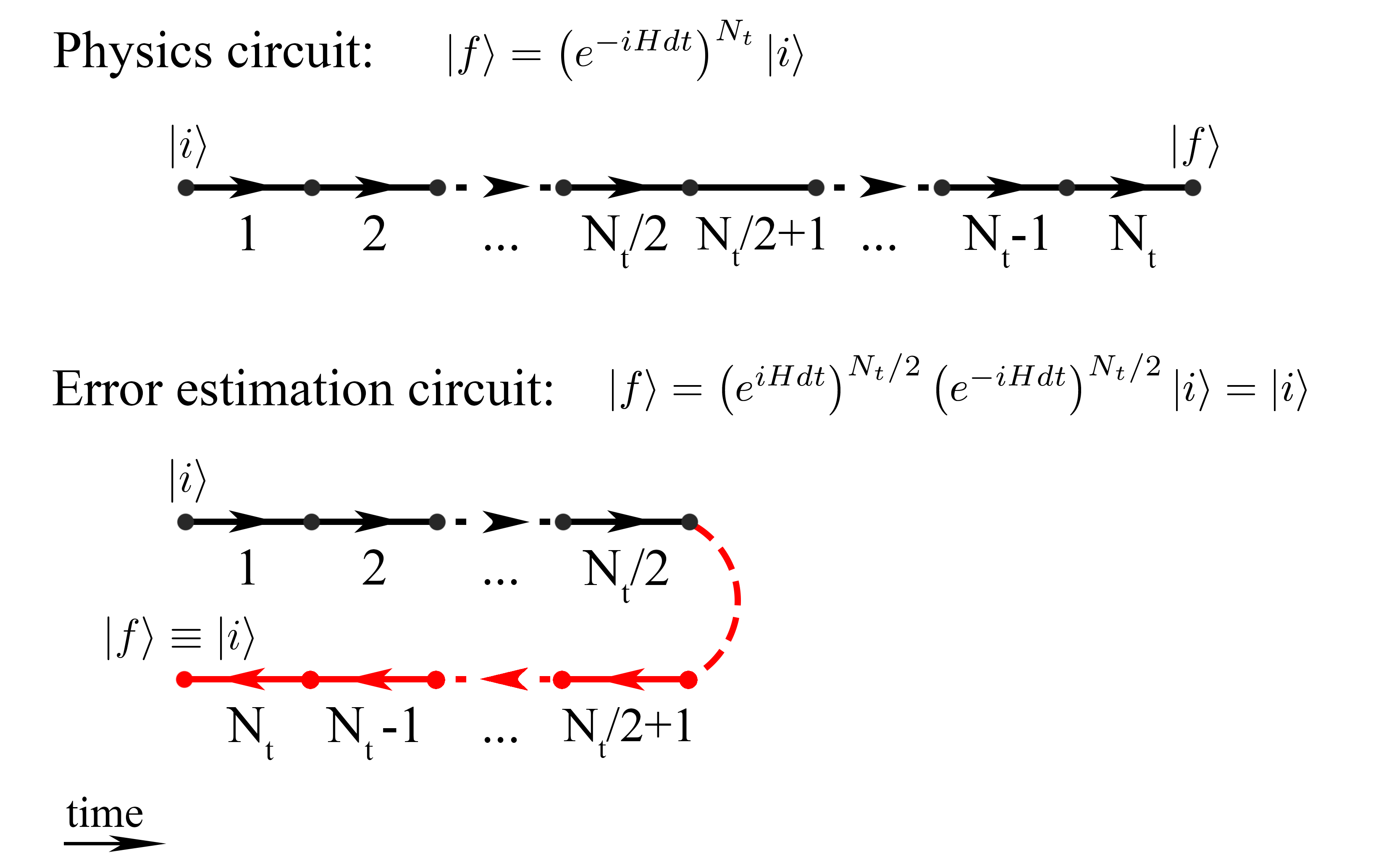}
\caption{Schematic representation of the error-estimation circuit compared to physics time evolution circuit used for self-mitigation. The physics time evolution circuit does  $N$ steps forward in time starting from an initial state $\left|i\right>$ ending to the final state $\left|f\right>$. The error-estimation circuit starts from the same initial state $\left|i\right>$ but it does $N/2$ time steps forward in time followed by $N/2$ steps backward returning to the initial state.}
\label{fig:Fig_self_mit}
\end{figure}
The physics circuit does $N$ time steps forward in time, while the error-estimation circuit does $N/2$ time steps forward and $N/2$ backward, returning to the initial state.
Therefore, the error-estimation circuit has the same gates in the same order and the identical or opposite variable inside the rotation gates, and accordingly it closely reproduces the physics circuit noise.
The hardware error can be estimated in the disparity between the final state of the error-estimation circuit and the chosen initial state.\\
\\
Due to the variability of the hardware the two circuits have to run back-to-back to ensure the hardware noise affecting them is similar. Therefore, the error mitigated final result for the physics circuit is obtained through the self-mitigation equation:
\begin{equation}
\left.\frac{P_{\rm true}-\tfrac{1}{2}}{P_{\rm computed}-\tfrac{1}{2}}\right|_{\rm physics\,run}
=
\left.\frac{P_{\rm true}-\tfrac{1}{2}}{P_{\rm computed}-\tfrac{1}{2}}\right|_{\rm mitigation\,run}
\label{ed:SFM_EQ}
\end{equation}
\item	Zero-noise extrapolation:\\
The method developed by the authors of \citep{Li_Benjamin} studies the circuit noise by creating copies of the original circuits where the noise is artificially increased by replacing each CNOT gate by odd multiples (triplet, quintet etc.) as shown in Fig.~\ref{fig:Cnot_triplet}.

\begin{figure}[H]
\centering
\begin{subfigure}{.5\textwidth}
  \centering
  \includegraphics[width=0.6\linewidth]{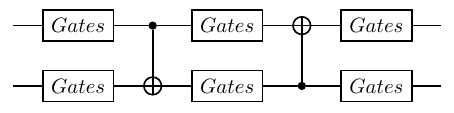}
  \label{fig:sub1}
\end{subfigure}%
\begin{subfigure}{.50\textwidth}
  \centering
  \includegraphics[width=0.9\linewidth]{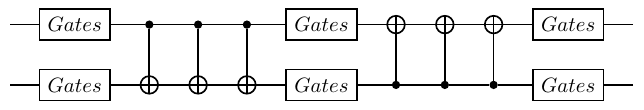}
  \label{fig:sub2}
\end{subfigure}
\caption{Example of how a circuit with artificially increased CNOT noise is created from the original circuit for the Zero-noise extrapolation technique. On the left the original circuit, while on the right the same circuit in which the CNOT gates are replaced by a CNOT triplet.}
\label{fig:Cnot_triplet}
\end{figure}

The zero-noise limit of the original circuit is extracted by fitting the circuits' measurements with a function of the CNOT multiples, as the intercept of the curve with the y-axis.\\
\end{itemize}

\section{SU(2) time evolution on IBM quantum computers}
The SU(2) pure gauge Hamiltonian considered in this chapter is the one presented in Chapter 1:
\begin{equation}\label{eq:H_SU2_Chap3}
\hat H = \frac{g^2}{2}\left(\sum_{i={\rm links}}\hat E_i^2-2x\sum_{i={\rm plaquettes}}\hat\square_i\right)
\end{equation}
The operators present in the Hamiltonian cannot be simply substituted with a combination of gates as can be done when considering an Ising model, therefore to encode the theory on the hardware it is necessary to consider the matrix representation of the systems we are interested in and to rewrite it using the available gates. This is the most general way to encode an operator on the hardware and gives the user the flexibility to choose which gates they want to use.\\

To illustrate how this encoding can be done consider the case of a 2-plaquette lattice with $j_{\rm max}=1/2$ with closed boundary conditions, whose Hamiltonian can be rewritten with the available gates from its matrix representation as:

\begin{eqnarray}\label{eq:2pla_IBM_decoposition}
\frac{2}{g^2}H =
\left(
\begin{array}{cccc}
0 & -2x & -2x & 0 \\
-2x & 3 & 0 & -x \\
-2x & 0 & 3 & -x \\
0 & -x & -x & \frac{9}{2}
\end{array}
\right) &=&
\frac{3}{8}\left(7 I_0 \otimes I_1  - 3Z_0 \otimes I_1 - Z_0 \otimes Z_1 - 3 I_0 \otimes Z_1\right) \nonumber \\
&&- \frac{x}{2}(3 X_0 \otimes  I_1 + X_0 \otimes Z_1) - \frac{x}{2}(3 I_0 \otimes X_1 +Z_0 \otimes X_1 )\nonumber \\
\end{eqnarray}
where the subscript at each gate denotes the qubit the gate acts on. It is fairly common in quantum information theory and in quantum computing literature to omit the Kronecker product symbols $\otimes$ and the identity gate $I$ for question of space.\\
\\
This decomposition in gates is obtained by considering only the matrix representation of the possible combinations of the identity and Pauli gates. For small matrices this can be done using pen and paper but a general way is presented in Appendix~\ref{sec:gate_decomposer}  where a gate decomposing algorithm that uses only the identity and Pauli gates is described in detail.\\
\\
The 4 states used to represent the Hamiltonian:
\begin{figure}[H]
\centering
\includegraphics[width=0.7\linewidth]{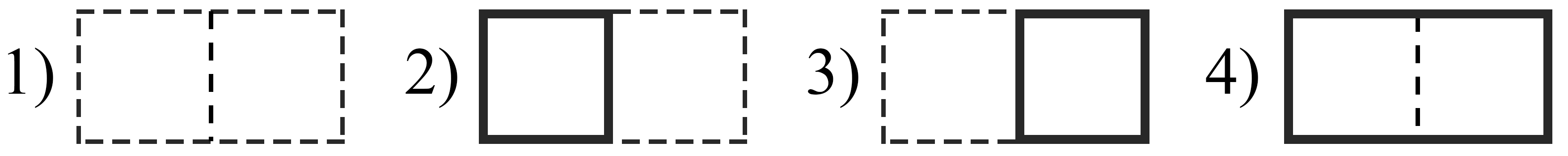}
\caption{The four possible states available on the 2-plaquette lattice with $j_{\rm max}=1/2$. Each solid line indicates a $j=1/2$ flux of energy is present, while the dashed line $j=0$ indicates no energy is present. If all the links of a plaquette have $j=1/2$ we label the plaquette as switched ``on'', otherwise the plaquette is switched ``off''.}
\label{fig:four_states_chap_3}
\end{figure}
These can be coded on the hardware using 2 qubits, one for each plaquette:
\begin{equation}\label{eq:4_states_IBM}
1) \longrightarrow \lvert 0 \rangle \lvert 0 \rangle,\qquad 2) \longrightarrow \lvert 0 \rangle \lvert 1 \rangle, \qquad  3) \longrightarrow \lvert 1 \rangle \lvert 0 \rangle, \qquad 4) \longrightarrow \lvert 1 \rangle \lvert 1 \rangle
\end{equation}
This is a very natural encoding that clearly represents the physics of the $j_{\rm max}=1/2$ gauge truncation with non periodic boundary conditions (NPBC) not only for the two plaquette lattice but for any lattice size. The motivation resides in the fact that with NPBC and for $j_{\rm max}=1/2$ to respect Gauss’ law the links in a plaquette have stringent restriction to their value. In fact, the top and bottom link of a plaquette are always equal with $j=0$ or $1/2$, while each vertical link shared by two plaquettes have a value of $j=0$ if the $j$ flux flowing in the two plaquettes is the same and $j=1/2$ otherwise. \\  
\\
Apart from this particular case, the labelling of the state using qubits is chosen in a completely arbitrary way and does not need to contain any physics meaning. In fact any encoding is acceptable as far as it matches the basis ordering used to represent the Hamiltonian.\\
\\
To access the time evolution of the theory one has to find a way to encode on the hardware the time evolution operator, that with the Hamiltonian rewritten in gates assumes the form:
\begin{equation}
e^{-iHt} = e^{-i\left( \frac{3}{8}\left(7 - 3Z_0 - Z_0Z_1 - 3Z_1\right) - \frac{x}{2}(3+Z_1)X_0 - \frac{x}{2}(3+Z_0)X_1\right) t}
\end{equation}
where for convenience time is in unit of $2/g^2$.\\ 
\\
Nevertheless, even in this new form there is no simple way to directly encode the exponential of gates on the hardware. Hence one possible way is to approximate it by discretizing the time into small time steps $dt$ and use a Suzuki-Trotter expansion \citep{Hatano2005}, keeping in mind that higher orders are more precise with an error scaling with higher power of $dt$ but have the downside of requiring too many gates, therefore a compromise has to  be found between the error introduced by the time discretization and the gate error introduced by a large number of gates.\\
\\
For the case of 2 plaquettes it was particularly convenient to use a second order Suzuki-Trotter expansion compared to a first order because after carefully reordering the gate terms, both orders require the same number of CNOT gates, but the second order is more precise and allows one to use a much larger time step $dt$. This means that we can access a wider time range with a circuit having the same number of CNOT gates; while an order larger than 2 has too many gates making the circuit too noisy for currently available hardware.\\ 
The expression for the second order expansion for $N_t$ time steps with time-step size $dt$ reads:
\begin{equation}\label{eq:2O_ST_expansion}
e^{-iHt} = e^{-i\sum_{j=1}^m H_jt}=\left(  \prod_{j=1}^m  e^{-i H_j \, dt/2} \: \prod_{j=m}^1  e^{-i H_j\, dt/2} \right)^{N_t}    + O\left( m^3 N_t \, dt^3 \right) 
\end{equation}
that in our case leads to:
\begin{eqnarray}
e^{-iHt} &=& e^{i(xt/4)Z_1Y_0}e^{i(3t/16)Z_0Z_1}e^{i(3xt/4)Y_0}e^{i(9t/16)Z_1} e^{i(9t/16)Z_0}e^{i(3xt/4)Y_1}e^{i(xt/4)Z_0Y_1} \nonumber \\
         & &e^{i(xt/4)Z_0Y_1} e^{i(3xt/4)Y_1} e^{i(9t/16)Z_0}e^{i(9t/16)Z_1}e^{i(3xt/4)Y_0} e^{i(3t/16)Z_0Z_1} e^{i(xt/4)Z_1Y_0} \,,        
\end{eqnarray}
where the Hamiltonian’s constant term $\frac{21}{8}$ was dropped because it will cancel out when calculating the probabilities of the time evolution. 
The expression can be expressed using single-qubit rotations and CNOT gates with the help of few a identities derived in Appendix~\ref{sec:exp_gate_identies}:
\begin{eqnarray}
e^{-i\theta Z_j} &=& RZ_j(2\theta)\,, \\
e^{-i\theta Y_j} &=& RY_j(2\theta) \,, \\
e^{-i\theta Z_jZ_k} &=& CX_{jk}RZ_k(2\theta)CX_{jk} \,, \\
e^{-i\theta Z_jY_k} &=& CX_{jk}RY_k(2\theta)CX_{jk} \,,
\end{eqnarray}\label{eq:2_gates_ identities}
\noindent where the CNOT gate for controlling qubit $j$ and target $k$ is labelled as $CX_{jk}$ and the rotation gate acting on the qubit $j$ along the $x$ and $y$ axis as $RX_j$ and $RY_j$ respectively. 
These identities lead to the following gates expression for the single step time evolution operator:
\begin{eqnarray}
&& CX_{10}\, RY_0(-\frac{1}{2}x t)\, RZ_0(-\frac{3}{8} t)\, CX_{10}\, RY_0(-\frac{3}{2}x t)\, RZ_0(-\frac{9}{4} t)\,    RZ_1(-\frac{9}{3} t)\, RY_1(-3 x t)\,    \nonumber \\
&&  CX_{01}\, RY_1(-x t)\, CX_{01}\, RY_0(-\frac{3}{2} x t)\, RZ_1(-\frac{9}{8} t)\, CX_{10}\,  RZ_0(-\frac{3}{8} t)\,  RY_0(-\frac{1}{2} x t)\, CX_{10}            
\end{eqnarray}
whose circuit realization is shown in Figure~\ref{fig:TE_2pla_j_1o2_circuit}:
\begin{figure}[H]
\centering
\includegraphics[width=0.7\linewidth]{ 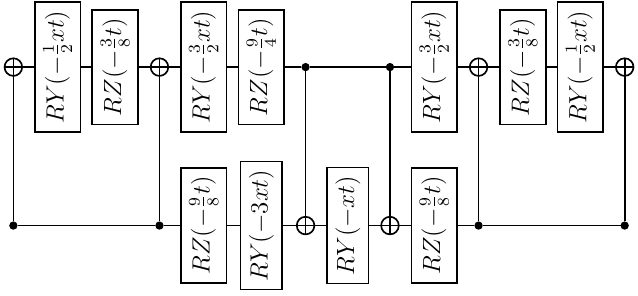}
\caption{Circuit encoding a single second-order Suzuki-Trotter step for the case of 2 plaquette lattice with $j_{\rm max}=1/2$ and non periodic boundary conditions. The top horizontal line represents the qubit 0 while the bottom one the qubit 1, and the gates are applied on the qubits from left to right. 
The circuit can be used for a forward time step with $t=dt$ and for a backward step with $t=-dt$. The circuit has 6 CNOT gates and 11 rotation gates, that are reduced to 4 CNOTs after cancellations with neighbouring steps are considered.}
\label{fig:TE_2pla_j_1o2_circuit}
\end{figure}
Nevertheless the circuit has a reasonably small number of gates, but with the goal of rationalizing the quantum resources the circuit can be optimized by reducing the number of CNOT gates by considering cancellations among the edge CNOT gates from neighbouring Suzuki-Trotter steps leading to only 4 CNOTs per Trotter-Suzuki step, with the only exception for the first and last CNOT of the full circuit. This has the consequence of exposing the rotation gates $RY(-\frac{1}{2} xt)$ that can now be combined into a single rotation with doubled angle $RY(-xt)$. These optimization operations are done inside the code written to submit the circuit to the hardware.\\
\\
The circuit with these extra optimizations can now be used to find the time evolution of the theory given an initial state by simply submitting to the hardware as many copies as needed of the circuit to reach the time range of interest and finally, obtaining the time evolution probabilities by measuring the final qubits state. An clear example is shown in the lower panel of Figure~\ref{fig:TE_2pla_j_1o2_circuit} where each point was obtained  by using the  quantum hardware {\tt ibm\_lagos} with a circuit containing two Trotter-Suzuki steps.
\begin{figure}[H]
\centering
\includegraphics[width=0.95\linewidth]{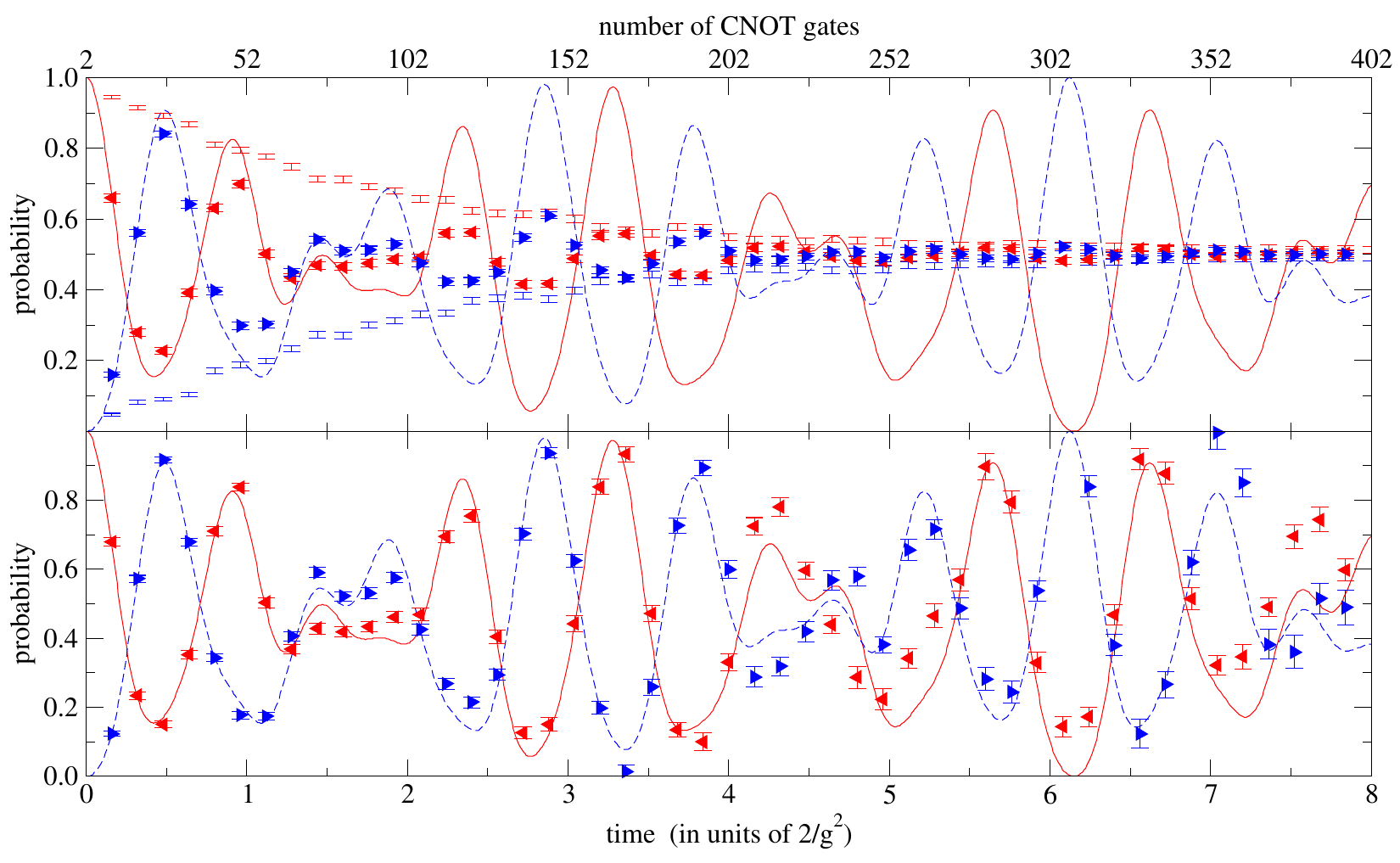}
\caption{Time evolution for the case of a 2-plaquette lattice at gauge coupling $x=2.0$ and time step $dt=0.08$ using the second state as the initial in Figure~\ref{fig:four_states_chap_3} in which the system has a $j=1/2$ flux of energy only in the left plaquette.  
The red solid (blue dashed) curve in both panels is the exact probability of the left (right) plaquette having an energy flux of $j=1/2$ over time. In the upper panel: The red left-pointing (blue right-pointing) triangles are the physics data from runs on the quantum hardware {\tt ibm\_lagos}. The red (blue) error bars without symbols are the error estimation circuit data from runs on {\tt ibm\_lagos}. All the hardware data in the upper panel are error mitigated using mitigation of measurement error and randomized compiling. In the lower panel the triangles are the final results of our error mitigation method that processes the upper panel data through the self-mitigation equation Eq.~\ref{ed:SFM_EQ}. Figure reproduced from Ref.~\citep{ARahman:2022tkr}.}
\label{fig:data_TE_2pla_IBM_self-m}
\end{figure}
It is important to notice that the probability that one plaquette has a flux of $j=1/2$, is the sum of two contributions, one when only that plaquette has a flux of $j=1/2$ and the other when both plaquettes have a flux of $j=1/2$. Therefore the probabilities displayed for the left and the right plaquette have a sum larger than 1.
The data points at each time in Figure~\ref{fig:TE_2pla_j_1o2_circuit} were obtained by submitting a two steps Trotter-Suzuki circuit using the 7 qubits quantum hardware {\tt ibm\_lagos} which allows on each job to run a sequence of 300 circuits back to back each with $10^4$ hits. The measurement protocol used is as follows:\\
\\
Of those 300 circuits, 4 circuits are used for the mitigation of measurement error; 148 circuits for randomized compiling of the error-estimation circuit and 148 for randomized compiling of the physics circuit.\\
Once the measurements of those 300 circuits are collected, the calibration matrix is applied to the measurements of the error-estimation circuits and the measurements of the physics circuits for mitigating the hardware measurement error.\\
The measurement of the physics circuits and the ones of the error-estimation circuits are combined inside the self-mitigation equation to mitigate the gate error present in the measurement of the physics circuits, by considering that the exact known result for the error-estimation circuit for the case considered has probability 1 for the left plaquette and 0 for the right plaquette.\\
Finally, the error bars for the error-estimation and the physics results are calculated as the sum in quadrature of the statistical error from the $10^4$ hits and the error from bootstrapping the data using 1480 samples.\\
\\
The effectiveness of self-mitigation can be appreciated by comparing the data in the upper panel where only mitigation of measurement error and randomized compiling is present with the self-mitigated data in the lower panel. In fact, in the upper panel for time larger than 1 the triangles are far away from the curves representing the exact classical results, and for time larger than 4 the signal looks completely lost into pure noise. By applying self-mitigation, the signal is extracted and it is overall consistent with the exact results even for points coming from circuits with up to 400 CNOT gates. The agreement of the data with the exact result could have been further improved by using zero-noise extrapolation on top of the error mitigation used.\\
\\
In conclusion, this approach can be used to find the time evolution operator given the theory Hamiltonian. Firstly, by decomposing the Hamiltonian in gates and secondly by approximating the time evolution operator by means of only CNOT and single qubit gates, leading to the final optimized time evolution circuit. This circuit can then be executed on a quantum hardware to calculate the time evolution using a few error mitigation techniques.\\
\\
In the next sections we will see how this approach to study time evolution is used to observe a traveling excitation across the 2-plaquette lattice and the time evolution for the case of 5-plaquettes.
\section{Traveling excitation on a two-plaquette lattice}
\begin{figure}[H]
\centering
\includegraphics[width=0.7\linewidth]{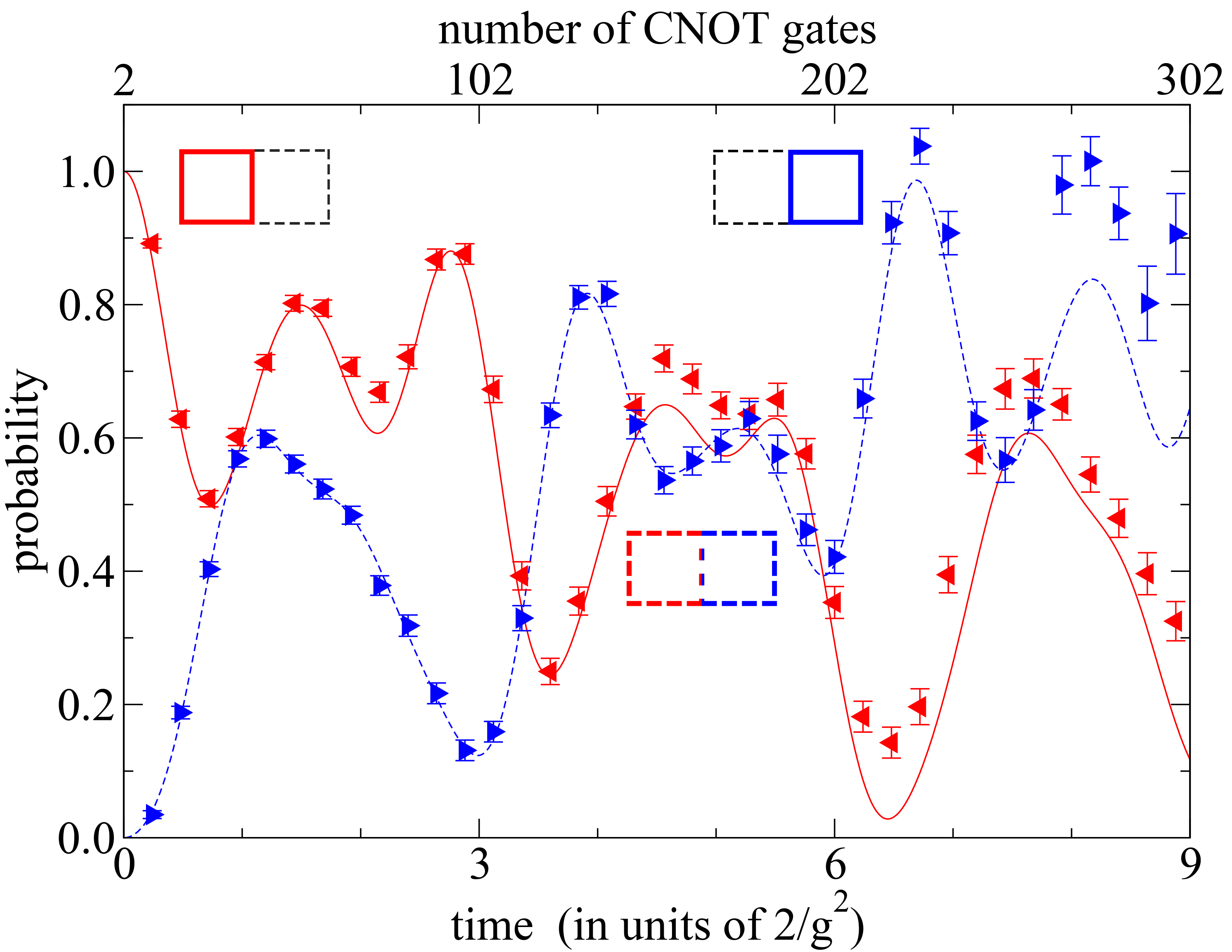}
\caption{Traveling excitation observed from the time evolution for the case of 2-plaquette lattice at gauge coupling $x=0.8$ and time step $dt=0.12$ using as an initial state a $j=1/2$ flux of energy only in the left plaquette.  
The red solid (blue dashed) curve in both panels is the exact probability of the left (right) plaquette having an energy flux of $j=1/2$ over time. The red left-pointing (blue right-pointing) triangles are the physics data from runs on the quantum hardware {\tt ibm\_lagos}. All the data are error mitigated using mitigation of measurement error, randomized compiling and self-mitigation. Figure adapted from Ref.~\citep{ARahman:2022tkr}.}
\label{fig:2pla_Traveling_Excitation_ibm}
\end{figure}

The robustness of the measurement protocol that uses three error mitigation techniques makes it possible to extract small variations among the probabilities giving the possibility to observe a travelling excitation across the 2-plaquette lattice from calculations done on a quantum hardware.\\
The code and the measurement protocol used to create Figure~\ref{fig:data_TE_2pla_IBM_self-m} is identical to the one of Figure~\ref{fig:2pla_Traveling_Excitation_ibm}. The only difference is in the physics parameters. The gauge coupling is now $x=0.8$ necessary so that for the theory to show traveling oscillations and the time step size $dt=0.12$.
It is important to notice that the probability that one plaquette has a flux of $j=1/2$, is the sum of two contributions, one when only that plaquette has a flux of $j=1/2$ and the other when both plaquettes have a flux of $j=1/2$. Therefore the probabilities displayed for the left and the right plaquette have a sum larger than 1.\\
\\
The agreement of the mitigated data with the exact classical result is remarkable up to $t=4$ where up to 100 CNOT gates are present. For larger time, when the number of CNOTs get much larger the discrepancy is particularly pronounced for the right plaquette. The extra systematic error present in the right part of the plot at large time can be in principle improved by using zero-noise extrapolation, but the large quantity of CNOT gates already present in the time evolution circuits make it not possible to adds extra CNOTs on the current hardware.\\
\\ 
Overall, the physics phenomenon is well distinguishable, the system is initially prepared with the left plaquette switched on (red triangles) and the right one switched off. With time the left plaquette probability drops, equating the right plaquette probability around the time range $4-5$ and after that it further drops close to zero, while the probability of the right plaquette rises close to 1. A similar study for a compact $U(1)$ lattice gauge theory was previously done in \citep{Lewis:2019wfx}, where the propagation and collision of travelling excitation real-time were observed with calculations coming from a quantum simulator.\\
In conclusion, this calculation clearly shows that the available IBM quantum hardware can be used to investigate interesting non-perturbative phenomena on a small lattice, and clearly indicates the future possibility of quantum hardware with larger numbers of qubits to investigate the same phenomena on larger lattices which a classical computer is not able to do, due to the exponentially large number of resources necessary.    
\section{Five-plaquette lattice with $j_{\rm max}=1/2$}
\begin{figure}[H]
\centering
\includegraphics[width=0.9\linewidth]{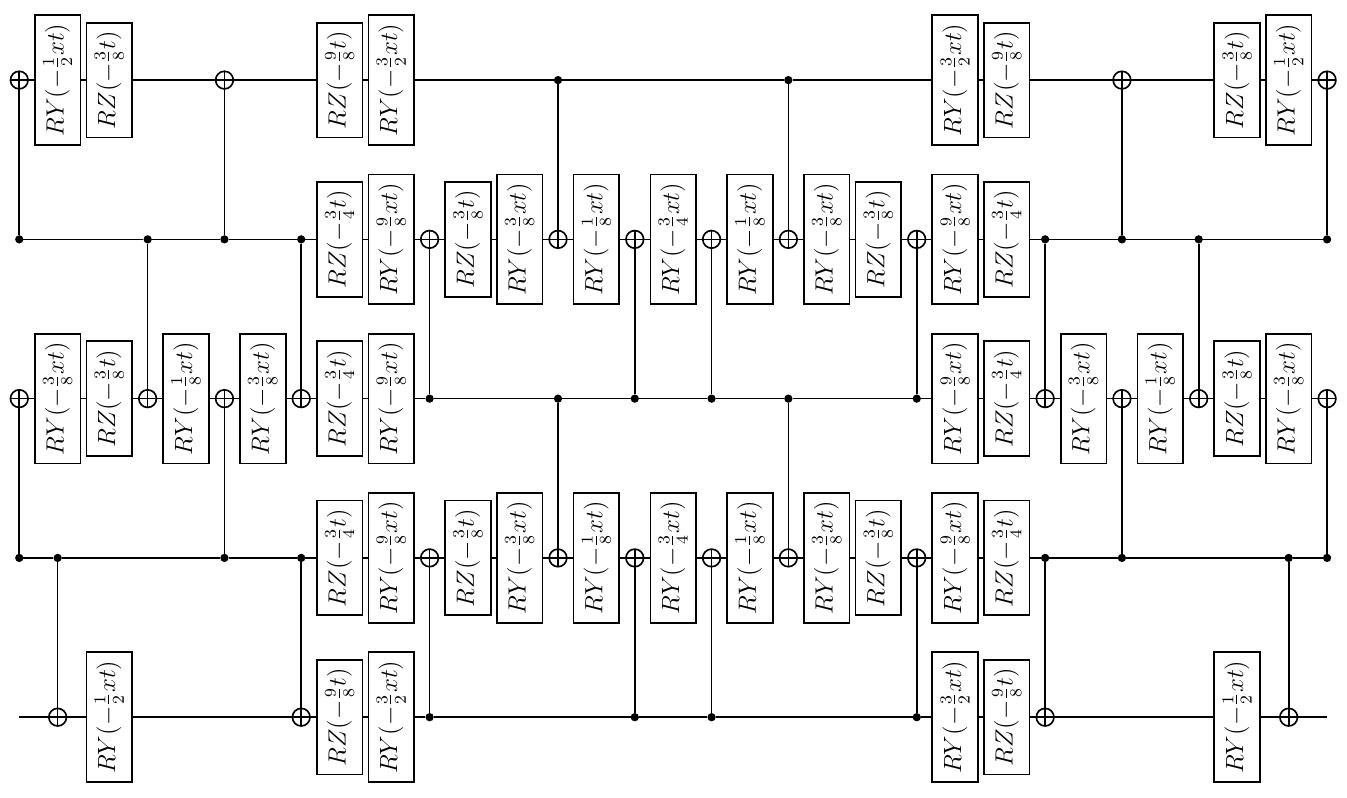}
\caption{Circuit encoding a single second-order Suzuki-Trotter step for the case of 5-plaquette lattice with $j_{\rm max}=1/2$ and non periodic boundary conditions. Each horizontal line represents a qubit, from top to bottom from qubit 0 to qubit 4 and, the operators are applied on the qubits from left to right. 
The circuit can be used for a forward time step with $t=dt$ and for a backward step with $t=-dt$. The circuit has 28 CNOT gates and 48 rotation gates, that are reduced to 22 CNOTs after cancellations with neighbouring steps are considered.}
\label{fig:TE_circuit_5pla}
\end{figure}

\begin{figure}[H]
\centering
\includegraphics[width=0.8\linewidth]{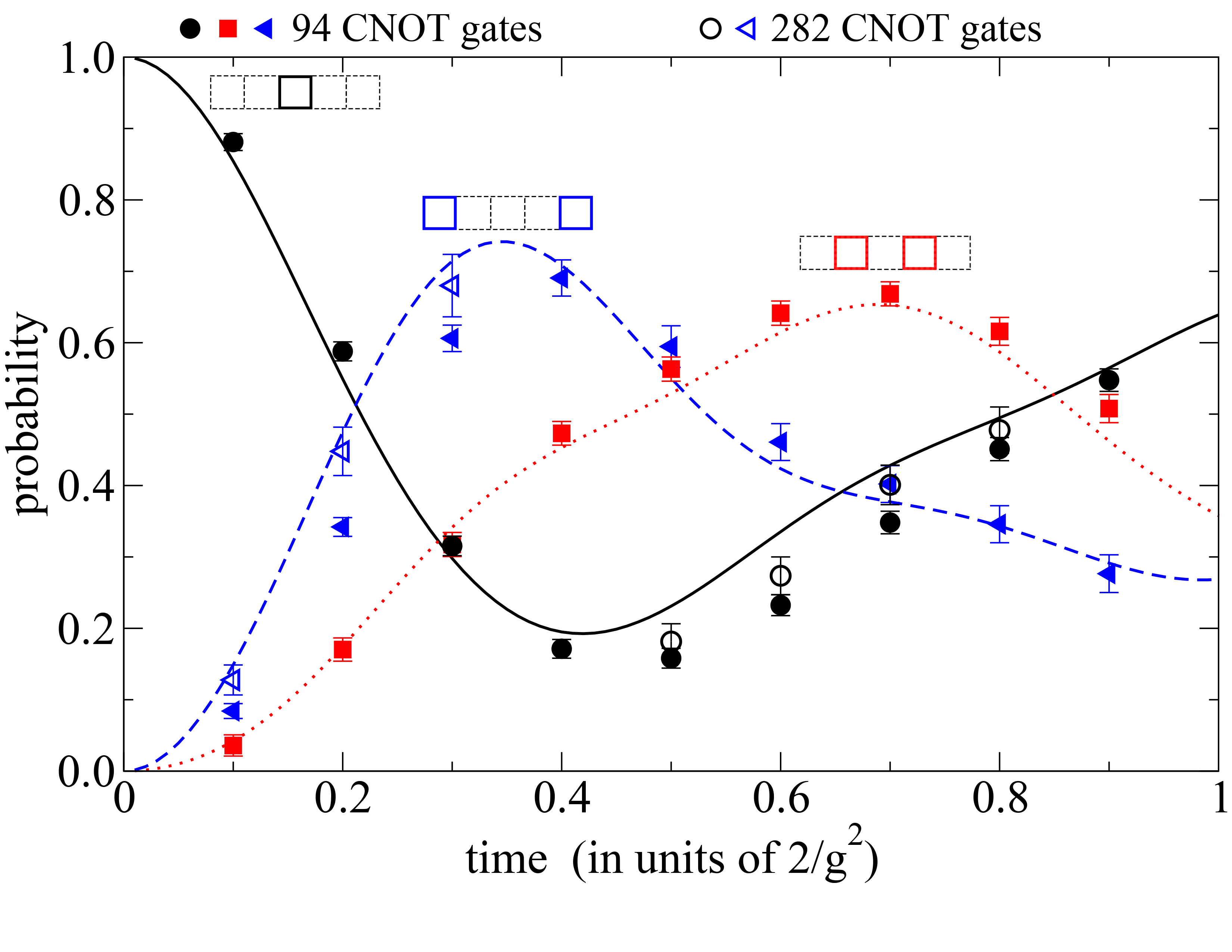}
\caption{Time evolution for the case of a 5-plaquette lattice with $j_{\rm max}=1/2$ at gauge coupling $x=2.0$, for only the centre plaquette with a flux of energy $j=1/2$ (switched on) as initial state, obtained by 4 Trotter-Suzuki steps per point at different values of $dt$. The black solid, blue dashed, and the red dotted curves are respectively the exact probability that the flux of energy $j=1/2$ is in the centre plaquette, the two neighbours to the centre plaquette and the two edge plaquettes. The solid black circle, blue triangle, and red square are the corresponding data from the quantum hardware {\tt ibm\_lagos} after self-mitigation.  The open circle, triangle, and square are quantum data from {\tt ibm\_lagos} after using both self-mitigation and zero-noise extrapolation. There are 94 CNOTs in each filled symbol while 282 in the empty symbols. Figure adapted from Ref.~\citep{ARahman:2022tkr}.}\label{fig:TE_5plagate}
\end{figure}

In the standard study of lattice gauge theory using simulations a crucial role is played by the lattice size, in fact the larger the lattice size the smaller the physics scale at which the physics is resolved, opening the possibility to measure quantities that can be extrapolated to the continuum limit and be relevant for comparison with experiments. Therefore, it is significant to explore the possibility to use a longer lattice and investigate the feasibility of the current quantum hardware.\\
\\
Inspired by the possibility of doing future studies on significant, large lattices, the Hamiltonian for a two-plaquette lattice with $j_{\rm max}=1/2$ presented in Equation~\ref{eq:2pla_IBM_decoposition} can be generalized to the case of N-plaquette with $j_{\rm max}=1/2$ as:
\begin{eqnarray}
H &=& \frac{g^2}{2}(h_E + h_B) \,, \\
\nonumber \\
h_E &=& \frac{3}{8}(3N+1) - \frac{9}{8}(Z_0+Z_{N-1}) - \frac{3}{4}\sum_{n=1}^{N-2}Z_n \nonumber \\
&& - \frac{3}{8}\sum_{n=0}^{N-2}Z_nZ_{n+1} \,, \\
\nonumber \\
h_B &=& -\frac{x}{2}(3+Z_1)X_0 - \frac{x}{2}(3+Z_{N-2})X_{N-1} \nonumber \\
&& - \frac{x}{8}\sum_{n=1}^{N-2}(9+3Z_{n-1}+3Z_{n+1}+Z_{n-1}Z_{n+1})X_n \,, \\
\nonumber
\label{eq:H5B}
\end{eqnarray}
which requires $N$ qubits to encode the $2^N$ physics states, clearly showing a polynomial scaling with the number of plaquettes on a quantum computer, instead of the $2^N$ on a classical computer.\\  
\\
With the present NISQ hardware, we considered the 5-plaquette case because on the 7 qubits hardware {\tt ibm\_lagos}, the longest qubit chain with nearest-neighbour connectivity has 5 qubits. This let us avoid the use of swap gates that are noisy while using 1 qubit for each plaquette, therefore the largest implementable lattice has 5 plaquettes.\\
\\
The time evolution operator can be constructed following the same approach used for the case of 2-plaquette lattice, by using the second order Trotter-Suzuki expansion in Equation~\ref{eq:2O_ST_expansion} and in addition to the identities presented in Equation~\ref{eq:2_gates_ identities}  a new one that contains the exponential of  three gates:
\begin{equation}
e^{-i\theta Z_jY_kZ_l} = CX_{lk}CX_{jk}RY_k(2\theta)CX_{jk}CX_{lk} \,.
\end{equation}
The time evolution operator for a single second order Trotter-Suzuki step displayed in Figure~\ref{fig:TE_circuit_5pla} has 28 CNOT gates that are further reduced to 22 after cancellations with neighbour steps, with the only exception for the first and last CNOTs of the full circuit. Consequently, three different pairs of rotation gates are combined into three rotation gates with a doubled angle. These optimizations are done inside the 5-plaquette code written to submit the circuit to the hardware, and whose structure is similar to the one used for the 2-plaquette case that can be found here \cite{duet_code}.
The time evolution circuit can be submitted to the hardware to study how a chosen initial state evolves in real-time as shown in Figure~\ref{fig:TE_5plagate}, where each point represents the time evolution of four Trotter-Suzuki steps each for different step sizes $dt$, obtained by averaging the result from 4 different jobs which showed variation within the statistical error.\\
\\
Similarly, to the measuring protocol of the 2 plaquette case, each job submitted to the hardware {\tt ibm\_lagos} by the 5-plaquette code, has a list with 300 circuits each executed $10^4$ times. Of those 300 circuits, 32 circuits are used for the mitigation of measurement-error, and the remaining 268 circuits are divided between the error estimation circuits necessary for randomize compiling and self-mitigation, and the physics calculation. Hence, the total statistics of each point is slightly larger than $5\times 10^5$.
The data labelled by the solid symbols were obtained by using three error mitigation techniques, mitigation of measurement error, randomize compiling and self-mitigation. The data labelled with the open symbols were obtained by using zero-noise extrapolation on top of the three error mitigation techniques with CNOTs replaced by triplets and performing a linear extrapolation.\\
\\
Looking to the data displayed in Figure~\ref{fig:TE_5plagate} we can notice that the quality of the agreement between the data and the exact curve is not the same for all the probabilities considered. The probability for the two neighbours to the centre plaquette, solid red squares, agrees with the exact result. The probability for the center plaquette, solid black circles, is consistent with the exact curve except for the time windows $0.4-0.8$. Similar behaviour is seen for the probability of the two edge plaquettes, solid blue triangles, with discrepancy in the time range $0.1-0.3$. For these two probabilities their discrepancy with the exact curve can be significantly reduced by using the zero-noise extrapolation, as clearly shown by the open symbols.\\
Motivated by this last improvement one could be tempted to consider a zero-noise extrapolation using higher-orders like quintet of CNOTs or even more, but this would create for the case of quintet a circuit with 470 CNOT gates, which will end up being too noisy making it useless to mitigate the hardware errors.\\
\\
Finally, it is important to remark that the result presented cannot be described as a travelling oscillation because at gauge coupling $x=2.0$ the system is in a regime in which the chromomagnetic term dominates mixing the states used to represent the Hamiltonian. We could have used the 5-plaquette lattice to observe a travelling oscillation at a small value of the gauge coupling $x$, but the phenomena happens at a much larger time than the 2-plaquette case. That would require a large number of Trotter-Suzuki steps, and the present noisy hardware  makes it unobservable.  
\section{Two-plaquette lattice with $j_{\rm max}>1/2$}
The first natural step to extend the previous works is to investigate the theory on 2-plaquettes with a larger energy truncation ($j_{\rm max}>1/2$).
Considering a larger truncation is motivated by the fact that the theory present in chapter~\ref{chap:Chapter_1} has an infinite dimensional Hilbert space, therefore using an angular momentum basis the full theory is reachable only by considering larger and larger truncations.
From the pure practical point of view, a larger truncation allows more states and therefore the time evolution processes that can be studied are much wider.\\
The encoding strategy used so far is to rewrite the Hamiltonian in gates and represent the states by $n$ qubits.\\ 
In this respect it is informative to consider how many states there are for each value of $j_{\rm max}$, and the corresponding number of qubits necessary for encoding the theory on a quantum hardware. The number of states for each value of $j_{\rm max}$ can be obtained using the algorithm presented in Appendix~\ref{appendix:state_algorithm}, and the number of qubits is the first integer greater than $\log_{2}(N^{°} states)$. These values are collected in Table~\ref{Tab:jmax_nstate}:
\begin{table}[H]
\centering
\begin{tabular}[t]{ccc}
\hline
\hline
$j_{\rm max}$	& $N°$ states	&  Min $N°$ qubits \\
\hline
1/2		&	4		&2\\
1		&	11		&4\\
3/2		&	23		&5\\
2		&	42		&6\\
\hline
\hline
\end{tabular}
\caption{Number of states and the minimum number of qubits necessary for each value of $j_{\rm max}$ for the 2-plaquette case with non periodic boundary conditions.}
\label{Tab:jmax_nstate}
\end{table}
There are no conceptual limitations to the value of $j_{\rm max}$ to choose, but in practice for a successful simulation of the system physics the $j_{\rm max}$ value is constrained by the quantum hardware number of qubits and its level of internal noise.
Considering the performance of the quantum hardware used so far, a wiser choice is to consider a Hamiltonian with a size of few dozen and maximum $5-6$ qubits to represent the states.\\
In planning the simulation it is important to attempt an efficient use of the number of qubits too. Since each qubit can represent 2 states, $n$ qubits can represent up to $2^n$ states. It is therefore reasonable to consider a truncation of the theory that has a number of states equal to $2^n$.\\
As we can see form the Table~\ref{Tab:jmax_nstate} only for the case $j_{\rm max}=1/2$ the number of states coincides with the maximum number of states representable with $N$ qubits. In all the other cases the number of states is smaller than the number of possible states representable with $N$ qubits. This means that there is a waste of resources and the presence of unphysical states in the system.\\
The presence of extra unphysical states is not a problem, but during the measurement process these states have to be recognized and manually eliminated. Furthermore, since these states should not be present, their measurement should always result in a zero count on an error-free hardware, therefore on the current NISQ hardware these states can be used as an error estimator and possibly as a source on which to build error mitigation techniques.\\
\\
A more elegant solution is to consider a truncation of the Hamiltonian that is not strictly bounded to a particular value of $j_{\rm max}$ and has a number of states exactly representable with $n$ qubits. To be more clear, if we are interested in a particular set of states present only to a certain value of $j_{\rm max}$, we can consider a larger truncation of the theory that contains these particular states plus other high energy states and at the same time it is representable using exactly $n$ qubits. This choice is more than reasonable even if it contains extra states with larger energies compared to the ones we are interested in. The reason is that the theory per se (by itself) is not organized by the value of $j_{\rm max}$, this distinction was introduced to better describe and understand the theory.\\
Finally, in the process of measuring the states of the qubits we are free to consider only the state transitions that are of our interest, exactly the same as we would do with a theory described by a finite-dimensional representation.

\subsection{Two-plaquette using a $16 \times 16$ truncation of the $j_{\rm max}=3/2$ case}
The first step beyond the $j_{\rm max}>1/2$ case is to consider the case $j_{\rm max}=1$ which has in total of 11 states of which 4 states are from the case $j_{\rm max}=1/2$. Using 11 states makes the encoding in qubits quite inefficient because 4 qubits can represent up to 16 states, therefore a wise decision is to have 16 states by including the 5 lowest energy states from the case $j_{\rm max}=3/2$. This means that we are considering a $16 \times 16$ truncation of the $j_{\rm max}=3/2$ case.\\
\\
The 16 states used to represent the 16$\times$16 truncation of the Hamiltonian are:
\begin{figure}[H]
\centering
\includegraphics[width=1.0\linewidth]{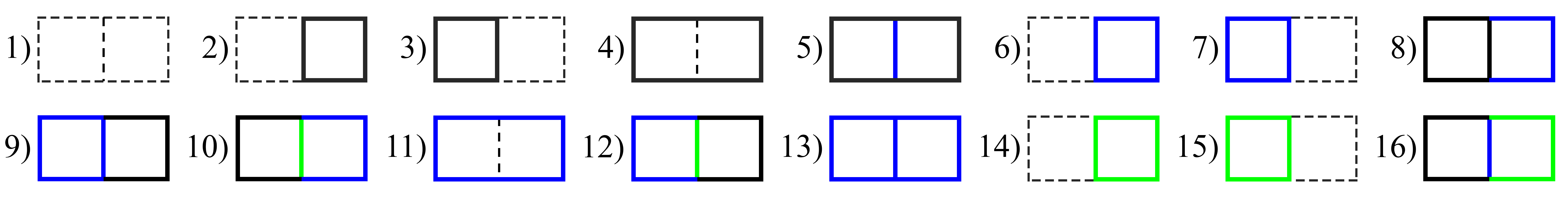}
\caption{The 16 states available on the 2-plaquette lattice after a 16$\times$16 truncation of the theory Hamiltonian for the case $j_{\rm max}=3/2$.
The states are ordered by growing chromoelectric energy. The color code is: dashed line for $j=0$, black solid line for $j=1/2$, blue solid line for $j=1$ and green solid line for $j=3/2$.}
\label{fig:four_states}
\end{figure}

\noindent and the Hamiltonian representation is:

\begin{equation}\label{eq:H_16x16}
\resizebox{0.90\hsize}{!}{%
$
H =\frac{g^2}{2}
\left(
  \begin{smallmatrix}
   \vrule width 0pt height 0pt
 0 & -2 x & -2 x & 0 & 0 & 0 & 0 & 0 & 0 & 0 & 0 & 0 & 0 & 0 & 0 & 0 \\
 -2 x & 3 & 0 & -x & -\sqrt{3} x & -2 x & 0 & 0 & 0 & 0 & 0 & 0 & 0 & 0 & 0 & 0 \\
 -2 x & 0 & 3 & -x & -\sqrt{3} x & 0 & -2 x & 0 & 0 & 0 & 0 & 0 & 0 & 0 & 0 & 0 \\
 0 & -x & -x & \tfrac{9}{2} & 0 & 0 & 0 & -\sqrt{3} x & -\sqrt{3} x & 0 & 0 & 0 & 0
   & 0 & 0 & 0 \\
 0 & -\sqrt{3} x & -\sqrt{3} x & 0 & \tfrac{13}{2} & 0 & 0 & -\tfrac{x}{3} &
   -\tfrac{x}{3} & -\tfrac{4 \sqrt{2} x}{3} & 0 & -\tfrac{4 \sqrt{2} x}{3} & 0 & 0 &
   0 & 0 \\
 0 & -2 x & 0 & 0 & 0 & 8 & 0 & -\tfrac{2 x}{\sqrt{3}} & 0 & -2 \sqrt{\tfrac{2}{3}}
   x & 0 & 0 & 0 & -2 x & 0 & 0 \\
 0 & 0 & -2 x & 0 & 0 & 0 & 8 & 0 & -\tfrac{2 x}{\sqrt{3}} & 0 & 0 & -2
   \sqrt{\tfrac{2}{3}} x & 0 & 0 & -2 x & 0 \\
 0 & 0 & 0 & -\sqrt{3} x & -\tfrac{x}{3} & -\tfrac{2 x}{\sqrt{3}} & 0 & 9 & 0 & 0 &
   -\tfrac{2 x}{\sqrt{3}} & 0 & -\tfrac{4 x}{3} & 0 & 0 & -\tfrac{4 \sqrt{2} x}{3} \\
 0 & 0 & 0 & -\sqrt{3} x & -\tfrac{x}{3} & 0 & -\tfrac{2 x}{\sqrt{3}} & 0 & 9 & 0 &
   -\tfrac{2 x}{\sqrt{3}} & 0 & -\tfrac{4 x}{3} & 0 & 0 & 0 \\
 0 & 0 & 0 & 0 & -\tfrac{4 \sqrt{2} x}{3} & -2 \sqrt{\tfrac{2}{3}} x & 0 & 0 & 0 &
   12 & 0 & 0 & -\tfrac{\sqrt{2} x}{3} & 0 & 0 & -\tfrac{x}{6} \\
 0 & 0 & 0 & 0 & 0 & 0 & 0 & -\tfrac{2 x}{\sqrt{3}} & -\tfrac{2 x}{\sqrt{3}} & 0 &
   12 & 0 & 0 & 0 & 0 & 0 \\
 0 & 0 & 0 & 0 & -\tfrac{4 \sqrt{2} x}{3} & 0 & -2 \sqrt{\tfrac{2}{3}} x & 0 & 0 & 0
   & 0 & 12 & -\tfrac{\sqrt{2} x}{3} & 0 & 0 & 0 \\
 0 & 0 & 0 & 0 & 0 & 0 & 0 & -\tfrac{4 x}{3} & -\tfrac{4 x}{3} & -\tfrac{\sqrt{2}
   x}{3} & 0 & -\tfrac{\sqrt{2} x}{3} & 14 & 0 & 0 & 0 \\
 0 & 0 & 0 & 0 & 0 & -2 x & 0 & 0 & 0 & 0 & 0 & 0 & 0 & 15 & 0 &
   -\sqrt{\tfrac{3}{2}} x \\
 0 & 0 & 0 & 0 & 0 & 0 & -2 x & 0 & 0 & 0 & 0 & 0 & 0 & 0 & 15 & 0 \\
 0 & 0 & 0 & 0 & 0 & 0 & 0 & -\tfrac{4 \sqrt{2} x}{3} & 0 & -\tfrac{x}{6} & 0 & 0 &
   0 & -\sqrt{\tfrac{3}{2}} x & 0 & \tfrac{31}{2}
       \vrule width 0pt depth 0pt
  \end{smallmatrix} \right)
$%
}
\end{equation}

The large size of the Hamiltonian forced us to develop an analytical algorithm to express any Hamiltonian of even size using only Pauli gates and the identity $(X,Y,Z,I)$. The decomposition is unique for each particular representation of the Hamiltonian, but by permuting the order of the states used to represent the Hamiltonian, different gate decompositions can be found creating the possibility to choose the one with the smallest number of gate terms. The method is described in Appendix~\ref{sec:gate_decomposer}.\\
\\
For this Hamiltonian with 16 states, there are $(16!)$ permutations that cannot be accessed with a reasonable amount of time. A possible heuristic solution is to randomly access some of them and find one that is decomposable with a reasonably small number of gates.
The optimal Hamiltonian found so far requires 68 gate-terms and it should be one of the matrix representations with the lowest number of gate-terms, since it requires less gate-terms than the number of non-zero elements in the matrix.\\
Using the gate decomposition method introduced in Appendix~\ref{sec:gate_decomposer} the optimal Hamiltonian written in gates is:
\begin{eqnarray}
H &=& \frac{g^2}{2}(h_E + h_B) \,, \\
\nonumber \\
h_E &=& \frac{1}{32} (293 \, I I I I + 31\, I I I Z -3\, I I Z I -13\, I I Z Z -13\, I Z I I +21\, I Z I Z \nonumber \\
					&& - 65\, I Z Z I -19\, I Z Z Z - 67\, Z I I I + 47\, Z I I Z +49\, Z I Z I + 31\, Z I Z Z \nonumber \\
					&& +31\, Z Z I I +25\, Z Z I Z -49\, Z Z Z I - 43\, Z Z Z Z)
\,, \\
\nonumber \\
h_B &=& x\,\Big( -\frac{17}{24} \, I I I X -\frac{9+5\sqrt{2}+2\sqrt{3}}{24}\, I I X I - \frac{-9+3\sqrt{2}-2\sqrt{3}}{24}\, I I X Z - \frac{5}{24}\, I I Z X   \nonumber \\
&&-\frac{29}{48}\, I X I I -\frac{1}{48}\, I X I Z -\frac{12+9\sqrt{2}+2\sqrt{3}}{24}\, I X X X -\frac{12+\sqrt{2}-2\sqrt{3}}{24}\, I X Y Y  \nonumber \\
&&-\frac{5}{16}\, I X Z I -\frac{11}{48}\, I X Z Z -\frac{-\sqrt{2}+2\sqrt{3}}{24}\, I Y X Y -\frac{-7\sqrt{2}+2\sqrt{3}}{24}\, I Y Y X +\frac{1}{8}\, I Z I X
\nonumber \\
&& +\frac{-3+3\sqrt{2}+2\sqrt{3}}{24}\, I Z X I +\frac{3+5\sqrt{2}-2\sqrt{3}}{24}\, I Z X Z -\frac{3}{8}\, I Z Z X  -\frac{7}{8\sqrt{3}}\, X I I X
\nonumber \\
 &&-\frac{-6\sqrt{3}+3\sqrt{6}}{48}\, X I X Z -\frac{7}{8\sqrt{3}}\, X I Z X -\frac{\sqrt{3}}{4}\, X X X X -\frac{\sqrt{3}}{4}\, X X Y Y -\frac{\sqrt{3}}{8}\, X Z I X \nonumber \\
&&-\frac{6\sqrt{3}-3\sqrt{6}}{48}\, X Z X I -\frac{-6\sqrt{3}+5\sqrt{6}}{48}\, X Z X Z -\frac{\sqrt{3}}{8}\, X Z Z X -\frac{\sqrt{3}}{8}\, Y I I Y
\nonumber \\
&& -\frac{-6\sqrt{3}-11\sqrt{6}}{48}\, Y I Y I -\frac{-6\sqrt{3}-3\sqrt{6}}{48}\, Y I Y Z -\frac{\sqrt{3}}{8}\, Y I Z Y -\frac{\sqrt{3}}{4}\, Y X X Y 
\nonumber \\
&&-\frac{\sqrt{3}}{4}\, Y X Y X  +\frac{1}{8\sqrt{3}}\, Y Z I Y -\frac{6\sqrt{3}+3\sqrt{6}}{48}\, Y Z Y I -\frac{-6\sqrt{3}-5\sqrt{6}}{48}\, Y Z Y Z 
\nonumber \\
&&+\frac{1}{8\sqrt{3}}\, Y Z Z Y -\frac{7}{24}\, Z I I X -\frac{9-5\sqrt{2}-2\sqrt{3}}{24}\, Z I X I -\frac{-9-3\sqrt{2}+2\sqrt{3}}{24}\, Z I X Z
\nonumber \\
&&+\frac{5}{24}\, Z I Z X -\frac{7}{48}\, Z X I I -\frac{11}{48}\, Z X I Z +\frac{12-9\sqrt{2}-2\sqrt{3}}{24}\, Z X X X 
\nonumber \\
&&-\frac{12-\sqrt{2}+2\sqrt{3}}{24}\, Z X Y Y -\frac{1}{16}\, Z X Z I -\frac{25}{48}\, Z X Z Z -\frac{\sqrt{2}-2\sqrt{3}}{24}\, Z Y X Y 
\nonumber \\
&&-\frac{7\sqrt{2}-2\sqrt{3}}{24}\, Z Y Y X -\frac{1}{8}\, Z Z I X -\frac{-3-3\sqrt{2}-2\sqrt{3}}{24}\, Z Z X I 
\nonumber \\
&&-\frac{3-5\sqrt{2}+2\sqrt{3}}{24}\, Z Z X Z -\frac{5}{8}\, Z Z Z X
\Big)\,,
\nonumber \\ \label{eq:h16x161gates}
\end{eqnarray}
where the operation among the gates is not a standard matrix multiplication but the Kronecker product, whose symbol $\otimes$ was omitted for reasons of space. This is a fairly common notation in quantum information theory and in quantum computing literature, (i.e. $Y Z I Y \longrightarrow Y\otimes Z\otimes I\otimes Y$).\\
The gate decomposition in Equation~\ref{eq:h16x161gates} is the gate decomposition of the optimal Hamiltonian representation obtained by exchanging rows and columns in the original Hamiltonian in Equation~\ref{eq:H_16x16} as: 
\begin{equation*}
\scalebox{0.89}{
$
(1,2,3,4,5,6,7,8,9,10,11,12,13,14,15,16) \longmapsto (6,2,8,4,9,1,6,10,13,11,12,16,14,5,3,15)
$}
\end{equation*}

The gate decomposition in Equation~\ref{eq:h16x161gates} is the one used to calculate the time evolution operator, and due to the large number of gates present only the first order Suzuki-Trotter expansion \citep{Hatano2005}:

\begin{equation*}\label{eq:ST_expansion}
e^{-iHt} = e^{-i\sum_{j=1}^m H_jt}=\left(  \prod_{j=1}^m  e^{-i H_j \, dt} \right)^{N_t}  + O\left( m^2 N_t \, dt^2 \right) 
\end{equation*}
where $N_t$ is the number of time steps and $dt$ the time-step size.
The time evolution operator circuit is made by the union of the circuits obtained by exponentiating each of the 68 gate terms necessary to decompose the optimal Hamiltonian.\\
This was done with using the procedure described in \ref{sec:exp_gate_identies}.\\
Since the CNOT gates are particularly noisy, these 68 circuits have been reorganized to minimize the number of CNOT gates present by cancellation, leading to what we call the optimal time evolution circuit.
The following table summarizes the optimization of the final circuits done by reorganizing the 68 circuits:
\begin{table}[htb]  
\centering
\begin{tabular}{ccc}
\hline
\hline
 Gate	 & Original & Optimal \\ \hline
 CNOT   & 234   & 112 \\ 
 RX RZ  & 67    & 67 \\ 
 H 	 & 114   & 52  \\ 
 S		 & 66  	 & 34\\
\hline
\hline
\end{tabular}
\caption{Table showing the different number of gates necessary for encoding the time evolution circuit for  the 2-plaquette lattice using the $16 \times 16$ truncation with $j_{max}=3/2$. 
In the left column there is the number of gates needed for each gate type without reorganizing the 68 circuits, while in the right column the number of gates for the optimal circuit obtained after reordering the terms.}\label{table:gamma_over_nucolor}
\end{table}

The corresponding time evolution circuit of a single first order Suzuki-Trotter step is organized in 14 distinct circuits:
\begin{equation}
\scalebox{0.89}{
$
T_0 \, W_1 \, W_2 \, W_4 \, W_5 \, T_{12} \, T_{17} \, T_{15} \, T_{18} \, T_{20} \, SWAP_{0,1} \, W_3 \, T_{21} \, T_{7} \, SWAP_{1,2} \,  T_{19} \,  SWAP_{2,1} \,  SWAP_{1,0}
$}
\end{equation}
\\
where each circuit labelled with $T$ and $W$ can be found in Appendix~\ref{sec:2pla_circuit16x16}, while the three $\rm SWAP$ gates are necessary to deal with the qubit connectivity required by the circuit.
\\
There are many interesting time evolution processes like multiple transitions among states, traveling excitations on the lattice, but unfortunately many of them require a large time span to be see for example Fig.~\ref{fig:Trave_exi}:
\begin{figure}[H]
\centering
\includegraphics[width=1.0\linewidth]{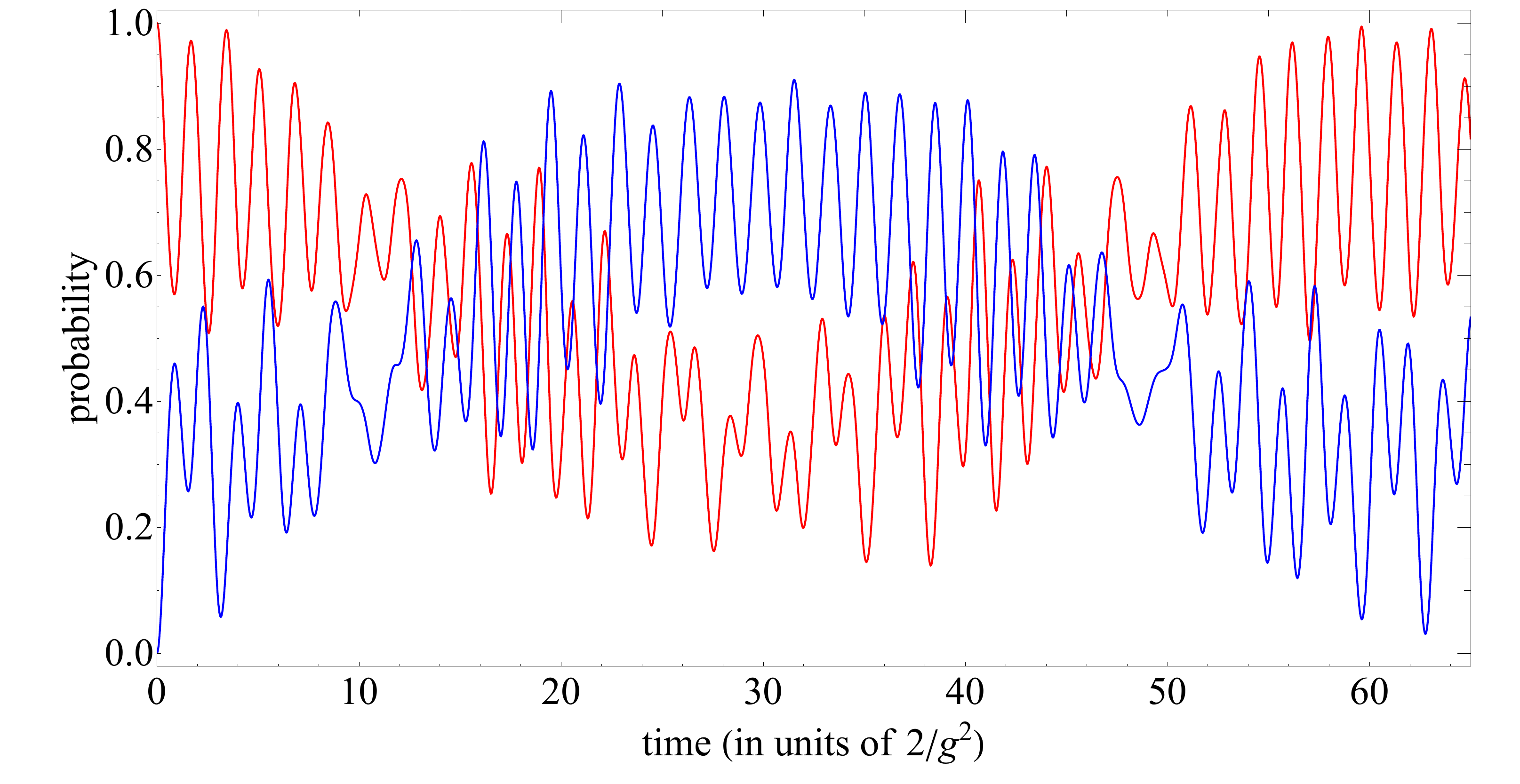}
\caption{Example of energy excitation moving from left to right for a 2-plaquette lattice for the $16 \times 16$  truncation of the Hamiltonian $j_{\rm max}=3/2$ case with $x=0.65$.
The red (blue) solid curve is the exact probability of the left (right) plaquette being switched on, having an energy flux of $j\geq 0$ over time. The traveling excitation transition point is around $t \approx 16$ clearly out of reach with the current hardware, because using a reasonably small $dt$ will require a large number of time steps, leading to a signal full of noise.}
\label{fig:Trave_exi}
\end{figure}
As we have shown in our previous work, \citep{Rahman:2022rlg} with the current NISQ hardware a time evolution circuit with 100 CNOT can be used only for a few time steps before the noise cannot be reasonably mitigated.\\
\\
One possible reasonable challenge is to consider the system initially prepared with the left plaquette switched ``on'' and time evolve it until it reaches a point of equilibrium between the states for which $j=1/2$, $j=1$ and $j=3/2$, at which the probability of finding the initial state in any group of states with higher $j$ value is close to equal, as shown in Figure~\ref{fig:IS_fj}:  
\begin{figure}[H]
\centering
\includegraphics[width=1.0\linewidth]{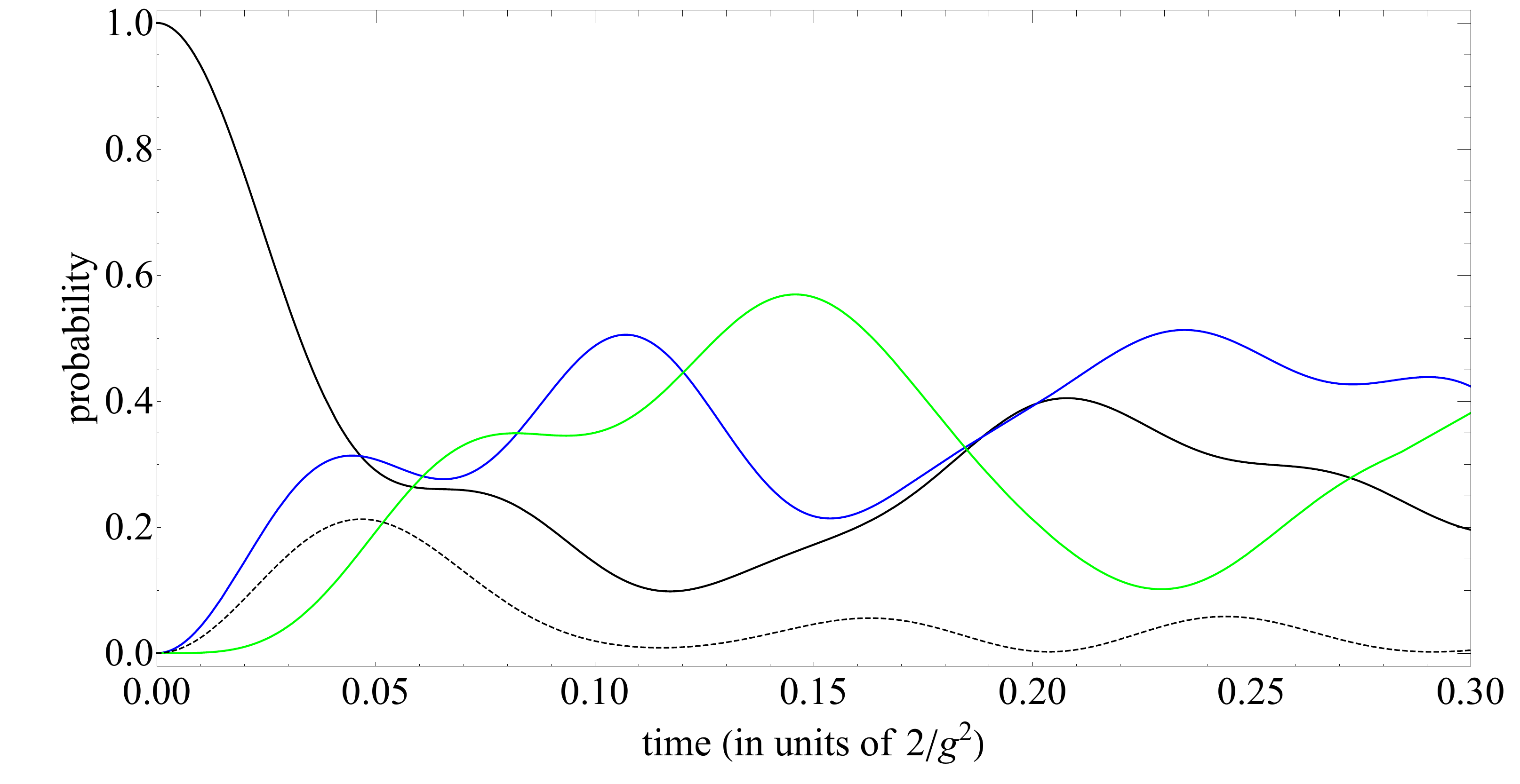}
\caption{Exact calculation of the time evolution of a 2-plaquette lattice for the $16 \times 16$ truncation of the Hamiltonian $j_{\rm max}=3/2$ case lattice from the initial state in which the left plaquette is switched on with gauge coupling x=8.0. The black solid, blue, green and black dashed lines are the exact probability over time of finding the time evolution of the left plaquette in state with a $j$ flux of $j=1/2$, $j=1$, $j=3/2$ and $j=0$ respectively.}
\label{fig:IS_fj}
\end{figure}
\noindent where the peculiar time evolution of equal balance between excited states is the point around the time instant $t=0.185$.\\
\\
The goal of the upcoming simulations on the IBM quantum hardware is to clarify the feasibility of this approach to study larger energy truncation and test the capabilities of the self-mitigation technique to mitigate the hardware error when dealing with theories with larger and more complex circuit implementation.\\
In the attempt to make clear the capabilities of the error mitigation techniques implemented so far, the code prepared for these new simulations is an adaptation of the one used in \citep{Rahman:2022rlg} for the case of 2-plaquette with $j_{\rm max}=1/2$ and whose public source is present on \citep{duet_code}.
\section{Conclusions and looking forward}
In this chapter we have presented the main features of the IBM gate-based quantum computer and shown the results obtained by using it to study the time evolution of a SU(2) pure gauge lattice theory, proving that the hardware can be successfully used in this scope. In fact the real-time evolution study was extended to a time range much larger than any previous study using non-Abelian lattice gauge theory \citep{Klco:2019evd, ARahman:2021ktn, Ciavarella:2021nmj, Illa:2022jqb}.\\
\\
Being able to perform a long time evolution made it possible to report the first observation of a excitation moving across a small lattice in real-time from a calculation done on quantum hardware, a dynamic phenomena that happens further away from the strong coupling regime of the theory. This has an important conceptual value, because this simple phenomenon cannot be observed using a classical computer once a much larger lattice is considered due to the exponential scaling of the resources. Therefore, this calculation even if done on a tiny lattice makes it clear that when the quantum hardware will have more quantum resources, they will be able to study phenomena not accessible on classical computers.\\
\\
Those results were possible thanks to the implementation of well-known error mitigation techniques, readout error mitigation, randomized compiling and zero-noise extrapolation, but an essential role was played by our error mitigation technique named self-mitigation developed for this study. In developing this method, we were inspired by a similar idea used in \citep{Urbanek:2021oej} in which the error estimation circuit is obtained from the physics circuit by removing all single gates, leaving only the CNOT gates. This was motivated by the fact that the largest contribution to the hardware error comes from the CNOT gate which is the noisiest. Instead, to better estimate the hardware error globally affecting the physics circuit, self-mitigation uses an error estimation circuit in which the physics circuits is used forward and backward in time, having the same number of gates, in the same order and even the angles of the rotation gates are the same or the opposite value. Therefore we think it better quantifies the hardware error affecting the physics circuit. In fact, self-mitigation made it possible to extend the time evolution study using close to 100 Suzuki-Trotter steps represented by circuits containing up to 400 CNOT gates.\\
\\
The method was successfully used by the authors of \citep{Atas:2022dqm} in which they simulate Quantum Chromodynamics in one spatial dimension with the goal of investigating the real-time evolution of a hadron system evolving from common hadrons to tetraquarks and pentaquarks. Furthermore, it was inspiring for the authors of \citep{Farrell:2022wyt} in developing a similar approach for the study of real-time dynamics from the vacuum to excited states of Quantum Chromodynamics in one spatial dimension.\\
\\
The idea behind self-mitigation can be applied to other operators not only the time evolution one, by simply creating an error estimation circuit that contains the operator circuit followed by its inverse circuit, and adapt the self-mitigation relation accordingly in Equation~\ref{ed:SFM_EQ} to take into account the different number of gates.\\
\\
In the last section we have presented an extension of this study to the case of 2-plaquette lattice with a $16 \time 16$ truncation of the $j_{\rm max}$ case. The future results of the calculation on the IBM hardware will help understand if self-mitigation can mitigate the hardware error on much larger and more complex circuits and since the time evolution circuit complexity comes from the Wilson approach, (in which the system is described with a global basis). Consequently it will make it clear if it is a feasible way to study the dynamics of SU(2) lattice gauge theory with larger gauge truncations $j_{\rm max} >1/2$. \\
\\
In this work the use of quantum resources was rationed by analytically calculating the time evolution operator instead of letting the Qiskit library calculate it from the Hamiltonian, and by optimizing it in two ways: analytically and at the code level.\\
Analytically a matrix decomposition is chosen that requires the least number of gates and, by rearranging the gate terms inside each single Suzuki-Trotter step to cancel CNOT gates and combine rotation gates into one. At the code level CNOTs are eliminated between consecutive time steps and Pauli gates are absorbed into rotation gates.  
\\
Despite, the many particle physics studies that have already been performed on the IBM gate-based quantum hardware, future hardware extensions dealing with the hardware errors are necessary to transform what is now an interesting and promising prototype in a tool able to perform calculations that cannot be done with classical computers. In this direction a crucial turning point is represented by increasing the qubit connectivity limited to 3. Increasing the connectivity to a value larger than 10 will help not only to encode more systems with more complex interactions but will make it possible to implement well known error correction algorithms like for example the Shor’s one which requires 9 qubits for each qubit to keep error free. Otherwise, increasing the qubits connectivity is challenging because it requires rethinking the internal structure of the quantum processing unit (QPU). The gates error rate should be reduced at a point that using many $\rm SWAP$ gates does not impact the hardware noise level, so that one can increase the qubit's connectivity by using $\rm SWAP$ gates.\\
\\
Indeed, the IBM quantum development roadmap for the next half of the decade is quite promising for lattice gauge theories because it goes towards the development of hardware with more than 4000 qubits with important actions to reduce hardware noise and to developing built-in error correction \citep{IBM_roadmap}.
\\
Looking forward, another important advancement is expected with the introduction of multilevel quantum computational units called qudits of which the qubit is a 2-level unit. Qudits will not only increase the computational capability to a much higher level but will make it possible to increase the complexity of the system that can be studied. For example, a qutrit, a quantum bit with three levels, will be perfectly suited to study Quantum Chromodynamics because it will make it natural to encode a quark or a gluon with their three-valued colour charge on a qutrit making the encoding scaling perfectly polynomial with the number of particles. 

%
%
%

\chapter{Conclusion}
In this final chapter we first discuss the overall content of the thesis by recalling the main rationale from the introduction that led to the use of two different quantum hardwares, the SU(2) lattice gauge theory presented in chapter 2, then we summarize the results presented in chapter 3 and 4 obtained with the D-Wave quantum annealer and the IBM gate-based quantum computer. A brief critical comparison of the two hardwares is provided, and we conclude by setting out some general perspectives on the future use of quantum hardware.
\\

\section{Synopsis of the thesis}
In the Introduction~\ref{chap:Introduction} we briefly discussed how despite the success in understanding quantum chromodynamics and other lattice gauge theories in the last 50 years of investigation, the sign problem is, in fact, an insurmountable obstacle that prevents the study of the real-time evolution of a theory, its phase diagram and other non-perturbative phenomena from being explored. In this scenario, a unique opportunity comes from the ability of quantum computers to store the theory Hilbert space in an exponentially more efficient way compared to a classical computer, making it possible to use the Hamiltonian formulation for direct numerical calculation, which is free from the sign problem, in place of the standard lattice gauge theory path integral formulation based on numerical calculation using the Monte Carlo method with importance sampling.\\

In Chapter~\ref{chap:Chapter_1}, we introduced the SU(2) pure gauge lattice theory given by Kogut
and Susskind \citep{PhysRevD.11.395}, describing in detail how to derive the matrix representation of the Hamiltonian for a lattice made by a row of plaquettes with a fixed value of the gauge truncation $j_{\rm max}$. To facilitate the reader in the following chapters where quantum hardware are used to extract the energy spectrum and the time evolution, the exact classical calculations of those observables for the 2-plaquette case are given. In Figures~\ref{fig:Te_x<1}, \ref{fig:Te_x>1} we showed how the two different regimes of the theory with the gauge coupling $x$ effected the time evolution. In Figure~\ref{fig:E_2pla_1o2} we showed how the energy spectrum of the theory for the 2-plaquette case as a function of the gauge coupling $x$ changes when considering different values of the gauge truncation $j_{\rm max}$.  Finally, in Figure~\ref{fig:to_jmax_2} the phenomena of a traveling excitation for the $j_{\rm max}=2$ case is shown.\\

In Chapter~\ref{chap:Chapter_2}, we first described the main feature of the D-Wave quantum annealer \citep{D-Wave} and then we  discussed in details how to use the hardware to find the ground state of a generic Hamiltonian by implementing the QAE algorithm Ref.~\citep{Teplukhin_2019, Teplukhin_2020, D0CP04272B, teplukhin2021sampling, Teplukhin2021} and the extension we developed AQAE \citep{ARahman:2022tkr}. The necessary algorithm steps were shown in Figures~\ref{fig:Fig_x_0.9__2_pla_jmax_1}, \ref{fig:Fig_x_0.9__2_pla_jmax_1_chain} and \ref{fig:Fig_2_pla_jmax_1_AQAE}. These algorithms were then used to calculate the energy spectrum of the theory as shown in Figures~\ref{fig:E_2pla_j_1o2_dw}, \ref{fig:E_2pla_j_1_dw}, \ref{fig:E_2pla_j_3o2_dw}, \ref{fig:E_4pla_j_1o2_dw} and \ref{fig:E_6pla_j_1o2_dw}, and those results were used to calculate the vacuum expectation value of the Hamiltonian as a function of the gauge coupling $x$ for different values of the gauge truncation $j$ and for longer lattices as shown in Figures~\ref{fig:0_H_0_jmax} and \ref{fig:0_H_0_latt}, respectively. Finally, using the Kitaev-Feynman clock states \citep{Feynman_85,McClean_2013, Tempel_2014, Kitaev_2022} the probability of oscillations between two excited states in real-time were obtained as shown in Figure~\ref{fig:TE_non_stoquastic} and \ref{fig:TE_stoquastic}.\\

In Chapter~\ref{chap:Chapter_3}, we first introduced the essential features of the IBM gate-based quantum computer \citep{IBM} needed to encode a generic operator on the hardware. Then, we addressed the hardware errors and consequently the necessary error mitigation techniques, showing the one we introduced self-mitigation \ref{item_Self_mitigation}, along with the other well-known ones. How to compute on the hardware the real-time evolution of the theory was discussed in details by presenting the result in Figure~\ref{fig:data_TE_2pla_IBM_self-m} for the 2-plaquette case with $j_{\rm max}=1/2$ and, an example of real-time traveling excitation is shown in Figure~\ref{fig:2pla_Traveling_Excitation_ibm}. The study of the real-time evolution was extended to the 5-plaquette case with $j_{\rm max}=1/2$ in Figure~\ref{fig:TE_5plagate}. Finally, the essential elements for calculating the time evolution for the case of 2-plaquette lattice considering a $16 \times 16$ truncation of the $j_{\rm max} = 3/2$  are shown with the targeted time evolution in Figure~\ref{fig:IS_fj}.
\section{Critical comparison between D-Wave and IBM}
Having studied the SU(2) lattice gauge theory on two different NISQ quantum hardwares, even though we did not study the exact same observables, it provides us with a unique opportunity to compare the two quantum hardwares and discuss the advantage and disadvantages.\\
\\
The main advantage of using the D-Wave annealer is that the Hamiltonian or any operator matrix representation can be encoded directly on the hardware as opposed to the IBM gate-based quantum computer where it has to be first decomposed in the available quantum gates. Secondly the D-Wave hardware can produce precise and accurate results without using any error mitigation techniques. Nevertheless, there are two main drawbacks that limit its usability: firstly, it is not a universal quantum annealer, in fact only operators with non-positive off-diagonal elements, called a stoquastic matrix, can be studied efficiently. Additionally there is a mild sign problem that severely limits the accuracy of the results. Secondly, the lack of freedom to directly interact with the qubit states makes it impossible to exploit qubit superposition and qubit entanglement, resulting in difficulties in designing computational algorithms that have polynomial scaling with the size of the system.\\
\\
In contrast, the IBM gate-based quantum hardware is a universal quantum computer, giving freedom to encode any operators on the hardware once it is decomposed in gates. Unlike D-Wave, on IBM's hardware it is possible to directly access and change the state of each qubit and one can formulate algorithms with polynomial scaling, which perform better than the classical one.
The major disadvantage of IBM hardware is the difficulty to extract results in good agreement with the exact result without the use of many error mitigation techniques due to the large plethora of different hardware errors. In fact, the hardware is still noisy, and in the interlude until a definitive error correction protocol is introduced, many problem-specific error mitigation techniques need to be formulated and adopted, but often with the drawback of requiring a large amount of quantum resources, causing the full encoding to lose its initial polynomial scaling with the system size.\\
\\ So ultimately, if D-Wave’s functionality is not extended, it will not be as useful as the IBM hardware in envisioning new ways to investigate unanswered particle physics questions. 
%
%
%

\section{Overall prospects for the future}
The real era of quantum computers has not yet begun, we are just at the beginning of the preparatory phase, what is called the noisy intermediate-scale quantum era, where the usability of the few prototypes of quantum hardware limited by the few noisy resources is under exploration. Therefore at least for the next decades classical computations will retain their predominant role in understanding nature.\\
\\
However, in keeping with the possibilities of the NISQ era, over the last four years many theories based on the U(1), SU(2) and SU(3) groups relevant to particle physics have been encoded on the available NISQ quantum hardware such as quantum annealing and gate-based quantum computers, but all those theories have been studied on a small one-dimensional lattice.
Thus, the first natural challenge is to extend those previous studies to longer one-dimensional spatial lattices and, later, to two-dimensional and three-dimensional lattices. The second challenge is to extend those previous studies to theories with terms describing phenomena that cannot be studied on large lattices using a classical computer due to the sign problem and/or an exponential growth of computational resources, for example by adding a chemical potential.\\
These will be the subject of research in the next few years or even decade, because every time the theory is extended to larger dimensions there is a need to reformulate its coding respecting the characteristics of the hardware, and noisy hardware will require adapting error mitigation techniques previously formulated and more will likely develop new error mitigation techniques that can handle much larger noise from a more complex system that will inevitably require more qubits and larger circuits than in the one-dimensional case. This future study will be greatly simplified if quantum hardware has a reduced error rate and increased qubit connectivity in the coming years.\\
\\
We should bear in mind that these future developments should be regarded as a preparatory exploration of quantum resources in the meantime that quantum hardware becomes fault-tolerant. Indeed it is insufficient to think of quantum computing as a "brute force" approach to look for a solution to problems that cannot be solved using the classical computer due to the highly intensive computational resources required.\\
Instead, quantum computers should induce a new paradigm in the study not only of particle physics but more generally in the understanding of nature, in which new formulations and new approaches are developed by exploiting quantum information properties. This is similar to what happened with the introduction of the classical computer, which led Stanislaw Ulam \citep{Metropolis_Ulam} in 1948 to invent the Markov chain Monte Carlo method that was implemented on the ENIAC computer \citep{Metropolis_memoir}, and later in the field of LGT led Michael Creutz in 1979 to study a SU(2) pure Yang-Mills theory. It paved the way for other researchers to use the 1974 Kenneth Wilson discretization of quantum chromodynamics and gauge theory from first principles to perform numerical studies on the classical computers available at that time, giving rise to Lattice QCD, something that was never thought before.\\
This will inevitably require a collective creative effort, but that is what scientists are here for, to introduce new ideas and/or new points of view on nature, and fortunately there are many who will be harnessing the quantum computing era in the decades to come.



\begin{appendices}

\chapter{Explanation on how to derive the SU(2) pure gauge lattice Hamiltonian}\label{appendix:HSU2_derivation}

In this appendix we present an explanation on how to obtain the 2-dimensional SU(2) pure gauge theory Hamiltonian present in Eq.~$\ref{eq:H_c2}$:
\begin{equation}\label{eq:H_DT}
\hat H = \frac{g^2}{2}\left(\sum_{i={\rm links}}\hat E_i^2-2x\sum_{i={\rm plaquettes}}\hat\square_i\right)
\end{equation}
starting from the Lagrangian density in the continuum.
Before doing that, it is important to remind the reader that the original formulation can be found in "Hamiltonian formulation of Wilson's lattice gauge theories" by John Kogut and Leonard Susskind \citep{PhysRevD.11.395}, where an analogy with the quantum rigid rotator is used to justify the introduction of the pure gauge-field term.\\
\\
Following the Yang-Mills approach a theory can be built by requiring the fields describing the particles to be gauge invariant under the chosen group. Therefore for our case the continuum SU(2) non-Abelian Lagrangian density in presence of fermions is:
\begin{equation}
\label{L_SU(2)_initial}
\mathcal{L}= - \dfrac{1}{2 N_c g^2}F^c_{ \mu \nu }(x)F^{c \, \mu \nu}(x) + \sum_f\overline{\psi_f}(x)(i\,\gamma^\mu \,D_\mu-m_f)\psi_f(x)
\end{equation}
where $g$ is the gauge coupling, $N_c$ is a numeric constant indicating the number of colors that for SU(2) is 2, and\\
\begin{equation}\label{eq:G_mn}
F_{\mu \nu}^c(x) = \partial_\mu A^{c}_\nu(x) - \partial_\nu A^{c}_\mu(x) +ig[A^{c}_\mu(x), A^{c}_\nu(x)]
\end{equation}
is the field strength tensor,  $A_\mu(x)$ is the gauge boson field and $c$ is the color index with value $c =1,2$.\\
$\psi_f(x)$ is a fermion quark field with mass $m_f$, $f$ specifies the fermion flavour, and $D_\mu = \partial_\mu - ig\sigma^{c} A_\mu^{c}$ is the covariant derivative.\\
\\
The gauge invariance of the Lagrangian under SU(2) is guaranteed by the following gauge transformation of its elements:
\begin{eqnarray}\label{con_local_transformation}
\psi(x) &\rightarrow& \psi'(x)=\Omega(x)\psi(x)\,, \\
\overline{\psi}(x)' &\rightarrow& \overline{\psi}(x)\Omega^\dagger (x)\,, \\
D_\mu &\rightarrow & D_\mu' \psi'(x)=\Omega(x)D_\mu \Omega(x)^\dagger \,, \\
A_{\mu}^c(x) &\rightarrow & A_{\mu}^c(x)'=\Omega(x) A_{\mu}^c(x) \Omega(x)^\dagger +i \left( \partial_\mu  \Omega(x)  \right)  \Omega(x)^\dagger \,.
\end{eqnarray}
where $\Omega(x)$ belongs to the fundamental representation of SU(2), and can be written as $\Omega(x) \equiv e^{-i\sigma^b\theta^b(x)}$ where $\omega^b$ are the 3 gauge transformation parameters, and $\sigma^b$ with $b=1,2,3$ are the Lie algebra generators, that for SU(2) are the Pauli matrices.\\
\\
Since we are interested in a pure-gauge theory we put the fermion field, $\psi_f$ to zero, therefore our starting point is:
\begin{equation}
\label{L_SU(2)}
\mathcal{L} = - \dfrac{1}{4}F^c_{\mu \nu}(x)F^{c \, \mu \nu}(x)
\end{equation}
The Hamiltonian density can be obtained by simply using the definition and the Euler-Lagrange equations leading to:

\begin{eqnarray}
\label{h_SU(2)}
\mathcal{H} &=& \dfrac{\delta \mathcal{L} }{\delta (\partial_0 A^c_\mu)(x)}\partial_0 A^c_\mu (x) - \mathcal{L} = \dfrac{\delta \mathcal{L} }{\delta (F^{ c \,\alpha \beta})} \dfrac{\delta F^{c \, \alpha \beta} }{\delta (\partial_0 A^c_\mu(x))}   \partial_0 A^c_\mu (x) - \dfrac{1}{4}F^c_{\mu \nu}(x)F^{c \, \mu \nu}(x) \nonumber \\
&=& \left(-\frac{1}{2}  \partial_0 A_\mu(x) \partial^0 A^\mu (x)  + \frac{1}{4}  F^c_{\mu \nu}(x)F^{c \, \mu \nu}(x) \right) 
\end{eqnarray}

Therefore the Hamiltonian is obtained by spatially integrating the Hamiltonian density:
\begin{equation}
H= \int d^Dx \, \mathcal{H}  = \int d^Dx \, \left(-\frac{1}{2}  \partial_0 A^c_\mu(x) \partial^0 A^{c \,\mu} (x)  + \frac{1}{4}  F^c_{\mu \nu}(x)F^{c \, \mu \nu}(x) \right)
\end{equation}
where $D$ is the dimension of the chosen spacetime.\\
\\
We can now expand the field strength tensor so that we can separate the terms containing time derivative of the fields $A_\mu (x)$ from the one without.

\begin{equation}
H= \int d^Dx -\frac{1}{2} \partial_0 A^c_\mu(x) \partial^0 A^{c \, \mu} (x) + \frac{1}{4} \left(  F^c_{00}(x)F^{c \, 00}(x) + F^c_{0j}(x)F^{c \, 0j}(x) + F^c_{i0}(x)F^{c \, i0}(x) + F^c_{ij}(x)F^{c \, ij}(x)\right) 
\end{equation}

where in the second passage we used the antisymmetry of the strength tensor $F^c_{\mu \nu}(x)=- F^c_{\nu \mu}(x)$ to prove that $ F^c_{0 j}(x) F^{c \, 0 j}(x)= F^c_{i 0}(x) F^{c \, i 0}(x)$, where $i$ and $j$ are dummy indices.\\
\\
Since we are free to fix the gauge, we can use it to simplify the calculations. We decide to work in the temporal in which $A_0(x)$ is set to zero. Even if we are always free to choose the gauge, the conjugate field of $A^c_0(x)$, $\Pi_0=\delta \mathcal{L} / \delta (\partial_0 A^c_0(x))$ is zero, therefore $A^c_0(x)$  is not a true canonic field, therefore setting it to zero by choosing the temporal gauge solves this conceptual problem as well.\\
\\
Furthermore, the electric field can be written using the potentials as $E_i(x)=-\nabla \phi (x) \cdot \hat{i} -\partial A_i(x)/\partial t$, but with no present charges, it reduces to $E_i(x)=-\partial_0 A_i(x)$.\\
\\
In the temporal gauge, the strength tensor components $F_{00} =0 $ and  $F_{0i} = \partial_i A_0(x) - \partial_i A_0(x) = \partial_0 A_i(x)$. This leads to $F_{0i}F^{0i}=E_i^2$ and therefore the Hamiltonian is:
\begin{equation}
H=  \int d^Dx \, \frac{1}{2} E_i^2(x) + \frac{1}{4} F^c_{ij}(x)F^{c \, ij}(x)
\end{equation}
Since in the following discussion the color index $c$ does not play a crucial role, we drop it to simplify the notation.
We can start noticing that the main distinction between the two terms is that the first term contains time derivative of the field $A_\mu(x)$ while the second not.\\
\\
It is interesting to notice that in the Abelian case the second term would be equal to the square of the magnetic flux field $B(x)$ because $F^{ij} =-\epsilon^{i j k} B^k$. It should be now evident why it is common to call the Hamiltonian terms even of the non-Abelian case as chromoelectric and chromomagnetic respectively for the first and second term.
\\
Before proceeding to introduce the lattice and discretize the Hamiltonian, it is necessary to change from the Minkowski to the Euclidian space and perform the Wick rotation that changes the time from real to imaginary. This means:
\begin{eqnarray}
x^0 & \rightarrow &  -ix_0\,, \\
A^0(x) & \rightarrow &  i A_0(x)\,, \\
\partial_0 & \rightarrow & i\partial_0 \,. 
\end{eqnarray}
This has the consequence that  $E_i(x)=-i\partial_0 A_i(x)$ but no effect on $E_i(x)^2$, therefore the Hamiltonian finally reads:
\begin{equation}
H= \int d^Dx  \,  \dfrac{1}{2} E_i^2(x) +  \dfrac{1}{4} F_{i j}(x) F^{i j}(x)
\end{equation}
The hypercubic lattice with linear sizes $L_\mu$ and lattice and spacing $a$ that discretizes the spacetime is formally defined as:
\begin{equation}
\Lambda=\left\lbrace x_\mu =an_\mu ~ \vert ~ n_\mu = 0,1, \ldots L_\mu/a  \right\rbrace 
\end{equation}\\
When we discretize the continuum spacetime with a D-dimensional lattice having lattice spacing $a$, the fundamental discretization prescription adopted are
\begin{eqnarray}
x_\mu &\rightarrow& n_\mu a\,, \\
\int d^D x \ldots &\rightarrow& a^D \sum_n \ldots, \\
\partial_\mu \psi(x) &\rightarrow& \dfrac{1}{2a} \left( \psi(n+\hat{\mu}) -  \psi(n-\hat{\mu}) \right) +\mathcal{O}(a^2) \,, \label{LQCD_derivative} 
\end{eqnarray}
which are very handy for showing that the lattice theory reproduces the continuum theory.\\
\\
A gauge field as element of SU(2) is attached at each link of the discretized D-dimensional spacetime. This element is called link variable and indicated with $U_\mu(n)$, which connects the points $n$ and $n+\hat{\mu}$  along the lattice direction $\mu.$\\
\\
The formal connection between the continuum and the discretized lattice theory is expressed in the relation between the link variable $U_\mu(n)$ and the gauge boson field $A_\mu(n)$:
\begin{equation}
\label{U_mu}
U_{\mu} (n) = exp(i a g A_\mu(n))
\end{equation}
that is the lattice version of the Schwinger line integral in the continuum theory $e^{i \mathrm{P}\int_x ^{x+a\hat{\mu}} dx_\mu A_{\mu}(x) }$
Assuming that in the limit $a \to 0$ the field $A_\mu (n)$ became sufficiently smooth, so that it can be expanded as:
\begin{equation}
\label{eq:A_mu}
A_{\nu}(n+\hat{\mu}) = A_{\nu}(n) + a \partial_\mu A_\nu(n) +\mathcal{O}(a^2)
\end{equation}
Since we are working in the temporal gauge, from equation Eq.~\ref{U_mu} implies that all the link variable along the temporal direction, $U_0(n)$ are set to the identity, therefore they do not play any role.\\
\\
In the pure gauge lattice theory an important role is played by the smallest product of link variables along a square, called plaquette variable:
\begin{equation} \label{eq:plaquette}
U_{\mu\nu}(n) =U_\mu(n)U_\nu(n+\hat{\mu})U_{\mu}^\dagger(n+\hat{\nu}) U_{\nu}^\dagger(n)
\end{equation}
its importance can be found in its connection with the discrete form of the strength tensor, which can be obtained by substituting in its definition the expression of the link variable in Eq.~\ref{U_mu}, obtaining:
\begin{equation}
\label{eq:plaq_expansion}
U_{\mu\nu}(n) = e^{i a g A_\mu(n)} e^{i a g A_\nu(n + \hat{\mu})}
e^{-i a g A_\mu(n + \hat{\nu})} e^{-i a g A_\nu(n)}
\end{equation}
to expand the product, we have to keep in mind that there are non-commuting terms, therefore we will use the Baker–Campbell–Hausdorff formula:
\begin{equation}
\label{eq:BCH_formula}
e^A e^B \approx e^{A + B +\frac{1}{2} [A,B] +\frac{1}{12} \left([A,[A,B]]+ [B,[A,B]] \right) + \cdots }
\end{equation}

For our specific case that has four operators can be generalized to:
\begin{equation}
\label{eq:BCH_4_oper}
e^A e^B e^C e^D  \approx e^{A + B + C + D +\frac{1}{2} \left( [A,B] +[A,C] + [A,D] + [B,C] + [B,D] + [C,D] \right)  + \cdots }
\end{equation}

Using it in Eq.~\ref{eq:plaq_expansion} we obtain:
\begin{equation}
\begin{aligned}
\label{h_SU(2)}
U_{\mu \nu}(n) &= e^{i a g A_\mu(n)} e^{i a g A_\nu(n + \hat{\mu})} e^{-i a g A_\mu(n + \hat{\nu})} e^{-i a g A_\nu(n)}  \\
&\approx   \exp( \, i a g \left\lbrace  A_\mu (n) + A_\nu (n+\hat{\mu}) - A_\mu (n+\hat{\nu}) - A_\nu (n)\right\rbrace + \\
&+ \dfrac{1}{2}  i^2 a^2 g^2 \{
\left[ A_\mu(n), A_\nu(n+ \hat{\mu}) \right] - \left[ A_\mu(n), A_\mu(n+\hat{\nu})\right] 
-\left[ A_\mu(n), A_\nu(n) \right] + \\
&- \left[ A_\nu(n+\hat{\mu}), A_\mu(n+ \hat{\nu}) \right] - \left[ A_\mu(n+\hat{\mu}), A_\nu(n) \right] +\left[ A_\mu(n), A_\nu(n) \right]  \} + \\
&+\mathcal{O}(a^3)  )
\end{aligned}
\end{equation}

where the terms with shifted arguments of the form $A_\mu(n+\hat{\nu})$ can be expanded with the use of Eq.~\ref{eq:A_mu}, therefore after few simplifications we obtain:
\begin{equation}
\begin{aligned}
\label{h_SU(2)_sempl}
U_{\mu \nu}(n) &\approx \exp( \, i a^2 g \left\lbrace  \partial_\mu A_\nu (n) -\partial_\nu A_\mu (n) \right\rbrace + \\
&- \dfrac{1}{2}  a^2 g^2 \{ 2 \left[ A_\mu(n), A_\nu(n+ \hat{\mu}) \right] + a \left[ A_\mu(n), A_\nu(n) \right] + a^2 \left[ A_\mu(n), A_\nu(n) \right]\} +\mathcal{O}(a^3) )=\\
&=\exp( i a^2 g \left\lbrace \partial_\mu A_\nu (n) -\partial_\nu A_\mu (n) + i g \left[ A_\mu(n), A_\nu(n) \right] \right\rbrace  +\mathcal{O}(a^3) )
\end{aligned}
\end{equation}
where the terms inside the curly brackets is the lattice field strength tensor $F_{\mu \nu}(n)$ and the relation between the plaquette variable and the lattice field strength tensor reads:
\begin{equation}
\label{U_Fmnu_rel}
U_{\mu \nu}(n) = \exp( i a^2 g F_{\mu \nu}(n) +\mathcal{O}(a^3) 
\end{equation}
where the lattice field strength tensor is simply:
\begin{equation}
\label{eq:F_mu_nu_lat}
F_{\mu \nu}(n)=\partial_\mu A_\nu (n) -\partial_\nu A_\mu (n) + i g \left[ A_\mu(n), A_\nu(n) \right] 
\end{equation}
The field strength can be seen as a generalized curl of the vector potential, and its volume integration can be viewed as the integration of the vector potential in a closed loop. This is the underlying Wilson's reasoning used in his proposal for a pure gauge lattice Lagrangian as the sum of product of gauge fields $U_\mu(n)$ around a square plaquette.\\
\\
This expression can be used to derive the lattice electric field, that is
\begin{equation}
F_{0 i} (n) = \partial_0 A_i (n) -\partial_i A_0 (n) + i g \left[ A_0(n), A_i(n) \right] = \partial_0 A_i (n) =- E_i(n)
\end{equation}
were in the second step we use the fact that we have chosen the temporal gauge, $A_0=0$.\\
\\
The lattice chromoelectric part of the Hamiltonian can be now written as:
\begin{equation}
\int d^Dx \, \frac{1}{2} E_i^2(x) 
\rightarrow H_E= a^D \sum_{n} E_i^2(n)
\end{equation}
where the sum is the discretization of the D-dimensional integral. \\
\\ 
We can now proceed to write the second term in the Hamiltonian, $F^{i j} F_{i j}$, using the plaquette operator defined as trace of a plaquette variable:
\begin{equation}
\hat{U}_{\mu \nu} = Tr \left[ U_{\mu \nu}  \right]= Tr \left[ \exp \left(i a^2 g F_{\mu \nu} +\mathcal{O}(a^3) \right)   \right]
\end{equation}
where the trace ensure that the operator is gauge invariant, in the SU(2) color space.\\
At the lowest order in the lattice spacing $a$ the following relation holds:
\begin{equation}
\label{U_Fmnu_rel}
\hat{U}_{\mu \nu}(n) + \hat{U}_{\mu \nu}^\dagger (n) = Tr\left[ 2 \mathds{1} -a^4 g^2 F_{\mu \nu}(n) F^{\mu \nu} (n) + \mathcal{O}(a^6) \right]
\end{equation}
therefore:
\begin{equation}
\label{U_Fmnu_rel}
Tr\left[ F_{\mu \nu}(n) F^{\mu \nu}(n)  \right] = \dfrac{4 - \left(  \hat{U}_{\mu \nu}(n) + \hat{U}_{\mu \nu}(n)^\dagger \right)}{a^4 g^2} + \mathcal{O}(a^6) 
\end{equation}
The term $F_{i j}(n) F^{i j}(n)$ can now be written as:
\begin{equation}
\label{U_Fmnu_rel}
Tr\left[ F_{i j}(n) F^{i j}(n)  \right] = \dfrac{4 - \left(  \hat{U}_{i j}(n) + \hat{U}_{i j}(n)^\dagger \right)}{a^4 g^2} + \mathcal{O}(a^6) = \dfrac{4 - 2   \hat{\square}(n) }{a^4 g^2} + \mathcal{O}(a^6)
\end{equation}
where in the second step we use the fact that for $SU(2)$ $\hat{U}_{\mu \nu}(n) = \hat{U}_{\mu \nu}^\dagger(n)$ and called $\hat{U}_{i j}(n)$ as $\hat\square(n)$, which is a only-spatial plaquette operator, made by only link variable laying on the spatial directions.\\
\\
The lattice chromomagnetic part of the Hamiltonian can be written as:
\begin{equation}
\int d^Dx \, \frac{1}{4} F_{ij}(x)F^{ij}(x) 
\rightarrow H_B= a^D \sum_{n} - \frac{1}{4}
\dfrac{2}{a^4 g^2} \hat{\square}(n)
\end{equation}
where we dropped the numeric constant, $4/(a^4 g^2)$, because it contributes only a constant energy shift.
\\

Finally, adding $H_E$ and $H_B$ the SU(2) lattice Hamiltonian is:
\begin{equation}
\begin{aligned}
H&= a^D \sum_{n} \left(  \dfrac{1}{2} E_i^2(n) - \dfrac{ 1 }{2 a^4 g^2} \hat{\square}(n) \right)  = a^{D-2}  \dfrac{ g^2 }{2} \sum_{n} \left(  \dfrac{a^2}{g^2} E_i^2(n) - \dfrac{ 1 }{a^2 g^4} \hat{\square}(n) \right) =\\
&=  a^{D-2}  \dfrac{1}{4} \dfrac{ g^2 }{2} \sum_{n} \left(  \dfrac{4 a^2}{g^2} E_i^2(n) - \dfrac{ 4 }{a^2 g^4} \hat{\square}(n) \right) =  a^{D-2} \dfrac{1}{4} \dfrac{ g^2 }{2} \sum_{n} \left(  a^2 \left( \dfrac{2}{g} E_i^2(n)\right) - \dfrac{ 2 x }{a^2} \hat{\square}(n)\right) 
\end{aligned}
\end{equation}
where in the first step we putted in highlighted $g^2/2$ while in the second step we multiplied and divided by 4 so that in the second term there is a $4/g^4$ that is $2x$ with $x=2/g^4$.\\
\\
Rescaling the chromoelectric field as $\left( \dfrac{2}{g} E_i(n)\right) \longrightarrow E_i(n)$, dropping the numeric multiplicative constant $1/4$, set the dimension to $D=2$, and setting the lattice spacing $a$ to 1, we finally obtain the expression used for the lattice SU(2) Hamiltonian:
\begin{equation}
\label{eq:H_final}
H= \dfrac{ g^2 }{2} \sum_{n}  E_i^2(n) - 2x \, \hat{\square}(n)
\end{equation}
where one has to keep in mind that for each node $n$ one has to sum the contributions of the electric field from each link composing the space. \\
\\
Even if an extra dimensionless numerical constant in the field is not important and the notation is clear, a final comment about the rescaling of the chromoelectric field is needed. In this discussion we have tried to obtain the final expression of $H$ in Eq.~\ref{eq:H_final} starting from its continuum formulation and limiting the ad hoc constant fixing only on the last step, as we did for the rescaling of the field. It is common in the literature to avoid rescaling the field as:
\begin{equation}
\int d^2x \, \frac{1}{2} E_i^2(x) 
\rightarrow H_E= \frac{g^2}{2} a^2 \sum_{n} E_i(n)^2 \,.
\end{equation}


\chapter{Deriving the plaquette operator matrix elements}\label{appendix:plaquette_derivation}
In this appendix \footnote{This appendix is drawn from \href{https://doi.org/10.1103/PhysRevD.104.034501}{Phys. Rev. D 104, 034501} done in collaboration with  S. A Rahman, R. Lewis and S. Powell.} we present a derivation of the plaquette operator matrix elements.\\
\\
In deriving the action of the plaquette operator on the matrix element we consider a one-dimensional lattice made by a single row of plaquettes with periodic boundary conditions in the longitudinal direction. Since the derivation is quite cumbersome it is important to have a clear picture of the labelling of the links, nodes and plaquettes as denoted in Figure~\ref{fig:lattice_appendix}:
\begin{figure}[H]
\centering
\includegraphics[width=0.6\linewidth]{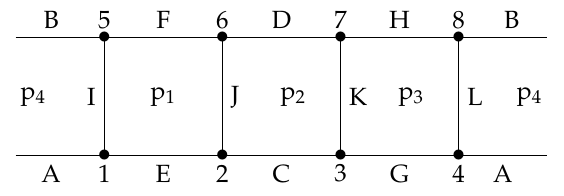}
\caption{A 4-plaquette lattice with periodic boundary condition in the longitudinal direction. The lattice is composed by 4 plaquettes made by the union of 12 gauge links which are joined by 8 3-link nodes. 
         Each plaquette is indicated by the label in the center of each square, each link by the letter at the middle point, and nodes by the close number.\label{fig:lattice_appendix}}
\end{figure}

To derive Eq.~(\ref{fig:lattice_appendix}), begin with the constraint of Gauss's law at lattice site 1 of Figure~\ref{fig:lattice},
which says the three gauge links form a colour singlet, $\left|j,m\right>_1=\left|0,0\right>_1$, giving
\begin{eqnarray}
\left|0,0\right>_1
&=& \sum_{m_I}\left|j_I,-m_I,j_I,m_I\right>_1\left<j_I,-m_I,j_I,m_I|0,0\right>_1 \\ \nonumber
&=& \sum_{m_I}\left|j_I,-m_I,j_I,m_I\right>_1\frac{(-1)^{j_I+m_I}}{\sqrt{2j_I+1}} \\ \nonumber
&=& \sum_{m_I}\frac{(-1)^{j_I+m_I}}{\sqrt{2j_I+1}}\sum_{m_A^\prime}\sum_{m_E}\left|j_A,m_A^\prime,j_E,m_E,j_I,m_I\right>_1 
\times \left<j_A,m_A^\prime,j_E,m_E,j_I,m_I|j_I,-m_I,j_I,m_I\right>_1 \\ 
&=& (-1)^{j_A-j_E+j_I}\sum_{m_A^\prime}\sum_{m_E}\sum_{m_I}\left|j_A,m_A^\prime,j_E,m_E,j_I,m_I\right>_1
    \left(\begin{array}{ccc} j_A & j_E & j_I \\ m_A^\prime & m_E & m_I \end{array}\right)
\end{eqnarray}
where in the third line we introduced an identity written as:
\begin{equation*}
\sum_{m_A^\prime}\sum_{m_E}\left|j_A,m_A^\prime,j_E,m_E,j_I,m_I\right>_1 \left<j_A,m_A^\prime,j_E,m_E,j_I,m_I \right|
\end{equation*}
and used the following relation connecting the Clebsch–Gordan coefficients to the Wigner 3-j symbols:
\begin{equation*}
\left(\begin{array}{ccc} j_1 & j_2 & j_3 \\ m_1 & m_2 & m_3 \end{array}\right) \equiv \frac{(-1)^{j_1-j_2-m3}}{\sqrt{2j_3+1}} \left<j_1,m_1,j_2,m_2|j_3,-m_3\right>
\end{equation*}
Our convention is to use $m$ for the left or bottom end of a link and to use $m^\prime$ for the right or top end of a link.
The other seven vertices have similar expressions:
\begin{eqnarray}
\left|0,0\right>_2
&=& (-1)^{j_C-j_E+j_J}\sum_{m_C}\sum_{m_E^\prime}\sum_{m_J}\left|j_C,m_C,j_E,m_E^\prime,j_J,m_J\right>_2
    \left(\begin{array}{ccc} j_C & j_E & j_J \\ m_C & m_E^\prime & m_J \end{array}\right) \,, \\
\left|0,0\right>_3
&=& (-1)^{j_C-j_G+j_K}\sum_{m_C^\prime}\sum_{m_G}\sum_{m_K}\left|j_C,m_C^\prime,j_G,m_G,j_K,m_K\right>_3
    \left(\begin{array}{ccc} j_C & j_G & j_K \\ m_C^\prime & m_G & m_K \end{array}\right) \,, \\
\left|0,0\right>_4
&=& (-1)^{j_B-j_G+j_L}\sum_{m_A}\sum_{m_G^\prime}\sum_{m_L}\left|j_A,m_A,j_G,m_G^\prime,j_L,m_L\right>_4
    \left(\begin{array}{ccc} j_A & j_G & j_L \\ m_A & m_G^\prime & m_L \end{array}\right) \,, \\
\left|0,0\right>_5
&=& (-1)^{j_B-j_F+j_I}\sum_{m_B^\prime}\sum_{m_F}\sum_{m_I^\prime}\left|j_B,m_B^\prime,j_F,m_F,j_I,m_I^\prime\right>_5
    \left(\begin{array}{ccc} j_B & j_F & j_I \\ m_B^\prime & m_F & m_I^\prime \end{array}\right) \,, \\
\left|0,0\right>_6
&=& (-1)^{j_D-j_F+j_J}\sum_{m_D}\sum_{m_F^\prime}\sum_{m_J^\prime}\left|j_D,m_D,j_F,m_F^\prime,j_J,m_J^\prime\right>_6
    \left(\begin{array}{ccc} j_D & j_F & j_J \\ m_D & m_F^\prime & m_J^\prime \end{array}\right) \,, \\
\left|0,0\right>_7
&=& (-1)^{j_D-j_H+j_K}\sum_{m_D^\prime}\sum_{m_H}\sum_{m_K^\prime}\left|j_D,m_D^\prime,j_H,m_H,j_K,m_K^\prime\right>_7
    \left(\begin{array}{ccc} j_D & j_H & j_K \\ m_D^\prime & m_H & m_K^\prime \end{array}\right) \,, \\
\left|0,0\right>_8
&=& (-1)^{j_B-j_H+j_L}\sum_{m_B}\sum_{m_H^\prime}\sum_{m_L^\prime}\left|j_B,m_B,j_H,m_H^\prime,j_L,m_L^\prime\right>_8
    \left(\begin{array}{ccc} j_B & j_H & j_L \\ m_B & m_H^\prime & m_L^\prime \end{array}\right) \,.
\end{eqnarray}
The product of the eight vertex states defines the state of the entire lattice.
Notice that we always list the gauge links $A$ through $L$ in alphabetical order
so the calculation will be self-consistent and have the correct Clebsch-Gordan phases.
The labeling of gauge links chosen in Figure~\ref{fig:lattice} (even horizontal, then odd horizontal, then vertical) is not required,
but it does maintain a convenient pattern among the four plaquette operators during the derivations.
The first plaquette operator is
\begin{equation}
\square_1 = \sum_{s_1}\sum_{s_2}\sum_{s_6}\sum_{s_5}(-1)^{s_1+s_2+s_6+s_5}U^E_{-s_1,s_2}U^J_{-s_2,s_6}U^F_{s_5,-s_6}U^I_{s_1,-s_5}
\end{equation}
where each sum includes only the two terms $s_i=\pm\tfrac{1}{2}$.
Notice that the subscripts on $U^F$ and $U^I$ have been interchanged because going counterclockwise around the plaquette
means we are going from the $m^\prime$ end to the $m$ end on those two links.
The effect of an operator $U$ on a link's state in the angular momentum notation is \cite{Byrnes_Yamamoto,Klco:2019evd}
\begin{eqnarray}
U_{s,s^\prime}\left|j,m,m^\prime\right> &=& \sum_{J=\left|j-\tfrac{1}{2}\right|}^{j+\tfrac{1}{2}}\sqrt{\frac{2j+1}{2J+1}}
\sum_M\sum_{M^\prime}\left<J,M|j,m;\tfrac{1}{2},s\right>\left<J,M^\prime|j,m^\prime;\tfrac{1}{2},s^\prime\right>
\left|J,M,M^\prime\right> \nonumber \\
&=& \sum_{J=\left|j-\tfrac{1}{2}\right|}^{j+\tfrac{1}{2}}\sqrt{2j+1}\sqrt{2J+1}
   \sum_M\sum_{M^\prime}(-1)^{1-2j+M+M^\prime}\left(\begin{array}{ccc} j & \tfrac{1}{2} & J \\ m & s & -M \end{array}\right) \nonumber \\    
&&\times   \left(\begin{array}{ccc} j & \tfrac{1}{2} & J \\ m^\prime & s^\prime & -M^\prime \end{array}\right)
   \left|J,M,M^\prime\right>
\end{eqnarray}
where the sums over $M$ and $M^\prime$ contain only a single nonzero term because the Clebsch-Gordan coefficients
vanish unless $M=m+s$ and $M^\prime=m^\prime+s^\prime$.
Applying $\square_1$ to our initial state gives
\begin{eqnarray} \label{eq:plaq_Istate}
\square_1\left|\psi_{\rm initial}\right>
&=&\sum_{M_E}\sum_{M_E^\prime}\sum_{M_J}\sum_{M_J^\prime}\sum_{M_F}\sum_{M_F^\prime}\sum_{M_I}\sum_{M_I^\prime}
   \sum_{m_A}\sum_{m_A^\prime}\ldots\sum_{m_L}\sum_{m_L^\prime}
   \sum_{J_F}\sum_{J_E}\sum_{J_I}\sum_{J_J}\sum_{s_1}\sum_{s_2}\sum_{s_6}\sum_{s_5} \nonumber \\
&& (-1)^{s_1+s_2+s_6+s_5}
   (-1)^{-2j_E-2j_J-2j_F-2j_I+M_E+M_E^\prime+M_J+M_J^\prime+M_F+M_F^\prime+M_I+M_I^\prime} \nonumber \\
&& \sqrt{2j_E+1}\sqrt{2J_E+1}
   \sqrt{2j_J+1}\sqrt{2J_J+1}
   \sqrt{2j_F+1}\sqrt{2J_F+1}
   \sqrt{2j_I+1}\sqrt{2J_I+1} \nonumber \\
&& \left(\begin{array}{ccc} j_A & j_E & j_I \\ m_A^\prime & m_E & m_I \end{array}\right)
   \left(\begin{array}{ccc} j_C & j_E & j_J \\ m_C & m_E^\prime & m_J \end{array}\right)
   \left(\begin{array}{ccc} j_C & j_G & j_K \\ m_C^\prime & m_G & m_K^\prime \end{array}\right)
   \left(\begin{array}{ccc} j_A & j_G & j_L \\ m_A & m_G^\prime & m_L \end{array}\right) \nonumber \\
&& \left(\begin{array}{ccc} j_B & j_F & j_I \\ m_B^\prime & m_F & m_I^\prime \end{array}\right)
   \left(\begin{array}{ccc} j_D & j_F & j_J \\ m_D & m_F^\prime & m_J^\prime \end{array}\right)
   \left(\begin{array}{ccc} j_D & j_H & j_K \\ m_D^\prime & m_H & m_K^\prime \end{array}\right)
   \left(\begin{array}{ccc} j_B & j_H & j_L \\ m_B & m_H^\prime & m_L^\prime \end{array}\right) \nonumber \\
&& \left(\begin{array}{ccc} j_E & \tfrac{1}{2} & J_E \\ m_E & -s_1 & -M_E \end{array}\right)
   \left(\begin{array}{ccc} j_E & \tfrac{1}{2} & J_E \\ m_E^\prime & s_2 & -M_E^\prime \end{array}\right)
   \left(\begin{array}{ccc} j_J & \tfrac{1}{2} & J_J \\ m_J & -s_2 & -M_J \end{array}\right)
   \left(\begin{array}{ccc} j_J & \tfrac{1}{2} & J_J \\ m_J^\prime & s_6 & -M_J^\prime \end{array}\right) \nonumber \\
&& \left(\begin{array}{ccc} j_F & \tfrac{1}{2} & J_F \\ m_F & s_5 & -M_F \end{array}\right)
   \left(\begin{array}{ccc} j_F & \tfrac{1}{2} & J_F \\ m_F^\prime & -s_6 & -M_F^\prime \end{array}\right)
   \left(\begin{array}{ccc} j_I & \tfrac{1}{2} & J_I \\ m_I & s_1 & -M_I \end{array}\right)
   \left(\begin{array}{ccc} j_I & \tfrac{1}{2} & J_I \\ m_I^\prime & -s_5 & -M_I^\prime \end{array}\right) \nonumber \\
&& \left|j_A,m_A,m_A^\prime\right>\left|j_B,m_B,m_B^\prime\right>\left|j_C,m_C,m_C^\prime\right>\left|j_D,m_D,m_D^\prime\right>
   \left|J_E,M_E,M_E^\prime\right>\left|J_F,M_F,M_F^\prime\right> \nonumber \\
&& \left|j_G,m_G,m_G^\prime\right>\left|j_H,m_H,m_H^\prime\right>
   \left|J_I,M_I,M_I^\prime\right>\left|J_J,M_J,M_J^\prime\right>\left|j_K,m_K,m_K^\prime\right>\left|j_L,m_L,m_L^\prime\right>.
\end{eqnarray}
Applying a final state to that result allows all sums to be performed and the answer simplifies to
\begin{eqnarray}
&&\left<\psi_{\rm final}\right|\square_1\left|\psi_{\rm initial}\right>
=\nonumber \\
&&(-1)^{j_A+j_B+j_C+j_D+2J_E+2J_F+2j_I+2j_J} \nonumber \\
& & \times \sqrt{2j_E+1}\sqrt{2J_E+1}\sqrt{2j_J+1}\sqrt{2J_J+1}\sqrt{2j_F+1}\sqrt{2J_F+1}\sqrt{2j_I+1}\sqrt{2J_I+1} \nonumber \\
& & \times \left\{\begin{array}{ccc} j_A & j_E & j_I \\ \tfrac{1}{2} & J_I & J_E \end{array}\right\}
    \left\{\begin{array}{ccc} j_B & j_F & j_I \\ \tfrac{1}{2} & J_I & J_F \end{array}\right\}
    \left\{\begin{array}{ccc} j_C & j_E & j_J \\ \tfrac{1}{2} & J_J & J_E \end{array}\right\}
    \left\{\begin{array}{ccc} j_D & j_F & j_J \\ \tfrac{1}{2} & J_J & J_F \end{array}\right\}
\end{eqnarray}
where the four curly brackets, one for each node, are the 6j symbols whose role is to enforce the Gauss's law at each node. The numerical values of the 6j symbols are typically a square roots of ratios, and can be obtained from:
\begin{eqnarray}
\left\{\begin{array}{ccc} j_0 & j_1 & j_2 \\ \tfrac{1}{2} & j_2+\tfrac{1}{2} & j_1+\tfrac{1}{2} \end{array}\right\}
&=& (-1)^{1+j_0+j_1+j_2}\sqrt{\frac{(1-j_0+j_1+j_2)(2+j_0+j_1+j_2)}{(2j_1+1)(2j_1+2)(2j_2+1)(2j_2+2)}} \,, \\
\left\{\begin{array}{ccc} j_0 & j_1 & j_2 \\ \tfrac{1}{2} & j_2-\tfrac{1}{2} & j_1+\tfrac{1}{2} \end{array}\right\}
&=& (-1)^{j_0+j_1+j_2}\sqrt{\frac{(1+j_0+j_1-j_2)(j_0-j_1+j_2)}{(2j_1+1)(2j_1+2)2j_2(2j_2+1)}} \,, \\
\left\{\begin{array}{ccc} j_0 & j_1 & j_2 \\ \tfrac{1}{2} & j_2+\tfrac{1}{2} & j_1-\tfrac{1}{2} \end{array}\right\}
&=& (-1)^{j_0+j_1+j_2}\sqrt{\frac{(1+j_0-j_1+j_2)(j_0+j_1-j_2)}{2j_1(2j_1+1)(2j_2+1)(2j_2+2)}} \,, \\
\left\{\begin{array}{ccc} j_0 & j_1 & j_2 \\ \tfrac{1}{2} & j_2-\tfrac{1}{2} & j_1-\tfrac{1}{2} \end{array}\right\}
&=& (-1)^{j_0+j_1+j_2}\sqrt{\frac{(1+j_0+j_1+j_2)(-j_0+j_1+j_2)}{2j_1(2j_1+1)2j_2(2j_2+1)}} \,,
\end{eqnarray}
but a 6j symbol is zero unless all four of its triangle conditions are satisfied \cite{Varshalovich:1988ye,Thompson1994AngularMA}.\\
\\
Results for $\left<\psi_{\rm final}\right|\square_i\left|\psi_{\rm initial}\right>$ with $i=2,3,4$ can be obtained simply by translation symmetry from the $i=1$ result or by explicit calculation. It is important to notice that this can be quite a challenging and error prone calculation, therefore a more practical notation can be useful to this scope.
Let's consider a generic plaquette $(\square_p)$ on a generic lattice with links labelled as in Figure~\ref{fig:plaquette_lattice}:
\begin{figure}[H]
\centering
\includegraphics[width=0.4\linewidth]{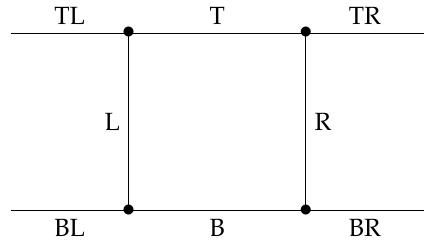}
\caption{A single plaquette with link labelled for an easy generalization of $\left<\psi_{\rm final}\right|\square_p\left|\psi_{\rm initial}\right>$ to any plaquette inside a 1-dimensional lattice. Each link  position is labelled with a letter that is a short notation for: top left (TL), top (T), top right (TR), left (L), right (R), bottom left (BL), bottom (B) and bottom right (BR).}\label{fig:plaquette_lattice}
\end{figure}
\noindent with this generic labelling the plaquette operator reads:

\begin{eqnarray}\label{eq:pla_coding_notation}
&&\left<\psi_{\rm final}\right| \hat \square_p\left|\psi_{\rm initial}\right>
=\nonumber \\
&& (-1)^{j_{BL}+j_{TL}+j_{BR}+j_{TR}+2J_B+2J_T+2j_L+2j_R} \nonumber \\
&& \times \sqrt{2j_B+1}\sqrt{2J_B+1}\sqrt{2j_R+1}\sqrt{2J_R+1}\sqrt{2j_T+1}\sqrt{2J_T+1}\sqrt{2j_L+1}\sqrt{2J_L+1} \nonumber \\
&& \times \left\{\begin{array}{ccc} j_{BL} & j_B & j_L \\ \tfrac{1}{2} & J_L & J_B \end{array}\right\} 
    \left\{\begin{array}{ccc} j_{TL} & j_T & j_L \\ \tfrac{1}{2} & J_L & J_T \end{array}\right\} 
    \left\{\begin{array}{ccc} j_{BR} & j_B & j_R \\ \tfrac{1}{2} & J_R & J_B \end{array}\right\} 
    \left\{\begin{array}{ccc} j_{TR} & j_T & j_R \\ \tfrac{1}{2} & J_R & J_T \end{array}\right\}
\end{eqnarray}

\chapter{State generator algorithm}\label{appendix:state_algorithm}
In this appendix we describe the algorithms developed and used to derive the physical states and the matrix representation of an SU(2) lattice gauge theory with Hamiltonian as:
\begin{equation}\label{eq:H_appendix}
\hat H = \frac{g^2}{2}\left(\sum_{i={\rm links}}\hat E_i^2-2x\sum_{i={\rm plaquettes}}\hat\square_i\right)
\end{equation}
In the angular momentum formulation, a generic state of a 2-dimensional lattice made by $N$ links arranged in plaquettes can be represented as:
\begin{equation}
\left|\psi\right> = \left|j_1,m_1,m_1^\prime\right>\left|j_2,m_2,m_2^\prime\right>\ldots\left|j_N,m_N,m_N^\prime\right> \,.
\end{equation}
We have shown in section \ref{section_H_representation} that the matrix element of the operators inside the Hamiltonian do not depend on the two projection $(m, m^\prime)$ at the end of each link, therefore a state is simply labelled as:\\
\begin{equation}
\left|\psi\right> = \left|j_1,j_2, \ldots, j_N \right>
\end{equation}
This state is a physical state only if it respects the SU(2) Gauss’s law at each node. This is expressed by forcing the links joining at a node to respect the angular momentum composition rule.\\
To better express the restriction at a code we should consider that on a generic 2-dimensional lattice there are two types of nodes depending on how many links touch it:
\begin{figure}[H]
\centering
\includegraphics[width=0.4\linewidth]{
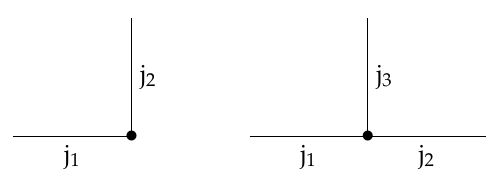}
\caption{The two possible type of nodes in a 1-dimensional lattice made by a row of plaquettes. On the left a 2-link node while on the right a 3-link node where $j_1$, $j_2$ and $j_3$ are generic links joining at the nodes.}\label{fig:2_3_links}
\end{figure}
We can label them as a 2-link node if two links touch it and a 3-link node if instead three links touch it. The angular momentum composition rule can be expressed for each node type as:
\begin{eqnarray}\label{eq:2LN_3LN}
\textrm{2-link node} \qquad &j_1==j_2 \\ \nonumber
\textrm{3-link node} \qquad &j_3 \in \lbrace \vert j_1 - j_2 \vert, \vert j_1 - j_2 \vert +1, \ldots, j_1 + j_2  \rbrace
\end{eqnarray}
These equations are sufficient to generate the physical states of the system once they are applied at each node of the lattice.\\
\\
The algorithm finds the physical states by first generating all the possible candidate states  $|phy>$ giving each link a j value from 0 up to a fixed $j_{max}$ by step of $1/2$ and secondly by checking that each link's value satisfies the angular momentum composition rule at each node they touch. For a lattice with N links organized in plaquettes the algorithm in pseudocode reads:
\begin{algorithm}[H]
\caption{Calculate physical states SU(2) Hamiltonian Eq.~\ref{eq:H_appendix}}
\begin{algorithmic} 
\FOR{$(j_1, j_2,\ldots,j_N)$ each with value from 0 to $j_{max}$ by step of 1/2}
\IF {(all 2LN are true) and (all 3LN are true)}
\STATE {$(j_1, j_2, … ,j_N)$  belong to the Hilbert space of the theory}
\ELSE
\STATE{it is not a physical state, discard it.}
\ENDIF
\ENDFOR
\end{algorithmic}
\end{algorithm}
\noindent where 2LN and 3LN are the 2-link and 3-link node function respectively of Eq.~\ref{eq:2LN_3LN}.\\
\\
There are two final observations: The algorithm could have been written without using the result for which the matrix element of the Hamiltonian does not depend on the value of the two projections $m,m^\prime$. Therefore a more general algorithm could use the more cumbersome notation $\left|\psi\right> = \left|j_A,m_A,m_A^\prime\right>\left|j_B,m_B,m_B^\prime\right>\ldots\left|j_L,m_L,m_L^\prime\right> $ and impose the SU(2) Gauss’s law by using for each node a 3j symbol as a restriction on each node. In this way the computation is massive and consequently the running time will be extremely large, because the algorithm has to explore all the possible values for $m,m^\prime \in (-j,-j+1/2,…,j)$ for each value $j$ of each link in the lattice. \\
\\
Secondly the algorithm could be easily extended to 2 and 3 dimensional lattices by introducing 4-link and n-link nodes by simply generalizing the link function and by deriving the plaquette contribution $\left<\psi_{\rm final}\right|\square_{p}\left|\psi_{\rm initial}\right>$ for a generic plaquette surrounded by 4 and n plaquettes respectively.

\section{Using the algorithm for 3 plaquettes}
In this section we show an example of how to use the algorithm to derive the states and the Hamiltonian representation of a 3-plaquette lattice with closed boundary condition (CBC) and $j_{max}=1/2$ truncation.\\
\\
The labeling of the lattice spatial element, link node and plaquette is presented in figure Figure~\ref{fig:3_pla_lattoce}:
\begin{figure}[H]
\centering
\includegraphics[width=0.5\linewidth]{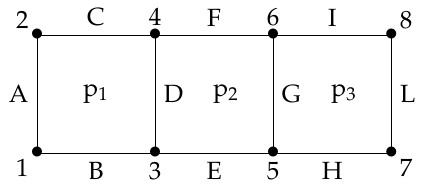}
\caption{A 3-plaquette lattice with closed boundary condition. 
Each plaquette is indicated by the label in the center of each square, each link by the letter at the middle point and, nodes by the closest number.}
\label{fig:3_pla_lattoce}
\end{figure}
\noindent Hence a global state on this lattice is represented using the following notation: \\
\begin{equation}
\left|\psi\right> = \left|j_a, j_b, j_c, j_d, j_e, j_f, j_g, j_h, j_i, j_l\right>\end{equation}
The selected lattice has 8 nodes, 4 2-link nodes (1,2,7,8) and 4 3-link nodes (3,4,5,6) and for each of them we impose the Gauss’s law.
The algorithm for this case reads:\\
\begin{algorithm}[H]
\caption{Calculate physical states SU(2) Hamiltonian 3-plaquette with CBC}
\begin{algorithmic} 
\FOR{$(j_a, j_b, j_c, j_d, j_e, j_f, j_g, j_h, j_i, j_l)$ each with value from 0 to $j_{max}$ by step of 1/2}
\IF {(~Ia==Ib and Ia==Ic and Il==Ii and Il==Ih) and\\ 
       3LN(Ic,Id,If) and 3LN(Ib,Id,Ie) and 3LN(Ie,Ig,Ih) and 3LN(If,Ig,Ii))}
\STATE {$(j_a, j_b, j_c, j_d, j_e, j_f, j_g, j_h, j_i, j_l)$  belong to the Hilbert space of the theory}
\ELSE
\STATE{discard it.}
\ENDIF
\ENDFOR
\end{algorithmic}
\end{algorithm}
Running this code with $j_{max}=1/2$ we get the following physical states:
\begin{eqnarray}
\nonumber
&&\left| 0, 0, 0, 0, 0, 0, 0, 0, 0, 0 \right> \\ \nonumber
&&\left| 0, 0, 0, 0, 0, 0, \frac{1}{2}, \frac{1}{2}, \frac{1}{2}, \frac{1}{2} \right> \\  \nonumber
&&\left| 0, 0, 0, \frac{1}{2}, \frac{1}{2}, \frac{1}{2}, \frac{1}{2}, 0, 0, 0 \right> \\  \nonumber
&&\left| \frac{1}{2}, \frac{1}{2}, \frac{1}{2}, \frac{1}{2}, 0, 0, 0, 0, 0, 0 \right> \\ 
&&\left| 0, 0, 0, \frac{1}{2}, \frac{1}{2}, \frac{1}{2}, 0, \frac{1}{2}, \frac{1}{2}, \frac{1}{2} \right> \\  \nonumber
&&\left| \frac{1}{2}, \frac{1}{2}, \frac{1}{2}, 0, \frac{1}{2}, \frac{1}{2}, \frac{1}{2}, 0, 0, 0 \right> \\  \nonumber
&&\left| \frac{1}{2}, \frac{1}{2}, \frac{1}{2}, 0, \frac{1}{2}, \frac{1}{2}, 0, \frac{1}{2}, \frac{1}{2}, \frac{1}{2} \right> \\  \nonumber
&&\left| \frac{1}{2}, \frac{1}{2}, \frac{1}{2}, \frac{1}{2}, 0, 0, \frac{1}{2}, \frac{1}{2}, \frac{1}{2}, \frac{1}{2} \right> \\  \nonumber
\end{eqnarray}
where we can see that starting from the vacuum (state 1), we see a state with the right plaquette switched ``on'' (state 2), then the center plaquette ``on'' (state 3) and so on until the last one where left and right plaquette are switched ``on''.\\

The matrix representation can be calculated using these states and the equation for the matrix element of the operators $\hat E_i^2$ and $\hat\square_{pi}$.\\
\\
The chromoelectric part is simply:
\begin{equation}
\left<\psi_{ \rm final}\right|\sum_i\hat E_i^2\left|\psi_{ \rm initial}\right> = \sum_{i=a }^lj_i(j_i+1) \delta_{\rm final, \rm initial} \,.
\end{equation}
\\
For the contribution of the chromomagnetic part we have to obtain the expression for each plaquette on this specific lattice configuration. To do that we use the notation in Appendix \ref{appendix:plaquette_derivation} Eq.~\ref{eq:pla_coding_notation}, and we get that the first plaquette is:\\
\begin{eqnarray}
\left<\psi_{\rm final}\right|\square_{p1}\left|\psi_{\rm initial}\right>
&&= (-1)^{j_f+2J_b+2J_c+2j_a+2j_d} \nonumber \\
&& \sqrt{2j_b+1}\sqrt{2J_b+1}\sqrt{2j_d+1}\sqrt{2J_d+1}\sqrt{2j_c+1}\sqrt{2J_c+1}\sqrt{2j_a+1}\sqrt{2J_a+1} \nonumber \\
&& \left\{\begin{array}{ccc} 0 & j_b & j_a \\ \tfrac{1}{2} & J_a & J_b \end{array}\right\} \nonumber
    \left\{\begin{array}{ccc} 0 & j_c & j_a \\ \tfrac{1}{2} & J_a & J_c \end{array}\right\} \nonumber
    \left\{\begin{array}{ccc} j_e & j_b & j_d \\ \tfrac{1}{2} & J_d & J_b \end{array}\right\} \nonumber
    \left\{\begin{array}{ccc} j_f & j_c & j_d \\ \tfrac{1}{2} & J_d & J_c \end{array}\right\}\\ \nonumber
&&   \delta_{j_e Je} \, \delta_{j_f Jf} \,  \delta_{j_g Jg} \,  \delta_{j_i Ji} \,  \delta_{j_h Jh} \,  \delta_{j_l Jl}\, \nonumber
\end{eqnarray}\label{eq:appendix_p1}
\noindent the second plaquette is:\\
\begin{eqnarray}
\left<\psi_{\rm final}\right|\square_{p2}\left|\psi_{\rm initial}\right>
&&= (-1)^{j_b+j_c+j_h+j_i+2J_e+2J_f+2j_d+2j_g} \nonumber \\
&& \sqrt{2j_e+1}\sqrt{2J_e+1}\sqrt{2j_g+1}\sqrt{2J_g+1}\sqrt{2j_f+1}\sqrt{2J_f+1}\sqrt{2j_d+1}\sqrt{2J_d+1} \nonumber \\
&& \left\{\begin{array}{ccc} j_b & j_e & j_d \\ \tfrac{1}{2} & J_d & J_e \end{array}\right\} \nonumber
    \left\{\begin{array}{ccc} j_c & j_f & j_d \\ \tfrac{1}{2} & J_d & J_f \end{array}\right\} \nonumber
    \left\{\begin{array}{ccc} j_h & j_e & j_g \\ \tfrac{1}{2} & J_g & J_e \end{array}\right\} \nonumber
    \left\{\begin{array}{ccc} j_i & j_f & j_g \\ \tfrac{1}{2} & J_g & J_f \end{array}\right\}\\ \nonumber  
&&   \delta_{j_a J_a} \, \delta_{j_b J_b} \,  \delta_{j_c J_c} \,  \delta_{j_h J_h} \,  \delta_{j_i J_i} \,  \delta_{j_l J_l}\, \nonumber
\end{eqnarray}\label{eq:appendix_p2}

\noindent and finally the third plaquette is:
\begin{eqnarray}
\left<\psi_{\rm final}\right|\square_{p3}\left|\psi_{\rm initial}\right>
&&= (-1)^{j_e+j_f+2J_h+2J_i+2j_g+2j_l} \nonumber \\
&& \sqrt{2j_h+1}\sqrt{2J_h+1}\sqrt{2j_l+1}\sqrt{2J_l+1}\sqrt{2j_i+1}\sqrt{2J_i+1}\sqrt{2j_g+1}\sqrt{2J_g+1} \nonumber \\
&& \left\{\begin{array}{ccc} j_e & j_h & j_g \\ \tfrac{1}{2} & J_g & J_h \end{array}\right\} \nonumber
    \left\{\begin{array}{ccc} j_f & j_i & j_g \\ \tfrac{1}{2} & J_g & J_i \end{array}\right\} \nonumber
    \left\{\begin{array}{ccc} 0 & j_h & j_l \\ \tfrac{1}{2} & J_l & J_h \end{array}\right\} \nonumber
    \left\{\begin{array}{ccc} 0 & j_i & j_l \\ \tfrac{1}{2} & J_l & J_i \end{array}\right\}\\ \nonumber
&&   \delta_{j_a J_a} \, \delta_{j_b J_b} \,  \delta_{j_c J_c} \,  \delta_{j_d J_d} \,  \delta_{j_e J_e} \,  \delta_{j_f J_f}\, \nonumber
\end{eqnarray}
where the extra deltas at the end of each plaquette are there to remind us that the plaquette contribution is non-zero only if the links in $\left|\psi_{\rm initial}\right>$ and $\left|\psi_{\rm final}\right>$ external to the acting plaquette have the same values. This is important if we use a code to perform the calculation.\\
\\
Now, a simple code can ease us from the burden of computing this series of tedious calculations. This code joined together with the one implementing the physical state algorithm form the 3-plaquette code for an SU(2) lattice gauge theory with closed boundary condition and tunable $j_{max}$.\\
\\
Finally using this code we get the Hamiltonian:
\begin{equation} 
H = \frac{g^2}{2}
\left(
\begin{array}{cccccccc}
 0 & -2 x & -2 x & -2 x & 0 & 0 & 0 & 0 \\
 -2 x & 3 & 0 & 0 & -x & 0 & 0 & -2 x \\
 -2 x & 0 & 3 & 0 & -x & -x & 0 & 0 \\
 -2 x & 0 & 0 & 3 & 0 & -x & 0 & -2 x \\
 0 & -x & -x & 0 & \frac{9}{2} & 0 & -x & 0 \\
 0 & 0 & -x & -x & 0 & \frac{9}{2} & -x & 0 \\
 0 & 0 & 0 & 0 & -x & -x & 6 & -\frac{x}{2} \\
 0 & -2 x & 0 & -2 x & 0 & 0 & -\frac{x}{2} & 6 \\
\end{array}
\right)
\begin{array}{c}
\left| 1_1^1 1_1^1 1_1^11 \right>  \\
\left| 1_1^1 1_1^1 2_2^22 \right>  \\
\left| 1_1^1 2_2^2 2_1^1 1 \right>  \\
\left| 2_2^22_1^11_1^11 \right> \\
\left| 1_1^12_2^21_2^22 \right>  \\
\left| 2_2^21_2^22_1^1 1 \right>  \\
\left| 2_2^2 1_2^21_2^22 \right>  \\
\left| 2_2^22_1^12_2^22 \right> \\
\end{array} \,
\end{equation}%
where to make it easier to read and recognize the states we use a compact notation that closely resembles a miniature of the lattice: $\left|\psi\right> =\left|A_B^C D_E^F G_H^I L\right>$ where the capital letters are $A=2j_a+1$ and are used to get rid of the fractions.\\
\\
The code can be used to generate the Hamiltonian for the case of $j_{max}=1$ that has 49 states therefore “this margin is too narrow to contain”.


\chapter{Decomposing an operator in Pauli gates}\label{sec:gate_decomposer}
In this section we discuss a simple way to decompose a generic operator using the identity and the Pauli gates $I, X, Y, Z$ in a way that is easy to code and therefore calculate by software. Finally, we present an algorithm that finds the gate decompositions with a small number of gates. 
Let’s consider a generic observable $O$ represented by a $2 \times 2$ matrix:
\begin{equation}
\hat{O} \doteq \left(
\begin{array}{cc}
a & b \\
c & d
\end{array}
\right)
\end{equation}
How can we decompose this observable using only I,X,Y and Z?
One could be tempted to create a system of equations and solve it. However, there is a simpler way that can be easily generalized to any observables represented by a $4 \times 4$ matrix or a $2^n \times 2^n$.
Starting from the $2 \times 2$ case we can simply notice that the observable is decomposable as:
\begin{equation}
\hat{O} \doteq 
a \left(
\begin{array}{cc}
1 & 0 \\
0 & 0
\end{array}
\right) +
b \left(
\begin{array}{cc}
0 & 1 \\
0 & 0
\end{array}
\right) +
c \left(
\begin{array}{cc}
0 & 0 \\
1 & 0 
\end{array}
\right) +
d \left(
\begin{array}{cc}
0 & 0 \\
0 & 1
\end{array}
\right)   
\end{equation}
where each element of the basis can be simply labelled based on the position of the 1 as:
\begin{equation}
TL = \left(
\begin{array}{cc}
1 & 0 \\
0 & 0
\end{array}
\right)\,,
\quad
TR=\left(
\begin{array}{cc}
0 & 1 \\
0 & 0
\end{array}
\right)\,,
\quad
BL= \left(
\begin{array}{cc}
0 & 0 \\
1 & 0 
\end{array}
\right)\,,
\quad
BR= \left(
\begin{array}{cc}
0 & 0 \\
0 & 1
\end{array}
\right) 
\end{equation}
Looking to the matrix representation of the chosen gates:
\begin{equation}
I = \left(
\begin{array}{cc}
1 & 0 \\
0 & 1
\end{array}
\right)\,,
\quad
X= \left(
\begin{array}{cc}
0 & 1 \\
1 & 0
\end{array}
\right)\,,
\quad
Y= \left(
\begin{array}{cc}
0 & - i \\
i & 0 
\end{array}
\right)\,,
\quad
Z = \left(
\begin{array}{cc}
1 & 0 \\
0 & -1
\end{array}
\right) 
\end{equation}
the basis can be rewritten using the chosen gates as:
\begin{equation}
TL =\dfrac{I + Z}{2}
\,,\quad
TR =\dfrac{X + i Y}{2}
\,,\quad
BL =\dfrac{X - i Y}{2}
\,,\quad
BR= \dfrac{I - Z}{2}
\end{equation}
We then define what we call the gate decomposer labelled as:
\begin{equation}
GD_{2 \times 2} =  \left[
\begin{array}{cc}
TL & TR \\
BL & BR
\end{array}
\right]
=
\left[
\begin{array}{cc}
(I + Z)/2 & (X + i Y)/2 \\
(X - i Y)/2 & (I - Z)/2
\end{array}
\right]
\end{equation}
where the square brackets in place of more common round brackets are used to make clear that it is not a common matrix but a matrix of matrices.
Finally the operator can be decomposed using the gates as: 
\begin{equation}
\hat{O} \doteq GD_{2 \times 2} \odot \hat{O} = \sum_{i,j}  (GD_{2 \times 2})_{i,j} \; O_{i,j}
\end{equation}
where $\odot$ is the element-wise product or Hadamard  product.\\
Executing all of the products, the operator reads:
\begin{equation}
\hat{O} \doteq \frac{1}{2}\left( (b+c) X + i (b - c) Y + (a-d) Z + (a+d)I\right) 
\end{equation}
To extend this approach to a generic operator represented by a $4 \times 4$ matrix, it is enough to rewrite the basis in gates. It is here that this method is quite effective, because there is a simple way to obtain the basis element of the $4 \times 4$ from the one of the $2 \times 2$. 
The first element of the $4 \times 4$ can be written as:
\begin{eqnarray}
\left(
\begin{array}{cccc}
1 & 0 & 0 & 0 \\
0 & 0 & 0 & 0 \\
0 & 0 & 0 & 0 \\
0 & 0 & 0 & 0 \\
\end{array}
\right)&=& \left(
\begin{array}{cc}
1 & 0 \\
0 & 0
\end{array}
\right)
\otimes
\left(
\begin{array}{cc}
1 & 0 \\
0 & 0
\end{array}
\right)
=
TL \otimes TL \\
&=& \dfrac{1}{4} \left( I\otimes I +I\otimes Z + Z \otimes I + Z \otimes Z \right) 
\end{eqnarray}
and similarly, the other elements of the basis are simply made by space ordered products, for example:
\begin{eqnarray}
\left(
\begin{array}{cccc}
0 & 0 & 0 & 0 \\
0 & 0 & 1 & 0 \\
0 & 0 & 0 & 0 \\
0 & 0 & 0 & 0 \\
\end{array}
\right) &=& \left(
\begin{array}{cc}
0 & 1 \\
0 & 0
\end{array}
\right)
\otimes
\left(
\begin{array}{cc}
0 & 0 \\
1 & 0
\end{array}
\right)
=
TR \otimes BL \\
&=& \dfrac{1}{4} \left( X \otimes X -i X \otimes Y + i Y \otimes X + Y \otimes Y \right) 
\end{eqnarray}
The $GD_{4 \times 4} $ is built using $GD_{2 \times 2}$ as follows:\\
\begin{equation}
GD_{4 \times 4}= GD_{2 \times 2} \otimes GD_{2 \times 2} =\left[
\begin{array}{cccc}
 \text{TL} \otimes \text{TL} & \text{TL} \otimes \text{TR} & \text{TR} \otimes \text{TL} &
   \text{TR} \otimes \text{TR} \\
 \text{TL} \otimes \text{BL} & \text{TL} \otimes \text{BR} & \text{TR} \otimes \text{BL} &
   \text{TR} \otimes \text{BR} \\
 \text{BL} \otimes \text{TL} & \text{BL} \otimes \text{TR} & \text{BR} \otimes \text{TL} &
   \text{BR} \otimes \text{TR} \\
 \text{BL} \otimes \text{BL} & \text{BL} \otimes \text{BR} & \text{BR} \otimes \text{BL} &
   \text{BR} \otimes \text{BR} \\
\end{array}
\right]
\end{equation}
by substituting the values of $TL, TR, BL, BR$ in terms of $I, X, Y, Z$ one has $GD_{4 \times 4} $ written in gates.
\\
The result can be easily extended to $2^n \times 2^n$ as:\\
\begin{equation}
GD_{2^n \times 2^n}= \underbrace{GD_{2 \times 2} \otimes GD_{2 \times 2} \otimes \cdots \otimes GD_{2 \times 2}}_{n}
\end{equation}
A few observations are necessary: first this strategy is meant to be used on a calculator, where one has pre-calculated the $ GD_{n \times n}$ and then use it to decompose the matrix representation of any observables with the same dimension. Secondly, in using this method one can optimize the calculation by using the fact that any operator is Hermitian therefore one can calculate only the upper triangular part and obtain the lower triangular part as the Hermitian conjugate of the upper one.
\\
This gate decomposer is part of an algorithm that finds the minimal possible gate decomposition of an observable by finding the matrix representation that requires the smallest number of gates. This search is done by permuting the states ordered in the basis used to represent the observables.\\
\begin{algorithm}[H]
\caption{}
\begin{algorithmic} 
\FOR{ all the permutation of the basis}
\STATE {decompose the operator representation in gates}
\IF {the number of gates needed is less that the original decomposition }
\STATE {save it}
\ENDIF
\ENDFOR
\end{algorithmic}
\end{algorithm}

This algorithm is computationally expensive because given a basis with $n$ states it needs to perform $n!$ checks. In case the number of needed permutations is too large one can opt to access a few of them randomly and stop the search once a reasonably small number of gates needed is reached.\\
Furthermore, whenever the system size allows access to all or a large part of permutations, it is better to choose among the ones with the same small number of gates as the permutation that contains the largest number of native gates present on the specific hardware, so that the circuit will be executed on the hardware without using extra gates necessary to convert it using native gates.\\

A final observation is that, this algorithm uses a large quantity of classical resources to minimize the circuit that encodes the theory on a quantum computer. This sounds like a contradiction to the use of quantum computers, which instead should have been used in places where classical computer needs large resources. This contradiction is only apparent because the quantum computers of the NISQ era are noisy prototypes of what an actual quantum computer will be. Therefore in using these prototypes we should consider any possible way to ration the use of the few quantum resources available.

\chapter{Exponential of combinations of Pauli gates}\label{sec:exp_gate_identies}
In this section we present a general way to exactly calculate the exponential of a combination of Pauli gates and show how to express it in terms of controlled Pauli X gates, a rotation gate and other single gates like Hadamard and $S$ gates. These tedious calculations are necessary not only to encode the time evolution operator on the hardware, but most importantly, to optimize the circuit by reducing the necessary number of gates, especially the noisiest ones: the CNOT and the $\rm SWAP$ gate.\\
\\
The most fundamental relation to prove is how to express the exponential of a single Pauli gate indicated as $A=X, Y, Z$, $exp(-i \theta A )$ in terms of $SU(2)$ rotation gates.\\
This can be simply proved by Taylor expanding the exponential as:
\begin{equation}
e^{-i \theta A } = I - i \theta A + \ldots +  {-i}^n\dfrac{\theta^n}{n!} A^n + \ldots 
\end{equation}
and  using the idempotent property of the operators $A^2=A$, the expression can be rearranged as:
\begin{equation} \label{eq:exp_A_pauli}
e^{-i \theta A } = I \left( \sum_{n \, \rm even } {-i}^n\dfrac{\theta^n}{n!} \right) - i \left( \sum_{n \, \rm odd } {-i}^{n-1}\dfrac{\theta^n}{n!} \right)= \cos(\theta) \, I - i \, sin(\theta) \, A
\end{equation}
where in the last line we recognise that the terms inside the parenthesis are the Taylor expansion of $\cos(\theta)$ and $\sin(\theta)$ respectively.\\
\\
To see how the previous expression leads to single qubit rotation gates, it is enough to substitute any of the Pauli operators in place of $A$ and perform the algebra as:
\begin{eqnarray}
e^{-i \theta X } &=&
\left(
\begin{array}{cc}
\cos(\theta) & - i \, sin(\theta) \\
- i \, sin(\theta) & \cos(\theta) \\
\end{array}
\right)
\equiv RX(2 \theta )  \,, \\ \nonumber
\\ 
e^{-i \theta Y } &=&
\left(
\begin{array}{cc}
\cos(\theta) &  sin(\theta) \\
 sin(\theta) & \cos(\theta) \\
\end{array}
\right)
\equiv RY(2 \theta )  \,, \\ \nonumber
\\
e^{-i \theta Z } &=&
\left(
\begin{array}{cc}
\cos(\theta) -i \, sin(\theta) &  0\\
 0 & \cos(\theta) + i \, sin(\theta) \\
\end{array}
\right)
\equiv RZ(2 \theta )  \,
\label{eq:Id_Hadamard_pauli}
\end{eqnarray}
where the rotations are not the common rotations of the real Euclidean space based on $SO(3)$ but are spin-1/2 rotations based on $SU(2)$.\\
\\
The result can be simply extended to a system with two qubits by simply following the same calculation in Equation~\ref{eq:exp_A_pauli}, obtaining: 
\begin{equation}\label{eq:exp_multigates}
e^{-i \theta B \otimes A } = cos(\theta) \, I \otimes I -  i \, sin(\theta) \,B \otimes A
\end{equation}
where the simple relation $(B \otimes A)^n =(B^n \otimes A^n)$ was used.\\
This expression can be rewritten using only a rotation gate and pairs of CNOT gates with the help of few identities, which are necessary to transform $B \otimes A$ into a pair of CNOT gates, a Pauli gate and one identity gate.\\
\\
The possible identities using only controlled Pauli X gates are:  
\begin{eqnarray}\label{eq:Id_CNOT_puali}
X_k \otimes X_j &=& CX_{jk} \, I_k \otimes X_j  \, CX_{jk} \,, \\ \nonumber
\\
Z_j \otimes Z_k &=& CX_{jk} \, I_j \otimes Z_k  \, CX_{jk} \,, \label{eq:Z_Z_cnot} \\ \nonumber
\\
X_k \otimes Y_j &=& CX_{jk} \, I_j \otimes Y_k  \, CX_{jk} \,, \label{eq:X_Y_cnot} \\ \nonumber
\\
Z_k \otimes Y_j &=& CX_{kj} I_j \otimes Y_k  \, CX_{kj} \,.
\end{eqnarray} 
\noindent which can be verified by performing the matrix multiplications.\\
\\
To see how this process works it is easier to explain with an example,
\begin{equation*}
e^{-i \theta X \otimes Y } = cos( \theta ) \, I \otimes I - i \, sin( \theta ) \, X \otimes Y
\end{equation*}
Using Equation~\ref{eq:X_Y_cnot} the two Pauli gates in the second term are rewritten as\\ $X_1  \otimes Y_0 = CX_{10} \, \left( I_1 \otimes Y_0 \right) \, CX_{10}$, leading to:
\begin{eqnarray}\label{eq:exp_XY_}
e^{-i \theta X_1 \otimes Y_0 } &=& cos(\theta) \, I_1 \otimes I_0 -  i \, sin(\theta) \, CX_{01} \, I_1 \otimes Y_0  \, CX_{01} =  \nonumber \\
&=& CX_{10} \, I_1 \otimes \left[ cos(\theta) \, I_0 -  i \, sin(\theta) \, Y_0 \right] CX_{01} = \nonumber \\
&=& CX_{01} \, I_1 \otimes RY_0(2 \theta) \, CX_{01} 
\end{eqnarray}
where in the second last equation we used the relation  $CX_{jk} (I_j \otimes I_k ) CX_{jk}= (I_j \otimes I_k )$ to bring the $CX$ pair outside, and finally recognized the $RY$ gate inside the square parenthesis. Finally its circuit is represented in Figure~\ref{fig:circuit_exp_X_Y}:
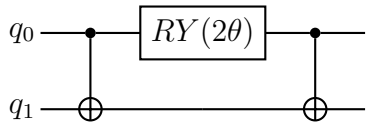
\begin{figure}[H]
\centering
\begin{quantikz}
&q_0 \, & \ctrl{1} & \gate{RY(2 \theta)}   & \ctrl{1}& \qw &  \\
&q_1 \, & \targ{} & \qw   & \targ{} & \qw &
\end{quantikz}
\caption{Two-qubit circuit representation for the operator $e^{-i \theta X_1 \otimes Y_0 }$ in Equation~\ref{eq:exp_XY_} expressed using only a pair of $CX$ and a rotation about the Y-axis $RY$. The two qubits used are labelled $q_0$ and $q_1$ respectively.}\label{fig:circuit_exp_X_Y}
\end{figure}
For the case there are two combinations of gates $X \otimes Z$ or $Y \otimes Y$ whose exponential cannot be rewritten using the relations in Equation~\ref{eq:Id_CNOT_puali}. For these two combinations one can first rewrite them using the following fundamental identities between Pauli gates and the Hadamard gate:
\begin{eqnarray}
X &=& H Z H\,, \\
Y &=&- H Y H\,,\\
Z &=&- H X H\,,
\label{eq:Id_Hadamard_pauli}
\end{eqnarray}
or the identities between Pauli gates and the $S$ gate:
\begin{eqnarray}
X &=& -S Y S^\dag\,, \\
Y &=& S X S^\dag\,,\\
Z &=& S X S^\dag\,,
\label{eq:Id_Sgate_pauli}
\end{eqnarray}
which both sets can be proved by simply performing the matrix multiplication.\\
\\
Using those relations $X \otimes Z$ or $Y \otimes Y$ can be rewritten for example in terms of two $Z$ gates:
\begin{eqnarray}
X \otimes Z &=& (H \otimes I ) \left( Z \otimes Z  \right)  (H \otimes I )\,, \\
Y \otimes Y &=& (S H \otimes S H ) \left( Z \otimes Z  \right)  (H S^\dag \otimes H S^\dag ) \,,
\end{eqnarray}
and now they can be exponentiated using the relations in Equation~\ref{eq:Z_Z_cnot} obtaining:
\begin{eqnarray}
e^{-i \theta X_j \otimes Z_k} &=& (H_j \otimes I_k ) (CX_{jk} \, I_j \otimes RZ_k( 2 \theta) \, CX_{jk})  (H_j \otimes I_k )\,,\\ \nonumber
\\
e^{-i \theta Y_j \otimes Y_k} &=& (S_j H_j \otimes S_k H_k ) (CX_{jk} \, I_j \otimes RZ_k( 2 \theta) \, CX_{jk}) (H_j S_j^\dag \otimes H_k S_k^\dag ) \,,
\end{eqnarray}
whose circuit representations are given in Figure~\ref{fig:circuit_XZ_YY}:
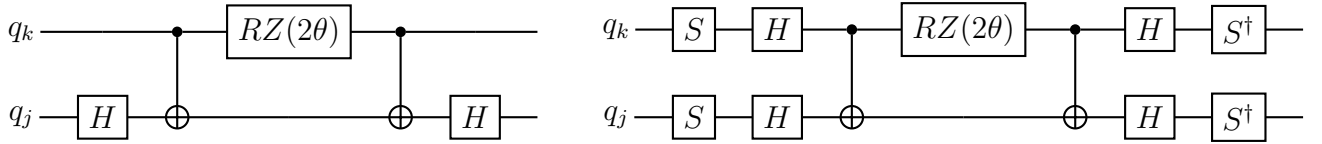
\begin{figure}[H]
\hspace{-0.8 cm} 
\begin{quantikz}
&q_k \, & \qw      & \ctrl{1} & \gate{RZ(2 \theta)}   & \ctrl{1}& \qw & \qw & \\
&q_j \, & \gate{H} & \targ{}  & \qw   				  & \targ{} &  \gate{H} & \qw &
\end{quantikz}
\hspace{-0.5 cm} 
\begin{quantikz}
&q_k \,& \gate{S} & \gate{H}      & \ctrl{1} & \gate{RZ(2 \theta)}   & \ctrl{1}& \gate{H} & \gate{S^\dag} & \qw & \\
&q_j \,& \gate{S} & \gate{H} & \targ{}  & \qw   				 	   & \targ{} &  \gate{H} & \gate{S^\dag} & \qw &
\end{quantikz}
\caption{Two-qubit circuit representation for the operators $e^{-i \theta X_j \otimes Z_k}$ on the left and $e^{-i \theta Y_j \otimes Y_k}$ on the right, expressed using a pair of $CX$,a rotation about the Z-axis $RY$ and pairs of Hadamard gates $H$ and $S$ gates. The first qubit is labelled $q_k$ while the second as $q_j$.}
\label{fig:circuit_XZ_YY}
\end{figure}
These results can be extended to a system with more than one qubit, in fact the exponential of a longer combination of Pauli gates $W \otimes  \dots \otimes B \dots A$, where $A$ acts on the first qubit and $W$ on the last one, can expressed as:
\begin{equation}\label{eq:exp_multigates}
e^{-i \theta W \otimes I \otimes \ldots \otimes A } = cos(\theta) \, I \otimes I \otimes \ldots \otimes I -  i \, sin(\theta) \, W \otimes I \otimes \ldots \otimes A
\end{equation}
This expression can be rewritten using only a rotation gate and pairs of $CX$ gates with the help of a few identities necessary to transform the gates in the second term into a combination of $CX$ gates and all the Pauli gates into identity gates except one.\\
\\
To perform this calculation one can use the relations in Equation~\ref{eq:Id_CNOT_puali} applying them to pair of consecutive gates or, use the following generalization to non-consecutive pairs:
\begin{eqnarray} \label{eq:Id_CNOT_puali_gen}
X_k \otimes [G] \otimes X_j &=& C_{jk} \, I_k \otimes [G] \otimes X_j  \, C_{jk} \,, \\ \nonumber
\\
Z_j \otimes [G] \otimes Z_k &=& C_{jk} \, I_j \otimes [G] \otimes Z_k  \, C_{jk} \,, \\ \nonumber
\\
X_k \otimes [G] \otimes Y_j &=& C_{jk} \, I_k \otimes [G] \otimes Y_j  \, C_{jk} \,, \\ \nonumber
\\
Z_k \otimes [G] \otimes Y_j &=& C_{kj} I_j \otimes [G] \otimes Y_k  \, C_{kj} \,.
\end{eqnarray}
\noindent where $[G]$ indicates a generic set of Pauli gates that is between the operators considered.\\
\\
To see how these identities are used it is worth considering an example $e^{-i \theta Z_2 \otimes Y_1 \otimes Z_0 }$. The three operators can be written by using the relation in Equation~\ref{eq:Id_CNOT_puali_gen}, contracting the outside operators first: 
\begin{equation}
\begin{aligned}
\wick{\c1 Z_2 \otimes Y_1 \otimes \c1 Z_0} &= \wick{ CX_{20} \left(I_2 \otimes \c1 Y_1 \otimes \c1 Z_0 \right)  CX_{20}}\\
&= CX_{20} \left(  I_2 \otimes \left[ CX_{01} \, Y_1 \otimes I_0 \, CX_{01} \right] \right)   CX_{20}
\end{aligned}
\end{equation}
inserting this expression into Equation~\ref{eq:exp_multigates} leads to:
\begin{equation}
e^{-i \theta Z_2 \otimes Y_1 \otimes Z_0 }= CX_{20} \left(  I_2 \otimes \left[ CX_{01} \, RY_1(2 \theta) \otimes I_0 \, CX_{01} \right] \right)   CX_{20}
\end{equation}
\\
Instead contracting consecutive pairs:
\begin{equation}
\begin{aligned}
\wick{ Z_2 \otimes \c1 Y_1 \otimes \c1 Z_0} &= \wick{ CX_{01} \c1 Z_2 \otimes \c1 Y_1 \otimes  I_0 \,  CX_{01}}\\
&= CX_{01} \, \left( CX_{21} \, I_2 \otimes Y_1 \, CX_{21} \right) \otimes I_0 \, CX_{01}
\end{aligned}
\end{equation}
\\
inserting this other expression into Equation~\ref{eq:exp_multigates} leads to:
\begin{equation}
e^{-i \theta Z_2 \otimes Y_1 \otimes Z_0 }= CX_{01} \, \left( CX_{21} \, I_2 \otimes RY_1 (2 \theta) \, CX_{21} \right) \otimes I_0 \, CX_{01}
\end{equation}
\\
The circuit representation of these two decompositions are shown in Figure~\ref{fig:circuit_ZYZ}:
\begin{figure}[H]
\centering
\begin{quantikz}
&q_0 \, & \targ{}  & \ctrl{1}   & \qw   			  & \ctrl{1}  & \targ{} &\qw \\
&q_1 \, & \qw  	   & \targ{}	& \gate{RY(2 \theta)} & \targ{}   & \qw & \qw\\
&q_2 \, & \ctrl{-2}   & \qw   		& \qw				  &  \qw & \ctrl{-2} &  \qw &
\end{quantikz}
\hspace{-0.5 cm} 
\begin{quantikz}
&q_0 \, & \ctrl{1}   & \qw          & \qw   			  & \qw        & \ctrl{1}  & \qw \\
&q_1 \, & \targ{}   & \targ{}		& \gate{RY(2 \theta)} & \targ{}   & \targ{}  & \qw\\
&q_2 \, & \qw   	 & \ctrl{-1}   	& \qw				  & \ctrl{-1}  & \qw 	   & \qw
\end{quantikz}
\caption{Three-qubit circuit representations for the operator $e^{-i \theta Z_2 \otimes Y_1 \otimes Z_0}$ on a three qubits system. On the left the circuit obtained by using external contraction, while on the right the circuit using consecutive contractions. Both circuits have the same number of gates and require the same qubit connectivity, therefore standing alone are computationally equivalent, but instead if part of a larger circuit, one could be better than the other if the number of gates is reduced when combined with other circuits.}
\label{fig:circuit_ZYZ}
\end{figure}
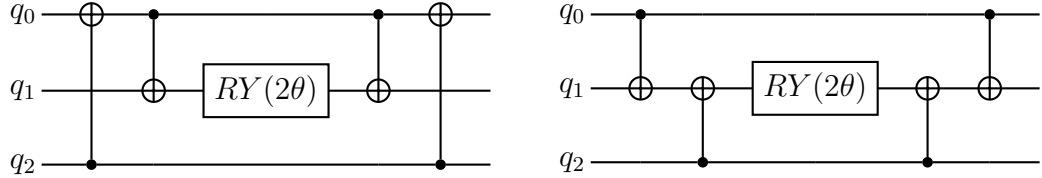
The circuits in Figure \ref{fig:circuit_ZYZ} are a clear examples showing that there can be more than one circuit represented for a given operator. This has important implications in terms of better use of the quantum resources, because one circuit can require the least number of gates; another can use only gates that are native for a specific hardware, making it computationally cheaper and less noisy to run; or another one uses only gates that respect the qubits connectivity present on the hardware, avoiding the use of $\rm SWAP$ gates. Furthermore, exploring different circuit representations is particularly valuable while using NISQ era hardware, because it can help reduce the number of gates by allowing cancellations among terms composing the full circuit. Therefore in designing a circuit one has to consider that not all those positive characteristics are shared by the same circuit, hence a trade-off between the different circuits has to be done, as it is necessary in developing the time evolution circuits presented in this chapter.

\chapter{ 2-plaquette using a $16\times16$ truncation of the $j_{\rm max}=3/2$}\label{sec:2pla_circuit16x16}

In this appendix a complete description of a single step first order Suzuki-Trotter real-time circuit for the 2-plaquette lattice using a $16\times16$ $j_{\rm max}=3/2$ truncation is given.\\
In listing each circuit a few notational conventions are used:
\begin{itemize}
\item The subscript following a rotation gate, an $H$, $S$ and $S^\dag$  indicates the qubit on which the gate acts; i.e. $RX_1(\theta)$ is a rotation along the x-axis of an angle $\theta$ that acts on the qubit 1.\\
\item Each controlled Pauli $X$, the common CNOT gate is labelled as $CX_{ij}$, where the first subscript is the control qubit while the second is the target qubit.
\end{itemize}
The complete circuit is made by 14 circuits and 4 $\rm SWAP$ gates:
\begin{equation}
\scalebox{0.85}{
$
T0 \, W1 \, W2 \, W4 \, W5 \, T12 \, T17 \, T15 \, T18 \, T20 \, SWAP_{0,1} \, W3 \, T21 \, T7 \, SWAP_{1,2} \,  T19 \,  SWAP_{2,1} \,  SWAP_{1,0}
$}
\end{equation}

where each term is a circuit whose gate contents is as follows:
\begin{eqnarray}
T0 &=& RZ_0(2 \, cd1 \,  dt) RX_0(2 \, ca1 \,  x dt) RZ_1(2 \, cd2 \, dt) RX_1(2 \, ca2  \, x dt)
\nonumber \\
&& RZ_2(2 \, cd4 \,  dt) RX_2(2 \, ca5 \,  x dt) RZ_3(2 \, cd8 \,  dt) 
\,, \\
\nonumber \\
W1 &=& CX_{3, 0} CX_{2, 0} RZ_0(2 \, cd13\,  dt) CX_{1, 0} RZ_0(2 \, cd15\,  dt) CX_{2, 0} 
\nonumber \\
&&RZ_0(2 \, cd11\,  dt) CX_{1, 0} RZ_0(2 \, cd9\,  dt) RX_3(2 \, ca17\,  x dt) H_1) CX_{1, 0} 
\nonumber \\
&& RZ_0(2 \, ca39\,  x dt) CX_{2, 0} RZ_0(2 \, ca51\,  x dt) CX_{3, 0} RZ_0(2 \, ca15\,  x dt) 
\nonumber \\
&&CX_{2, 0} RZ_0(2 \, ca3\,  x dt) CX_{1, 0} H_1 
\,, \\
\nonumber \\
W2 &=& H_3 CX_{3, 0} H_1 CX_{1, 0} RZ_0(2  \, ca19 \, x  dt) CX_{2, 0} RZ_0(2  \, ca25 \, x  dt)
\nonumber \\
&&CX_{1, 0} H_1 CX_{3, 0} H_3 RZ_0(2  \, cd5 \, dt) CX_{1, 0} RZ_0(2  \, cd7 \, dt) CX_{2, 0}
\nonumber \\
&&RZ_0(2  \, cd3 \, dt) H_2 CX_{2, 0} RZ_0(2  \, ca10 \, x  dt) CX_{3, 0} RZ_0(2  \, ca46 \, x  dt)
\nonumber \\
&&CX_{1, 0} RZ_0(2  \, ca42  \, x  dt) CX_{3, 0} RZ_0(2  \, ca6 \, x  dt) CX_{2, 0} H_2
\,, \\
\nonumber \\
W4 &=& CX_{0, 2} CX_{0, 1} RX_0(2  \, ca7 \, x  dt) CX_{0, 3} RX_0(2  \, ca21 \, x  dt) CX_{0, 1}
\nonumber \\
&&CX_{0, 2} H_0 H_3 CX_{1, 0} RZ_0(2  \, ca20 \, x  dt) CX_{2, 0} RZ_0(2  \, ca26 \, x  dt) 
\nonumber \\
&& CX_{1, 0} RZ_0(2  \, ca23 \, x  dt) CX_{3, 0} H_3 RZ_0(2  \, ca13 \, x  dt) CX_{1, 0} 
\nonumber \\
&& RZ_0(2  \, ca16 \, x  dt) CX_{3, 0} RZ_0(2  \, ca52  \, x  dt) CX_{1, 0} RZ_0(2  \, ca49 \, x  dt) 
\nonumber \\
&&  RZ_0(2  \, ca37 \, x  dt) CX_{1, 0} RZ_0(2  \, ca40 \, x  dt) CX_{3, 0} 
\nonumber \\
&&RZ_0(2  \, ca4 \, x  dt) CX_{1, 0} H_0
\,, \\
\nonumber \\ 
W5 &=& S_0 H_0 S_3 H_3 CX_{3, 0} CX_{1, 0} RZ_0(2 \, ca30\, x  dt) CX_{2, 0} RZ_0(2 \, ca36\, x  dt)
\nonumber \\
&& CX_{1, 0} RZ_0(2 \, ca33\, x  dt) CX_{2, 0} H_0 H_3 RX_0(2 \, ca27\, x  dt) CX_{0, 2} CX_{0, 1} 
\nonumber \\
&& RX_0(2 \, ca31 \, x  dt) CX_{0, 2} CX_{0, 3} S^\dag_3 S_2 CX_{0, 2} RX_0(2 \, ca11\, x  dt) CX_{0, 2}
\nonumber \\
&& S^\dag_2 CX_{0, 1} S^\dag_0
\,, \\
\nonumber \\
T12 &=& S_0 CX_{0, 2} S_1 CX_{0, 1} RX_0(2 \, ca8\,  x dt) CX_{0, 3} RX_0(2 \, ca22 \, x dt) CX_{0, 1}
\nonumber \\
&& S^\dag_1 CX_{0, 2} CX_{0, 3} S^\dag_0
\,, \\
\nonumber \\
T17 &=& S_0 H_0 CX_{3, 0} H_2 CX_{2, 0} S_1 H_1 CX_{1, 0} RZ_0(2 \, ca44 \, x  dt) CX_{1, 0} CX_{2, 0}
\nonumber \\ 
&&H_1 S^\dag_1 H_1 H_2 S_2 H_2 CX_{2, 0} CX_{1, 0} RZ_0(2 \, ca47 \, x  dt) CX_{1, 0} H_1 CX_{2, 0}
\nonumber \\ 
&& H_2 S^\dag_2 CX_{3, 0} H_0 S^\dag_0
\,, \\
\nonumber \\ 
T15 &=& S_3 H_3 CX_{3, 0} S_1 H_1 CX_{1, 0} RZ_0(2 \, ca29 \,  x dt) CX_{2, 0} RZ_0(2 \, ca35 \, x dt)
\nonumber \\
&& CX_{1, 0} H_1 S^\dag_1 CX_{2, 0} CX_{3, 0} H_3 S^\dag_3
\,, \\
\nonumber \\
T18 &=& H_0 CX_{3, 0} H_2 CX_{2, 0} H_1 CX_{1, 0} RZ_0(2 \,ca43 \, x  dt) CX_{1, 0} CX_{2, 0} \nonumber \\
&&H_1 S_1 H_1 H_2 S_2 H_2 CX_{2, 0} CX_{1, 0} RZ_0(2 \,ca48 \, x  dt) CX_{1, 0} H_1 S^\dag_1
\nonumber \\
&& CX_{2, 0} H_2 S^\dag_2 CX_{3, 0} H_0
\,,
\label{eq:16x16_circuit_p1}
\end{eqnarray}

\begin{eqnarray}%
T20 &=& S_1 CX_{0, 1} S_2 CX_{0, 2} RX_0(2 \,ca12 \, x  dt) CX_{0, 2} S^\dag_2 CX_{0, 2} S_3 CX_{0, 3} 
\nonumber \\
&& RX_0(2 \,ca32 \, x  dt) CX_{0, 3} S^\dag_3 CX_{0, 2} CX_{0, 1} S^\dag_1
\,, \\
\nonumber \\
W3 &=& H_2 CX_{2, 1} RZ_1(2 \, ca9 \, x \, dt) CX_{3, 1} RZ_1(2 \, ca45 \, x \, dt) CX_{2, 1} H_2
\nonumber \\
&& RZ_1(2 \, cd10 \, dt) CX_{2, 1} RZ_1(2 \, cd14 \, dt) CX_{2, 1} RX_3(2 \, ca18 \, x dt)
\nonumber \\
&& CX_{3, 1} CX_{2, 1} RZ_1(2 \, cd6 \, dt) CX_{2, 1}
\,, \\
\nonumber \\
T21 &=& S_1 S_3 CX_{1, 3} RX_1(2 \, ca28 \, x dt) H_1 H_3 CX_{2,1} RZ_1(2 \, ca34 \, x dt) CX_{2,1}
\nonumber \\
&& CX_{3,1} H_1 H_3 S^\dag_1 S^\dag_3
\,, \\
\nonumber \\
T7 &=& H_1 CX_{2, 1} H_3 CX_{3, 1} RZ_1(2 \, ca24\, x dt, 1) CX_{3, 1} H_3 RZ_1(2 \, ca14\, x dt, 1) 
\nonumber \\
&&CX_{3, 1} RZ_1(2 \, ca50\, x dt, 1) CX_{2, 1} RZ_1(2 \, ca38\, x dt, 1) CX_{3, 1} H_1
\,, \\
\nonumber \\
T19 &=& H_2 \: CX_{3,2} \: RZ_2(2\, ca41\, x dt)\: CX_{3,2}\: H_2 \: CX_{3,2}\: \: RZ_2(2\, cd12\, dt) \: CX_{3,2}
\,.\nonumber \\
\label{eq:16x16_circuit_p2}
\end{eqnarray}
where each rotation gate is a function of $x$, $dt$ and numerical coefficients whose values are listed in Equation~\ref{d_coeff} and Equation~\ref{offD_coeff}.
\begin{align}
cd0 &= 293/32  		&   &  & cd8 &= -67/32 	\nonumber \\
cd1 &= 31/32	  	&   &  & cd9 &= 47/32 	\nonumber \\
cd2 &= -3/32		&   &  & cd10 &= 49/32 	\nonumber \\
cd3 &= -13/32		&   &  & cd11 &= 31/32 	\nonumber \\ \label{d_coeff} 
cd4 &= -13/32		&   &  & cd12 &= 31/32  \\ 
cd5 &= 21/32		&   &  & cd13 &= 25/32 	\nonumber \\
cd6 &= -65/32		&   &  & cd14 &= -49/32 	\nonumber \\
cd7 &= -19/32		&   &  & cd15 &= -43/32 	\nonumber
\end{align}
Here is a list of the numerical coefficients present in the rotation gates $RX$ and $RZ$ coming from the diagonal term of the Hamiltonian.
\begin{align}
ca1 &= -17/24   								&   &  & ca27 &= -1/8 \sqrt{3} 					\nonumber \\
ca2 &= -1/24 (9 + 5 \sqrt{2} + 2 \sqrt{3})   	&   &  & ca28 &= -1/48 (6 \sqrt{3} - 11 \sqrt{6}) 	\nonumber \\
ca3 &= -1/24 (-9 + 3 \sqrt{2} - 2 \sqrt{3})   	&   &  & ca29 &= -1/48 (-6 \sqrt{3} - 3 \sqrt{6})	\nonumber \\
ca4 &= -5/24 								  	&   &  & ca30 &= -1/8 \sqrt{3}					 	\nonumber \\
ca5 &= -29/48   								&   &  & ca31 &= -1/4 \sqrt{3} 					\nonumber \\
ca6 &= -1/48								   	&   &  & ca32 &= -1/4 \sqrt{3}					 	\nonumber \\
ca7 &= -1/24 (12 + 9 \sqrt{2} + 2 \sqrt{3})		&   &  & ca33 &= 1/(8 \sqrt{3}) 					\nonumber \\
ca8 &= -1/24 (12 +\sqrt{2} - 2 \sqrt{3})    	&   &  & ca34 &= -1/48 (6 \sqrt{3} + 3 \sqrt{6})	\nonumber \\
ca9 &= -5/16	   								&   &  & ca35 &= -1/48 (-6 \sqrt{3} - 5 \sqrt{6}) 	\nonumber \\
ca10 &= -11/48								   	&   &  & ca36 &= 1/(8 \sqrt{3}) 				 	\nonumber \\
ca11 &= -1/24 (-\sqrt{2} + 2 \sqrt{3})			&   &  & ca37 &= -7/24			 					\nonumber \\
ca12 &= -1/24 (-7 \sqrt{2} + 2 \sqrt{3})	   	&   &  & ca38 &= -1/24 (9 - 5 \sqrt{2} - 2 \sqrt{3})\nonumber \\ \label{offD_coeff}
ca13 &= 1/8 	   								&   &  & ca39 &= -1/24 (-9 - 3 \sqrt{2} + 2 \sqrt{3})  \\ 
ca14 &= -1/24 (-3 + 3 \sqrt{2} + 2 \sqrt{3})   	&   &  & ca40 &= 5/24								 	\nonumber \\
ca15 &= -1/24 (3 + 5 \sqrt{2} - 2 \sqrt{3})		&   &  & ca41 &= -7/48 							\nonumber \\
ca16 &= -3/8								   	&   &  & ca42 &= -11/48						 	\nonumber \\
ca17 &= -7/(8 \sqrt{3})							&   &  & ca43 &= -1/24 (12 - 9 \sqrt{2} - 2 \sqrt{3}) 	\nonumber \\
ca18 &= -1/48 (6 \sqrt{3} + 11 \sqrt{6})	   	&   &  & ca44 &= -1/24 (12 -\sqrt{2} + 2 \sqrt{3}) 	\nonumber \\
ca19 &= -1/48 (-6 \sqrt{3} + 3 \sqrt{6})		&   &  & ca45 &= 1/16 					\nonumber \\
ca20 &= -7/(8 \sqrt{3})						   	&   &  & ca46 &= -25/48 	\nonumber \\
ca21 &= -1/4 \sqrt{3} 							&   &  & ca47 &= -1/24 (\sqrt{2} - 2 \sqrt{3})		\nonumber \\
ca22 &= 1/4 \sqrt{3}						   	&   &  & ca48 &= -1/24 (7 \sqrt{2} - 2 \sqrt{3}) 	\nonumber \\
ca23 &= -1/8 \sqrt{3}							&   &  & ca49 &= -1/8			 					\nonumber \\
ca24 &= -1/48 (6 \sqrt{3} - 3 \sqrt{6})		   	&   &  & ca50 &= -1/24 (-3 - 3 \sqrt{2} - 2 \sqrt{3}) 	\nonumber \\
ca25 &= -1/48 (-6 \sqrt{3} + 5 \sqrt{6})		&   &  & ca51 &= -1/24 (3 - 5 \sqrt{2} + 2 \sqrt{3})	\nonumber \\
ca26 &= -1/8 \sqrt{3}						   	&   &  & ca52 &= -5/8 									\nonumber
\end{align}
Here is a list of the numerical coefficients present in the rotation gates $RX$ and $RZ$ coming from the off-diagonal term of the Hamiltonian.

\end{appendices}



\addcontentsline{toc}{chapter}{\refname}
\printbibliography
\end{document}